\documentclass[twocolumn]{aastex63}
\usepackage{amsmath}
\usepackage{amssymb}
\usepackage{mathrsfs}
\usepackage[shortlabels]{enumitem}

\definecolor{purple}{rgb}{0.5,0,0.5}
\definecolor{darkgreen}{rgb}{0.1,0.6,0.1}
\definecolor{orange}{rgb}{1,0.6,0}

\newcommand{\brvs}{Br\"unt-V\"ais\"al\"a}

\submitjournal{ApJS}
\shorttitle{Atlas of Stellar Convection}
\shortauthors{Jermyn et al.}

\begin{document}

\title{An Atlas of Convection in Main-Sequence Stars}

\correspondingauthor{Adam S. Jermyn}
\email{adamjermyn@gmail.com}

\author[0000-0001-5048-9973]{Adam S. Jermyn}
\affiliation{Center for Computational Astrophysics, Flatiron Institute, New York, NY 10010, USA}

\author[0000-0002-3433-4733]{Evan H. Anders}
\affiliation{CIERA, Northwestern University, Evanston IL 60201, USA}

\author[0000-0002-7635-9728]{Daniel Lecoanet}
\affiliation{CIERA, Northwestern University, Evanston IL 60201, USA}
\affiliation{Department of Engineering Sciences and Applied Mathematics, Northwestern University, Evanston IL 60208, USA}

\author[0000-0001-5048-9973]{Matteo Cantiello}
\affiliation{Center for Computational Astrophysics, Flatiron Institute, New York, NY 10010, USA}
\affiliation{Department of Astrophysical Sciences, Princeton University, Princeton, NJ 08544, USA}

\begin{abstract}
Convection is ubiquitous in stars and occurs under many different conditions.
Here we explore convection in main-sequence stars through two lenses: dimensionless parameters arising from stellar structure and parameters which emerge from the application of mixing length theory.
We first define each quantity in terms familiar both to the 1D stellar evolution community and the hydrodynamics community.
We then explore the variation of these quantities across different convection zones, different masses, and different stages of main-sequence evolution.
We find immense diversity across stellar convection zones.
Convection occurs in thin shells, deep envelopes, and nearly-spherical cores; it can be efficient of inefficient, rotationally constrained or not, transsonic or deeply subsonic.
This atlas serves as a guide for future theoretical and observational investigations by indicating which regimes of convection are active in a given star, and by describing appropriate model assumptions for numerical simulations.
\end{abstract}


\tableofcontents

\section{Introduction}\label{sec:intro}

Convection is central to a wide range of mysteries in stars.
How are pulsations excited~\citep{1990ApJ...363..694G}?
How are magnetic fields  generated in the Sun and stars in general \citep{1955ApJ...122..293P,2005PhR...417....1B}?  What drives stellar activity and stellar spots \citep{2020NatAs...4..658L,2009A&ARv..17..251S}? How do stars spin down \citep{1967ApJ...150..551K}?
What sets the radii of M-dwarfs~\citep{2018AJ....155..225K,2019MNRAS.489.2615M}?
Where does stochastic low-frequency photometric variability come from~\citep{2021ApJ...915..112C,2020FrASS...7...70B}?
What causes mass loss in Luminous Blue Variables~\citep{2018Natur.561..498J}?

Convection occurs not only during core hydrogen burning,  but also in later evolutionary phases, including the red (super) giant phase \citep{2011ApJ...731...78T,2021arXiv211003261G,2009A&A...506.1351C}, during the helium flash \citep[e.g.][]{2006ApJ...639..405D,2009A&A...501..659M}, the thermally pulsating asymptotic giant branch \citep[e.g.][]{2005ARA&A..43..435H,2008A&A...483..571F,2010A&A...512A..10S},  in white dwarfs \citep[e.g.][]{1996A&A...313..497F,2020MNRAS.492.3540C}, classical novae \citep[e.g.][]{2013ApJ...762....8D}, late phases of nuclear burning \citep[e.g.][]{2006ApJ...637L..53M,2007ApJ...667..448M,2016ApJ...833..124M}, and during core collapse \citep[e.g.][]{2011ApJ...733...78A,2013RvMP...85..245B,2016PASA...33...48M}.

For some goals, such as modeling main-sequence stellar evolution, it is sufficient to use the steady-state one-dimensional convective formulation of Mixing Length Theory~\citep[MLT]{1958ZA.....46..108B,1965ApJ...142..841H,1999A&A...346..111L}, but for questions of dynamics~\citep{2016MNRAS.456.3475S}, rotation~\citep{Antia_2008}, waves~\citep{2019ApJ...880...13Z} and precision stellar evolution~\citep{2018ApJ...856...10J}, this approach is insufficient.

To that end, significant effort has gone into improved one-dimensional theories~\citep[e.g.][]{1977ApJ...214..196G,1982ApJ...262..330S,1986A&A...160..116K,1991ApJ...370..295C,Deng_2006,2015LRSP...12....8H,2018MNRAS.476..646J} as well as into dynamical numerical simulations of convection zones~\citep[e.g.][among many others]{
2011A&A...535A..22C,
2012ApJ...753L..13S,
2013ApJ...773..137G,2013ApJ...769...18T,
2014ApJ...785...90R,2014ApJ...786...24H,
2015ApJ...813...74J,2016ApJ...829...92A,
2016ApJ...833L..28Y,
2017MNRAS.471..279C,
2018ApJ...863...35S,
2019MNRAS.488.2503C,2019ApJ...876....4E,
2020ApJ...902L...3B,2020A&A...638A..15P,2020MNRAS.491..972A,2020A&A...641A..18H,
2021arXiv211011356A,2021ApJ...923...52K,
2022ApJ...924L..11S}.
A challenge in these efforts is that different convection zones can lie in very different parts of parameter space.
Just to name a few axes of variation: convection can  occur in thin shells or deep envelopes or nearly-spherical cores, it can carry most of the energy flux or very little, it can be rotationally constrained or not, and it can be transsonic or deeply subsonic.

Here we explore the dimensionless parameters that set the stage for convection in main-sequence stars, as well as those which emerge with the application of mixing length theory.
We begin in Section~\ref{sec:CZ} by describing the different convection zones which appear in main-sequence stars.
In Section~\ref{sec:params} we then define each quantity of interest in terms familiar both to the 1D stellar evolution community and the hydrodynamics community.
In Section~\ref{sec:methods} we explain how we construct our stellar models.
We then explore in Section~\ref{sec:discussion} how these quantities vary across different convection zones, different masses, and different stages of main-sequence evolution.
We finally discuss the different regimes which arise in main-sequence stellar convection, and the prospects for realizing these regimes in numerical simulations.

\section{Convection Zones}\label{sec:CZ}

\begin{figure*}
  \includegraphics[width=\textwidth]{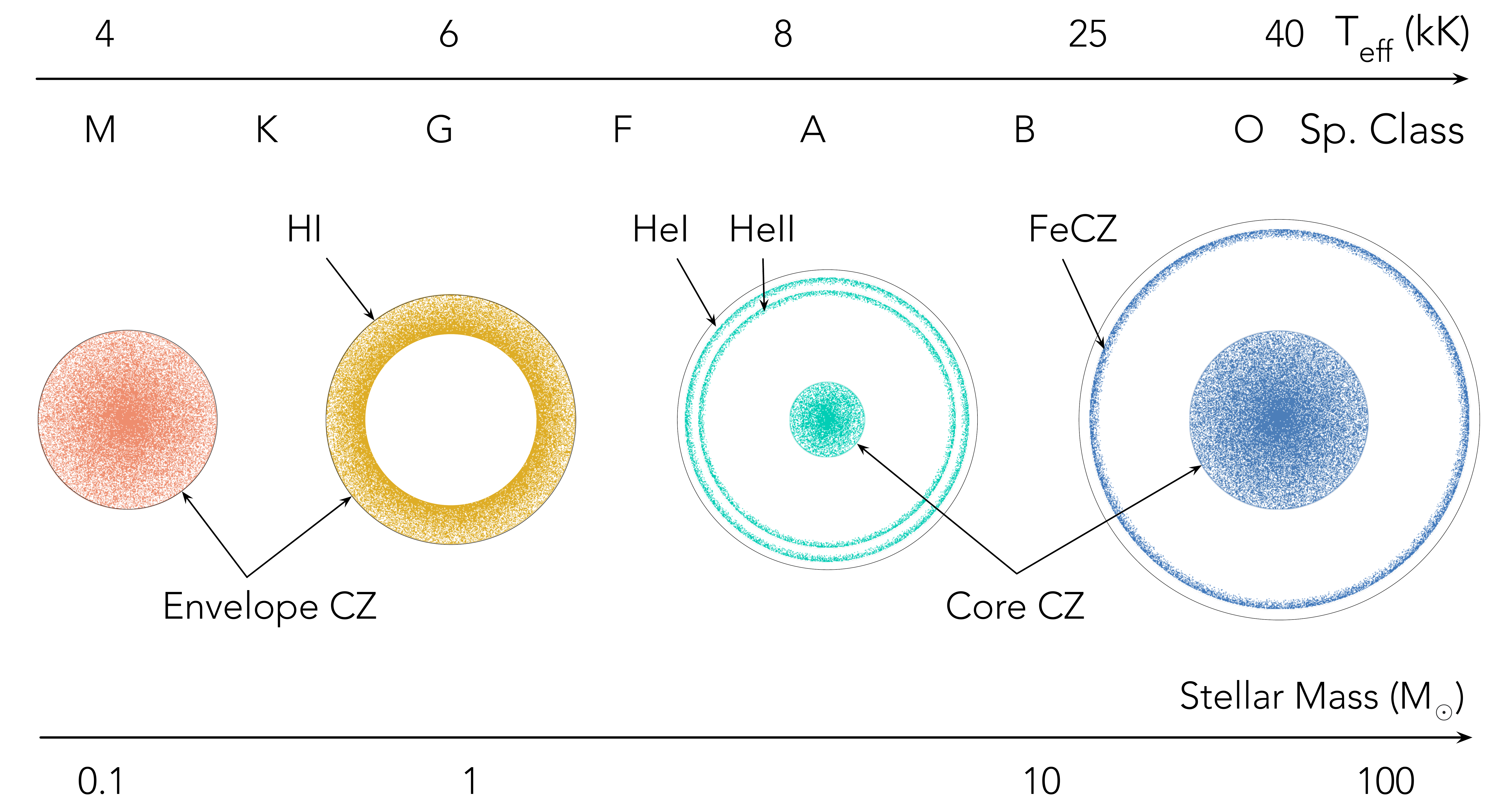}
  \caption{Landscape of convection in main sequence stars of different mass and spectral type. Large opacity in low-mass stars is responsible for deep envelope convection zones and fully convective stars. Thin ionization-driven convection zones can be caused by the recombination of hydrogen (HI), helium (HeI and HeII), and iron group elements (FeCZ) in early-type stars.  For masses above 1.1M$_\odot$ nuclear burning leads to steep-enough temperature gradients to drive core convection. This outline is informed by one-dimensional stellar evolution calculations, see Fig.~2 in \citet{2019ApJ...883..106C} for a more quantitative picture. We warn that some of the thin ionization-driven convection zones might have subcritical Rayleigh numbers, and so may not have convective motions in real stars \citep[see e.g.][]{2022arXiv220110567J}.}
   \label{fig:cz}
\end{figure*}

Convection occurs in different regions of stars on the main sequence, which we depict schematically in Figure~\ref{fig:cz} for stars of different mass and spectral types.
For a more quantitative picture obtained using one-dimensional stellar evolution calculations we refer to Fig.~2 in \citet{2019ApJ...883..106C}.

When we refer to a convection zone (CZ) below we mean a Schwarzschild-unstable layer, assuming that convection rapidly homogenizes the composition of the zone.

Generally speaking, convection in stars can occur in the presence of a steep temperature gradient due to nuclear energy generation (e.g. core convection), 
a large gradient in opacity, and/or a
high heat capacity and consequently low adiabatic gradient. An increase of the opacity and a decrease of the adiabatic gradient tend to occur in stellar envelopes, at temperatures corresponding to the ionization of different species \citep{2019ApJ...883..106C}.

We classify these convection zones first into three broad classes:
\begin{enumerate}[1)]
    \item Core CZs (e.g. the core of a $20 M_\odot$ star).
    \item Deep Envelope CZs (e.g. the Sun's envelope)
    \item Thin ionization-driven CZs (e.g. the HeII ionization zone in a $2M_\odot$ star)
\end{enumerate}
There is some ambiguity in assigning some convection zones to these classes. 
This arises primarily in two scenarios.
First, is the convection zone in an M-dwarf, which can span the entire star, a Deep Envelope CZ or a Core CZ?
Secondly, at what point does the Deep Envelope CZ seen in solar-like stars become a thin zone driven by HI/HeI/HeII ionization?

In the first instance we want to have as continuous a definition as possible across the Hertzsprung-Russel (HR) diagram.
So we say that an M dwarf has a Deep Envelope CZ, because the convection zone gradually retreats upwards with increasing mass.

In the case of the transition from Deep Envelope to thin ionization-driven convection, we make the decision based on the temperature at the base of the CZ, because ionization zones occur at nearly fixed temperatures.

In full then, our classification scheme is
\begin{enumerate}[1)]
\item Core CZ - If $m_{\rm inner}/M_\star < 0.03$ and $M_\star > M_\odot$.
\item Deep Envelope CZ - If $T_{\rm inner} > 500,000\,\mathrm{K}$ and $M_\star < 2 M_\odot$.
\item Thin ionization-driven CZ otherwise.
\end{enumerate}
Here $m_{\rm inner}$ and $T_{\rm inner}$ are respectively the mass coordinate and temperature at the inner boundary of the convection zone.
Note that this classification scheme does not distinguish if a thin ionization-driven CZ is driven by a decrease in the adiabatic gradient or an opacity increase.
Lower numbers in the above list receive priority if a CZ is validly described by multiple classes (e.g., a CZ that satifies \#1 and \#2 is classified as \#1). 
Next, if the zone is ionization-driven, we select the sub-class by:
\begin{enumerate}[3a)]
\item HI CZ - If $T_{\rm outer} < 11,000\,\mathrm{K}$.
\item HeI - If the CZ contains a point with $11,000\,\mathrm{K}<T<35,000\,\mathrm{K}$.
\item HeII - If the CZ contains a point with $35,000\,\mathrm{K}<T<100,000\,\mathrm{K}$.
\item FeCZ - If the CZ contains a point with $100,000\,\mathrm{K}<T<500,000\,\mathrm{K}$.
\end{enumerate}
This approach works because the ionization state is much more strongly controlled by temperature than by density, so temperature is actually a good proxy for ionization state.
If a zone meets multiple of these criteria we classify it by the hottest subclass it fits.

\section{Dimensionless Parameters}\label{sec:params}

\begin{deluxetable*}{lll}[t]
  \tablecolumns{3}
  \tablewidth{0.5\textwidth}
  \tablecaption{Dimensionless Input Parameters.\label{tab:inputs}}
  \tablehead{\colhead{Name} & \colhead{Description} & \colhead{Appears}}
  \startdata
$\mathrm{Pr}$ & Prandtl number & Eq.~\eqref{eq:Pr}\\
$\mathrm{Pm}$ & Magnetic Prandtl number & Eq.~\eqref{eq:Pm}\\
$\beta_{\rm rad}$ & Ratio of radiation to total pressure & Eq.~\eqref{eq:beta}\\
$\mathrm{Ra}$ & Rayleigh number & Eq.~\eqref{eq:Ra}\\
$\mathrm{A}$ & Aspect ratio & Eq.~\eqref{eq:A}\\
$\mathrm{D}$ & Density ratio & Eq.~\eqref{eq:D}\\
$\tau_{\rm CZ}$ & CZ optical depth & Eq.~\eqref{eq:tau_cz}\\
$\tau_{\rm outer}$ & Optical depth from surface to outer edge of CZ & Eq.~\eqref{eq:tau_surf}\\
$\Gamma_{\rm Edd}$ & Eddington ratio & Eq.~\eqref{eq:GammaEdd}\\
$\mathrm{Ek}$ & Ekman number & Eq.~\eqref{eq:Ek}\\
  \enddata
\end{deluxetable*}

\begin{deluxetable}{lll}
  \tablecolumns{3}
  \tablewidth{0.5\textwidth}
  \tablecaption{Output Parameters.\label{tab:outputs}}
  \tablehead{\colhead{Name} & \colhead{Description} & \colhead{Appears}}
  \startdata
$\mathrm{Re}$ & Reynolds number & Eq.~\eqref{eq:Re}\\
$\mathrm{Pe}$ & P{\'e}clet number & Eq.~\eqref{eq:Pe}\\
$\mathrm{Ma}$ & Mach number & Eq.~\eqref{eq:M}\\
$\Gamma_{\rm Edd}^{\rm rad}$ & Radiative Eddington ratio & Eq.~\eqref{eq:GammaRad}\\
$F_{\rm conv}/F$ & Ratio of convective to total energy flux & N/A\\
$\mathrm{S}$ & Stiffness & Eq.~\eqref{eq:S}\\
$\mathrm{Ro}$ & Rossby number & Eq.~\eqref{eq:Ro}\\
$t_{\rm conv}$ & Turnover time & Eq.~\eqref{eq:tconv}\\
  \enddata
\end{deluxetable}

We describe convection using two categories of dimensionless parameters: ``inputs'' (Table~\ref{tab:inputs}), which depend on the stellar models and microphysics, and ``outputs'' (Table~\ref{tab:outputs}), which include a measure of the convective velocity or mixed temperature gradient and therefore depend upon a theory of convection.
For the latter we employ Mixing Length Theory (MLT) following the prescription of~\citet{1968pss..book.....C}; we emphasize that this theory reflects a drastic simplification of reality and is in many regards unsatisfactory, but it does allow us to provide order-of-magnitude estimates for these quantities.
We note also that even the ``inputs'' in this paper depend upon our choice to employ MLT, because the stratification achieved in convective regions can subtly alter the stratification and values throughout the model.
Nonetheless, MLT reflects the current standard in stellar modelling, and points the way to many open questions, so it is valuable to use as a baseline.

\subsection{Input Parameters}

\subsubsection{Microphysics}

We begin with quantities set solely by the microphysics.
The Prandtl number is the ratio
\begin{align}\label{eq:Pr}
	\mathrm{Pr} \equiv \frac{\nu}{\chi},
\end{align}
where $\nu$ is the kinematic viscosity and $\chi$ is the thermal diffusivity.
The magnetic Prandtl number is similarly
\begin{align}\label{eq:Pm}
	\mathrm{Pm} \equiv \frac{\nu}{\eta},
\end{align}
where $\eta$ is the magnetic diffusivity.
Our prescriptions for these diffusivities are given in Appendix~\ref{sec:diffusivities}.

We further write the ratio of radiation pressure to total pressure as
\begin{align}\label{eq:beta}
	\beta_{\rm rad} \equiv \frac{P_{\rm rad}}{P}.
\end{align}

\subsubsection{Stellar Structure}

Next, we present quantities determined by stellar structure.
The first of these is the Rayleigh number, 
\begin{align}\label{eq:Ra}
    \mathrm{Ra} \equiv \delta r^4 \frac{\langle g\rangle \langle \nabla_{\rm rad}-\nabla_{\rm ad}\rangle }{\langle \nu\rangle\langle\chi\rangle}\left(\frac{1}{\langle h\rangle}\right)\left(\frac{4-\langle \beta_{\rm gas}\rangle}{\langle\beta_{\rm gas}\rangle}\right)
\end{align}
which measures how unstable a region is to convection.
When $\mathrm{Ra} < \mathrm{Ra}_{\rm crit} \approx 10^3$, convective motions are diffusively stabilized, so the convective velocity is zero~\citep{1961hhs..book.....C}.
Here, $g$ is the acceleration of gravity, $\delta r$ is the thickness of the convection zone, $h \equiv -dr/d\ln P$ is the pressure scale height, $\beta_{\rm gas}\equiv P_{\rm gas}/P=1-\beta_{\rm rad}$ is the ratio of gas pressure to total pressure,
\begin{align}
	\nabla_{\rm ad} \equiv \left.\frac{\partial \ln T}{\partial \ln P}\right|_s
\end{align}
is the adiabatic temperature gradient, $s$ is the entropy,
\begin{align}
	\nabla_{\rm rad} \equiv \frac{3\kappa L P}{64 \pi G M \sigma T^4}
\end{align}
is the radiative temperature gradient, $\kappa$ is the opacity, $L$ is the luminosity, $P$ is the pressure, $T$ is the temperature, $G$ is the gravitational constant, $M$ is the mass coordinate, and $\sigma$ is the Stefan-Boltzmann constant.
The factor in equation~\eqref{eq:Ra} depending on $\beta$ arises from the thermal expansion coefficient, and is equal to the ratio of the density susceptibility to the temperature susceptibility ($\chi_\rho / \chi_T$).
Note that the Rayleigh number is a global quantity defined over the whole convection zone.
Ra depends on the depth of the zone ($\delta r$), but the other quantities on the right-hand side of equation~\eqref{eq:Ra} are generally functions of radius and so must be averaged, which we denote with angled brackets.
The details of this averaging are described in Section~\ref{sec:average}.

Stellar structure provides the geometry of the CZ.
The aspect ratio is
\begin{align}\label{eq:A}
	\mathrm{A} \equiv \frac{r_{\rm outer}}{\delta r},
\end{align}
where $r_{\rm outer}$ is the radial coordinate of the outer boundary of the convection zone.
Thin shell CZs have large aspect ratios and can be adequately modeled with local Cartesian simulations, but CZs with small aspect ratios require global (spherical) geometry to capture realistic dynamics.

The density contrast is
\begin{align}\label{eq:D}
	\mathrm{D} \equiv \frac{\rho_{\rm inner}}{\rho_{\rm outer}},
\end{align}
where $\rho_{\rm inner}$ and $\rho_{\rm outer}$ are respectively the density on the inner and outer boundaries of the convection zone.
When $\mathrm{D}$ is small, density stratification can be neglected, as in the Boussinesq approximation.

The optical depth across the convection zone
\begin{align}\label{eq:tau_cz}
	\tau_{\rm CZ} \equiv \int_{\rm CZ} \kappa \rho dr,
\end{align}
and the optical depth from the outer boundary of the convection zone to infinity
\begin{align}\label{eq:tau_surf}
	\tau_{\rm outer} \equiv \int_{r_{\rm outer}}^{\infty} \kappa \rho dr,
\end{align}
together tell us whether or not radiation can be treated diffusively in the convection zone.
Note that our stellar models do not extend into the atmosphere of the star, which is handled separately as a boundary condition, so $\tau_{\rm outer}$ will never be less than the optical depth of the base of the atmosphere.
In our models this minimum surface optical depth is $\tau_{\rm min} = 2/3$.

Related, the Eddington ratio is
\begin{align}
	\Gamma_{\rm Edd} \equiv \frac{L}{L_{\rm Edd}},
	\label{eq:GammaEdd}
\end{align}
where
\begin{align}
	L_{\rm Edd} \equiv \frac{4\pi G M c}{\kappa},
\end{align}
and $c$ is the speed of light.
Here $M$ is the mass coordinate, and both $\kappa$ and $L_{\rm Edd}$ are functions of $M$.

Finally, we measure the ratio of the rotational and viscous timescales via the Ekman number
\begin{align}\label{eq:Ek}
	\mathrm{Ek} \equiv \frac{\nu}{2\Omega \delta r^2}.
\end{align}
Here $\Omega$ is the angular velocity of the CZ and we assume solid body rotation.
We choose $\Omega$ as a function of mass and evolutionary state to reproduce typical observed rotation periods.
Details on this choice are described in Appendix~\ref{sec:rotation}.

In the presence of rotation, Ek, Ra, and Pr combined determine the diffusive instability of a CZ.
Rotation dominates viscous effects when $\mathrm{Ek} \ll 1$, but the output Rossby number (below) more reliably describes how effectively rotation deflects convective flows.

\subsection{Outputs}

Using Mixing Length Theory we obtain the convection speed $v_c$ and temperature gradient $\nabla$~\citep{1958ZA.....46..108B,1968pss..book.....C}.
We define the Reynolds number
\begin{align}\label{eq:Re}
	\mathrm{Re} \equiv \frac{v_c \delta r}{\nu},
\end{align}
which is the ratio of the viscous timescale to the convective turnover time, so convection at large $\mathrm{Re}$ is turbulent.

We further define the P{\'e}clet number
\begin{align}\label{eq:Pe}
	\mathrm{Pe} \equiv \frac{v_c \delta r}{\chi},
\end{align}
which is the ratio of the thermal diffusion timescale across the zone to the turnover time, and so measures the relative importance of convective and radiative heat transfer.

Next we define the Mach number
\begin{align}\label{eq:M}
	\mathrm{Ma} \equiv \frac{v_c}{c_s},
\end{align}
where $c_s$ is the adiabatic sound speed.
$\mathrm{Ma}$ measures the magnitude of thermodynamic fluctuations, and sound-proof (e.g., anelastic) models are valid at low $\mathrm{Ma}$.
In radiation-dominated ($\beta_{\rm rad} \approx 1$), low Pe gases, the relevant fluctuations become isothermal and we should replace $c_s$ with the isothermal sound speed ${c_{s,\rm iso}^2 = \partial P/\partial \rho|_T}$~\citep{2015ApJ...808L..31G,2015ApJ...813...74J}.
For CZs where this is relevant we will show both the adiabatic and isothermal Mach numbers ($\mathrm{Ma}_{\rm iso} \equiv v_c/c_{s,\rm iso}$).

With $\nabla$ we obtain the ratio of convective flux to total flux $F_{\rm conv}/F$, which is closely related to both the Nusselt number $\mathrm{Nu}$ and to the efficiency parameter $\Gamma$ in Mixing Length Theory~\citep[see the discussion in][]{Jermyn_2022}.
We also construct the radiative Eddington ratio 
\begin{align}\label{eq:GammaRad}
\Gamma_{\rm Edd}^{\rm rad} \equiv \frac{L_{\rm rad}}{L_{\rm Edd}},
\end{align}
where $L_{\rm rad}$ is the radiative luminosity, which depends upon the steady-state temperature gradient $\nabla$ achieved by the convection, and is therefore distinct from the Eddington ratio, $\Gamma_{\rm Edd}$.

We further write the stiffness
\begin{align}\label{eq:S}
	\mathrm{S} = \frac{N_{\rm RZ}^2}{f_{\rm conv}^2}
\end{align}
of the convective-radiative boundary, which measures how difficult it is for convective motions to proceed past the boundary.
Here $N_{\rm RZ}$ is the \brvs\ frequency in the radiative zone and the convective frequency $f_{\rm conv} = v_c/h$.
Inner and outer boundaries in shell CZs are distinguished as $\mathrm{S}_{\rm inner/outer}$.
We only report values for convection zones which have the relevant boundaries\footnote{Core convection zones do not have inner boundaries. Likewise, when convection zones reach the surface of our models we cannot calculate an outer stiffness because there is no outer boundary inside of the model. Physically there ought to be a location where the energy transport becomes radiative, but this occurs in the atmosphere, which we treat with a simple boundary condition.}.

We measure rotation using the Rossby number
\begin{align}\label{eq:Ro}
	\mathrm{Ro} \equiv \frac{v_c}{2\Omega \delta r}.
\end{align}
We use the same uniform $\Omega$ to calculate Ro as we did for Ek (Equation~\ref{eq:Ek}).

Finally, we compute the dimensional turnover time
\begin{align}
	t_{\rm conv} \equiv \int_{\rm CZ} \frac{dr}{v_c},
	\label{eq:tconv}
\end{align}
which is useful for reasoning about the effects of rotation as a function of $\Omega$.

\section{Methods}\label{sec:methods}

\subsection{Stellar Evolution}

We calculated stellar evolutionary tracks for stars ranging from $0.3-60 M_\odot$ using release r21.12.1 of the Modules for Experiments in Stellar Astrophysics software instrument
\citep[MESA][]{Paxton2011, Paxton2013, Paxton2015, Paxton2018, Paxton2019}.
Details on the MESA microphysics inputs are provided in Appendix~\ref{appen:mesa}.
Our models were run at the Milky Way metallicity of $Z=0.014$, use convective premixing~\citep[][Section 5.2]{Paxton2019} and the Cox MLT option~\citep{1968pss..book.....C} with $\alpha_{\rm MLT}=1.6$, and determine the convective boundary using the Ledoux criterion.
All data and scripts used in compiling this atlas are available publicly.
See Appendix~\ref{appen:data} for details.

\subsection{Averaging}\label{sec:average}

We are interested in obtaining a global measure of each parameter (e.g., Re) across a convection zone, yet these parameters depend on quantities which themselves vary across the convection zone (e.g., $v_c$, $\nu$), so we require an averaging procedure.
We denote the radial average of a quantity $q$ over a convection zone by
\begin{align}
	\langle q \rangle \equiv \frac{1}{\delta r}\int_{\rm CZ} q dr.
\end{align}
With this, we calculate the averages
\begin{align}
	\mathrm{Pr} &= \frac{\langle \nu\rangle}{\langle \chi \rangle}\\
	\mathrm{Pm} &= \frac{\langle \nu\rangle}{\langle \eta\rangle}\\
	\beta_{\rm rad} &= \langle \frac{P_{\rm rad}}{P}\rangle\\
    \mathrm{Ra} &= \delta r^4 \frac{\langle g\rangle \langle \nabla_{\rm rad}-\nabla_{\rm ad}\rangle }{\langle \nu\rangle\langle\chi\rangle}\left(\frac{1}{\langle h\rangle}\right)\left(\frac{4-\langle \beta_{\rm gas}\rangle}{\langle\beta_{\rm gas}\rangle}\right) \label{eqn:avg_Ra}
\end{align}
\begin{align}
	\Gamma_{\rm Edd} &= \langle\frac{L}{L_{\rm Edd}}\rangle\\
	\Gamma_{\rm Edd}^{\rm rad} &= \langle\frac{L_{\rm rad}}{L_{\rm Edd}}\rangle\\
	\frac{F_{\rm conv}}{F} &= \langle \frac{F_{\rm conv(r)}}{F(r)}\rangle\\
	\mathrm{Ek} &= \frac{\langle \nu \rangle}{2\Omega \delta r^2}\\
	\mathrm{Re} &= \frac{\langle v_c\rangle}{\langle \nu\rangle}\delta r\\
	\mathrm{Pe} &= \frac{\langle v_c\rangle}{\langle \chi \rangle}\delta r\\
	\mathrm{Ma} &= \frac{\langle v_c\rangle}{\langle c_s\rangle}\\
	\mathrm{Ro} &= \frac{\langle v_c\rangle}{2\Omega \delta r}
\end{align}

To compute the stiffness we radially average quantities within one pressure scale height of the convective-radiative boundary\footnote{If the CZ or RZ does not extend for a full pressure scale height we average over as much as we can without crossing another convective-radiative boundary.}, where the scale height is measured at the boundary.
Once more denoting radial averages by $\langle ... \rangle$, we write
\begin{align}
	\mathrm{S} = \frac{\langle N\rangle_{\rm RZ}^2}{\langle (v_c/h)^2\rangle_{\rm CZ}},
\end{align}
where the subscript RZ denotes an average on the radiative side of the boundary, and CZ denotes an average on the convective side.

We chose to compute averages radially because convection involves radial heat transport, making averaging in one dimension (radially) seem more appropriate than a volume average, which would pick up much more contribution from low-density outer layers, or a mass average, which is dominated by motions very deep down.
Arguments could well be made for other choices, including mass, volume, or pressure weighting, and in specific applications it may be clearer which kind of average is most appropriate.

In Appendix~\ref{sec:avg} we examine the effects of averaging profiles of e.g., $\mathrm{Pr}$ directly (i.e. $\mathrm{Pr} = \langle \nu/\chi\rangle$) rather than averaging the diffusivities individually.
For most convection zones we find small ($\la 2\times$) differences between the two, but for Deep Envelope CZs quantities can differ by orders of magnitude depending on the averaging procedure because those zones are strongly stratified.
To better capture this variation we additionally report averages over the outer and inner pressure scale height of the Deep Envelope CZs.

\section{Discussion}\label{sec:discussion}

\subsection{Summary of Results}

In Appendix~\ref{sec:CZ_A} we present plots of the quantities in Tables~\ref{tab:inputs} and~\ref{tab:outputs} for each convection zone across the HR diagram.
We group quantities by the class and subclass of convection zone, as our aim is to provide as complete a picture as possible about the properties of each kind of convection zone.
Our aim is for each subsection of Appendix~\ref{sec:CZ_A} to stand on its own, and so the same points are often repeated between these.

For each zone we begin with a figure showing the aspect ratio to introduce the geometry of the zone.
We then study each of our other parameters in Tables~\ref{tab:inputs} and~\ref{tab:outputs}.
Along the way we discuss implications for studying these convection zones in simulations, as well as the prospects for answering specific science questions.

Table~\ref{tab:res_in} shows the ranges of key input parameters we obtain for each convection zone, and Table~\ref{tab:res_out} shows the same for output parameters.
For some convection zones we show both the full mass range and subsets meant to highlight properties that vary strongly with mass.
Moreover, the HeI convection zone has sub-critical Rayleigh numbers over certain mass ranges.
In those ranges the HeI zone has no convective motions because it is not unstable~\citep{2022arXiv220110567J}.
As such we have restricted the mass range for the HeI CZ in our table to just that which supports super-critical Rayleigh numbers\footnote{
    Here we have used the simplest possibile stability analysis: comparing the average Ra from Eqn.~\ref{eqn:avg_Ra} to the canonical critical value.
    A more robust stability analysis could be performed by solving for the eigenvalues associated with the stellar stratification, but such an analysis is beyond the scope of this Atlas and we encourage future authors to explore this.
}.
The same applies to the HeII CZ, but for it sub-critical Rayleigh numbers occur at high masses with large Eddington ratios which may reflect instability~\citep{2015ApJ...813...74J,2018Natur.561..498J}, so we retain the full mass range for the HeII CZ.

Tables~\ref{tab:res_in} and~\ref{tab:res_out} tell a story of enormous diversity.
For a single class of convection zone, many properties, like Ra, Re, and Pe span five or more orders of magnitude as a function of mass!
Across different classes of convection zones, the diversity is even more striking.
We see aspect ratios ranging from unity to $10^3$, density ratios from unity to $10^7$, optical depths from unity up to $10^{12}$, and Rossby numbers from $10^{-3}$ up to $10$.
In Sections~\ref{sec:regimes} and~\ref{sec:questions} we examine the implications of these parameter ranges for the choice of simulation techniques and for the scientific questions that each CZ poses.

\begin{deluxetable*}{lllllllllr}[t]
  \tablecolumns{10}
  \tablewidth{0.5\textwidth}
  \tablecaption{Ranges of input quantities obtained from 1D stellar models, rounded to the nearest integer $\log_{10}$. For quantities which span less than one decade we just list one value. All mass ranges are approximate and metallicity-dependent. To compute $\mathrm{Ek}$ we use $\Omega$ corresponding to a surface velocity of $150\,\mathrm{km\,s^{-1}}$ for O-F stars (down to $1.2 M_\odot$), and scale the surface velocity by $M^2$ below that point. Note that the properties of Deep Envelope CZs vary significantly in space, and we explore this variation in detail in Appendix~\ref{sec:inner_outer}.\label{tab:res_in}}
  \vspace{0.5em}
  \tablehead{\colhead{CZ} & \colhead{Mass Range} & \colhead{$\log_{10}\beta_{\mathrm{rad}}$} & \colhead{$\log_{10}\mathrm{Ra}$} & \colhead{$\log_{10}\mathrm{A}$} & \colhead{$\log_{10}\mathrm{D}$} & \colhead{$\log_{10}\tau_{\mathrm{CZ}}$} & \colhead{$\log_{10}\tau_{\mathrm{surf}}$} & \colhead{$\log_{10} \Gamma_{\mathrm{Edd}}$} & \colhead{$\log_{10}\mathrm{Ek}$}}
  \startdata
Deep Envelope & $0.3 M_\odot\la M \la 1.2 M_\odot$ & $(-5,-3)$ & $(25,34)$ & $(0,1)$ & $(4,7)$ & $(9,13)$ & $0$ & $(-2,0)$ & $(-16,-14)$\\
HI & $1.3 M_\odot\la M \la  3 M_\odot$ & $(-3,-1)$ & $(1,23)$ & $(1,3)$ & $(0,3)$ & $(0,8)$ & $0$ & $(-2,-1)$ & $(-14,-4)$\\
HI (low mass) & $1.3 M_\odot\la M \la 1.5 M_\odot$ & $(-3,-2)$ & $(9,23)$ & $(1,3)$ & $(0,3)$ & $(2,8)$ & $0$ & $(-2,-1)$ & $(-14,-7)$\\
HI (high mass) & $1.5 M_\odot\la M \la  3 M_\odot$ & $(-2,-1)$ & $(1,16)$ & $(2,3)$ & $(0,1)$ & $(0,5)$ & $0$ & $(-2,-1)$ & $(-11,-4)$\\
HeI & $ 2 M_\odot\la M \la 2.5 M_\odot$ & $(-2,-1)$ & $(3,4)$ & $3$ & $0$ & $1$ & $1$ & $-2$ & $(-6,-5)$\\
HeII & $1.5 M_\odot\la M \la 60 M_\odot$ & $(-2,0)$ & $(0,10)$ & $(2,3)$ & $0$ & $(0,4)$ & $(0,3)$ & $(-2,0)$ & $(-8,-3)$\\
FeCZ & $ 7 M_\odot\la M \la 60 M_\odot$ & $(-1,0)$ & $(5,11)$ & $(1,3)$ & $(0,1)$ & $(3,4)$ & $(1,4)$ & $(-1,0)$ & $(-7,-5)$\\
FeCZ (low mass) & $ 7 M_\odot\la M \la 30 M_\odot$ & $(-1,0)$ & $(4,9)$ & $(1,3)$ & $(0,1)$ & $(3,4)$ & $(2,4)$ & $(-1,0)$ & $(-7,-6)$\\
FeCZ (high mass) & $30 M_\odot\la M \la 60 M_\odot$ & $0$ & $(7,11)$ & $(1,2)$ & $(0,1)$ & $(3,4)$ & $(1,2)$ & $0$ & $(-7,-5)$\\
Core & $1.1 M_\odot\la M \la 60 M_\odot$ & $(-3,0)$ & $(24,27)$ & $0$ & $(0,1)$ & $(11,12)$ & $(10,12)$ & $(-3,0)$ & $(-15,-12)$\\
Core (low mass) & $1.1 M_\odot\la M \la 30 M_\odot$ & $(-3,0)$ & $(24,27)$ & $0$ & $(0,1)$ & $(11,12)$ & $(10,12)$ & $(-3,0)$ & $(-15,-13)$\\
Core (high mass) & $30 M_\odot\la M \la 60 M_\odot$ & $(-1,0)$ & $(24,25)$ & $0$ & $1$ & $11$ & $10$ & $0$ & $(-14,-12)$\\

  \enddata
\end{deluxetable*}

\begin{deluxetable*}{llllllllll}[t]
  \tablecolumns{10}
  \tablewidth{0.5\textwidth}
  \tablecaption{Ranges of output quantities obtained from 1D stellar models rounded to the nearest integer $\log_{10}$. For quantities which span less than one decade we just list one value. All mass ranges are approximate and metallicity-dependent. To compute $\mathrm{Ro}$ we use $\Omega$ corresponding to a surface velocity of $150\,\mathrm{km\,s^{-1}}$ for O-F stars (down to $1.2 M_\odot$), and scale the surface velocity by $M^2$ below that point. Note that the properties of Deep Envelope CZs vary significantly in space, and we explore this variation in detail in Appendix~\ref{sec:inner_outer}.\label{tab:res_out}}
  \vspace{0.5em}
  \tablehead{\colhead{CZ} & \colhead{Mass Range} & \colhead{$\log_{10}\mathrm{Re}$} & \colhead{$\log_{10}\mathrm{Pe}$} & \colhead{$\log_{10}\mathrm{Ma}$} & \colhead{$\log_{10}\mathrm{Ro}$} & \colhead{$\log_{10} t_{\rm conv}/\mathrm{d}$} & \colhead{$\log_{10}\frac{F_{\rm conv}}{F}$} & \colhead{$\log_{10}\mathrm{S}_{\rm outer}$} & \colhead{$\log_{10}\mathrm{S}_{\rm inner}$}}
  \startdata
Deep Envelope & $0.3 M_\odot\la M \la 1.2 M_\odot$ & $13$ & $(4,7)$ & $(-4,-2)$ & $(-3,-1)$ & $(1,2)$ & $0$ & $(-1,1)$ & $(4,8)$\\
HI & $1.3 M_\odot\la M \la  3 M_\odot$ & $(1,12)$ & $(-8,4)$ & $(-5,-1)$ & $(-3,1)$ & $(-1,3)$ & $(-16,0)$ & $(-1,0)$ & $(-1,9)$\\
HI (low mass) & $1.3 M_\odot\la M \la 1.5 M_\odot$ & $(8,12)$ & $(-1,4)$ & $(-2,-1)$ & $(-1,1)$ & $(-1,0)$ & $(-1,0)$ & $0$ & $(-1,3)$\\
HI (high mass) & $1.5 M_\odot\la M \la  3 M_\odot$ & $(1,10)$ & $(-8,2)$ & $(-5,-1)$ & $(-3,1)$ & $(-1,3)$ & $(-16,0)$ & $(-1,0)$ & $(-1,9)$\\
HeI & $ 2 M_\odot\la M \la 2.5 M_\odot$ & $(2,3)$ & $(-6,-5)$ & $-4$ & $(-3,-2)$ & $(1,3)$ & $(-14,-11)$ & $(4,8)$ & $(6,7)$\\
HeII & $1.5 M_\odot\la M \la 60 M_\odot$ & $(-1,8)$ & $(-8,0)$ & $(-5,-1)$ & $(-3,0)$ & $(-1,3)$ & $(-16,-1)$ & $(1,9)$ & $(1,9)$\\
FeCZ & $ 7 M_\odot\la M \la 60 M_\odot$ & $(5,7)$ & $(-2,1)$ & $(-3,-1)$ & $(-1,0)$ & $(0,2)$ & $(-6,-1)$ & $(2,4)$ & $(1,3)$\\
FeCZ (low mass) & $ 7 M_\odot\la M \la 30 M_\odot$ & $(5,7)$ & $(-2,1)$ & $(-3,-1)$ & $(-1,0)$ & $(0,2)$ & $(-6,-1)$ & $(2,4)$ & $(1,4)$\\
FeCZ (high mass) & $30 M_\odot\la M \la 60 M_\odot$ & $(5,7)$ & $(-1,1)$ & $(-2,-1)$ & $0$ & $(1,2)$ & $(-4,-1)$ & $(2,4)$ & $(1,2)$\\
Core & $1.1 M_\odot\la M \la 60 M_\odot$ & $(11,13)$ & $(6,7)$ & $(-4,-3)$ & $(-3,-1)$ & $(1,2)$ & $(-1,0)$ & $(5,8)$& N/A\\
Core (low mass) & $1.1 M_\odot\la M \la 30 M_\odot$ & $(11,13)$ & $(6,7)$ & $(-4,-3)$ & $(-3,-2)$ & $(1,2)$ & $(-1,0)$ & $(5,8)$& N/A\\
Core (high mass) & $30 M_\odot\la M \la 60 M_\odot$ & $11$ & $6$ & $-3$ & $(-2,-1)$ & $(1,2)$ & $0$ & $(5,6)$& N/A\\

  \enddata
\end{deluxetable*}

\subsection{Physical Regimes}\label{sec:regimes}

\begin{deluxetable*}{lllllll}[t]
  \tablecolumns{7}
  \tablewidth{0.5\textwidth}
  \tablecaption{Summary of important physical processes in main-sequence convection zones. ``Bound.'' refers to the outer boundary, ``Diff.'' refers to the diffusive approximation. All mass ranges are approximate and metallicity-dependent. We do not include mass ranges in which convection zones which have subcritical Rayleigh numbers as these zones do not undergo convective motions~\citep{2022arXiv220110567J}.\label{tab:recs}}
  \tablehead{\colhead{CZ} & \colhead{Mass Range} & \colhead{Geometry} & \colhead{Radiation Pressure?} & \colhead{Radiative Transfer?} & \colhead{Rotation?} & \colhead{Fluid Dynamics}}
  \startdata
  Deep Envelope		& $M \la 1.2 M_\odot$ & Global & Neglect & Diff. Bulk, Full Bound. & Include & Mixed\tablenotemark{a} \\
  HI (low mass)		& $1.3 M_\odot \la M \la 1.5 M_\odot$ 	& Local  & Neglect & Diff. Bulk, Full Bound. & Neglect & Compressible \\
  HI (high mass)	& $1.5 M_\odot \la M \la 2.5 M_\odot$	& Local  & Neglect & Mixed\tablenotemark{b} & Include & Boussinesq	\\
  HeI				& $2 M_\odot \la M \la 2.5 M_\odot$		& Local  & Neglect & Diffusive & Include & Boussinesq	\\
  HeII 				& $1.5 M_\odot \la M \la 9 M_\odot$		& Local  & Neglect & Diffusive & Include & Boussinesq	\\
  FeCZ (low mass)	& $7M_\odot \la M \la 30 M_\odot$		& Local  & Include & Diffusive & Include & Boussinesq	\\
  FeCZ (high mass)	& $M \ga 30 M_\odot$	& Global & Include & Full & Neglect & Compressible \\
  Core (low mass)	& $1.1 M_\odot \la M \la 30 M_\odot$	& Global & Neglect & Diffusive & Include & Boussinesq	\\
  Core (high mass)	& $M \ga 30 M_\odot$  & Global & Include & Diffusive & Include & Anelastic	\\
  \enddata
  \tablenotetext{a}{The anelastic approximation is appropriate in the deeper portions of these zones, where the Mach numbers are low. The compressible equations are needed near the surface, where the Mach numbers can approach unity.}
  \tablenotetext{b}{Diffusion is appropriate in the bulk, but full radiative transfer is important near the surface.}
\end{deluxetable*}

We have classified convection zones in solar metallicity main-sequence stars into six categories:
\begin{enumerate}
\item Deep Envelope ($M \la 1.2 M_\odot$)
\item HI ($1.3 M_\odot \la M \la 3 M_\odot$)
\item HeI ($2 M_\odot \la M \la 2.5 M_\odot$)
\item HeII ($M  \ga 1.5 M_\odot$)
\item FeCZ ($M \ga 7 M_\odot$)
\item Core ($M \ga 1.1 M_\odot$)
\end{enumerate}
Different physical processes dominate in each of these (Table~\ref{tab:recs}), which suggests using different kinds of numerical experiments for studying them.

\subsection{Scientific Questions}\label{sec:questions}

\begin{deluxetable*}{lll}[t]
  \tablecolumns{3}
  \tablewidth{0.5\textwidth}
  \tablecaption{Scientific topics in main-sequence convection zones. For more on fossil magnetic fields and convection see~\citet{2021ApJ...923..104J} and references therein. ``Rad.'' means radiation. \label{tab:topics}}
  \tablehead{\colhead{CZ} & \colhead{Mass Range} & \colhead{Topics}}
  \startdata
  Deep Envelope		& $M \la 1.2 M_\odot$ 					& Rotational constraints. Dynamo action. Impact of stratification.   									\\
  HI (low mass)		& $1.3 M_\odot \la M \la 1.5 M_\odot$ 	& Internal Gravity Waves. Wave Mixing. Impact of sonic motion. 											\\
  HI (high mass)	& $1.5 M_\odot \la M \la 2.5 M_\odot$	& Marginal $\mathrm{Ra}$. Weak turbulence. Interaction with fossil fields.	\\
  HeI				& $2 M_\odot \la M \la 2.5 M_\odot$		& Marginal $\mathrm{Ra}$. Weak turbulence.	Rotational constraints. Optically thin convection.	\\
  HeII 				& $1.5 M_\odot \la M \la 9 M_\odot$		& Marginal $\mathrm{Ra}$. Weak turbulence.	Interaction with fossil fields. Dynamo action.	\\
  FeCZ (low mass)	& $7M_\odot \la M \la 30 M_\odot$		& Interaction with fossil magnetic fields.	Low-stiffness boundaries.  Dynamo action.					\\
  FeCZ (high mass)	& $M \ga 30 M_\odot$					& Impact of Rad. Pressure. Sonic motion. Super-Eddington limit. Low $\mathrm{Pe}$.	 Dynamo action.	\\
  Core (low mass)	& $1.1 M_\odot \la M \la 30 M_\odot$	& Internal Gravity Waves. Dynamo action. Rotational constraints. Stiff boundaries.						\\
  Core (high mass)	& $M \ga 30 M_\odot$					& Internal Gravity Waves. Dynamo action. Impact of Rad. Pressure.									\\
  \enddata
\end{deluxetable*}

Different regimes of convection pose distinct scientific questions.
Table~\ref{tab:topics} summarizes some of the topics which are interesting to study for each class of convection zone.

\subsubsection{Rotation}

Rapid global rotation can stabilize the convective instability \citep{1961hhs..book.....C}.
This effect is most prominent in the regime of $\mathrm{Ek} \ll 1$ and $\mathrm{Pr} \gtrsim 1$; since stellar CZs have $\mathrm{Pr} \ll 1$, we do not expect rotation to substantially modify the value of the critical Rayleigh number.

Assuming convective instability, rotation influences convection by deflecting its flows via the Coriolis force.
The Rossby number Ro measures how large advection is compared to the Coriolis force, and when $\mathrm{Ro} \ll 1$, the Coriolis force dominates over the nonlinear inertial force.
In the regime of $\mathrm{Ro} \ll 1$ attained in Deep Envelope and Core CZs, rotation two-dimensionalizes the convection, deflecting flows into tall, skinny columns that align with the rotation axis \citep[discussed in the solar context in e.g.,][]{2016ApJ...830L..15F, 2021PNAS..11822518V}.
In the regime of $\mathrm{Ro} \gg 1$, rotation does not appreciably modify convection compared to a non-rotating system.

It is unclear how rotation influences convective dynamics in the transitional $\mathrm{Ro} \sim 1$ regime.
A brief and modern review describing the modern theory of rotating convection is provided by~\citet{2020PhRvR...2d3115A}; we note that the regimes of $\mathrm{Ro} \gg 1$ and $\mathrm{Ro} \ll 1$ are described in detail, but $\mathrm{Ro} \sim 1$ does not have a clean theoretical understanding.
On the scale of whole convection zones rotation can couple with convection to drive differential rotation, and the associated latitudinal shear can be crucial in driving dynamos \citep{2017ApJ...836..192B}.

\subsubsection{Magnetism}

Main-sequence CZs consist of strongly-ionized fluid, so magnetism should be universally important.

Dynamos are some of the most exciting applications of magnetoconvection~\citep{2005PhR...417....1B}.
Deep Envelope CZs are known to support complex dynamo cycles (as in the Sun), and a full understanding of these cycles remains elusive \citep[e.g.,][]{2010ApJ...711..424B, 2016Sci...351.1427H, 2022ApJ...926...21B}.
Dynamos are likewise of interest in the HeII CZ because it has the highest kinetic energy density of all subsurface CZs for most A/B stars~\citep{2019ApJ...883..106C,2020ApJ...900..113J}, making dynamo action there a good candidate to explain the observed magnetic fields of e.g. Vega and Sirius~A~\citep{2010A&A...523A..41P,2011A&A...532L..13P}.
The FeCZ is similarly of interest because FeCZ dynamos probably set the magnetic field strengths of most O stars~\citep{2011A&A...534A.140C,2019MNRAS.487.3904M}.
In Core CZs, dynamo-generated fields could be the progenitors of those observed in Red Giant cores and compact stellar remnants~\citep{2015Sci...350..423F,2016ApJ...824...14C}, so understanding their configuration and magnitude is important for connecting to observations.

We also note that Deep Envelope CZs uniformly have $\mathrm{Pm} \ll 1$, while each other CZ class has $\mathrm{Pm} \gg 1$ in some mass range, so these could exhibit very different dynamo behaviors and likely need to be studied separately.
Many fundamental studies into scaling laws and force balances in magnetoconvection employ the quasistatic approximation for magnetohydrodynamics~\citep[e.g.][]{yan_calkins_maffei_julien_tobias_marti_2019}.
The quasistatic approximation assumes that $\mathrm{Rm} = \mathrm{Pm} \mathrm{Re} \rightarrow 0$; in doing so, this approximation assumes a global background magnetic field is dominant and neglects the nonlinear portion of the Lorentz force. 
This approximation breaks down in convection zones with $\mathrm{Pm} > 1$ and future numerical experiments should seek to understand how magnetoconvection operates in this regime.
Understanding which lessons from these reduced experiments apply in the $\mathrm{Pm} > 1$ regime may help unravel some mysteries of dynamo processes in the HI/HeI/HeII/Fe/Core CZs.

Beyond dynamos, convection can interact with fossil magnetic fields~\citep{Zeldovich}, and it has been suggested that this can serve to erase or hide the near-surface fields of early-type stars~\citep{2020ApJ...900..113J,2021ApJ...923..104J}.
In the limit of $\nabla_{\rm rad} \gg \nabla_{\rm ad}$ and weak magnetic fields the result is likely that the fossil field is erased and replaced with the result of a convective dynamo~\citep{2001ApJ...549.1183T,2021MNRAS.503..362K}, though considerable uncertainties remain in this story~\citep{Featherstone:2009}.
In the limit of weak convection ($\nabla_{\rm rad} \sim \nabla_{\rm ad}$) and strong magnetic fields convection is actually stabilized and so prevented from occuring in the first place~\citep{1966MNRAS.133...85G,2019MNRAS.487.3904M,2020ApJ...900..113J}.
These interactions are least-understood and so most interesting in the intermediate limit where the kinetic and magnetic energy densities are comparable, which occurs only in the HeI, HeII, and Fe CZs.

\subsubsection{Internal Gravity Waves}

Internal Gravity Waves (IGW) are generated at the boundaries of convection zones~\citep{1990ApJ...363..694G}.
The power in these waves peaks in frequency near the convective turnover frequency, and the overall flux of IGW scales as $F_{\rm conv} \mathrm{Ma}^\alpha$ for some order-unity $\alpha$~\citep{1990ApJ...363..694G,Rogers_2013,2013MNRAS.430.2363L,2018JFM...854R...3C}.
This makes it particularly important to study wave generation at the low-mass end of the HI CZ, because the HI CZ has $F_{\rm conv} \sim F$ and $\mathrm{Ma} \sim 0.3$, so the wave flux is expected to be a substantial fraction of the total stellar luminosity.
This could produce substantial wave mixing~\citep{1991ApJ...377..268G} or even alter the thermal structure of the star.
Wave mixing may also be induced by Core convection zones~\citep{2017ApJ...848L...1R}, where the smaller Mach number is offset by the larger stellar luminosity, resulting in even larger IGW fluxes. 

Since g-modes are evenescent in convective regions, they have small predicted amplitudes at the surface of stars with convective envelopes, making them difficult to detect \citep{2010A&ARv..18..197A}. There are claims of detection of solar g-modes in the literature \citep{2007Sci...316.1591G,2017A&A...604A..40F}, but so far none has been fully verified \citep{2018SoPh..293...95S,2019ApJ...877...42S,2019A&A...629A..26B}.

Even in early-type stars is usually difficult to observe IGW with very long periods, because these require long observing campaigns.
Because the IGW excitation spectrum peaks near the convective turnover time, IGW from the Core, Fe, and low-mass HI convection zones should have the most readily observable time-scales, with typical periods less than ten days.
However, the transmission of waves to the surface may make IGW difficult to observe from core convection zones~\citep{2019ApJ...886L..15L,2021ApJ...915..112C}.

\subsubsection{Marginal Stability}

Convective instability requires a Rayleigh number greater than the critical value of $\sim 10^3$~\citep{1961hhs..book.....C}, and the character of marginally unstable convection can be quite different from that of high-$\mathrm{Ra}$ convection.
This is most relevant for the HeI, HeII, and high-mass HI convection zones, as these have Rayleigh numbers that cross the threshold from sub-critical (stable) to super-critical (unstable)~\citep{2022arXiv220110567J}.

The actual stability of a convection zone is determined by the radial profile of the Rayleigh number. For cases where the average Rayleigh number is close to the critical value, it is important to solve the global stability problem to determine if the CZ is stable or not. If the Rayleigh number is everywhere subcritical, the putative CZ should be stable.

Stellar convection at moderate Rayleigh numbers is likely dominated by thermal structures with aspect ratios near unity which are very diffusive. These will organize into ``roll'' structures with predominantly horizontal vorticity, which exhibit dynamics such as spiral defect chaos~\citep[e.g.][]{PhysRevFluids.5.093501}. 
Although the thermal structures may be diffusive, the flows are still highly turbulent~\citep[e.g.][]{2018NatCo...9.2118P,PhysRevFluids.6.100503} due to the values of $\mathrm{Pr} \ll 1$ and $\mathrm{Re} \gg 1$. 
This may be important for studying the convective zone boundary, mixing, and/or interaction with waves.

\subsubsection{Stratification}

Stratification effects are strongest in Deep Envelope CZs, where the density can vary by four to seven orders of magnitude across the CZ.
Fundamentally, density stratification breaks the symmetry of convective upflows and downflows, leading to intense, fast, narrow downflows and broad, slower, diffuse upflows.
Global simulations suggest that these broad upflows, named ``giant cells,'' should imprint on surface convection flows, but there is debate and disagreement regarding whether or not these flows are observed in the Sun \citep{2016AnRFM..48..191H}.
It is possible that rotation influences giant cells in a way that changes their observational signature \citep{2016ApJ...830L..15F, 2021PNAS..11822518V}.
It is also possible that high density stratification turns cold downflows into small, fast features called ``entropy rain'' \citep{2016ApJ...832....6B, 2019ApJ...884...65A}.
The very low diffusivities present in stars could allow small downflows to traverse the full depth of these stratified CZs without diffusing, but modern convection simulations generally do not have enough spatial resolution (or have too large diffusivity) to resolve these motions.
In summary, stratification breaks symmetry, but the precise consequences of this symmetry breaking on convective dynamics and observed phenomena is not clear.

\subsubsection{Super-Eddington Convection}

When the \emph{radiative} energy flux through the system is locally super-Eddington the hydrostatic solution develops a density inversion and becomes unstable to convection~\citep{1973ApJ...181..429J}.
In deep stellar interiors the thermal energy density is large enough that convection can carry most of the energy flux.
This lowers the radiative flux, allowing it to remain sub-Eddington and eliminating the density inversion~\citep{2015ApJ...813...74J}.
Convection is then sustained by the usual superadiabatic temperature gradient.
In this way the convective cores of very massive stars can remain nearly hydrostatic despite super-Eddington total energy fluxes. 

On the other hand in near-surface layers where convection becomes inefficient convection may not able to carry enough heat and the radiative flux can stay super-Eddington.
Radiative acceleration can then drive a density inversion in late-type stars, and in luminous stars can even drive outflows~\citep{Owocki:2012,2018Natur.561..498J}.
This is what happens in the FeCZ at high masses.

\subsubsection{Radiation Pressure}

Even in sub-Eddington systems radiation pressure can play a role by changing the effective polytropic index of the adiabat (i.e. $\gamma \rightarrow 4/3$ as $\beta_{\rm rad} \rightarrow 1$).
We are We are not aware of studies addressing effects of high-$\beta$ convection, which suggests it could be interesting to study this limit, and in particular to compare otherwise-identical systems both including and excluding radiation pressure to understand what differences arise.

\subsubsection{Radiative Transfer}

Radiation pressure and a high Eddington ratio can substantially affect the nature of convection~\citep{1973ApJ...181..429J,1998ApJ...494L.193S,Paxton2013}.
The resulting dynamics depends sensitively on the ratio of the photon diffusion time to the dynamical time~\citep{2015ApJ...813...74J}.
This determines whether calculations can use a simplified diffusive treatment of radiation, or if they need to use full radiative transfer.

In Core CZs the photon diffusion time is long, so while both $\beta_{\rm rad}$ and $\Gamma_{\rm Edd}$ approach unity there it is still suitable to treat radiation in the diffusion approximation \citep[e.g. as done by][]{2016ApJ...829...92A}.

By contrast, in the FeCZ in massive stars the photon diffusion time is comparable or shorter than the dynamical timescale.
Moreover, the FeCZ has ${\beta_{\rm rad}, \Gamma_{\rm Edd} \sim 1}$.
Such zones thus must be studied using full radiative transfer~\citep{2015ApJ...813...74J,2017ApJ...843...68J,2018Natur.561..498J,2020ApJ...902...67S,2022ApJ...924L..11S}.

\subsubsection{Stiffness and the P{\'e}clet number}

The stiffness of a convective boundary determines the extent of overshooting past the edge of the boundary.
All of the upper boundaries we examined are extremely stiff except for the HI CZ, which has $\mathrm{S}_{\rm outer} \sim 0.1-1$, and Deep Envelope CZs, which have $\mathrm{S}_{\rm outer} \sim 0.1-10$ when they don't extend all the way to the surface of the model.
This suggests that motions could carry a decent fraction of a scale-height past the upper boundary of the HI CZ, which might cause observable motions at the photosphere and so could conceivably explain some of the observed macro/microturbulence \citep{Landstreet:2009}. The connection between (near-)surface convection and surface velocity fields is well established in the regime of late-type stars \citep[e.g.][]{2000A&A...359..729A,2007A&A...469..687C,2011ApJ...741..119M,2012MNRAS.427...27B,2013MSAIS..24...37S,2013ApJ...769...18T},
but is still under investigation for early-type stars \citep{2009A&A...499..279C,2015ApJ...813...74J,2020ApJ...902...67S,2022ApJ...924L..11S}.
Most of the lower boundaries are similarly stiff, though the low-mass HI CZs and the FeCZ both have relatively low-stiffness lower boundaries with $\mathrm{S}_{\rm inner} \sim 10-100$.
This could be responsible for some chemical mixing beneath these CZs.

When Pe is small, radiative diffusion dominates over advection and the motion becomes effectively isothermal rather than adiabatic.
Overshooting convective flows feel a reduced stiffness, and so can extend further than the stiffness alone would suggest.
This may be why 3D calculations show that motions from the FeCZ extend far beyond 1D predictions and up to the stellar surface \citep{2015ApJ...813...74J,2018Natur.561..498J}, and could account for some of the observed stellar variability and surface turbulence in stars with near-surface thin ionization CZs~\citep{2021ApJ...915..112C,2022ApJ...924L..11S,2022MNRAS.509.4246E}. This said, other processes than near-surface convection could be important in driving the observed surface velocity fields \citep[see e.g.][]{2009A&A...508..409A,2013ApJ...772...21R}.

\subsubsection{Transonic Convection}

Thermodynamic perturbations in a convection zone scale like the square Mach number \citep{PhysRevFluids.2.083501}.
Flows with Mach numbers approaching one at the photospheres of Deep Envelope convection zones can therefore add to the stellar photometric variability, participating to the ``bolometric flicker'' observed in light curves \citep{1998mons.proc...59T,2006A&A...445..661L,2011ApJ...741..119M,2014A&A...567A.115C,van_kooten_etal_2021}.
Improved models of high-Mach number surface convection can therefore assist in cleaning photometric lightcurves and assist in exoplanet detection.

We also observe high Mach number flows in the opacity-driven HI and Fe CZs.
Local models of these regions have been studied by e.g., \citet{2015ApJ...813...74J} and \citet{2022ApJ...924L..11S}, who observed that the large thermal perturbations associated with this convection created low-density, optically-thin ``chimneys'' through which radiation could escape the star directly.
Turbulent motions driven by these high Mach number regions is a candidate for the origin of observed low-frequency variability \citep{2021ApJ...915..112C}.

In summary, transonic (high-Mach number) convection occurs in various convective regions near the surfaces of both early and late type stars, and the large thermodynamic perturbations associated with this convection can generate short-timescale variability of the stellar luminosity.

\subsection{The Road to 3D Stellar Evolution}

The study of stars has relied heavily on one-dimensional stellar evolution calculations.
Three-dimensional stellar evolution calculations are not yet feasible due to the formidable range of spatial and temporal scales characterizing the problem.
For example, the average Reynolds number in the solar convection zone is about $10^{12}$.
Assuming a Kolmogorov spectrum of turbulence, the range of scales that need to be resolved to achieve an accurate and resolved direct numerical simulation is then $\ell_{\rm max}/\ell_{\rm min} \sim \,$Re$^{3/4} \sim 10^9$.
Such a simulation would need about $10^{27}$ resolution elements, while the largest hydrodynamic simulations currently use $\sim 10^{12}$.

Similarly, simulations that aim to capture the full dynamics of convection on the longer timescales of stellar evolution (thermal and nuclear) would require an enormous number of timesteps.
For example the dynamical timescale in the solar convection zone is of order an hour, while the thermal and nuclear timescales are $10^7\,\mathrm{yr}$  and $10^{10}\,\mathrm{yr}$, respectively.
Assuming one timestep per dynamical timescale (a vast underestimate), this implies that $\sim 10^{11-14}$ timesteps would be needed to resolve the solar convection zone for a thermal (nuclear) timescale, exceeding the number of steps achieved in state-of-the-art multidimensional numerical simulations \citep[e.g. $\sim 4\times 10^8$,][]{2021arXiv211011356A} by many orders of magnitude. 

Even assuming that Moore's law will continue to hold, a fully resolved stellar turbulence calculation of solar convection will not be achievable for $1.5 \log_2 10^{15}  \sim $ 50 years \citep[e.g.][]{2008IAUS..252..439M}.
A similar calculation protracted for a full thermal (nuclear) timescale is  $1.5 \log_2 10^{20}  \sim $ 70 years ($1.5 \log_2 10^{23}  \sim $ 80 years)  away.

It is clear, then, that one-dimensional stellar evolution calculations will remain an important tool for decades to come.
At the same time, the last decade has marked a transition to a new landscape in the theoretical study of stars.

Driven by progress in both hardware and numerical schemes, multi-dimensional calculations are becoming common tools for studying stellar interiors. 
Even though these calculations do not resolve the full range of relevant scales, it is likely that many of the flow features they reveal are robust to varying resolution.
While this is not guaranteed to be true in all cases (e.g. in magnetohydrodynamic systems or other setups with inverse cascades), and so must be checked carefully, numerical simulations can provide valuable insight into real astrophysical situations.

As computers and numerical methods become more and more powerful, new problems in stellar convection become accessible to numerical study.
Our aim with this atlas is to provide a guide useful for the next generation of stellar physicists performing numerical simulations of stellar convection.
Our ``maps'' illuminate the range of parameters involved in such numerical efforts, and our hope is that they will be used by modellers to better navigate this landscape.

\acknowledgments

The Flatiron Institute is supported by the Simons Foundation.
This research was supported in part by the National Science Foundation under Grant No. PHY-1748958.
We are grateful to the Kavli Institute for Theoretical Physics and its staff for providing the uniquely engaging environment which enabled this work.
EHA thanks Ben Brown and Imogen Cresswell for many discussions on stellar convection broadly and magnetoconvection. 
EHA thanks CIERA and Northwestern University for CIERA Postdoctoral fellowship funding.
DL is supported in part by NASA HTMS grant 80NSSC20K1280.

\software{
\texttt{MESA} \citep[][\url{http://mesa.sourceforge.net}]{Paxton2011,Paxton2013,Paxton2015,Paxton2018,Paxton2019},
\texttt{MESASDK} \citep{mesasdk_linux,mesasdk_macos},
\texttt{matplotlib} \citep{hunter_2007_aa}, 
\texttt{NumPy} \citep{der_walt_2011_aa}
         }

\clearpage

\appendix

\section{Diffusivities}\label{sec:diffusivities}

\subsection{Viscosity}

Computing the viscosity of a plasma is complicated.
For simplicity we use the viscosity of pure hydrogen plus radiation, so that
\begin{align}
	\nu \approx \nu_{\rm ie} + \nu_{\rm rad}.
	\label{eq:nu}
\end{align}
The radiation component is~\citep{1962pfig.book.....S}
\begin{align}
	\nu_{\rm rad} = \frac{4 a T^4}{15 c \kappa \rho^2},
	\label{eq:nu_rad}
\end{align}	
where $a$ is the radiation gas constant.
We obtain the hydrogen viscosity using the Braginskii-Spitzer formula~\citep{osti_4317183,1962pfig.book.....S}
\begin{align}
	\nu_{\rm ie} \approx 2.21\times 10^{-15} \frac{(T/\mathrm{K})^{5/2}}{(\rho/\mathrm{g\,cm^{-3}}) \ln \Lambda}\mathrm{cm^2\,s^{-1}},
	\label{eq:nu_ie}
\end{align}
where $\ln \Lambda$ is the Coulomb logarithm, given for hydrogen by~\citep{1987ApJ...313..284W}
\begin{align}
	\ln \Lambda = -17.9 + 1.5\ln \frac{T}{\mathrm{K}} - 0.5 \ln\frac{\rho}{\mathrm{g\,cm^{-3}}}
\end{align}
for temperatures $T < 4.2\times 10^5\,\mathrm{K}$ and
\begin{align}
	\ln \Lambda = -11.5 + \ln \frac{T}{\mathrm{K}} - 0.5 \ln\frac{\rho}{\mathrm{g\,cm^{-3}}}
\end{align}
for $T > 4.2\times 10^5\,\mathrm{K}$.
Corrections owing to different compositions are generally small relative to the many orders of magnitude we are interested in here.
For instance the difference between pure hydrogen and a cosmic mixture of hydrogen and helium is under $30\%$~\citep{2008ApJ...674..408B}.

\subsection{Thermal Diffusion}

Thermal diffusion in main-sequence stars is dominated by photons, resulting in the radiative diffusivity
\begin{align}
	\chi \equiv \frac{16 \sigma T^3}{3\kappa c_p \rho^2}.
	\label{eq:chi}
\end{align}

\subsection{Electrical Conductivity}

We use routines built into MESA to compute the conductivity $\sigma$. These routines were provided by~\citep{SCY} and follow the formula of~\citet{1962pfig.book.....S}, with a Coulomb logarithm computed by~\citet{1987ApJ...313..284W}. Given $\sigma$, the magnetic diffusivity is computed via
\begin{align}
	\eta = \frac{c^2}{4\pi \sigma}.
	\label{eq:eta}
\end{align}

\section{MESA} \label{appen:mesa}

The MESA EOS is a blend of the OPAL \citep{Rogers2002}, SCVH
\citep{Saumon1995}, FreeEOS \citep{Irwin2004}, HELM \citep{Timmes2000},
PC \citep{Potekhin2010}, and Skye \citep{Jermyn2021} EOSes.

Radiative opacities are primarily from OPAL \citep{Iglesias1993,
Iglesias1996}, with low-temperature data from \citet{Ferguson2005}
and the high-temperature, Compton-scattering dominated regime by
\citet{Poutanen2017}.  Electron conduction opacities are from
\citet{Cassisi2007}.

Nuclear reaction rates are from JINA REACLIB \citep{Cyburt2010}, NACRE \citep{Angulo1999} and
additional tabulated weak reaction rates \citet{Fuller1985, Oda1994,
Langanke2000}.  Screening is included via the prescription of \citet{Chugunov2007}.
Thermal neutrino loss rates are from \citet{Itoh1996}.

Models were constructed on the pre-main sequence with $Z=0.014$, $Y=0.24+2Z$, and $X=1-Y-Z$ and evolved from there.
We neglect rotation and associated chemical mixing.

\section{Data Availability}\label{appen:data}

The inlists and run scripts used in producing the HR diagrams in this work are available in this other GitHub \href{https://github.com/adamjermyn/conv_trends}{repository} on the \texttt{atlas} branch in the commit with short-sha \texttt{701d74d9}. The data those scripts produced are available in~\citet{adam_s_jermyn_2022_6533097}, along with the plotting scripts used in this work.

\section{Rotation Law}\label{sec:rotation}

\begin{figure*}
\centering
\begin{minipage}{0.48\textwidth}
\includegraphics[width=\textwidth]{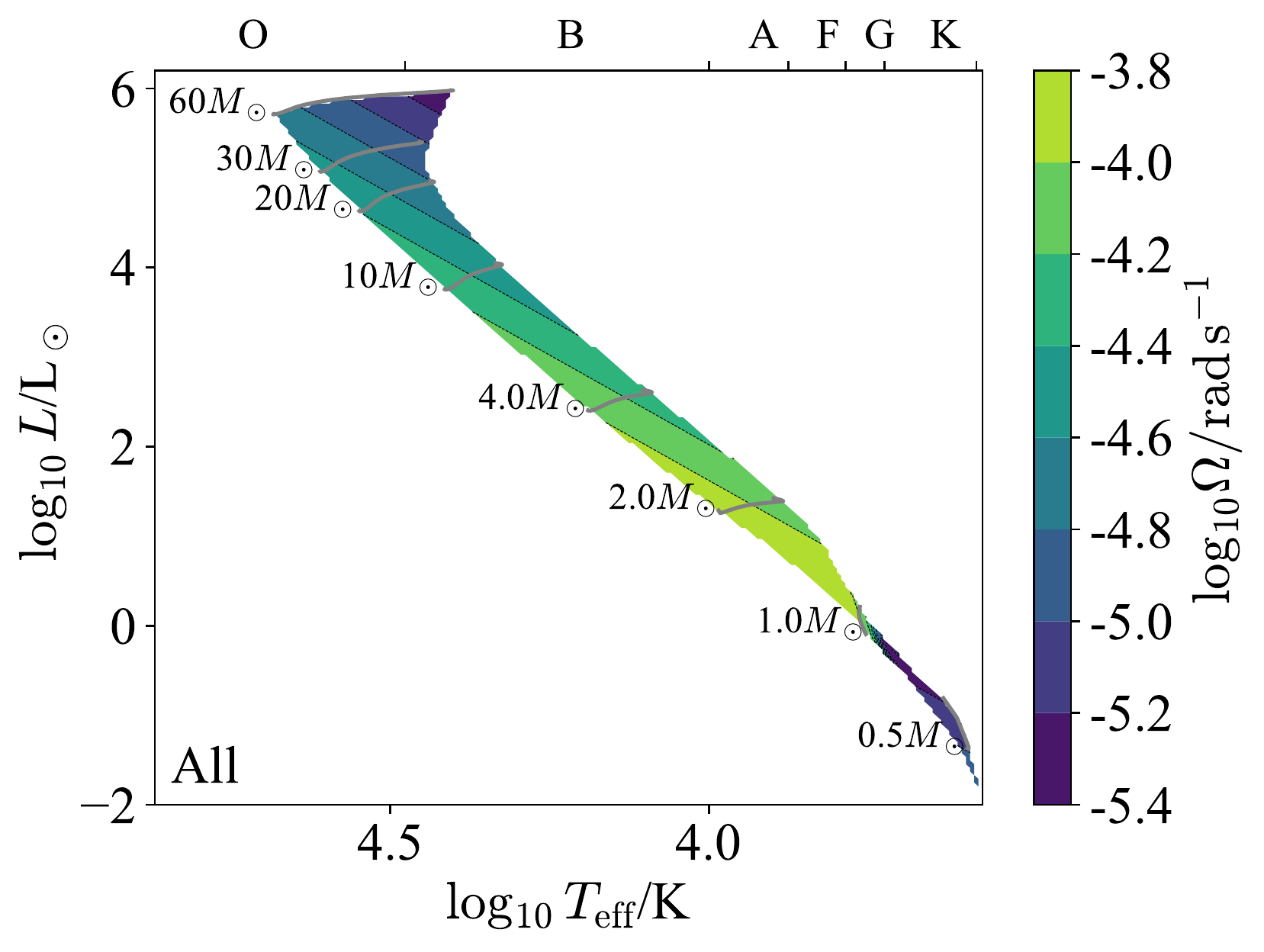}
\end{minipage}

\caption{The angular velocity $\Omega$ is shown in terms of $\log T_{\rm eff}$/spectral type and $\log L$ for stellar models with Milky Way metallicity $Z=0.014$. Note that for Sun-like stars we predict angular velocities of just under $10^{-4}\mathrm{rad\,s^{-1}}.$.}
\label{fig:envelope_diffusivities}
\end{figure*}

Throughout this work we choose a typical $\Omega$ corresponding to a surface velocity of $150\,\mathrm{km\,s^{-1}}$ for O-F stars (down to $1.2 M_\odot$).
From $1.2 M_\odot$ to $0.8 M_\odot$ we vary the surface velocity linearly down to $3\,\mathrm{km\,s^{-1}}$, and treat it as constant below this point.
This simple form approximately reproduces the inferred equatorial velocities from $v\sin i$ measurements by~\citet{2005yCat.3244....0G}, the typical equatorial velocities reported by~\citet{2013A&A...560A..29R}, and the rotation period inferred from Kepler observations by~\citep{2013A&A...557L..10N} for B-M stars.
Binary interactions can, of course, change the relevant velocities~\citep{2013ApJ...764..166D}, though in general we expect the dominant variation in rotation-related quantities to be due to variation in the properties of convection and not variation in the rotation periods.

\section{Averaging Sensitivity}\label{sec:avg}

\begin{figure*}[!ht]
\centering
\begin{minipage}{0.48\textwidth}
\includegraphics[width=\textwidth]{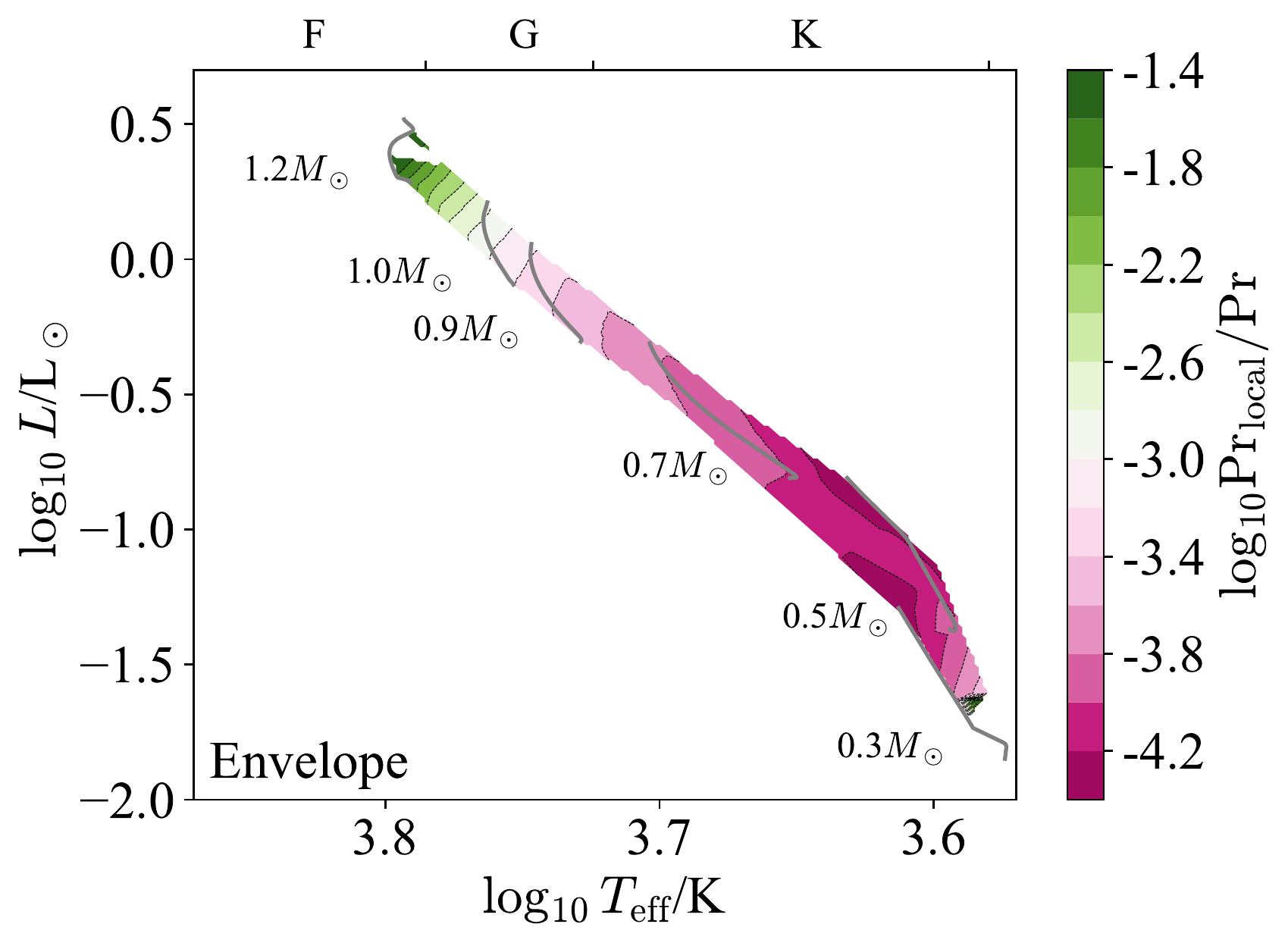}
\end{minipage}
\hfill
\begin{minipage}{0.48\textwidth}
\includegraphics[width=\textwidth]{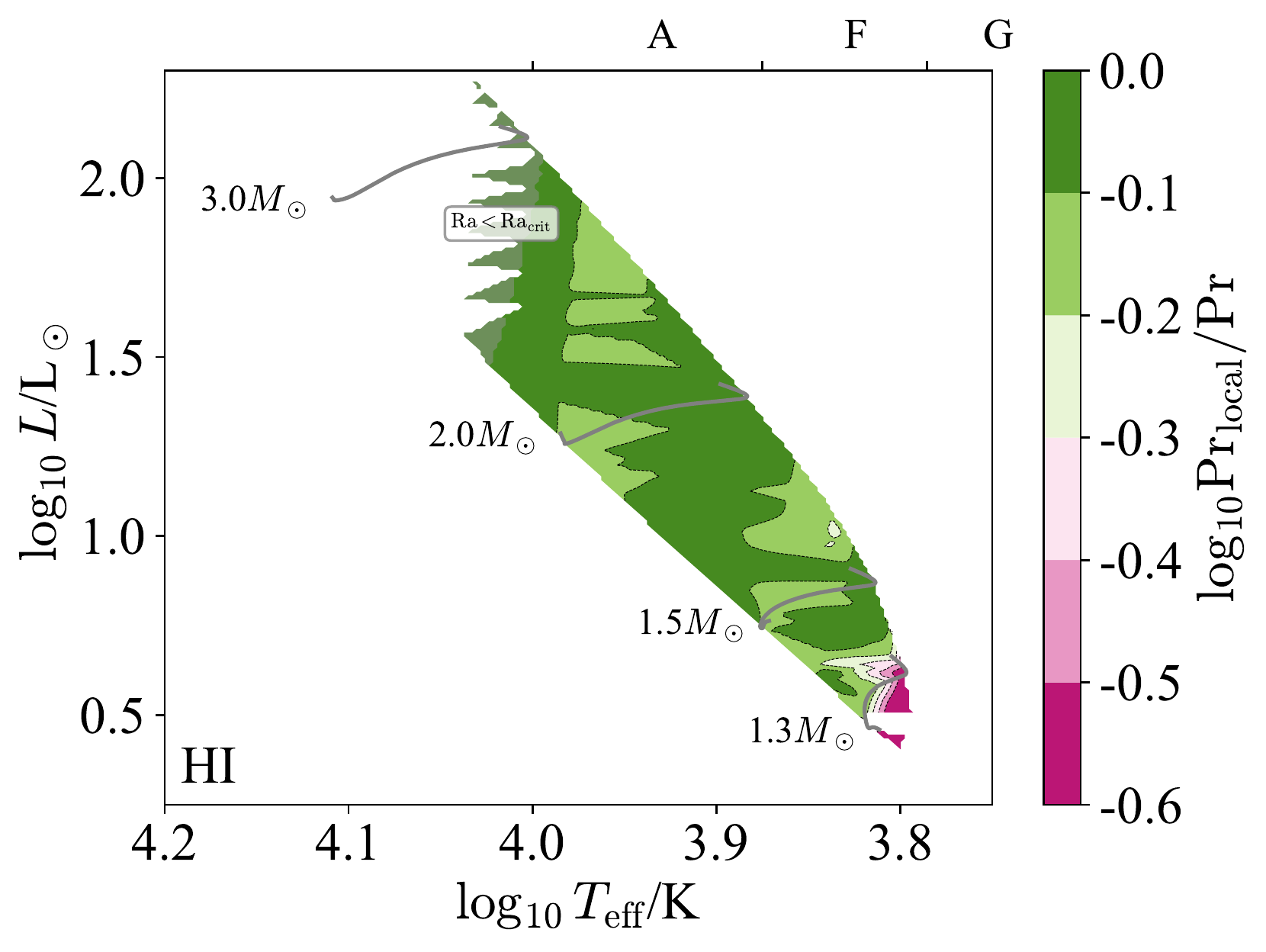}
\end{minipage}
\hfill
\begin{minipage}{0.48\textwidth}
\includegraphics[width=\textwidth]{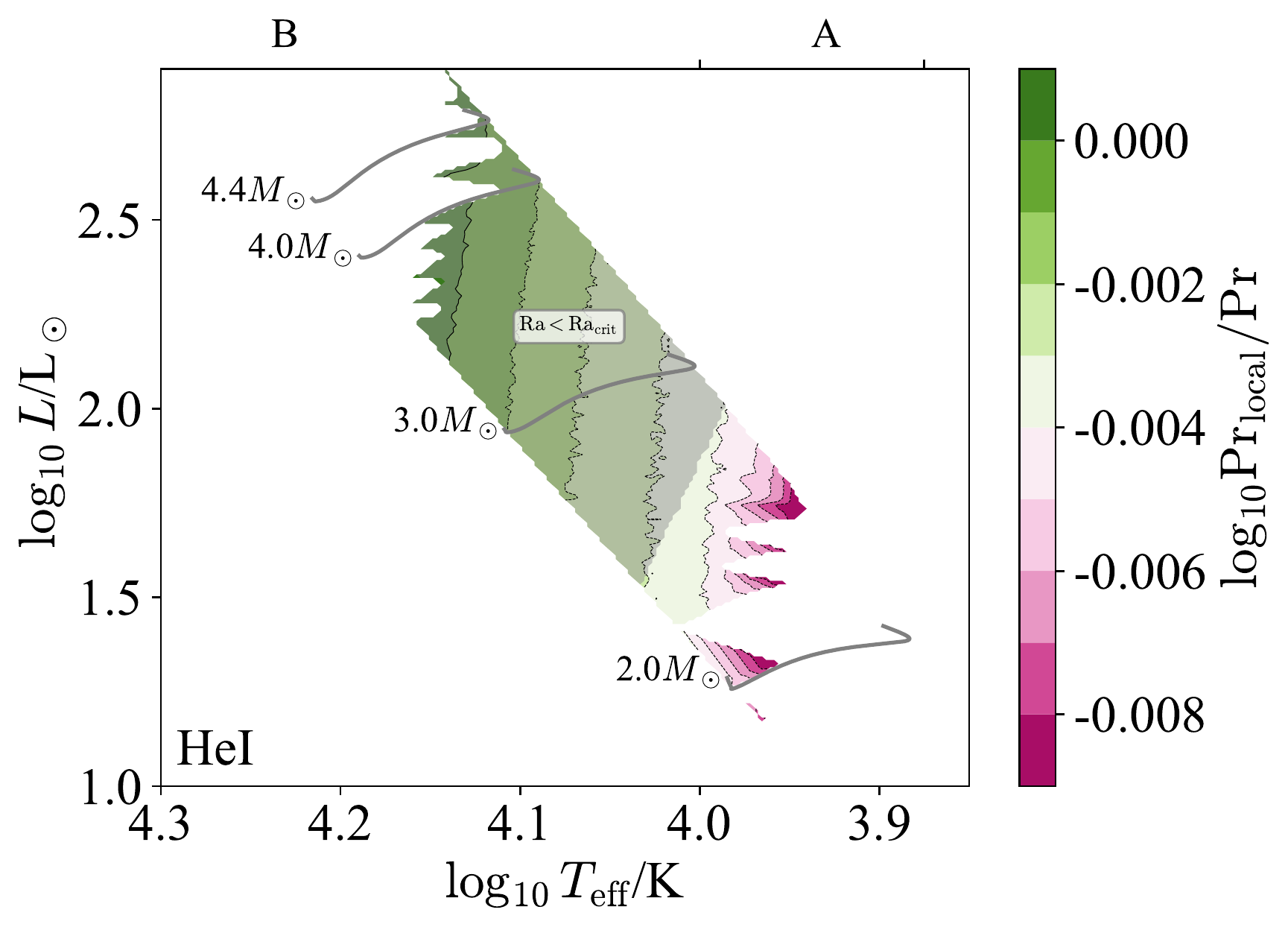}
\end{minipage}
\hfill
\begin{minipage}{0.48\textwidth}
\includegraphics[width=\textwidth]{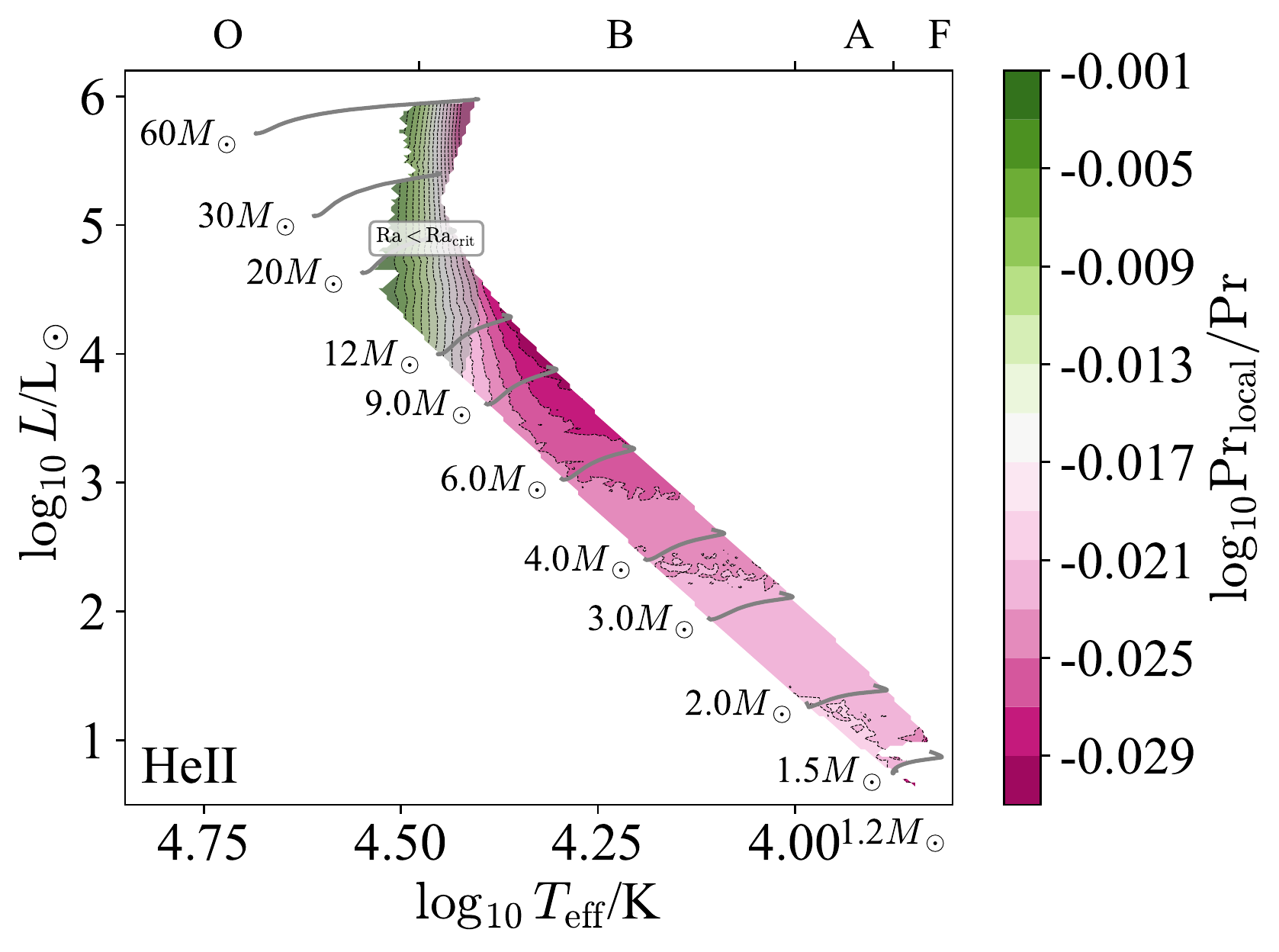}
\end{minipage}
\hfill
\begin{minipage}{0.48\textwidth}
\includegraphics[width=\textwidth]{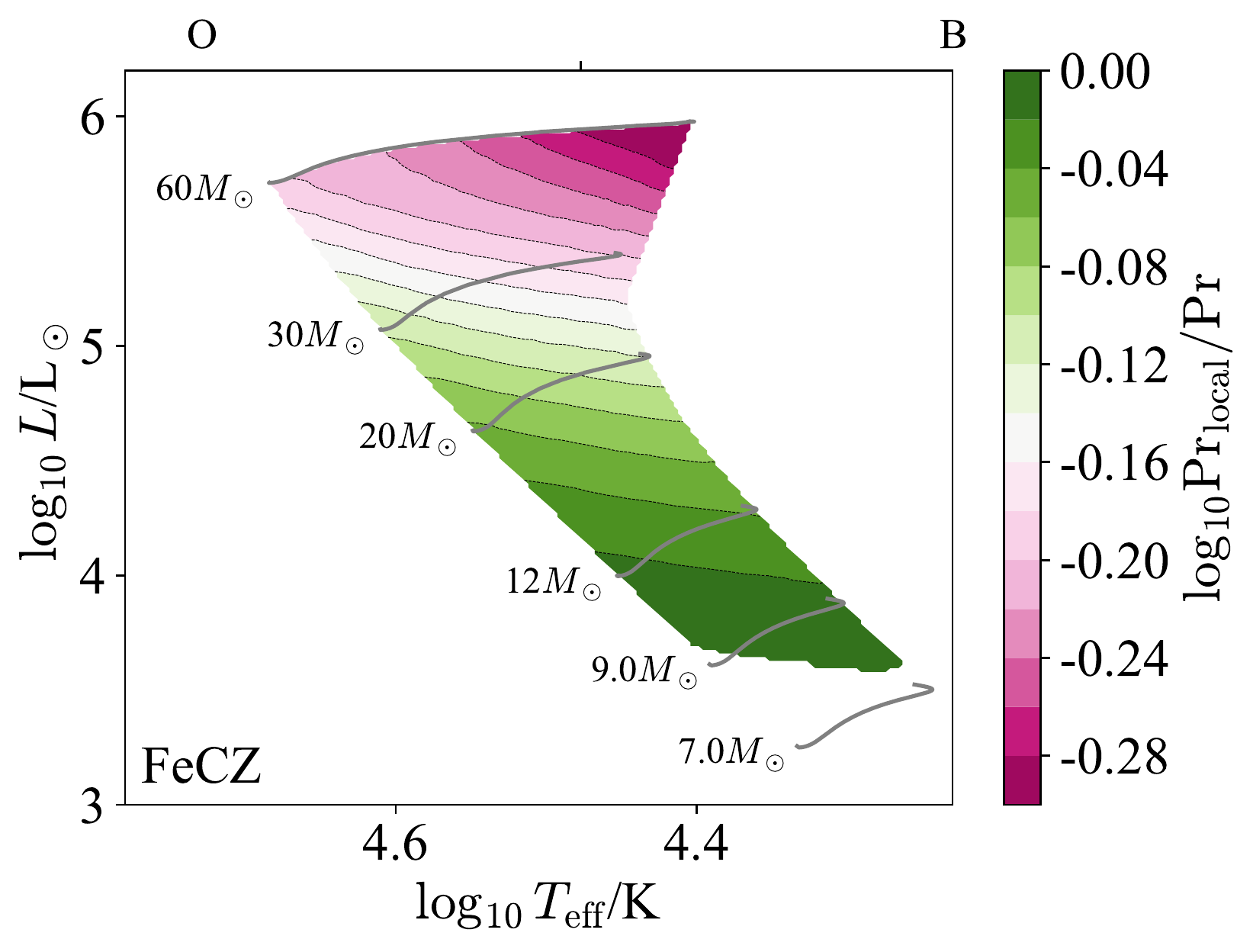}
\end{minipage}
\hfill
\begin{minipage}{0.48\textwidth}
\includegraphics[width=\textwidth]{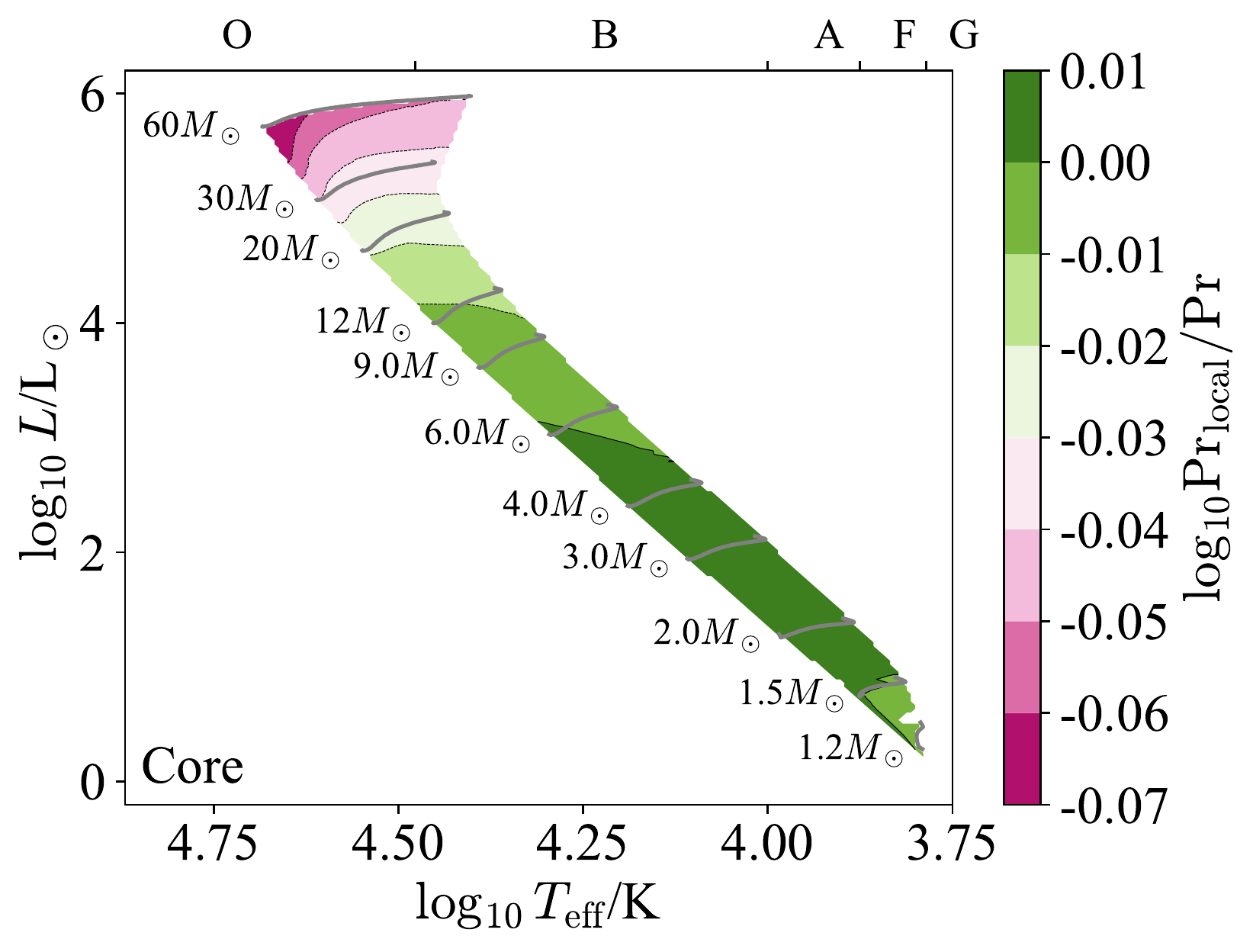}
\end{minipage}
\hfill

\caption{The difference between the Prandtl number computed by averaging $\nu$ and $\chi$ radially and that computed by radially averaging $\nu/\chi$ is shown in terms of $\log T_{\rm eff}$ and $\log L$ for stellar models with core convection zones and Milky Way metallicity $Z=0.014$. Different panels show different convection zones. For all but the Deep Envelope CZs the differences are small ($\la 2\times$), but for the Deep Envelope zones we see differences of up to $10^5$.}
\label{fig:averaging}
\end{figure*}

We had many choices in how to average different quantities, so it is worth performing a sensitivity analysis.
Figure~\ref{fig:averaging} shows the Prandtl number computed both by averaging $\nu$ and $\chi$ independently and by averaging $\nu/\chi$ directly.
This was chosen as a particularly simple quantity, and we show it for each convection zone.
For all but the Deep Envelope CZs the differences are small ($\la 2\times$), but for the Deep Envelope zones we see differences of up to $10^5$, motivating our choice to additionally show quantities near the inner and outer boundaries of Deep Envelope CZs.

\clearpage
\section{Convection Zone HR Diagrams}\label{sec:CZ_A}

\subsection{Deep Envelope CZ}

Here we examine Deep Envelope convection zones, which occur in low-mass stars ($M_\star \la 1.2 M_\odot$).
Because these zones are strongly density-stratified, we first examine averages over the entire CZ and then compare those with averages over just the innermost and outermost pressure scale heights.

\subsubsection{Whole-Zone Averages}

\begin{figure*}
\centering
\begin{minipage}{0.48\textwidth}
\includegraphics[width=\textwidth]{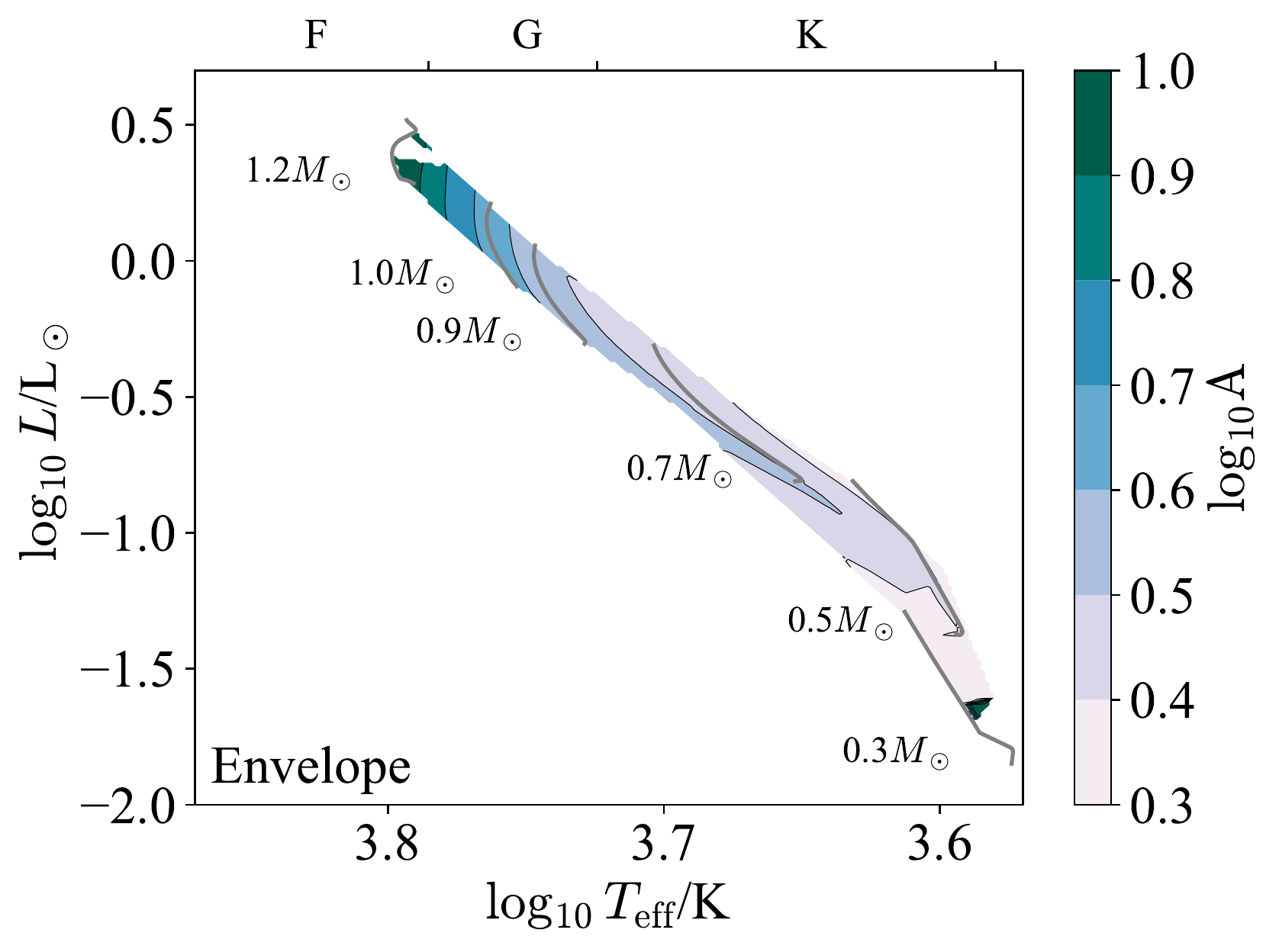}
\end{minipage}

\caption{The aspect ratio $\mathrm{A}$ is shown in terms of $\log T_{\rm eff}$/spectral type and $\log L$ for stellar models with deep envelope convection zones and Milky Way metallicity $Z=0.014$. Note that the aspect ratio is an input parameter, and does not depend on a specific theory of convection.}
\label{fig:envelope_structure}
\end{figure*}

We begin with the bulk structure of Deep Envelope CZs.
Figure~\ref{fig:envelope_structure} shows the aspect ratio $\mathrm{A}$, which ranges from $3-10$.
These small-to-moderate aspect ratios suggest that the global (spherical shell) geometry could be important.

\begin{figure*}
\begin{minipage}{0.48\textwidth}
\includegraphics[width=\textwidth]{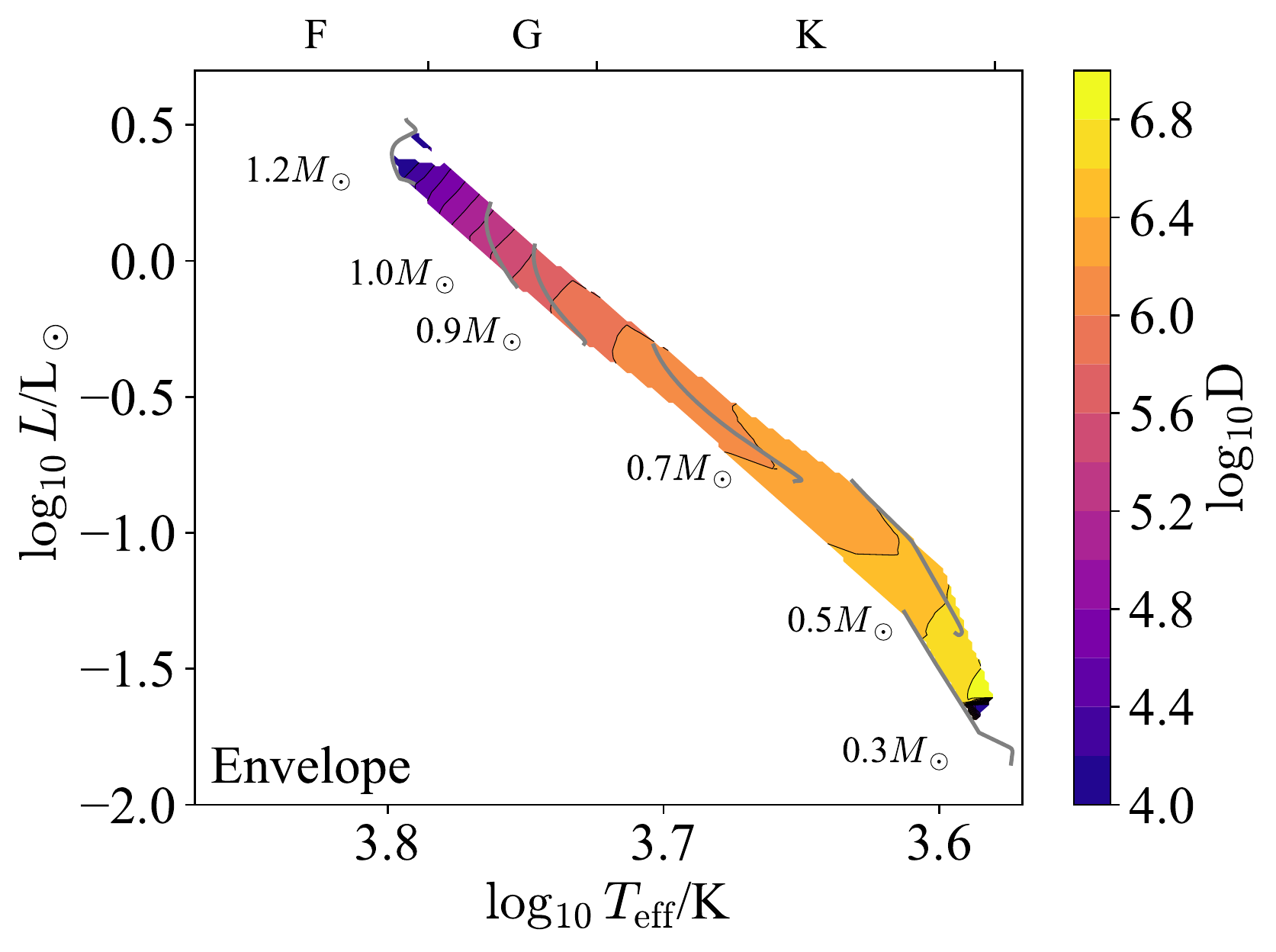}
\end{minipage}
\hfill
\begin{minipage}{0.48\textwidth}
\includegraphics[width=\textwidth]{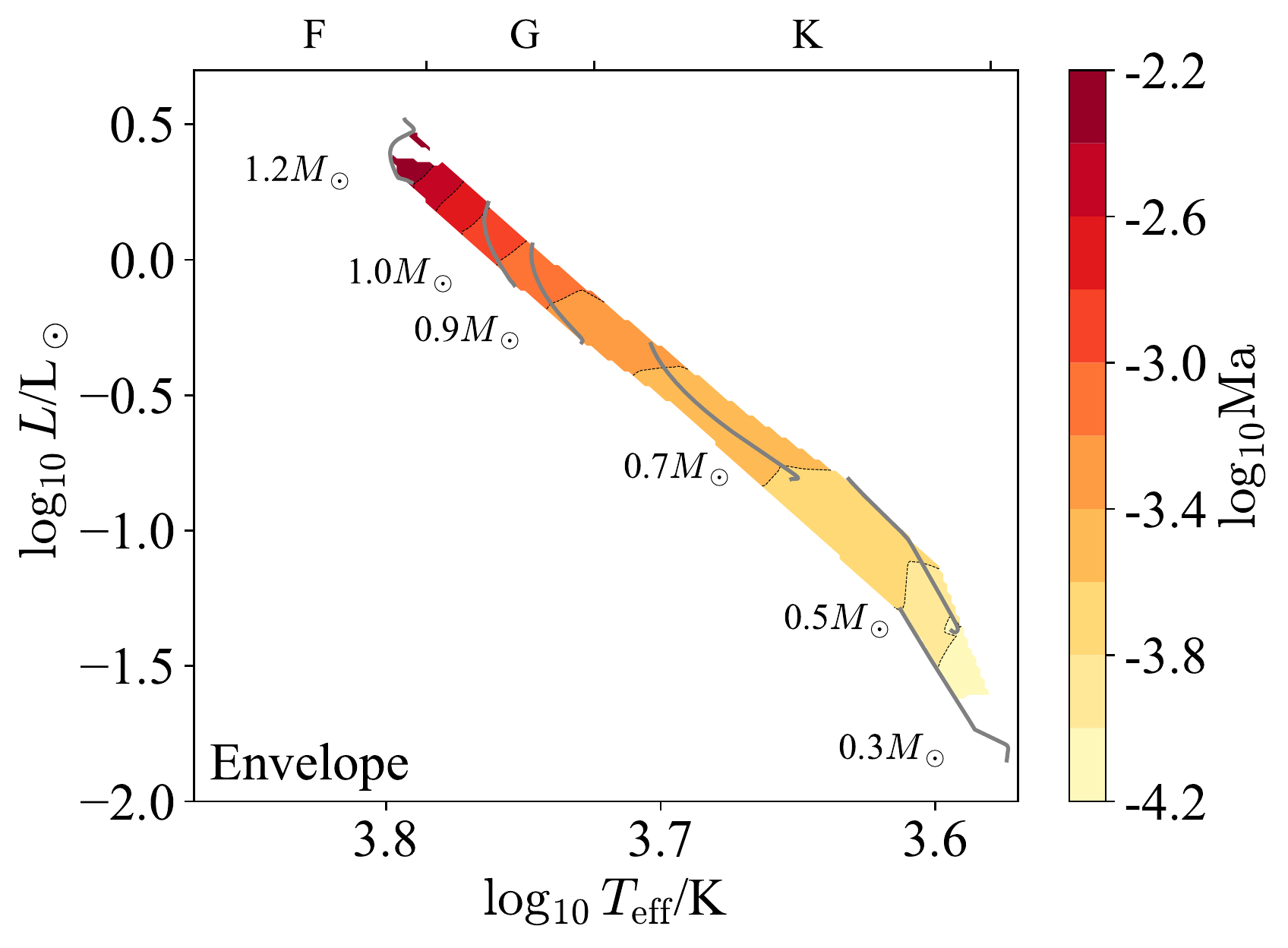}
\end{minipage}
\hfill

\caption{The density ratio $\mathrm{D}$ (left) and Mach number $\mathrm{Ma}$ (right) are shown in terms of $\log T_{\rm eff}$/spectral type and $\log L$ for stellar models with deep envelope convection zones and Milky Way metallicity $Z=0.014$. Note that while the density ratio is an input parameter and does not depend on a specific theory of convection, the Mach number is an output of such a theory and so is model-dependent.}
\label{fig:envelope_equations}
\end{figure*}

Next, the density ratio $\mathrm{D}$ (Figure~\ref{fig:envelope_equations}, left) and Mach number $\mathrm{Ma}$ (Figure~\ref{fig:envelope_equations}, right) inform which physics the fluid equations must include to model these zones.
Deep Envelope CZs are strongly density-stratified, so the Boussinesq approximation is likely insufficient to study them.
For the most part convection in these zones is also highly subsonic.
This, along with the density ratio, suggests it is appropriate to use the anelastic approximation.
However, near the surface $v_c$ becomes large and the fully compressible equations may be necessary (Section~\ref{sec:inner_outer}).

\begin{figure*}
\begin{minipage}{0.48\textwidth}
\includegraphics[width=\textwidth]{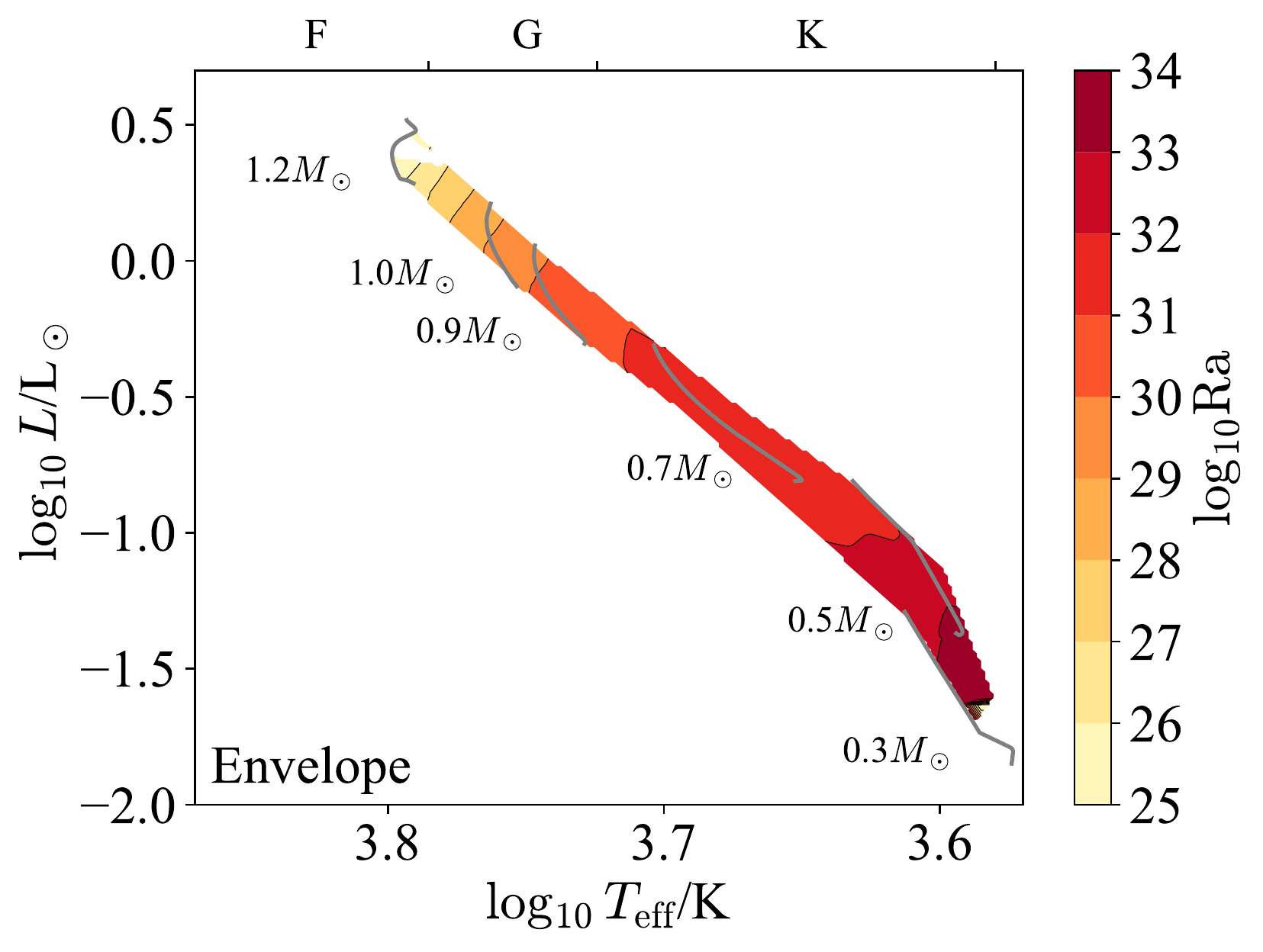}
\end{minipage}
\hfill
\begin{minipage}{0.48\textwidth}
\includegraphics[width=\textwidth]{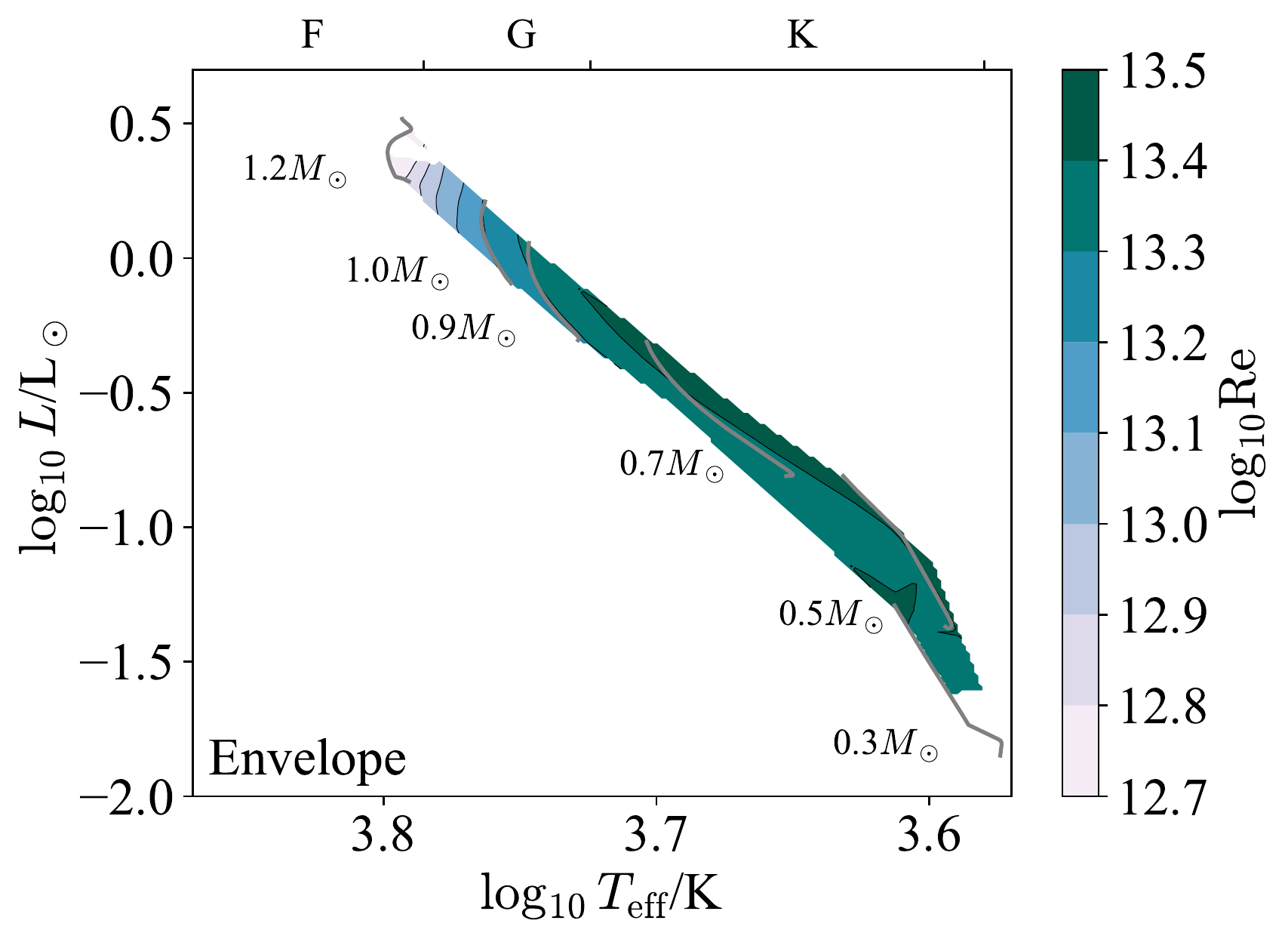}
\end{minipage}
\hfill

\caption{The Rayleigh number $\mathrm{Ra}$ (left) and Reynolds number $\mathrm{Re}$ (right) are shown in terms of $\log T_{\rm eff}$/spectral type and $\log L$ for stellar models with deep envelope convection zones and Milky Way metallicity $Z=0.014$.  Note that while the Rayleigh number is an input parameter and does not depend on a specific theory of convection, the Reynolds number is an output of such a theory and so is model-dependent.}
\label{fig:envelope_stability}
\end{figure*}

The Rayleigh number $\mathrm{Ra}$ (Figure~\ref{fig:envelope_stability}, left) determines whether or not a putative convection zone is actually unstable to convection, and the Reynolds number $\mathrm{Re}$ determines how turbulent the zone is if instability sets in (Figure~\ref{fig:envelope_stability}, right).
In these zones both numbers are enormous, so we should expect convective instability to result in highly turbulent flows.

\begin{figure*}
\centering
\begin{minipage}{0.48\textwidth}
\includegraphics[width=\textwidth]{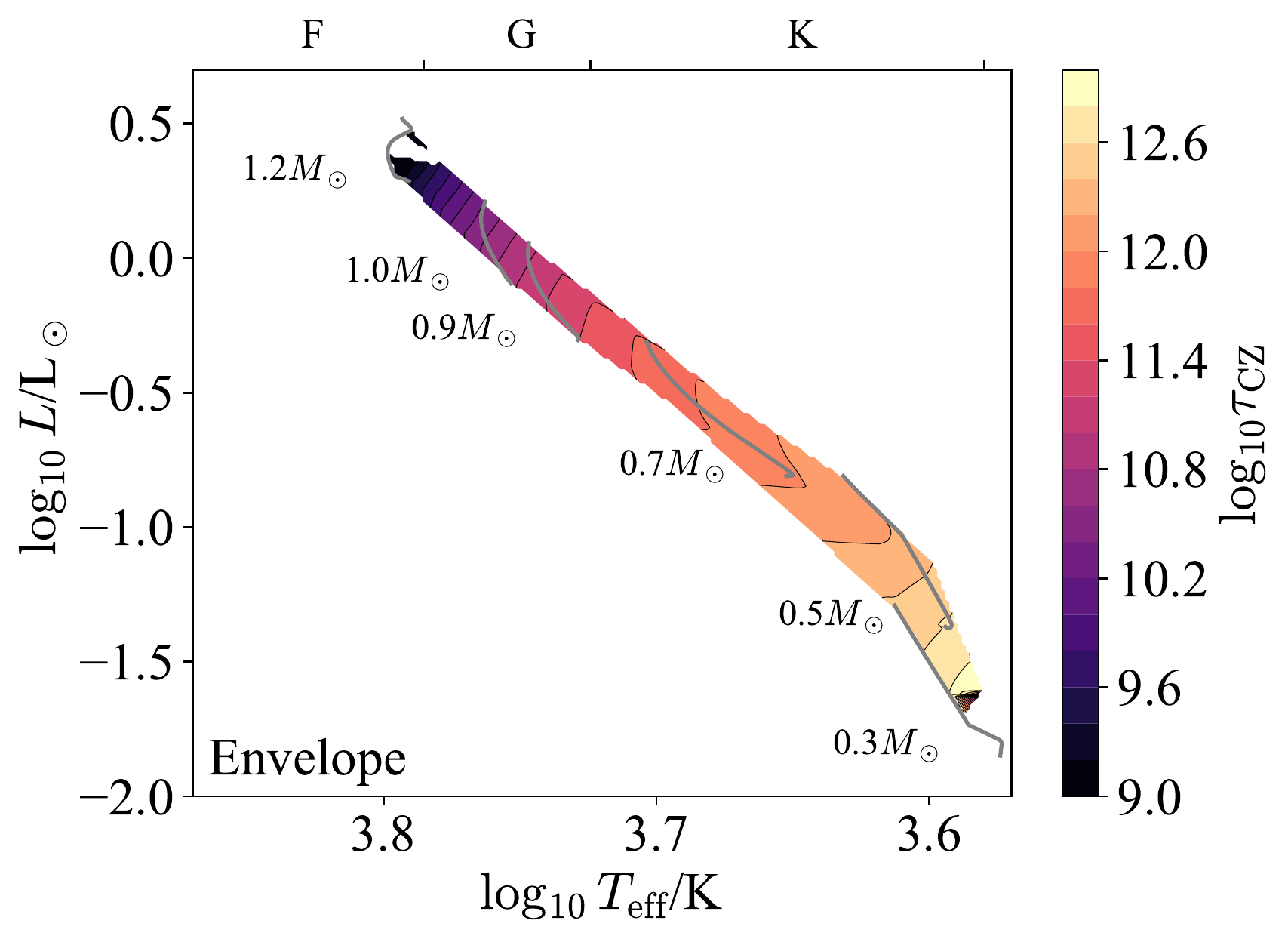}
\end{minipage}
\hfill
\begin{minipage}{0.48\textwidth}
\includegraphics[width=\textwidth]{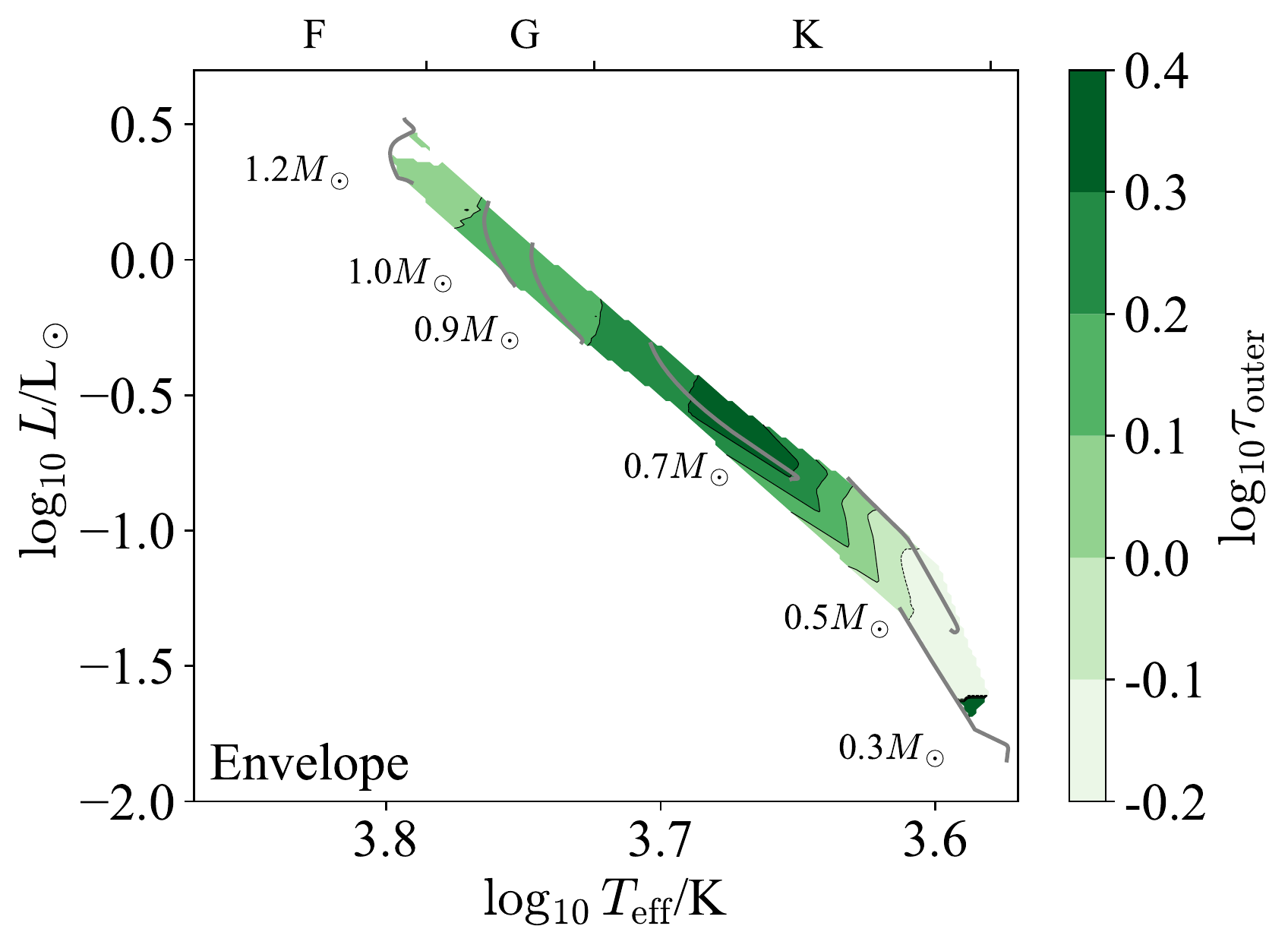}
\end{minipage}
\hfill
\caption{The convection optical depth $\tau_{\rm CZ}$ (left) and the optical depth to the surface $\tau_{\rm outer}$ (right) are shown in terms of $\log T_{\rm eff}$/spectral type and $\log L$ for stellar models with deep envelope convection zones and Milky Way metallicity $Z=0.014$. Note that both of these are input parameters, and do not depend on a specific theory of convection.}
\label{fig:envelope_optical}
\end{figure*}

The optical depth across a convection zone $\tau_{\rm CZ}$ (Figure~\ref{fig:envelope_optical}, left) indicates whether or not radiation can be handled in the diffusive approximation, while the optical depth from the outer boundary of the convection zone to infinity $\tau_{\rm outer}$ (Figure~\ref{fig:envelope_optical}, right) indicates the nature of radiative transfer and cooling in the outer regions of the convection zone.
We see that the optical depth across these zones is enormous ($\tau_{\rm CZ} \sim 10^{11}$) but their outer boundaries lie at very small optical depths ($\tau_{\rm outer} \sim 1$). This means that the bulk of the CZ can be modeled in the limit of radiative diffusion, but the dynamics of the outer regions likely require radiation hydrodynamics.

\begin{figure*}
\centering
\begin{minipage}{0.48\textwidth}
\includegraphics[width=\textwidth]{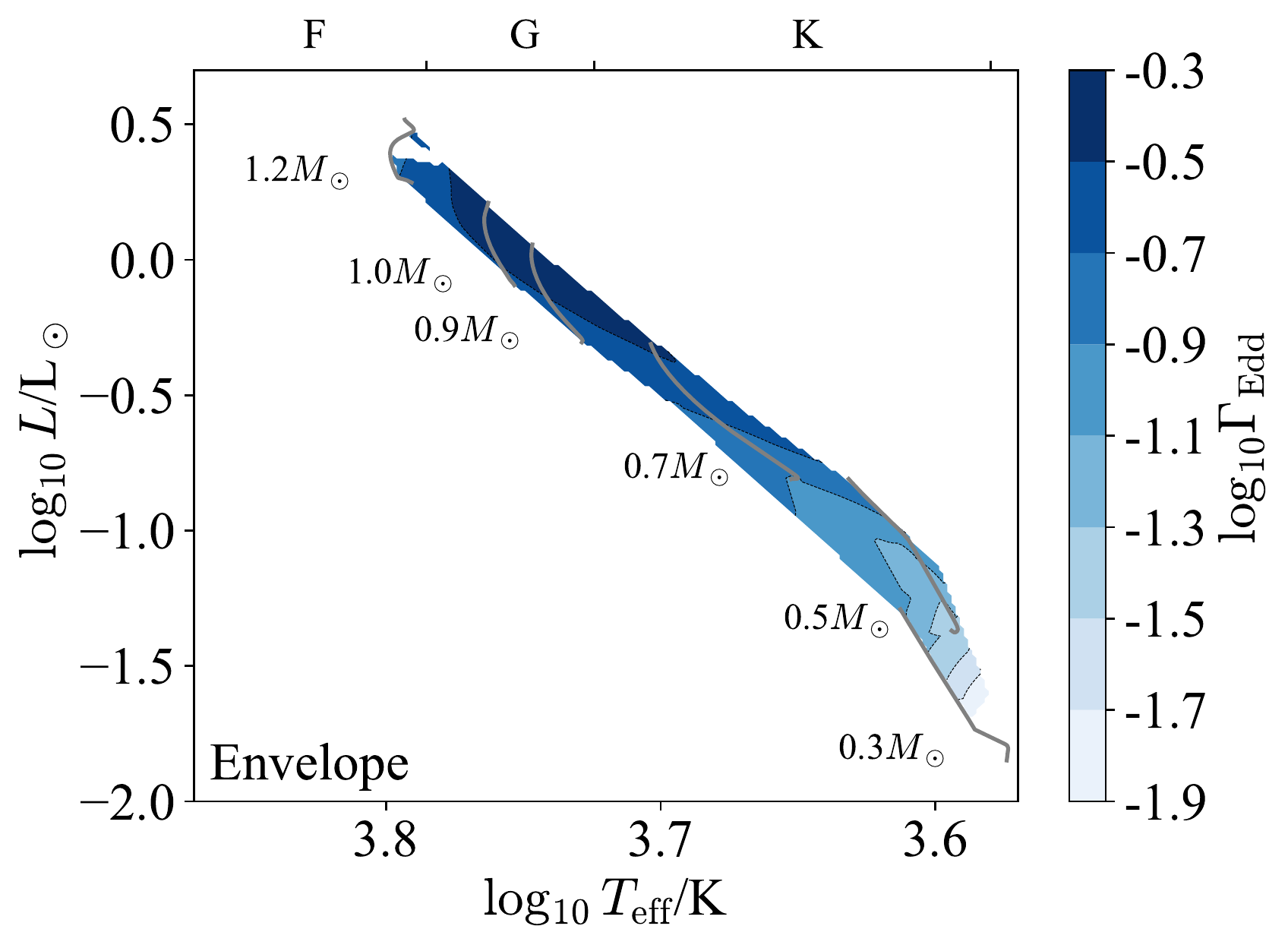}
\end{minipage}
\hfill
\begin{minipage}{0.48\textwidth}
\includegraphics[width=\textwidth]{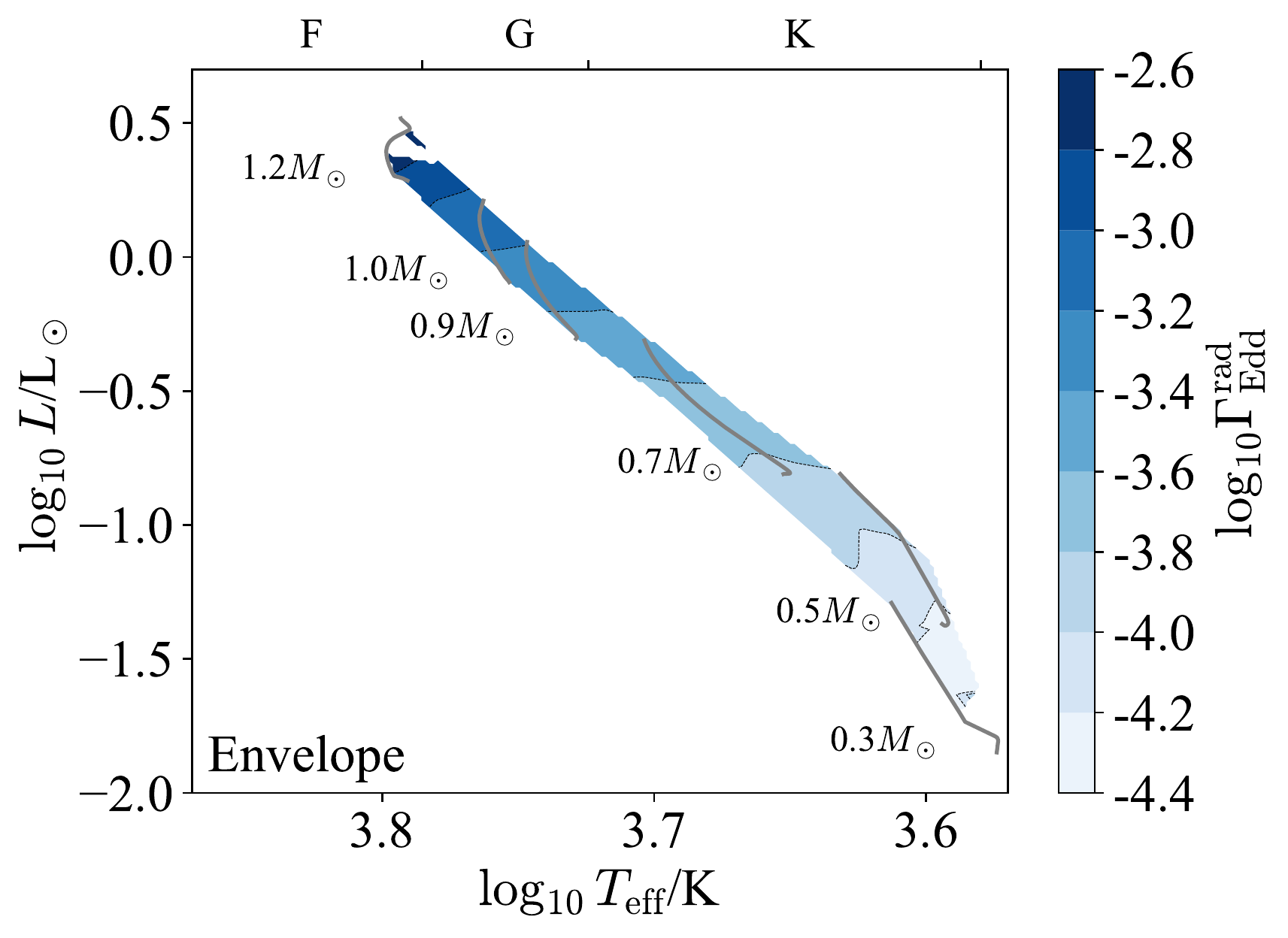}
\end{minipage}
\hfill
\caption{The Eddington ratio with the full luminosity $\Gamma_{\rm Edd}$ (left) and the radiative luminosity (right) are shown in terms of $\log T_{\rm eff}$/spectral type and $\log L$ for stellar models with deep envelope convection zones and Milky Way metallicity $Z=0.014$. Note that while $\Gamma_{\rm Edd}$ is an input parameter and does not depend on a specific theory of convection, $\Gamma_{\rm Edd}^{\rm rad}$ is an output of such a theory and so is model-dependent.}
\label{fig:envelope_eddington}
\end{figure*}

The Eddington ratio $\Gamma_{\rm Edd}$ (Figure~\ref{fig:envelope_eddington}, left) indicates whether or not radiation hydrodynamic instabilities are important in the non-convecting state, and the radiative Eddington ratio $\Gamma_{\rm Edd}^{\rm rad}$ (Figure~\ref{fig:envelope_eddington}, right) indicates the same in the developed convective state.
Here we see that in the absence of convection Deep Envelope CZs would reach moderate $\Gamma_{\rm Edd} \sim 0.3$, but because convection transports some of the flux this is reduced to $\la 0.1$ and radiation hydrodynamic instabilities are unlikely to matter.

The Prandtl number $\mathrm{Pr}$ (Figure~\ref{fig:envelope_diffusivities}, left) measures the relative importance of thermal diffusion and viscosity, and the magnetic Prandtl number $\mathrm{Pm}$ (Figure~\ref{fig:envelope_diffusivities}, right) measures the same for magnetic diffusion and viscosity.
We see that both are very small, so the thermal diffusion and magnetic diffusion length-scales are much larger than the viscous scale.

\begin{figure*}
\centering
\begin{minipage}{0.48\textwidth}
\includegraphics[width=\textwidth]{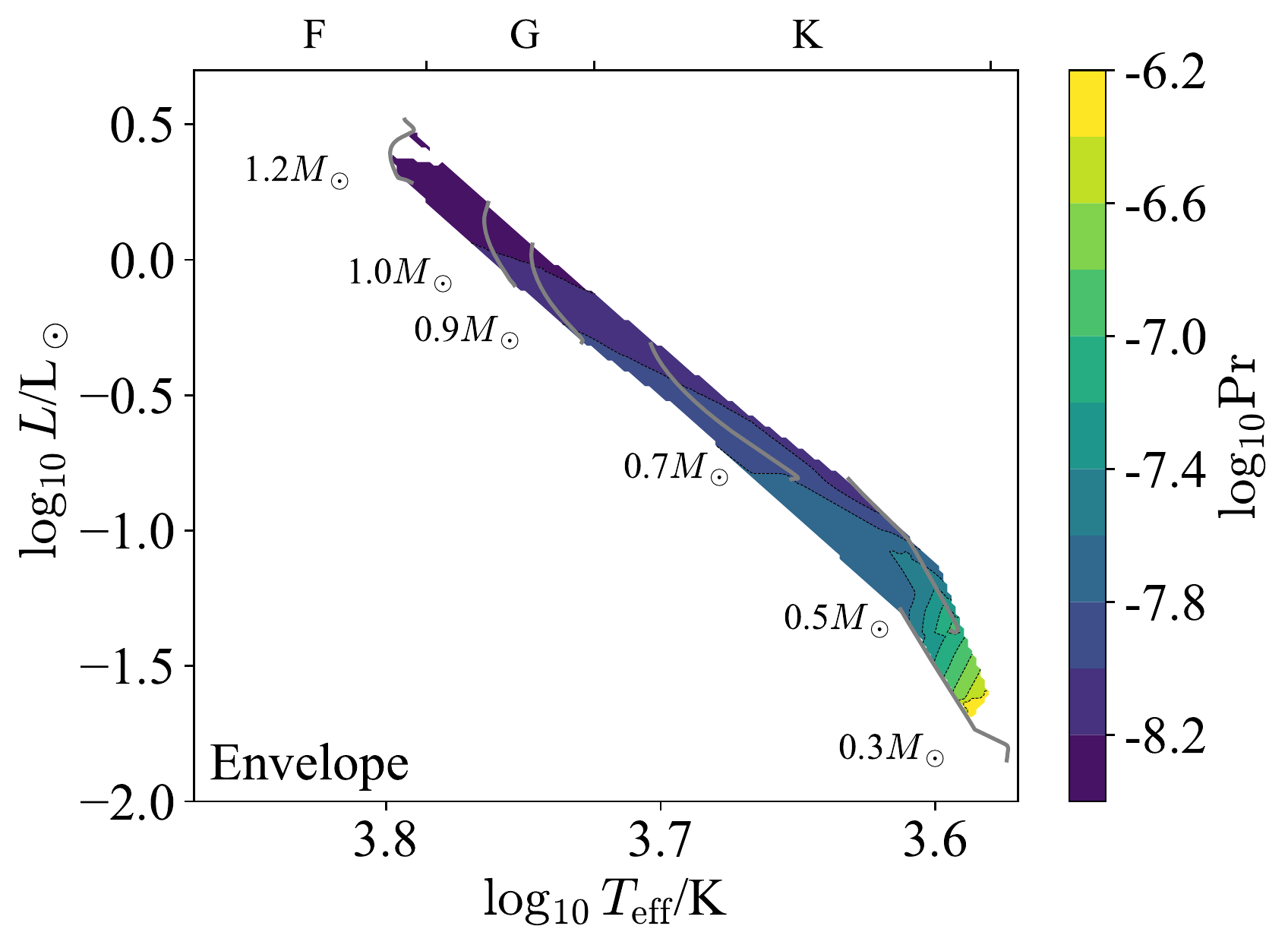}
\end{minipage}
\hfill
\begin{minipage}{0.48\textwidth}
\includegraphics[width=\textwidth]{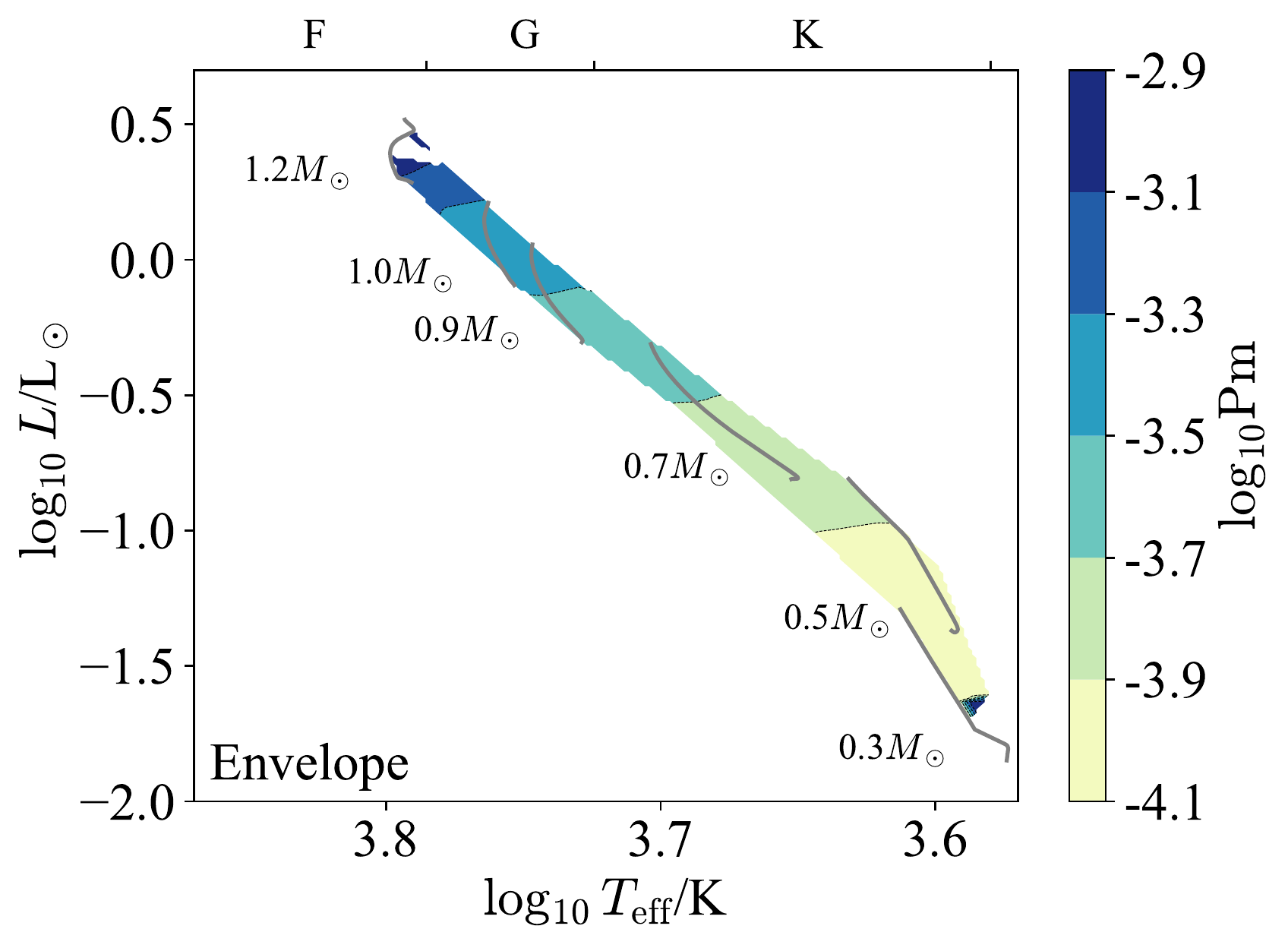}
\end{minipage}
\hfill

\caption{The Prandtl number $\mathrm{Pr}$ (left) and magnetic Prandtl number $\mathrm{Pm}$ (right) are shown in terms of $\log T_{\rm eff}$/spectral type and $\log L$ for stellar models with deep envelope convection zones and Milky Way metallicity $Z=0.014$. Note that both $\mathrm{Pr}$ and $\mathrm{Pm}$ are input parameters, and so do not depend on a specific theory of convection.}
\label{fig:envelope_diffusivities}
\end{figure*}

\begin{figure*}
\centering
\begin{minipage}{0.48\textwidth}
\includegraphics[width=\textwidth]{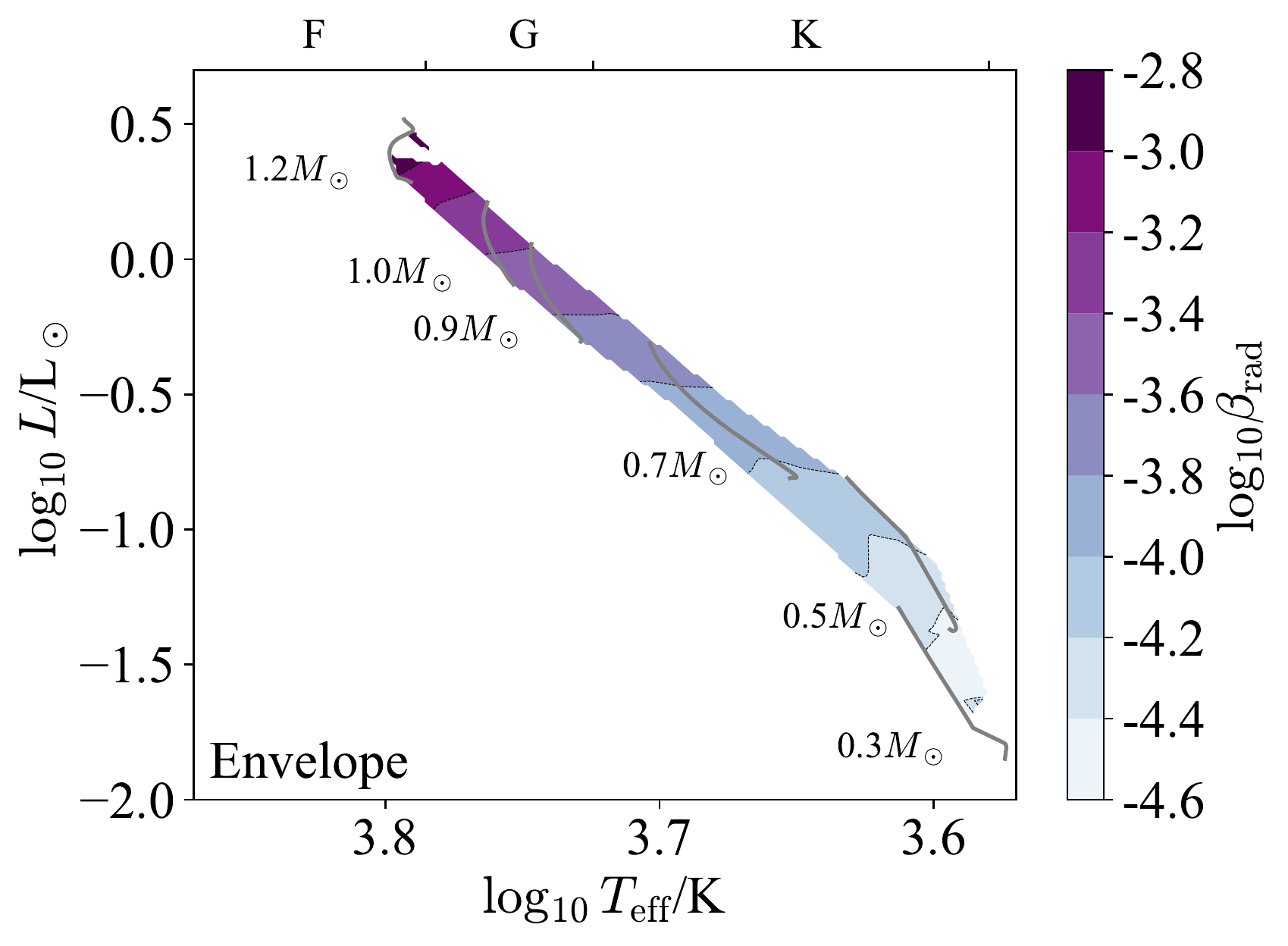}
\end{minipage}
\hfill
    \caption{The radiation pressure ratio $\beta_{\rm rad}$ is shown in terms of $\log T_{\rm eff}$/spectral type and $\log L$ for stellar models with deep envelope convection zones and Milky Way metallicity $Z=0.014$. Note that $\beta_{\rm rad}$ does not depend on the steady state convective velocity or temperature gradient, so it is not sensitive to the choice of convection theory.}
\label{fig:envelope_beta}
\end{figure*}

The radiation pressure ratio $\beta_{\rm rad}$ (Figure~\ref{fig:envelope_beta}) measures the importance of radiation in setting the thermodynamic properties of the fluid.
We see that this is uniformly small and so radiation pressure likely plays a sub-dominant role in these zones.

\begin{figure*}
\centering
\begin{minipage}{0.48\textwidth}
\includegraphics[width=\textwidth]{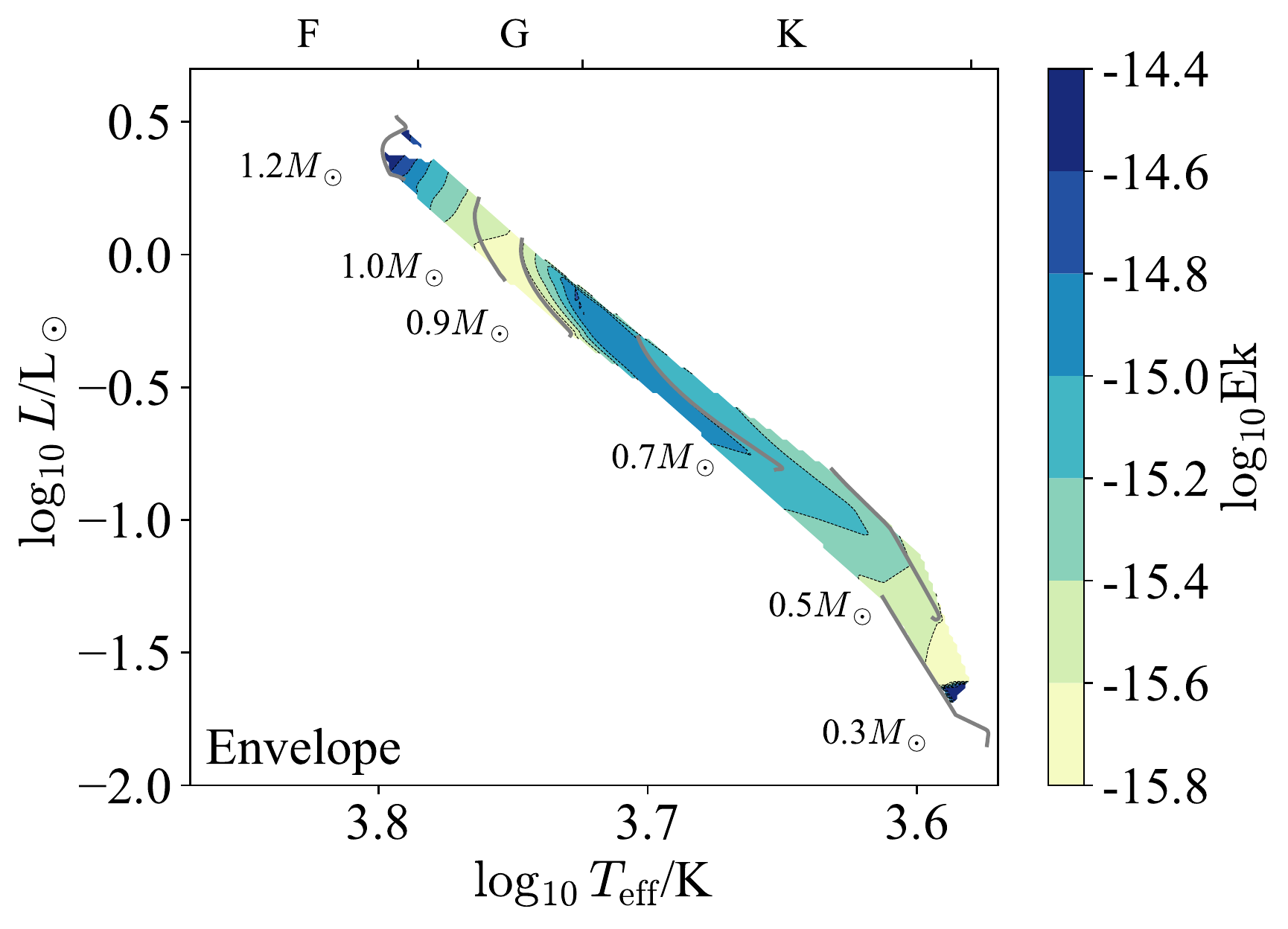}
\end{minipage}

\caption{The Ekman number $\mathrm{Ek}$ is shown in terms of $\log T_{\rm eff}$/spectral type and $\log L$ for stellar models with deep envelope convection zones and Milky Way metallicity $Z=0.014$. Note that the Ekman number is an input parameter, and does not depend on a specific theory of convection.}
\label{fig:envelope_ekman}
\end{figure*}

The Ekman number $\mathrm{Ek}$ (Figure~\ref{fig:envelope_ekman}) indicates the relative importance of viscosity and rotation.
This is tiny across the HRD, so we expect rotation to dominate over viscosity, except at very small length-scales.

\begin{figure*}
\centering
\begin{minipage}{0.48\textwidth}
\includegraphics[width=\textwidth]{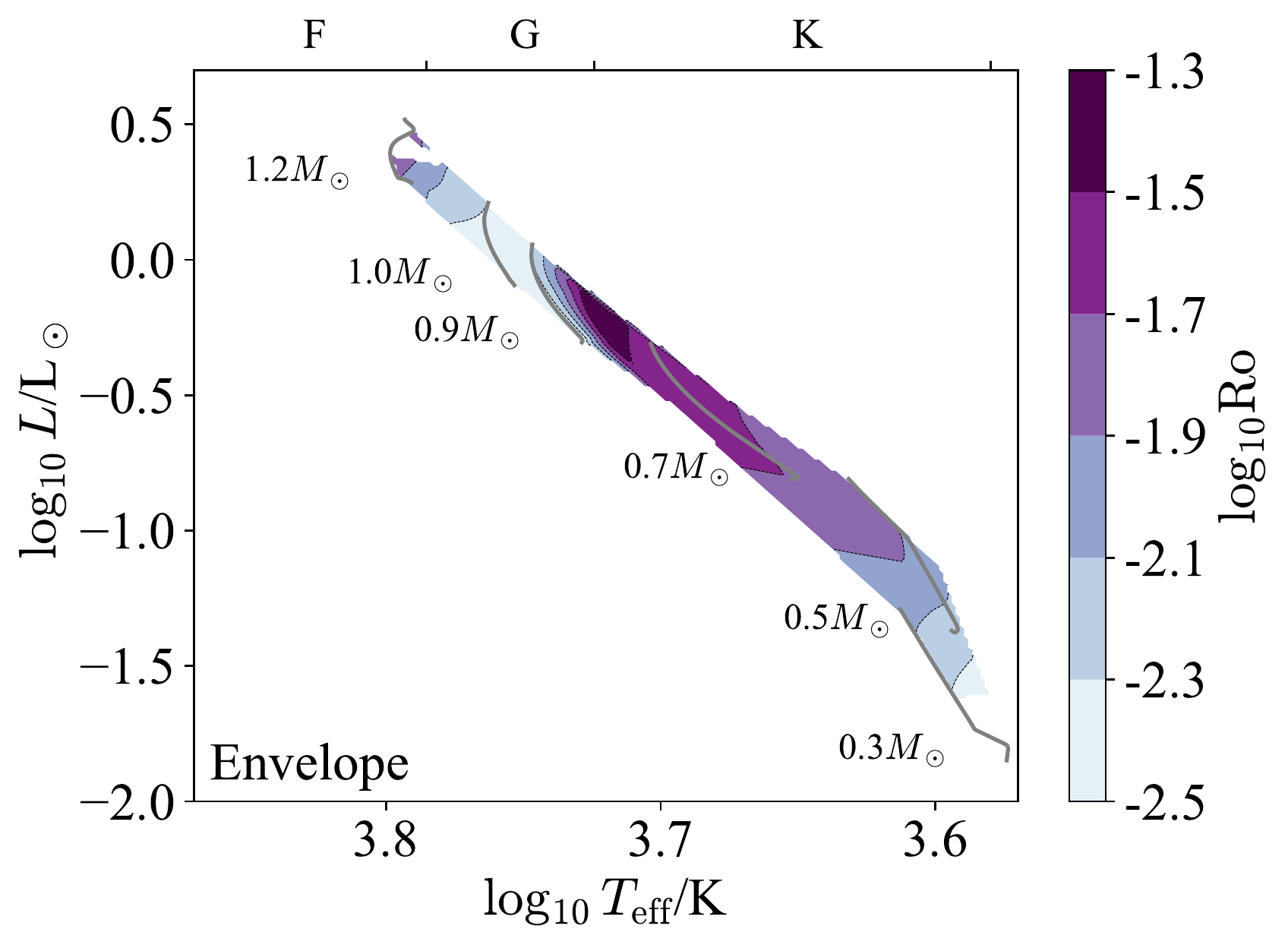}
\end{minipage}
\hfill
\begin{minipage}{0.48\textwidth}
\includegraphics[width=\textwidth]{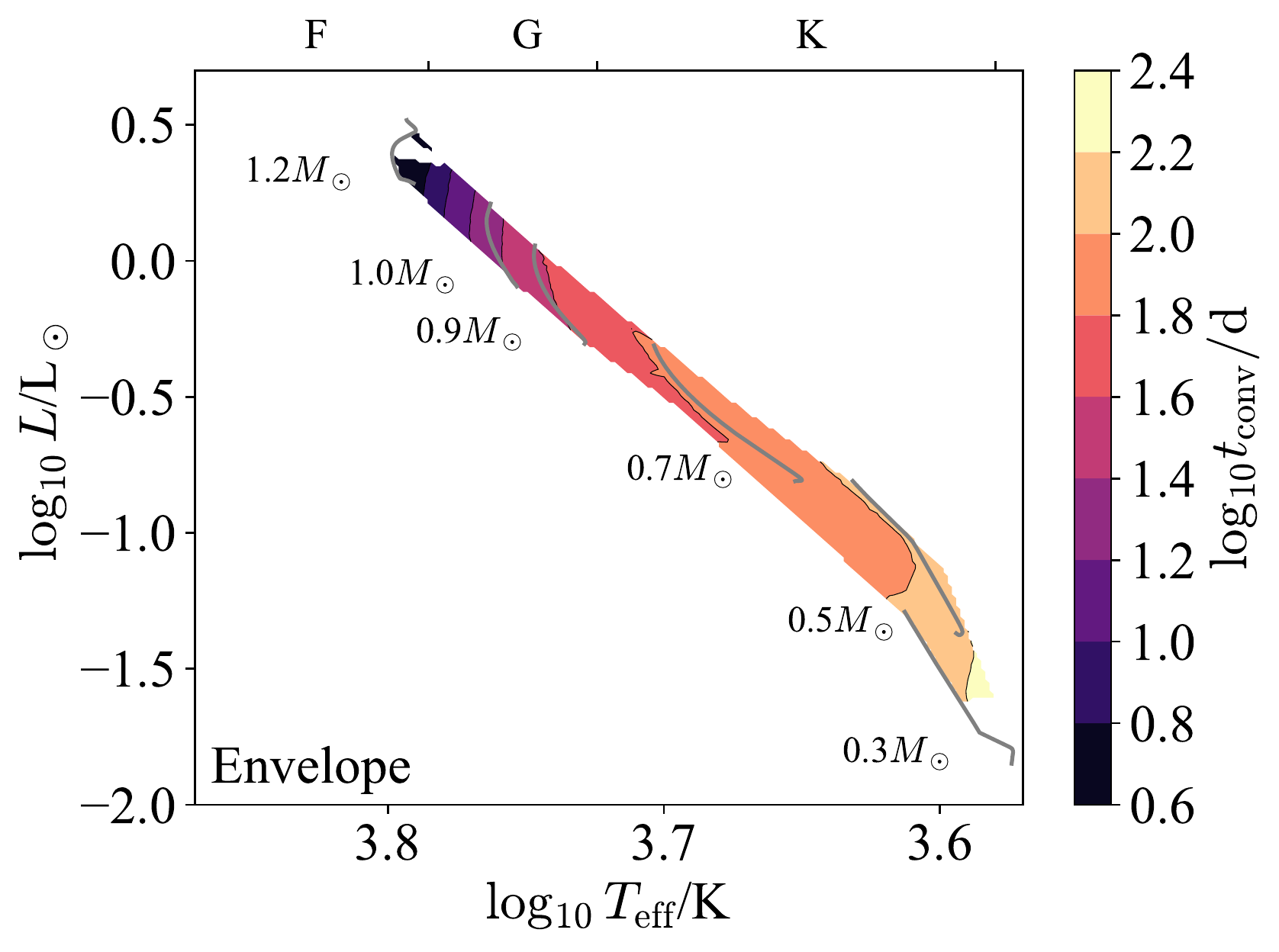}
\end{minipage}
\hfill

\caption{The Rossby number $\mathrm{Ro}$ (left) and turnover time $t_{\rm conv}$ (right) are shown in terms of $\log T_{\rm eff}$/spectral type and $\log L$ for stellar models with deep envelope convection zones and Milky Way metallicity $Z=0.014$. Note that both $\mathrm{Ro}$ and $t_{\rm conv}$ are outputs of a theory of convection and so are model-dependent.}
\label{fig:envelope_rotation}
\end{figure*}

The Rossby number $\mathrm{Ro}$ (Figure~\ref{fig:envelope_rotation}, left) measures the relative importance of rotation and inertia.
This is small, so these zones are strongly rotationally constrained for typical rotation rates~\citep{2013A&A...557L..10N}, though becoming less so towards higher masses. This is also strongly depth-dependent (Section~\ref{sec:inner_outer}), and near the surface $v_c$ becomes large and flows are typically not rotationally constrained.

We have assumed a fiducial rotation law to calculate $\mathrm{Ro}$.
Stars exhibit a variety of different rotation rates, so we also show the convective turnover time $t_{\rm conv}$ (Figure~\ref{fig:envelope_rotation}, right) which may be used to estimate the Rossby number for different rotation periods.

\begin{figure*}
\centering
\begin{minipage}{0.48\textwidth}
\includegraphics[width=\textwidth]{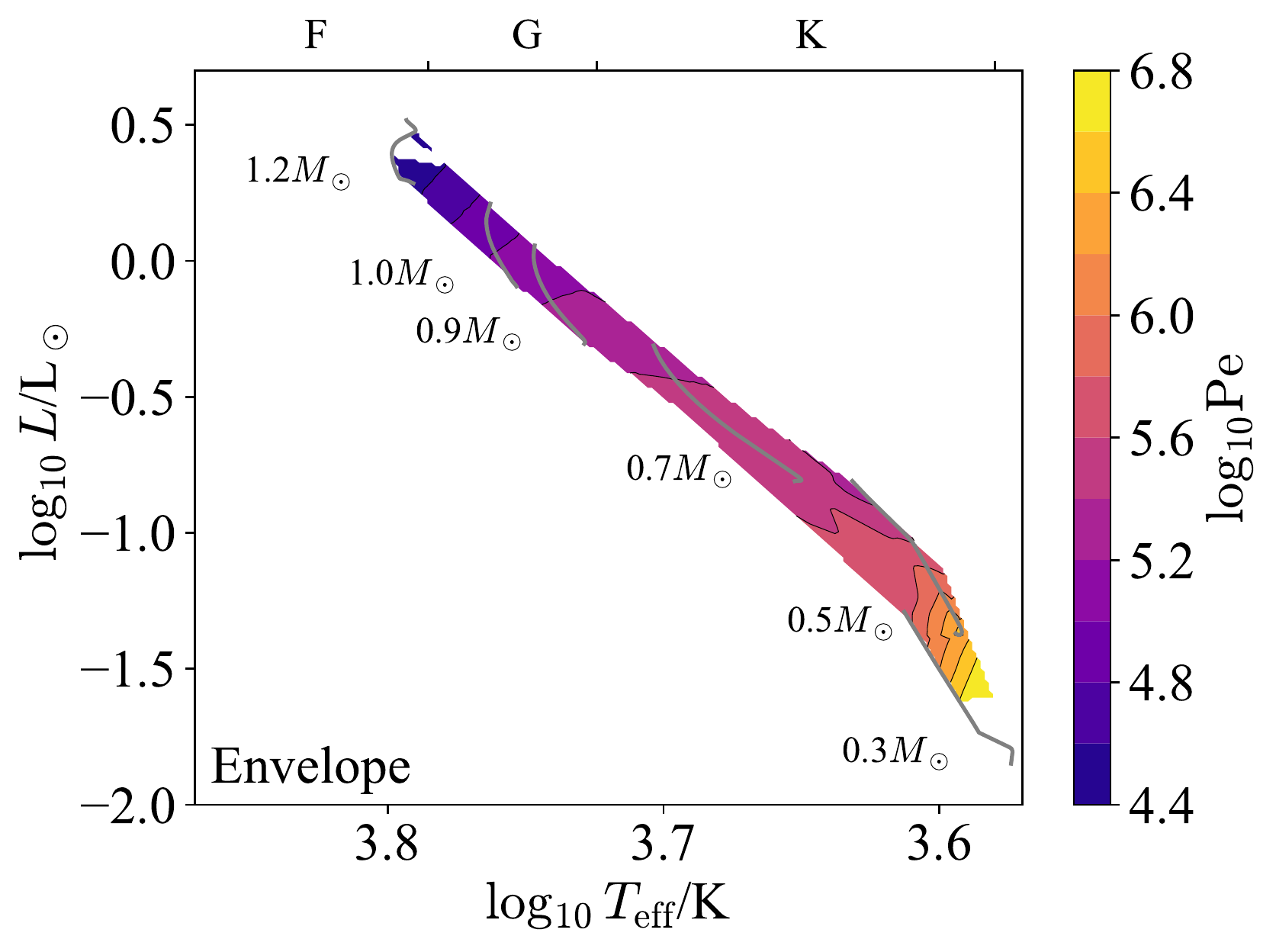}
\end{minipage}
\hfill
\begin{minipage}{0.48\textwidth}
\includegraphics[width=\textwidth]{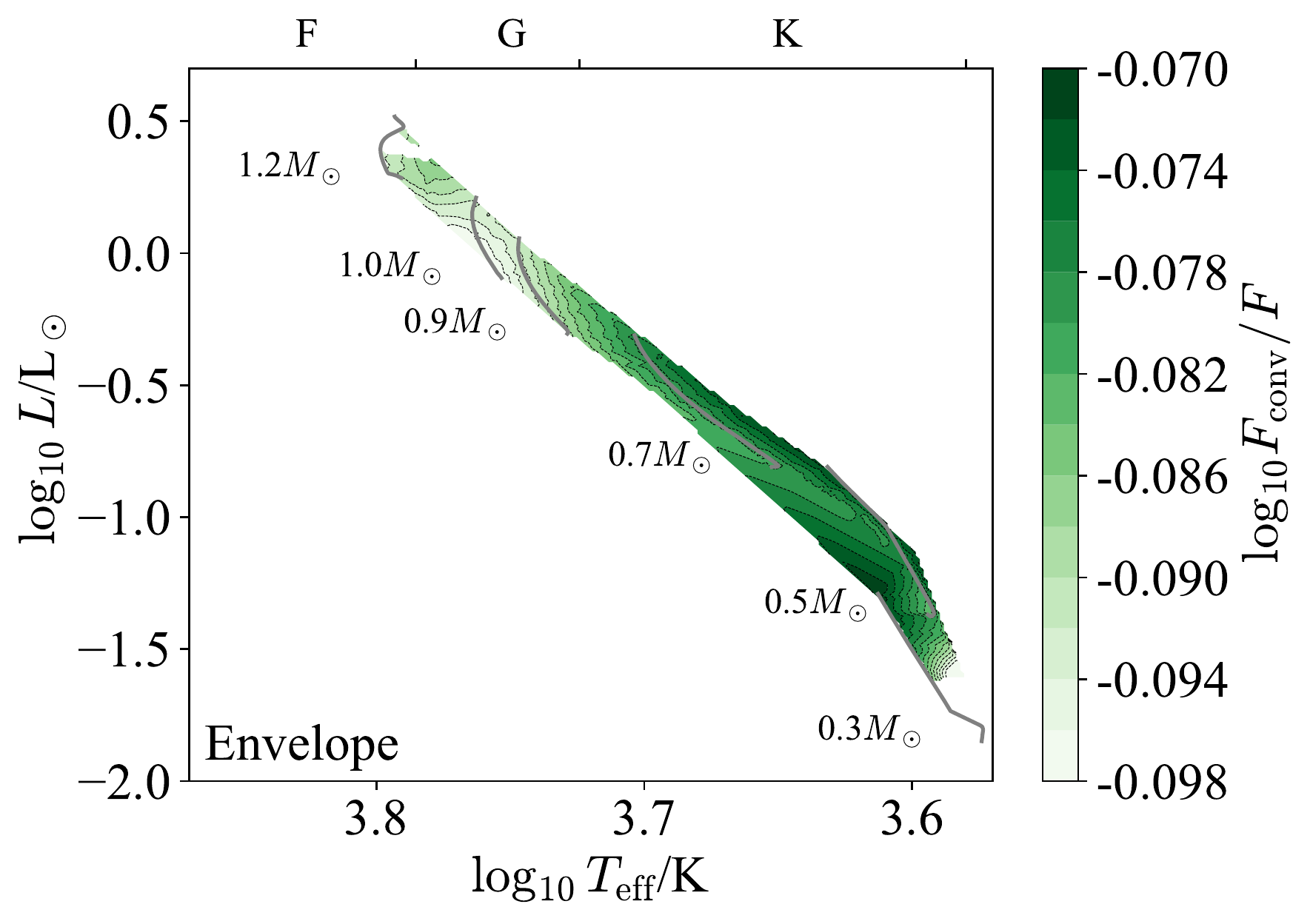}
\end{minipage}

\caption{The P{\'e}clet number $\mathrm{Pe}$ (left) and $F_{\rm conv}/F$ (right) are shown in terms of $\log T_{\rm eff}$/spectral type and $\log L$ for stellar models with deep envelope convection zones and Milky Way metallicity $Z=0.014$. Note that both $\mathrm{Pe}$ and $F_{\rm conv}/F$ are outputs of a theory of convection and so are model-dependent.}
\label{fig:envelope_efficiency}
\end{figure*}

The P{\'e}clet number $\mathrm{Pe}$ (Figure~\ref{fig:envelope_efficiency}, left) measures the relative importance of advection and diffusion in transporting heat, and the flux ratio $F_{\rm conv}/F$ (Figure~\ref{fig:envelope_efficiency}, right) reports the fraction of the energy flux which is advected.
The former is large and the latter is near-unity, so we conclude that convection in these zones is highly efficient, and heat transport is dominated by advection.

\begin{figure*}
\centering
\begin{minipage}{0.48\textwidth}
\includegraphics[width=\textwidth]{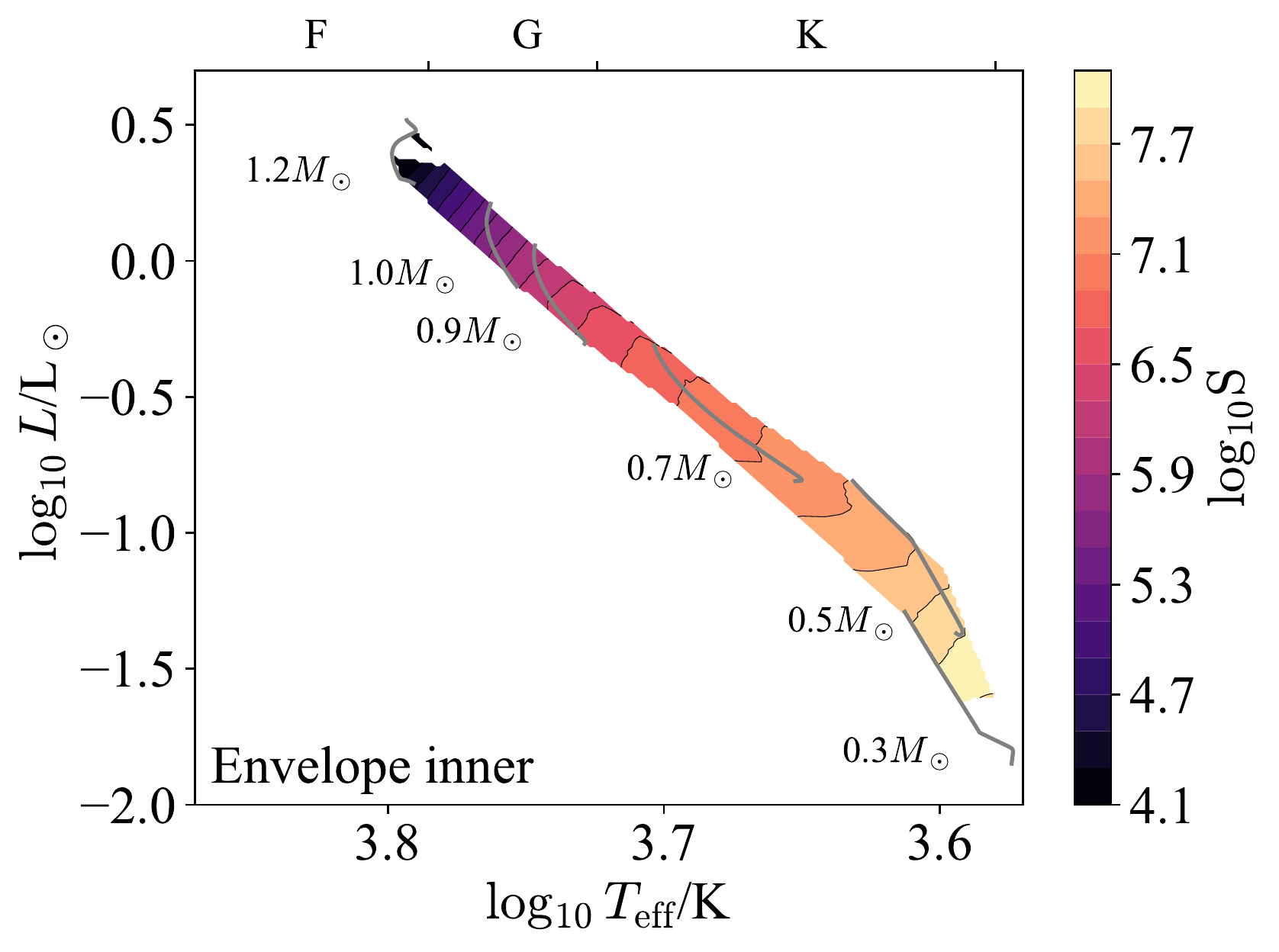}
\end{minipage}

\caption{The stiffness of the inner convective boundary is shown in terms of $\log T_{\rm eff}$ and $\log L$ for stellar models with deep envelope convection zones and Milky Way metallicity $Z=0.014$. Note that the stiffness is an output of a theory of convection and so is model-dependent.}
\label{fig:envelope_stiff}
\end{figure*}

Figure~\ref{fig:envelope_stiff} shows that the base of the envelope convection zone is always extremely stiff, with $S \sim 10^{4-8}$. We expect very little mechanical overshooting as a result, though there could still well be convective penetration~\citep{2021arXiv211011356A}.

\subsubsection{Inner and Outer Scale Heights}\label{sec:inner_outer}

Because Deep Envelope CZs are strongly stratified their properties vary tremendously with depth.
We now examine this variation by considering some of the same quantities averaged just over the innermost or outermost pressure scale heights.
For this section only we evaluate $\mathrm{Ek}$, $\mathrm{Re}$, $\mathrm{Pe}$, and $\mathrm{Ro}$ using the pressure scale height at the relevant boundary rather than the full size of the convection zone $\delta r$.
We do this to focus on the dynamics of motions near the boundary.

\begin{figure*}
\centering
\begin{minipage}{0.433\textwidth}
\includegraphics[width=\textwidth]{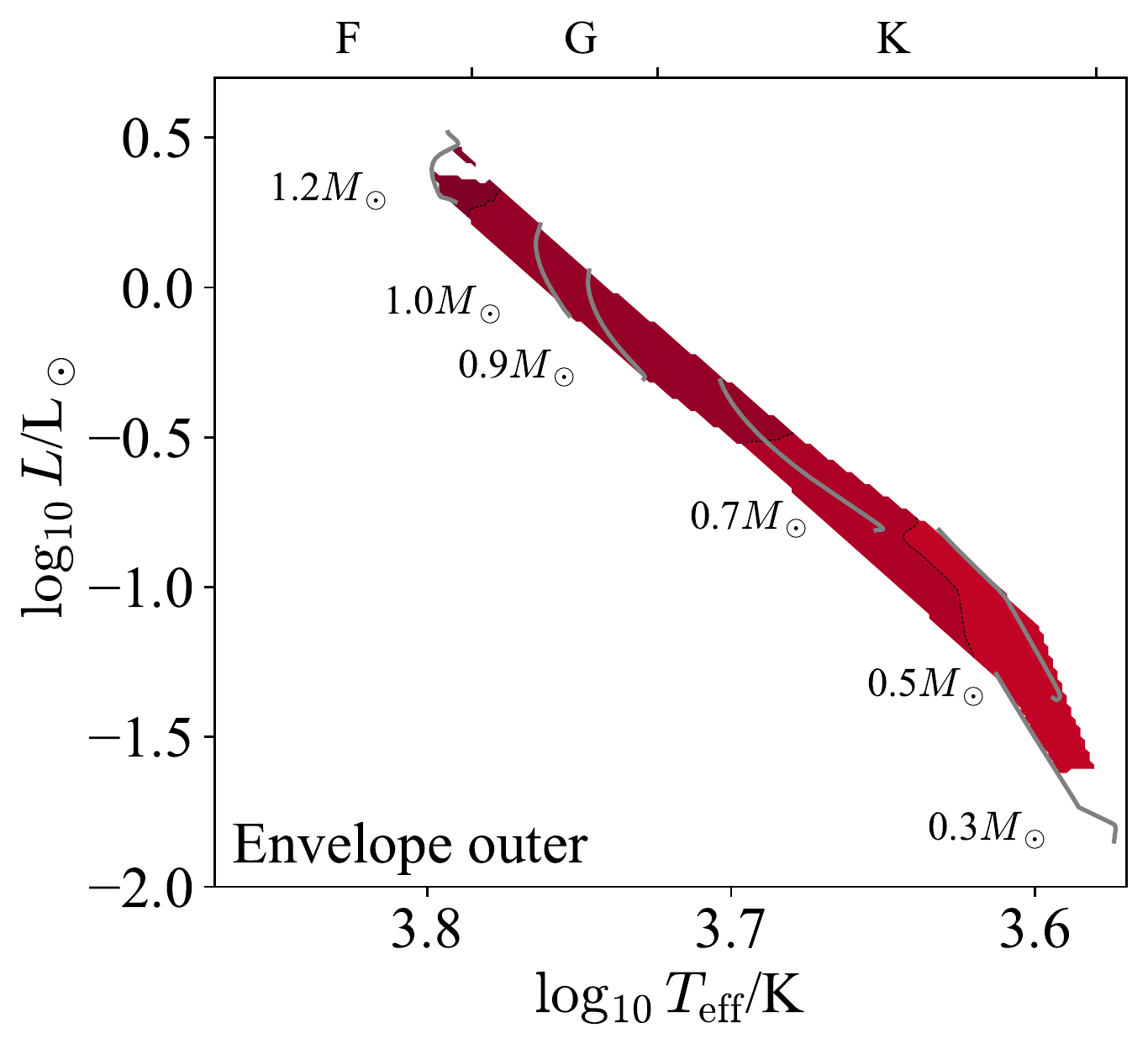}
\end{minipage}
\hfill
\begin{minipage}{0.537\textwidth}
\includegraphics[width=\textwidth]{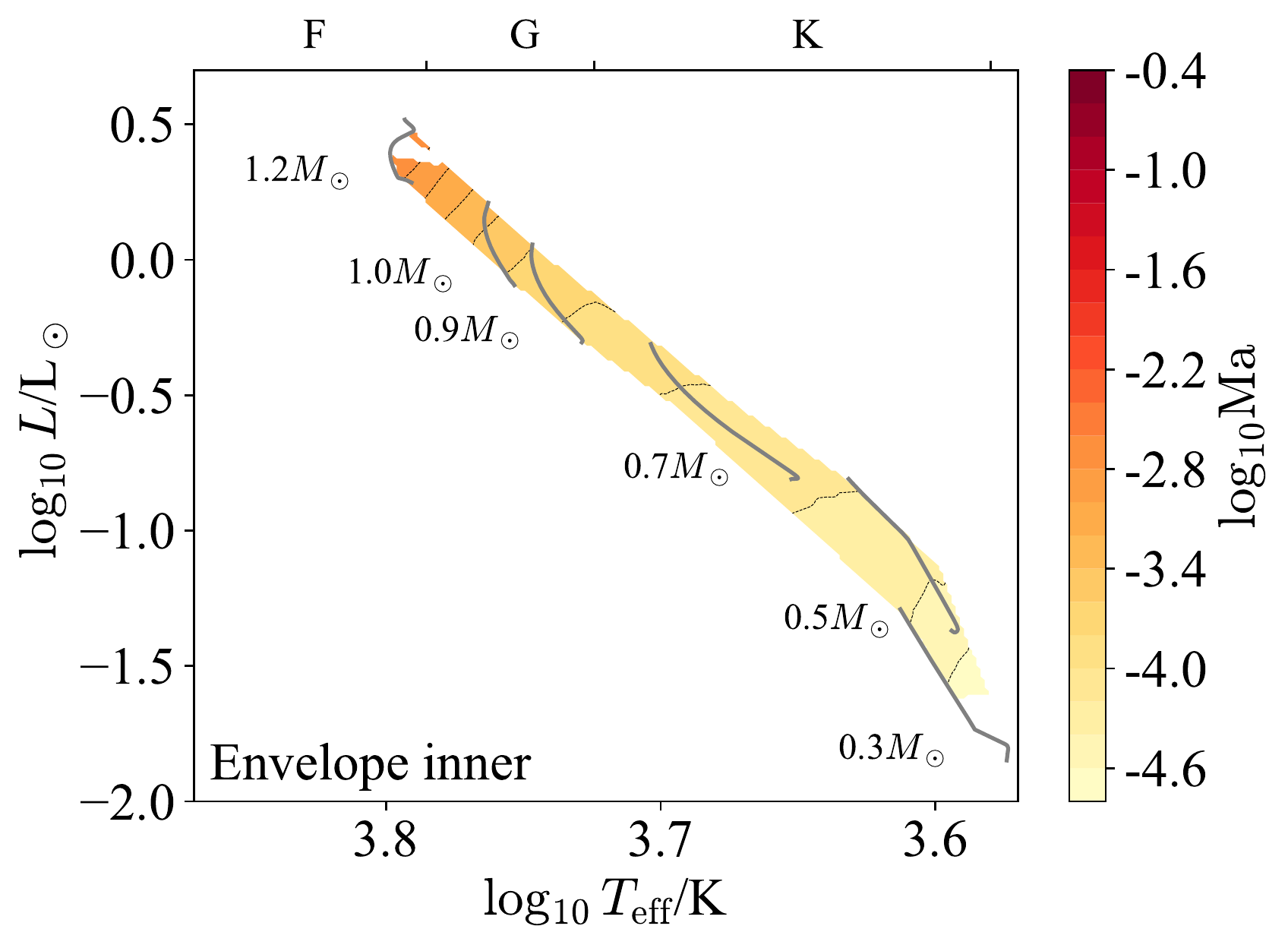}
\end{minipage}
\hfill

\caption{The Mach number $\mathrm{Ma}$ is shown averaged over the outermost pressure scale height (left) and innermost one (right) in terms of $\log T_{\rm eff}$/spectral type and $\log L$ for stellar models with deep envelope convection zones and Milky Way metallicity $Z=0.014$. Note that the Mach number is an output of a theory of convection and so is model-dependent.}
\label{fig:envelope_top_bottom_M}
\end{figure*}

Figure~\ref{fig:envelope_top_bottom_M} shows the Mach number $\mathrm{Ma}$.
The Mach number shows important differences, being much larger in the outermost pressure scale (left) than the innermost one (right), and reaches $0.3$ at the outer boundaries of more massive stars ($M \approx 1.1 M_\odot$).
So, while the innermost regions of these zones are well-modeled by the anelastic approximation, the near-surface regions likely require using the fully compressible equations.

\begin{figure*}
\centering
\begin{minipage}{0.433\textwidth}
\includegraphics[width=\textwidth]{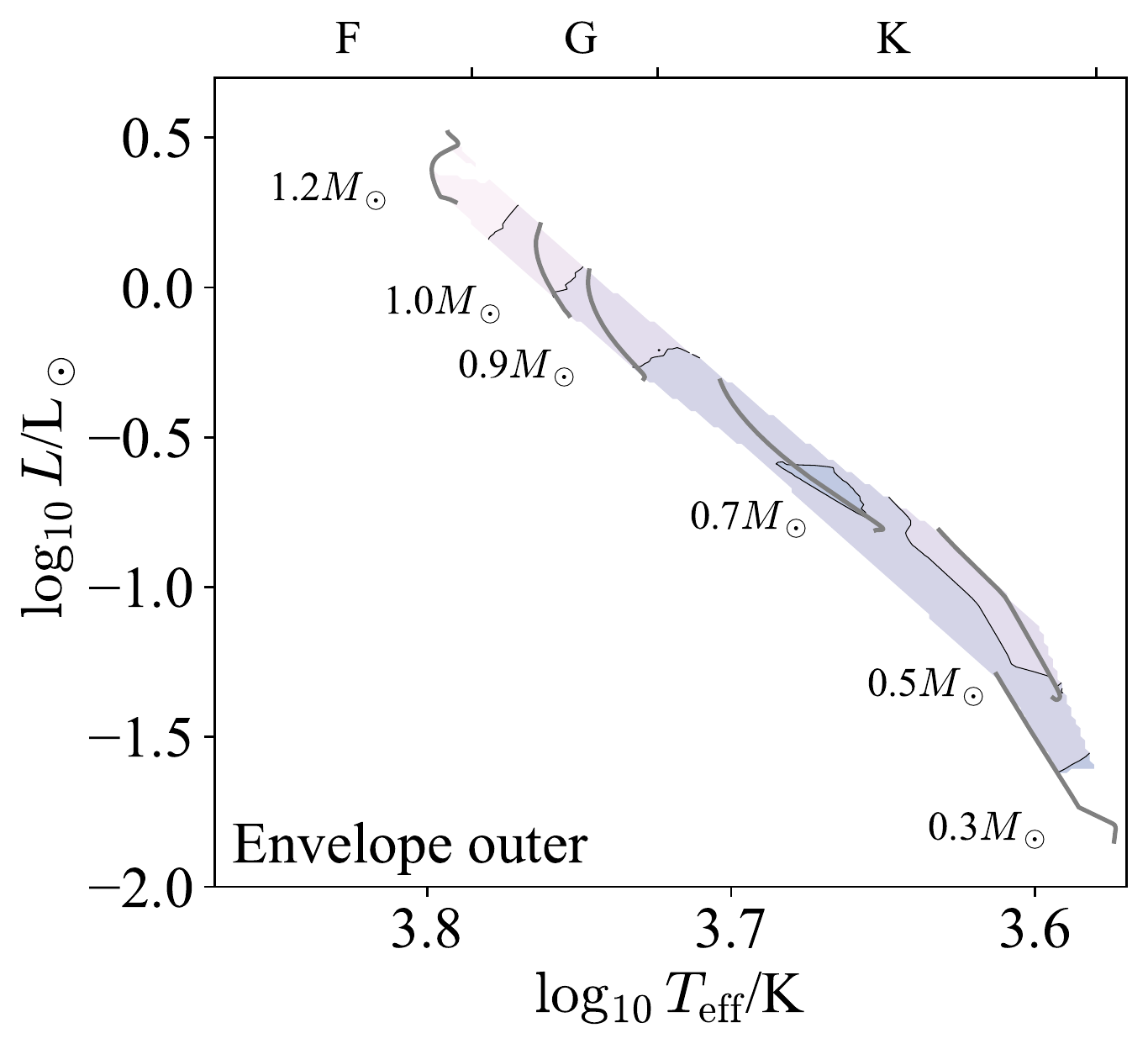}
\end{minipage}
\hfill
\begin{minipage}{0.537\textwidth}
\includegraphics[width=\textwidth]{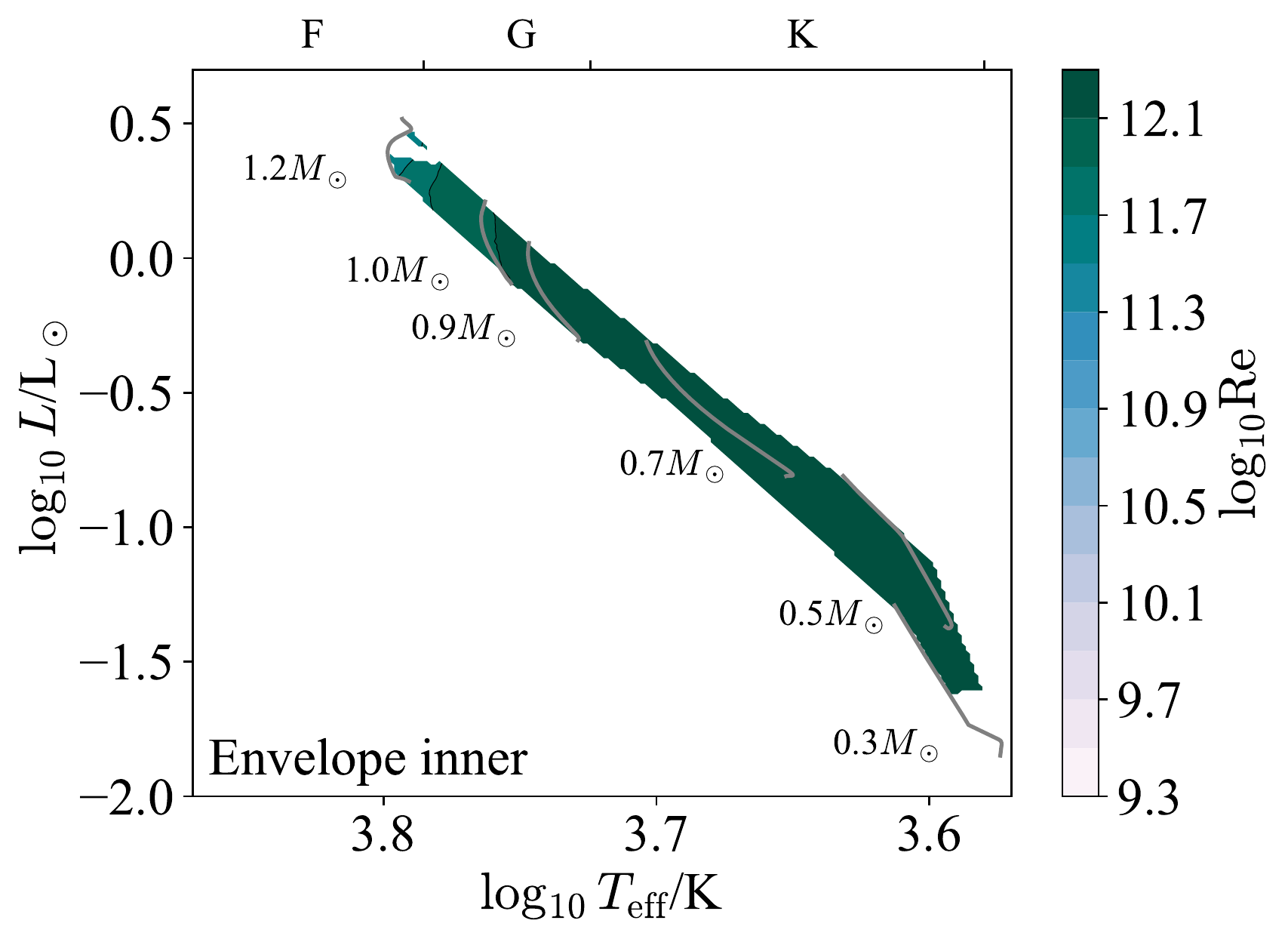}
\end{minipage}
\hfill

\caption{The Reynolds number $\mathrm{Re}$ is shown averaged over the outermost pressure scale height (left) and innermost one (right) in terms of $\log T_{\rm eff}$/spectral type and $\log L$ for stellar models with deep envelope convection zones and Milky Way metallicity $Z=0.014$. Note that the Reynolds number is an output of a theory of convection and so is model-dependent.}
\label{fig:envelope_top_bottom_Re}
\end{figure*}

Figure~\ref{fig:envelope_top_bottom_Re} shows the Reynolds number $\mathrm{Re}$.
The Reynolds number shows a larger range of values near the outer boundary than near the inner one, but centered on similar typical values of $\sim 10^{12}$.
This indicates well-developed turbulence near both boundaries.

\begin{figure*}
\centering
\begin{minipage}{0.433\textwidth}
\includegraphics[width=\textwidth]{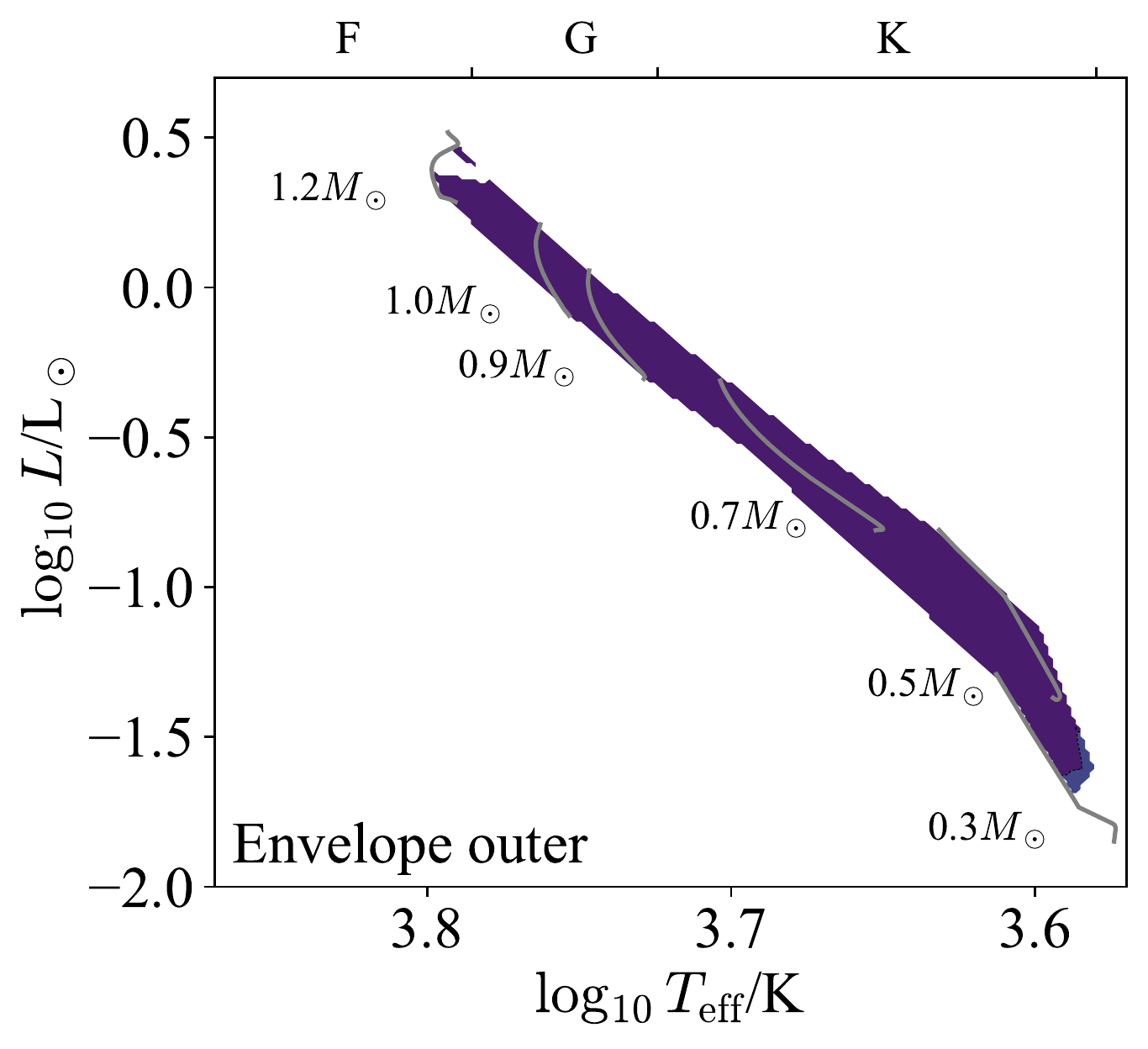}
\end{minipage}
\hfill
\begin{minipage}{0.537\textwidth}
\includegraphics[width=\textwidth]{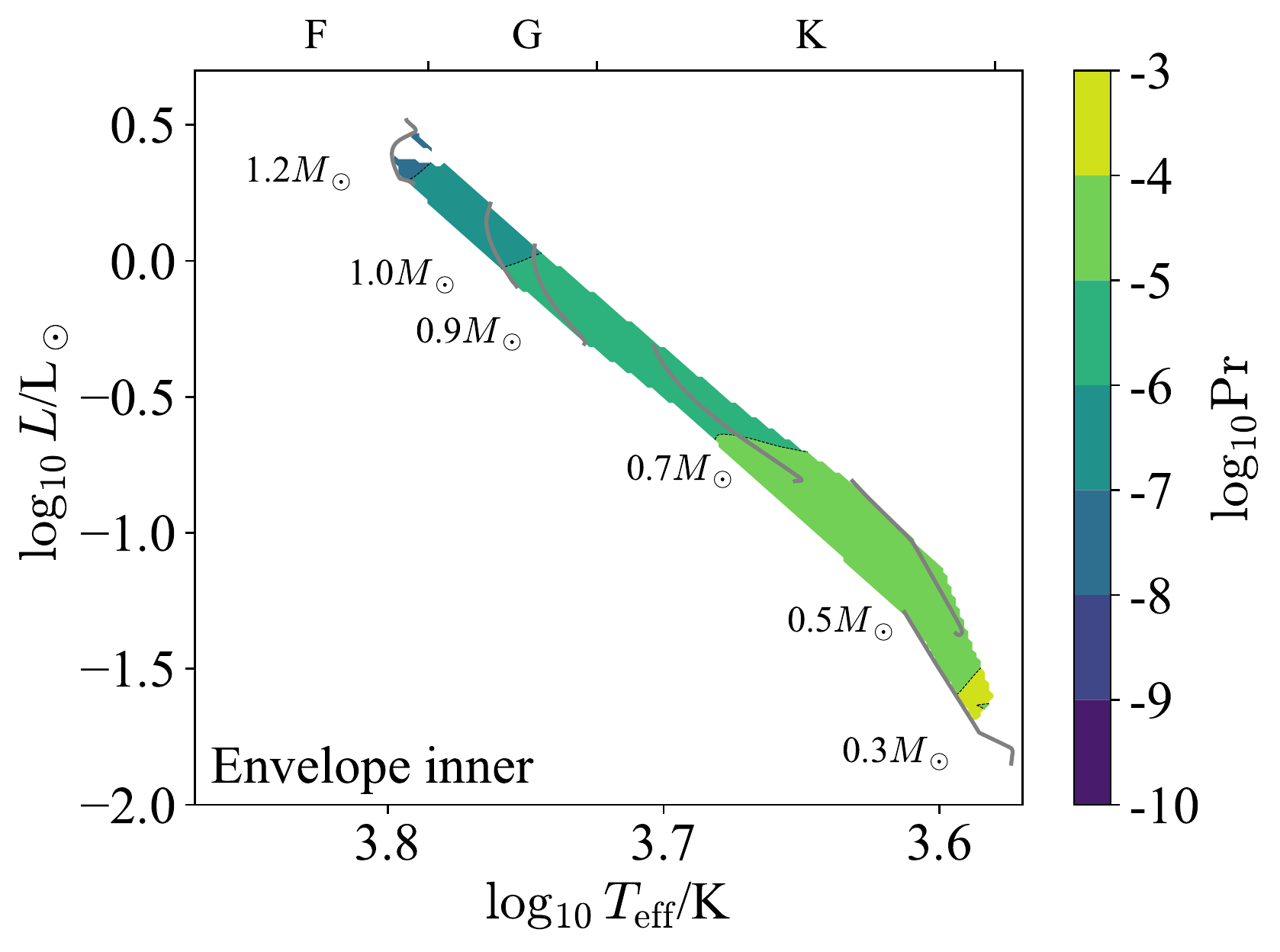}
\end{minipage}
\hfill
\begin{minipage}{0.433\textwidth}
\includegraphics[width=\textwidth]{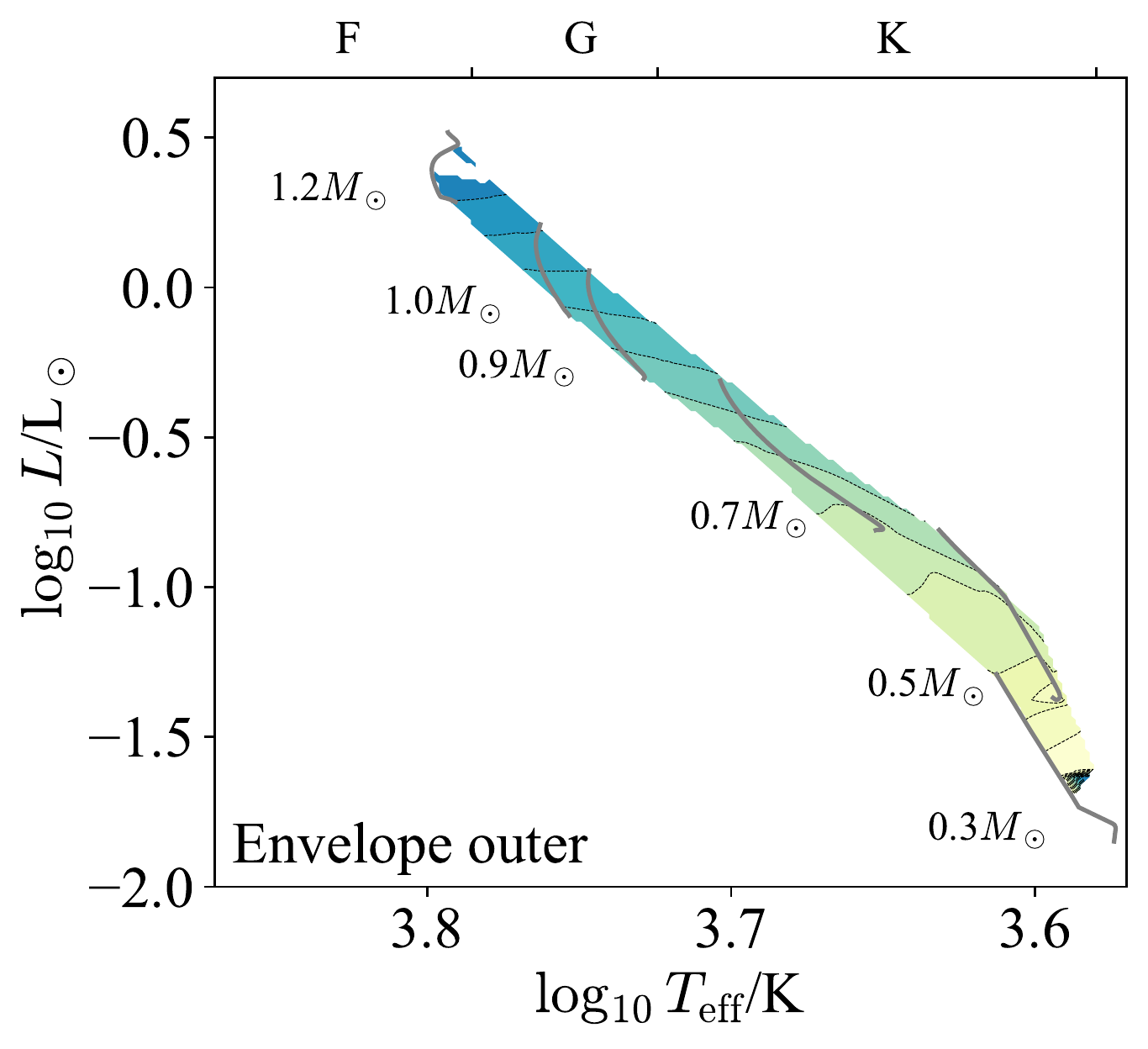}
\end{minipage}
\hfill
\begin{minipage}{0.537\textwidth}
\includegraphics[width=\textwidth]{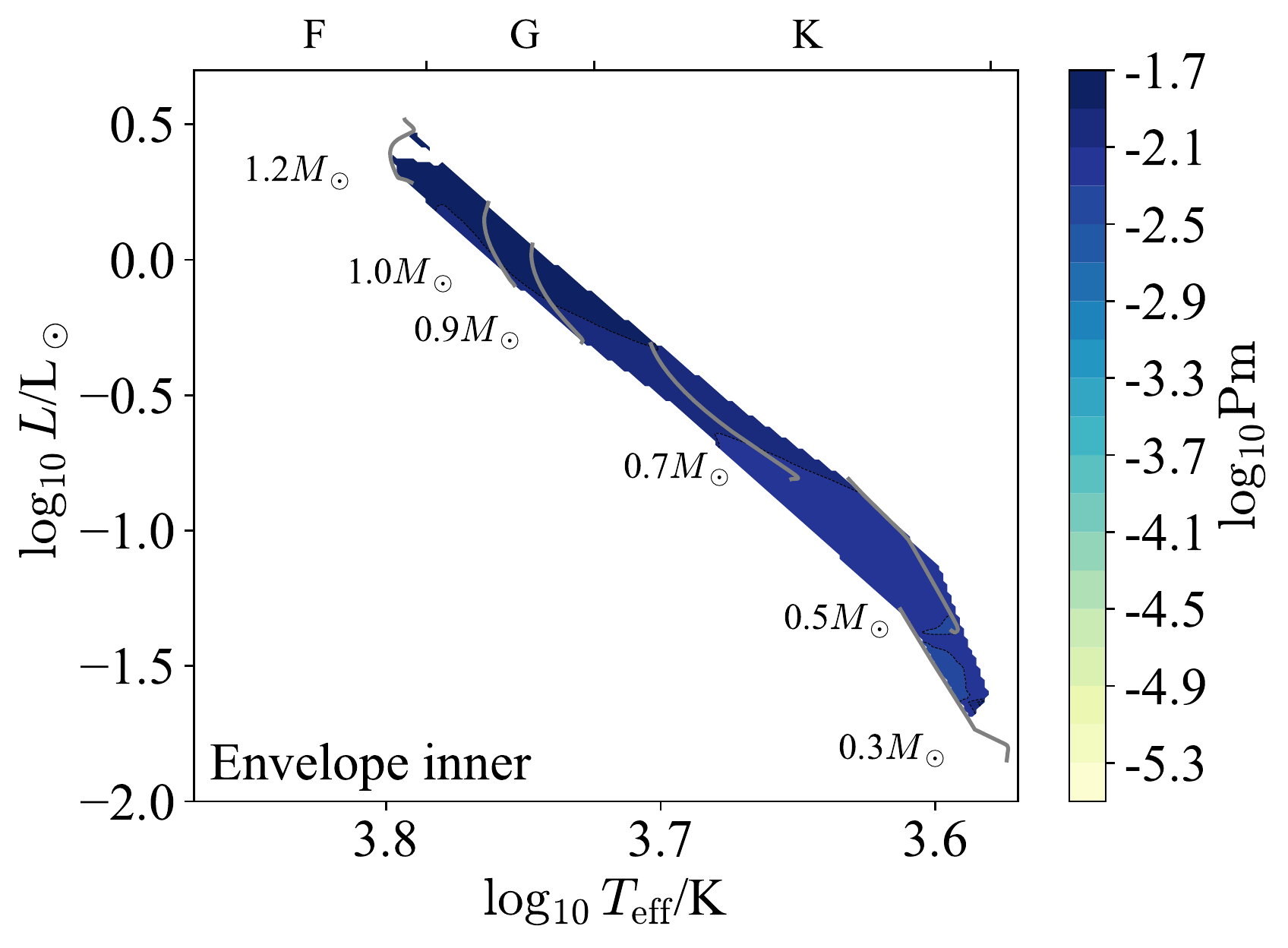}
\end{minipage}
\hfill

\caption{The Prandtl number $\mathrm{Pr}$ (upper) and magnetic Prandtl number $\mathrm{Pm}$ (lower) are shown averaged over the outermost pressure scale height (left) and innermost one (right) in terms of $\log T_{\rm eff}$/spectral type and $\log L$ for stellar models with deep envelope convection zones and Milky Way metallicity $Z=0.014$. Note that both $\mathrm{Pr}$ and $\mathrm{Pm}$ are input parameters, and so do not depend on a specific theory of convection.}
\label{fig:envelope_micro_top_bottom_inputs_1}
\end{figure*}

Figure~\ref{fig:envelope_micro_top_bottom_inputs_1} shows the Prandtl number $\mathrm{Pr}$ (upper) and magnetic Prandtl number $\mathrm{Pm}$ (lower).
Both are much smaller near the outer boundary of the convection zone (left) than the inner boundary (right), though the difference is much more stark ($\sim 10^{2}-10^{5}$) for the Prandtl number than for the magnetic Prandtl number ($\sim 10^{1}-10^{3}$).
Both remain small near both boundaries, however, so the qualitative features of convection that they reflect are unchanged through the zone.

\begin{figure*}
\centering
\begin{minipage}{0.433\textwidth}
\includegraphics[width=\textwidth]{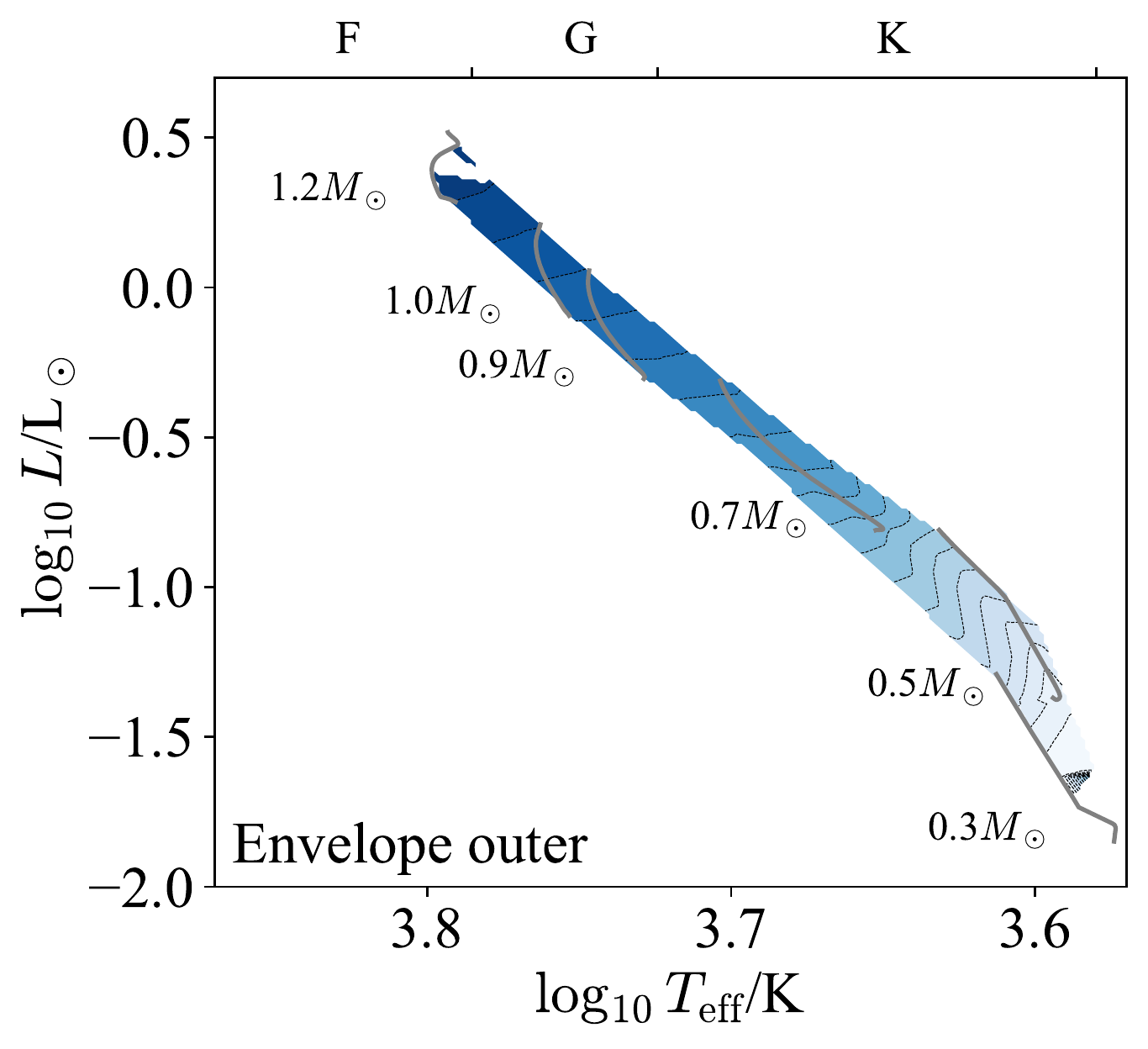}
\end{minipage}
\hfill
\begin{minipage}{0.537\textwidth}
\includegraphics[width=\textwidth]{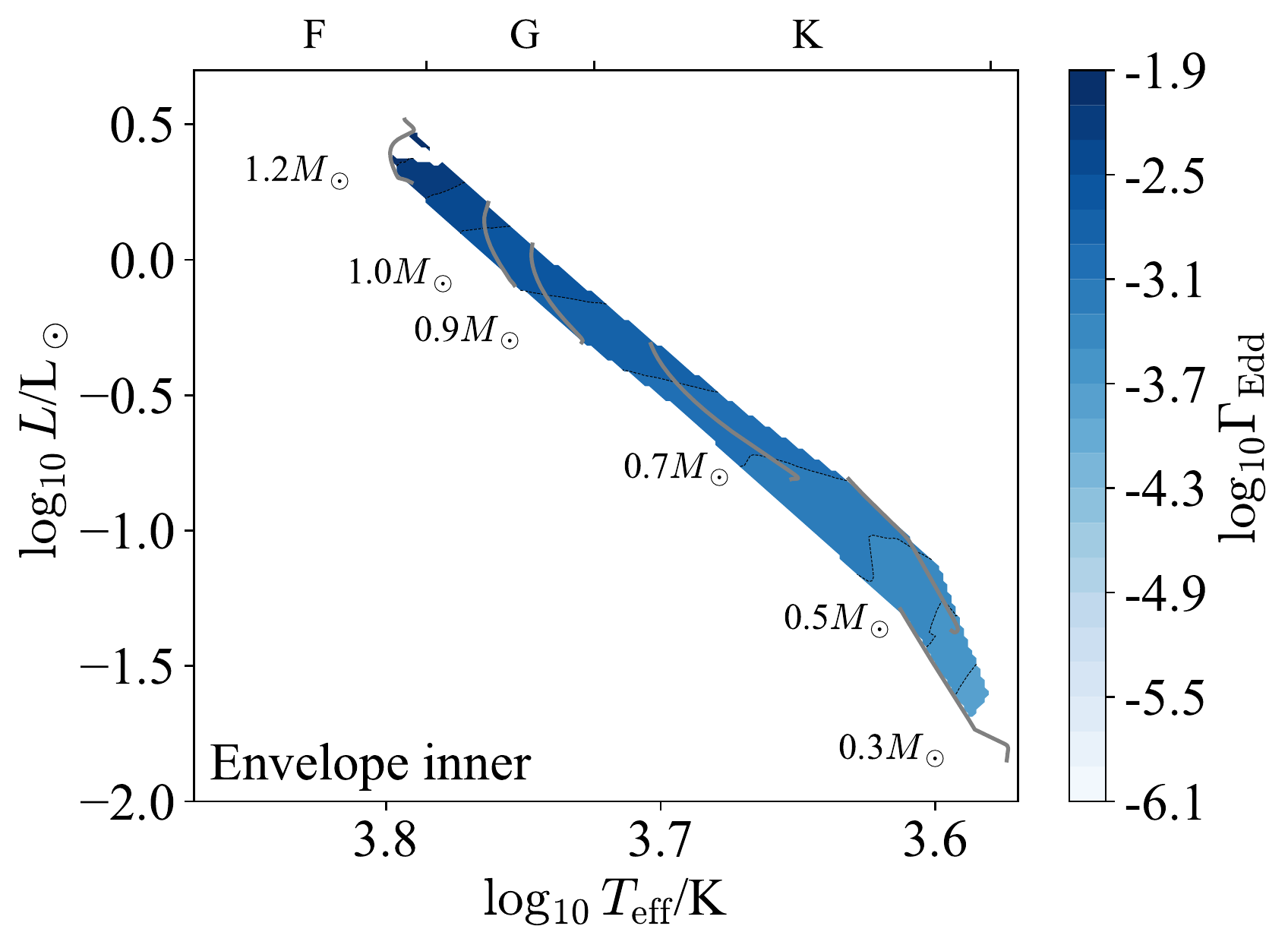}
\end{minipage}
\hfill
\begin{minipage}{0.433\textwidth}
\includegraphics[width=\textwidth]{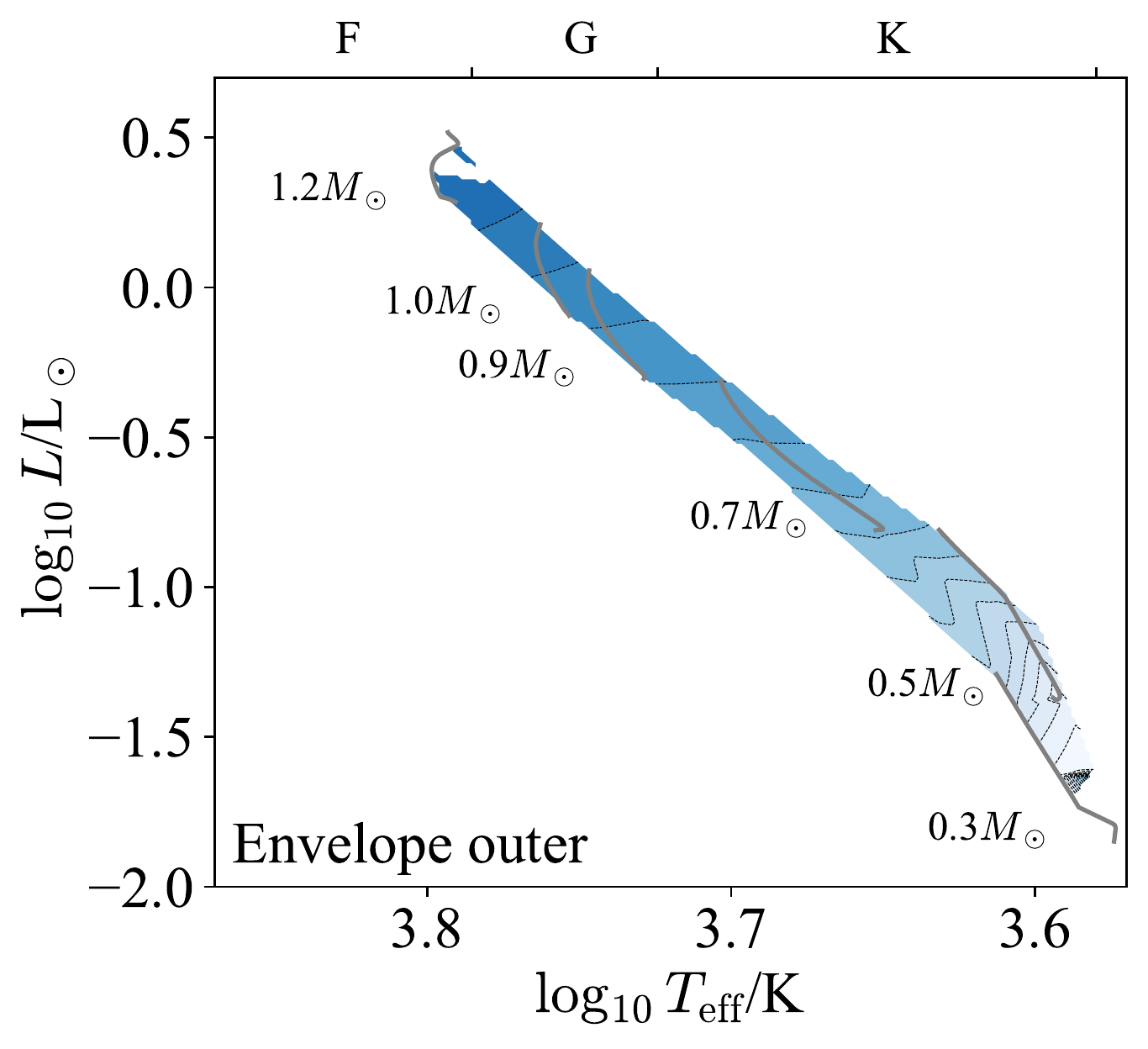}
\end{minipage}
\hfill
\begin{minipage}{0.537\textwidth}
\includegraphics[width=\textwidth]{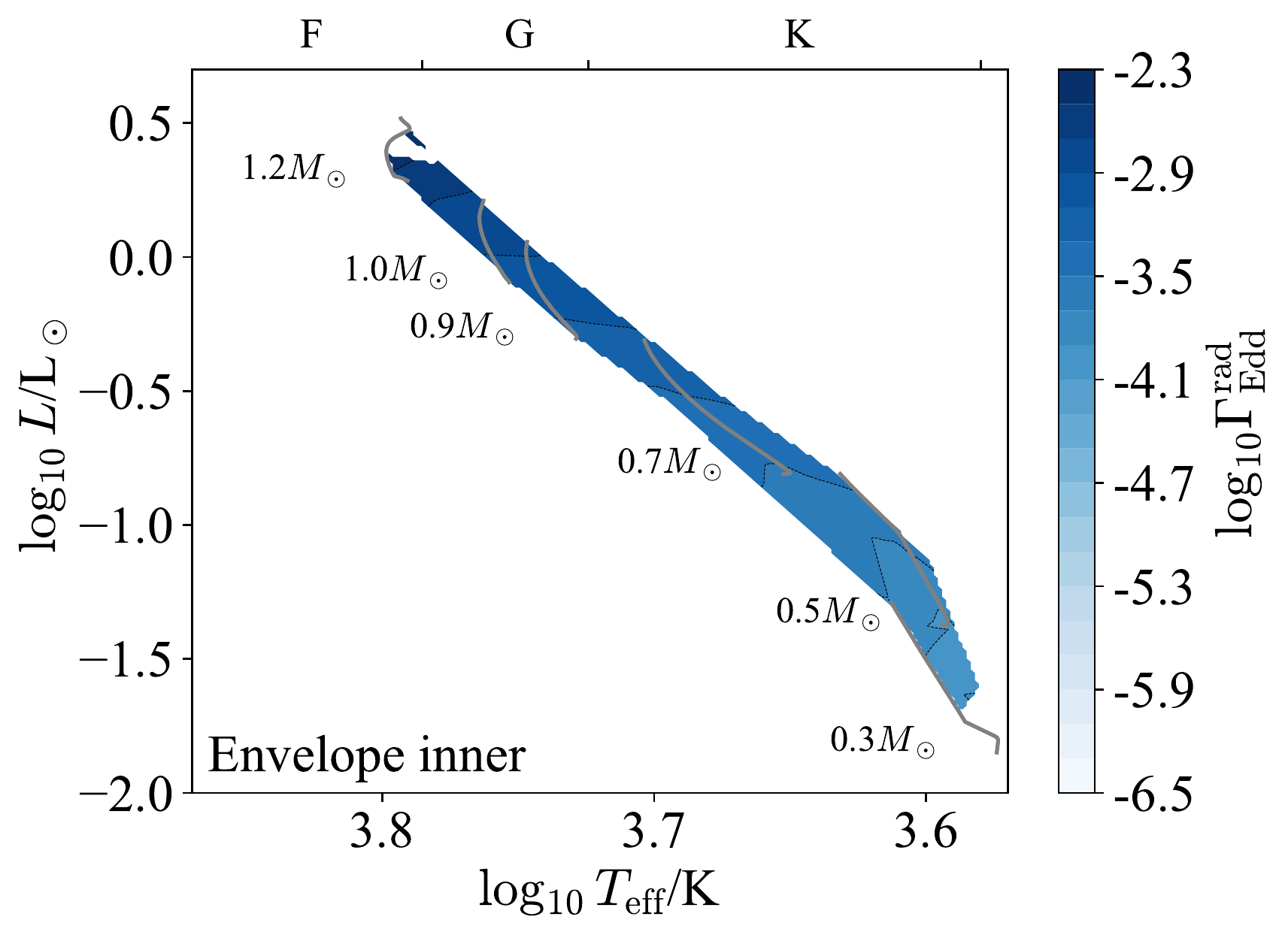}
\end{minipage}
\hfill

\caption{The Eddington ratio with the full luminosity $\Gamma_{\rm Edd}$ (upper) and the radiative luminosity (lower) are shown averaged over the outermost pressure scale height (left) and innermost one (right) in terms of $\log T_{\rm eff}$/spectral type and $\log L$ for stellar models with deep envelope convection zones and Milky Way metallicity $Z=0.014$. Note that while $\Gamma_{\rm Edd}$ is an input parameter and does not depend on a specific theory of convection, $\Gamma_{\rm Edd}^{\rm rad}$ is an output of such a theory and so is model-dependent.}
\label{fig:envelope_outputs_top_bottom_2}
\end{figure*}

Figure~\ref{fig:envelope_outputs_top_bottom_2} shows the Eddington ratios $\Gamma_{\rm Edd}$ (upper) and $\Gamma_{\rm Edd}^{\rm rad}$ (lower).
These are both small at both boundaries, so there is no significant difference in the regime of convection between the two boundaries.

\begin{figure*}
\centering
\begin{minipage}{0.433\textwidth}
\includegraphics[width=\textwidth]{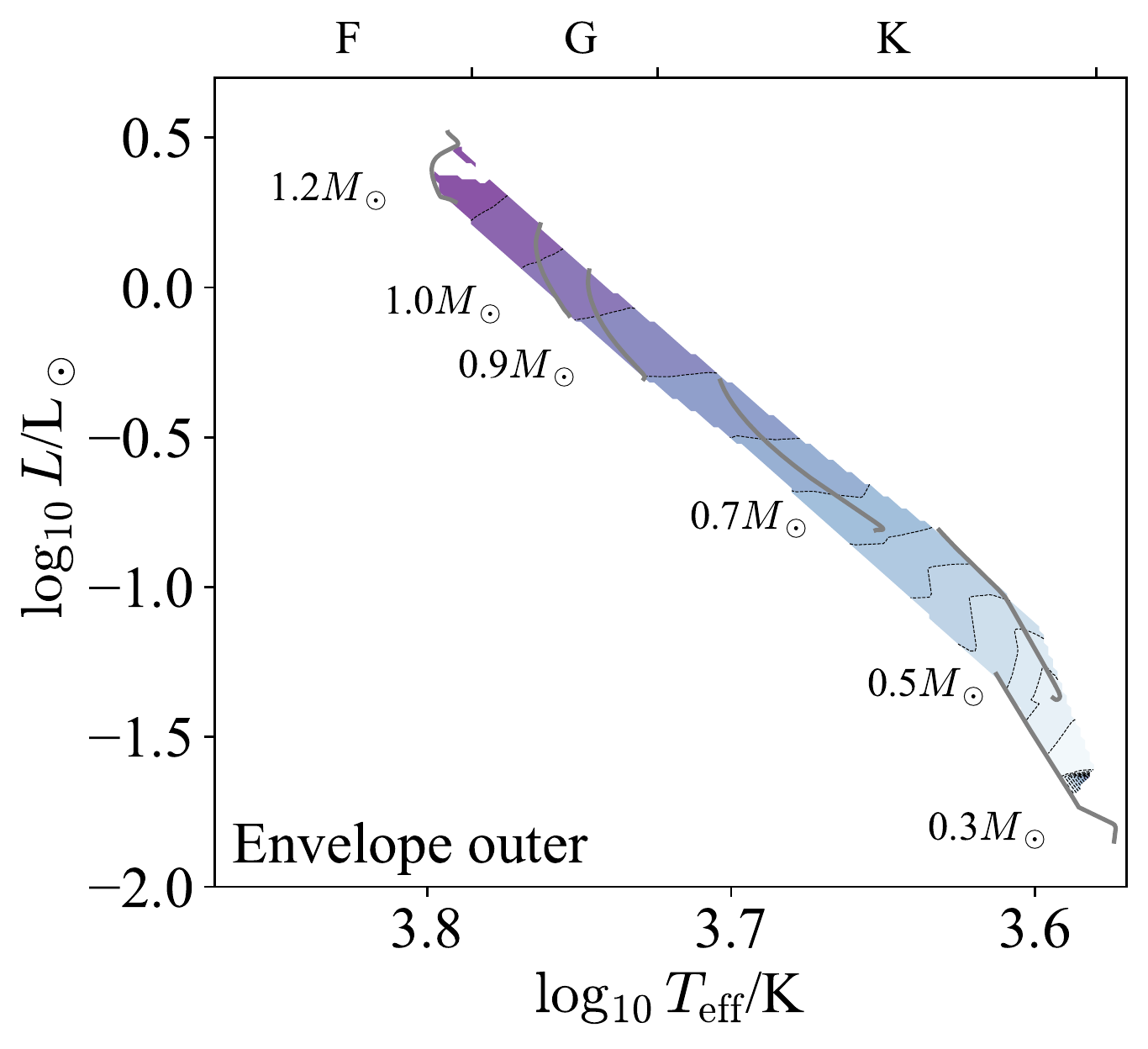}
\end{minipage}
\hfill
\begin{minipage}{0.537\textwidth}
\includegraphics[width=\textwidth]{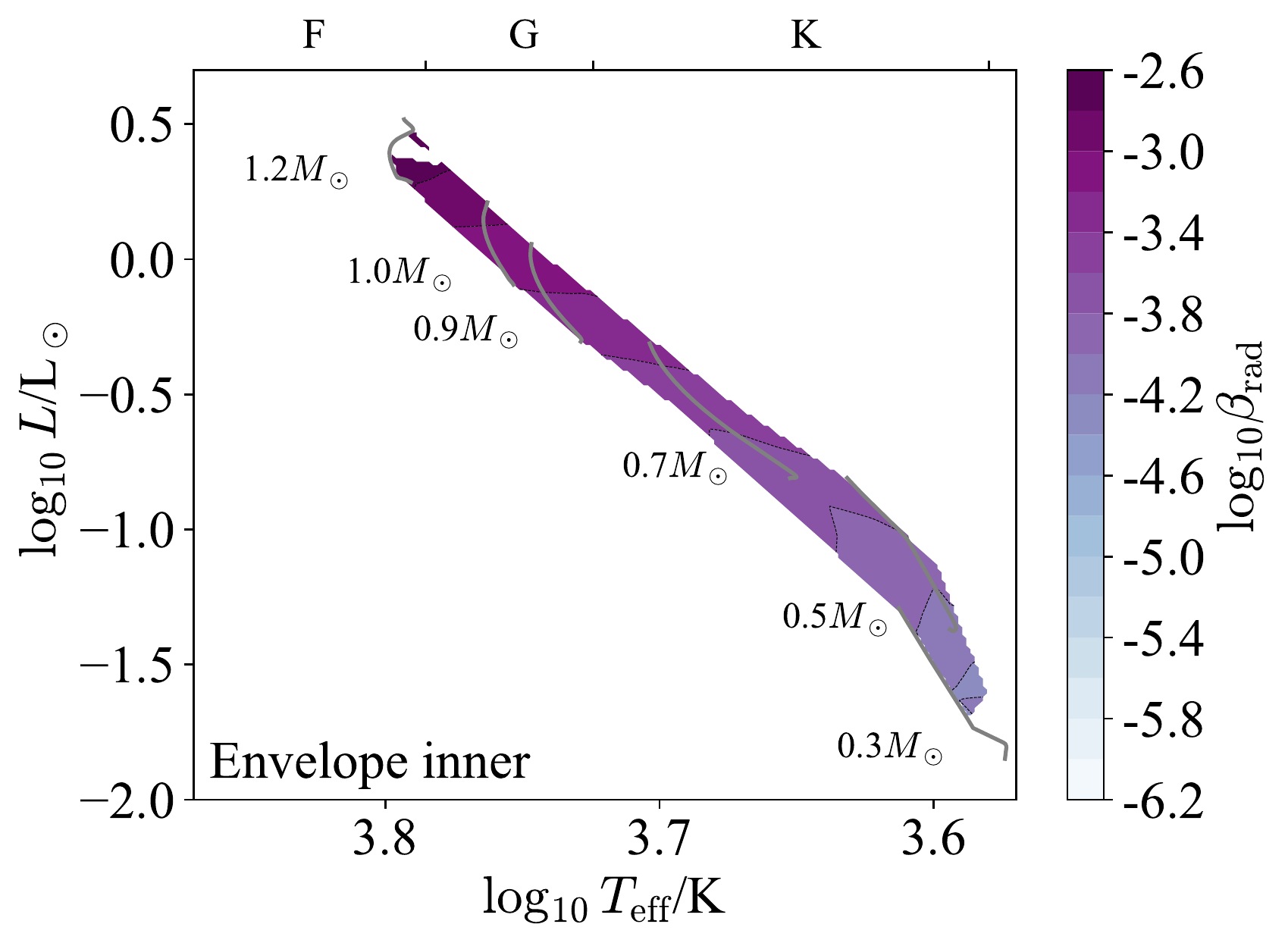}
\end{minipage}
\hfill

\caption{The radiation pressure ratio $\beta_{\rm rad}$ is shown averaged over the outermost pressure scale height (left) and innermost one (right) in terms of $\log T_{\rm eff}$/spectral type and $\log L$ for stellar models with deep envelope convection zones and Milky Way metallicity $Z=0.014$.}
\label{fig:envelope_micro_top_bottom_inputs_2}
\end{figure*}

Figure~\ref{fig:envelope_micro_top_bottom_inputs_2} shows the radiation pressure ratio $\beta_{\rm rad}$.
The radiation pressure ratio is smaller by a factor of $10-100$ near the outer boundary of the convection zone (left) than the inner boundary (right), though it remains less than one percent in both cases and so radiation pressure may be safely ignored in these convection zones.

\begin{figure*}
\centering
\begin{minipage}{0.433\textwidth}
\includegraphics[width=\textwidth]{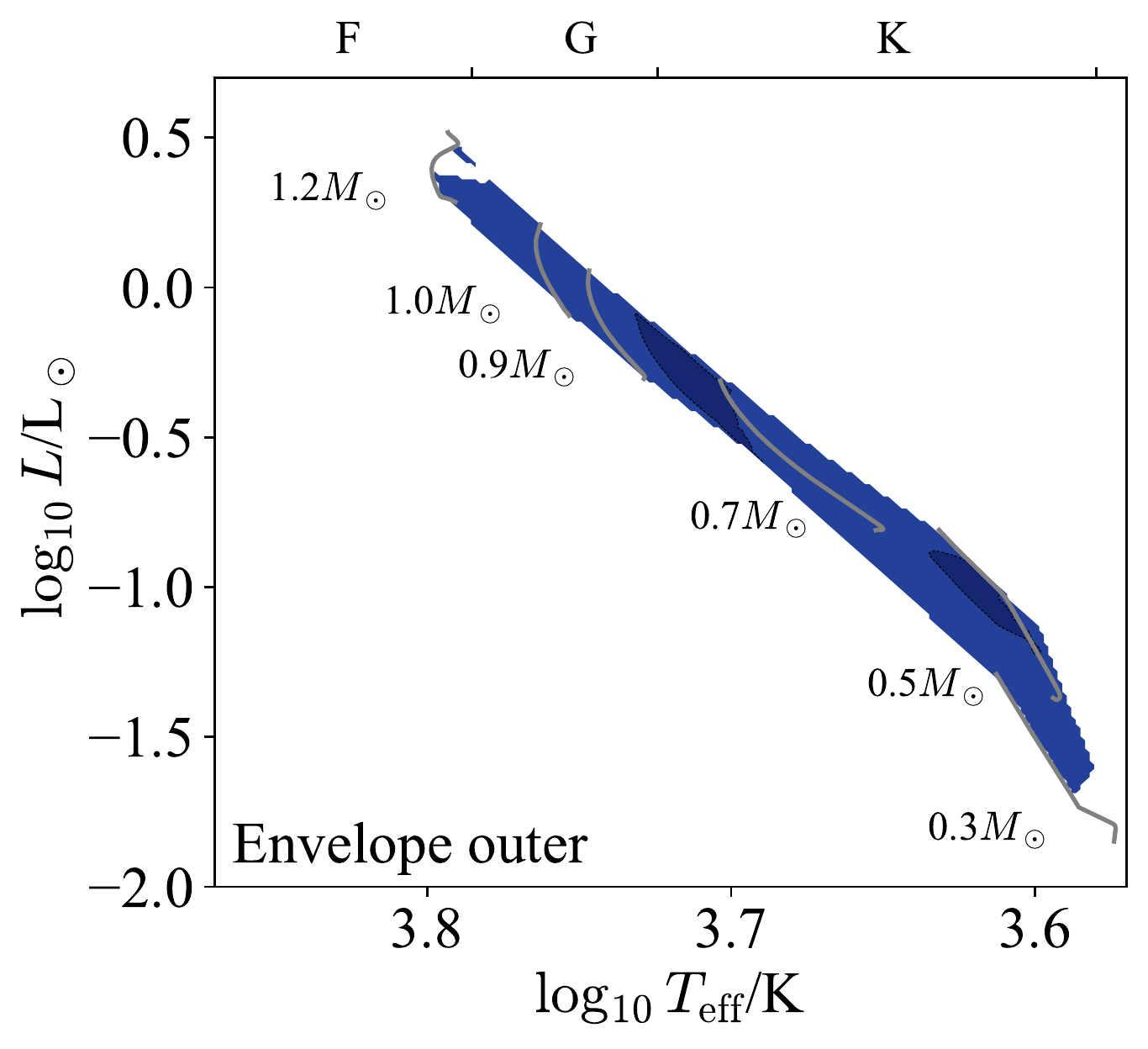}
\end{minipage}
\hfill
\begin{minipage}{0.537\textwidth}
\includegraphics[width=\textwidth]{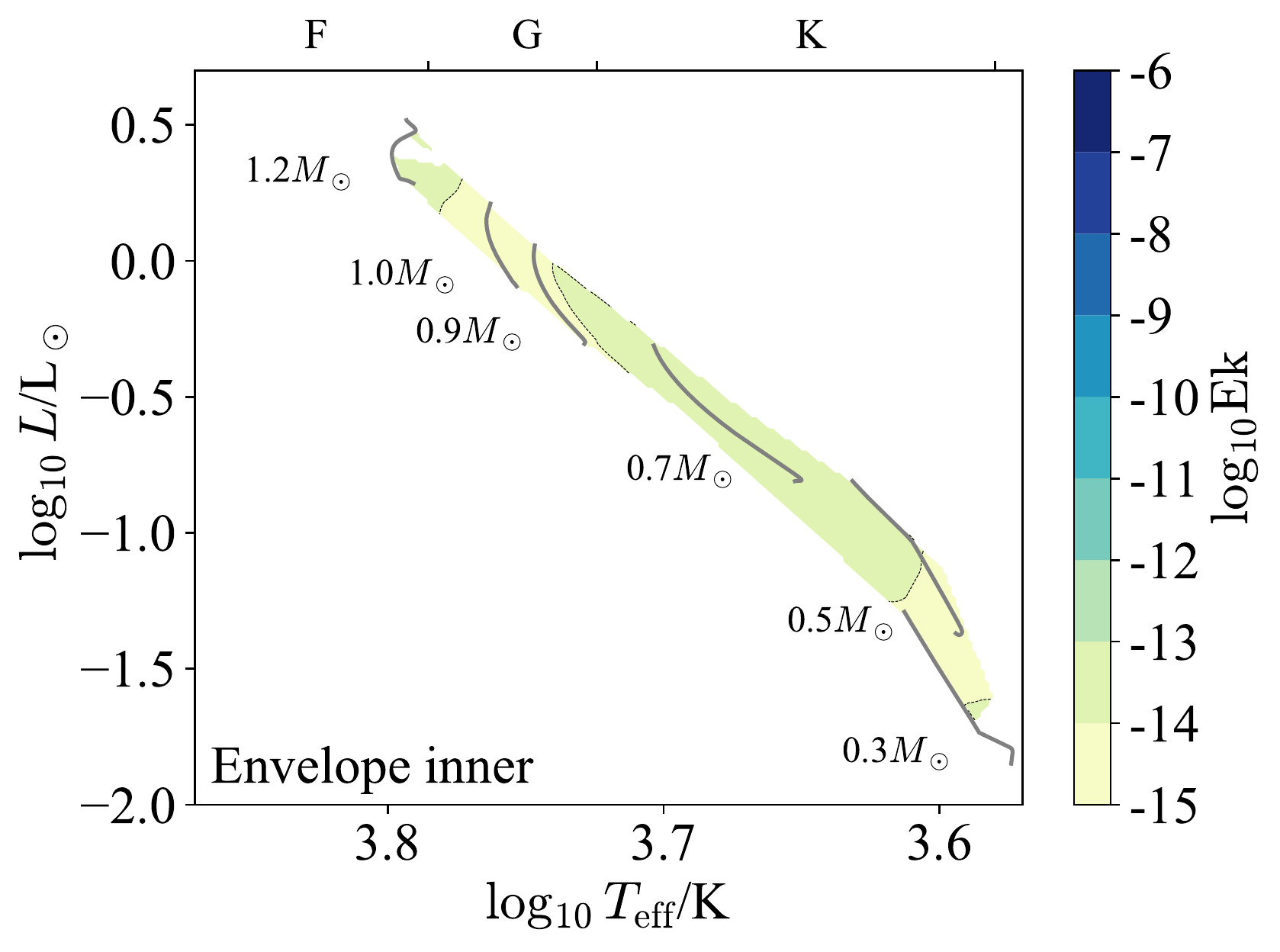}
\end{minipage}
\hfill
\begin{minipage}{0.433\textwidth}
\includegraphics[width=\textwidth]{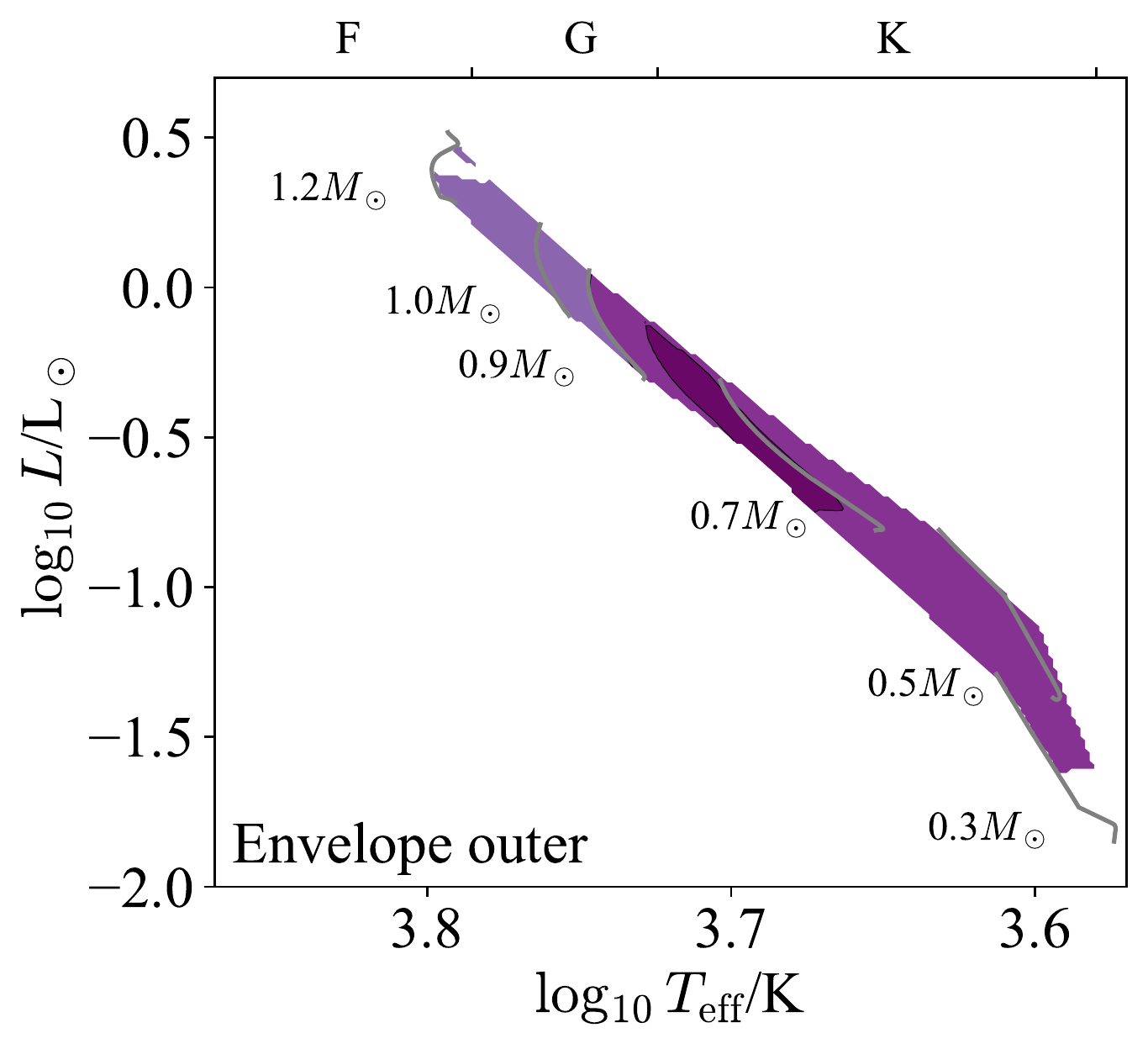}
\end{minipage}
\hfill
\begin{minipage}{0.537\textwidth}
\includegraphics[width=\textwidth]{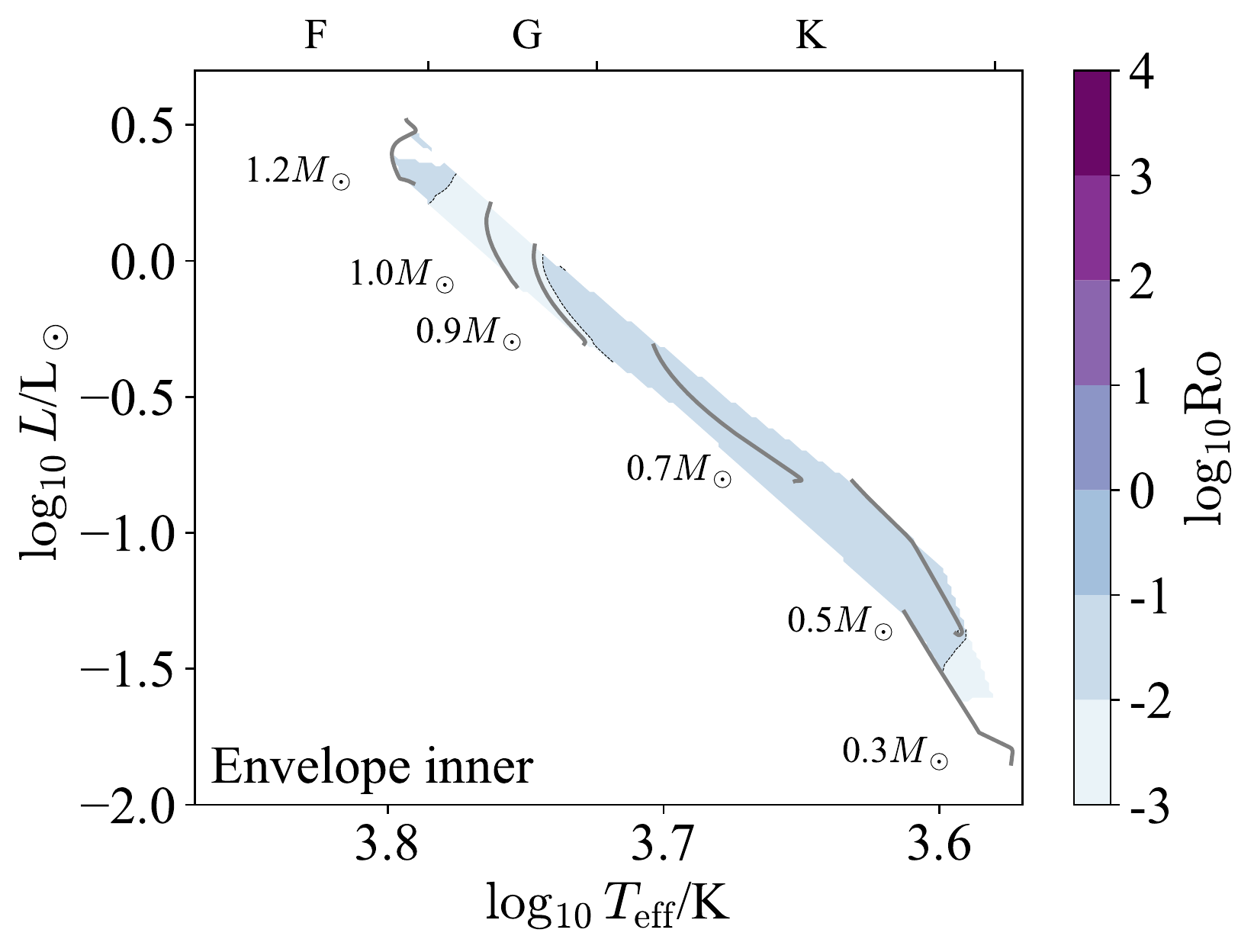}
\end{minipage}
\hfill

\caption{The Rossby number $\mathrm{Ro}$ (upper) is shown averaged over the outermost pressure scale height (left) and innermost one (right) in terms of $\log T_{\rm eff}$/spectral type and $\log L$ for stellar models with deep envelope convection zones and Milky Way metallicity $Z=0.014$. Note that both numbers were calculated here using the pressure scale height at the relevant boundary rather than the full width of the convection zone $\delta r$.}
\label{fig:envelope_outputs_top_bottom_3}
\end{figure*}

Figure~\ref{fig:envelope_outputs_top_bottom_3} shows the Rossby number $\mathrm{Ro}$ (upper) and Ekman number $\mathrm{Ek}$ (lower).
The Rossby number is somewhat larger near the outer boundary than near the inner one.
Importantly, the Rossby number near the outer boundary is larger than unity, meaning that the flows are not rotationally constrained.
This is in contrast to both the average and inner boundary values, which have $\mathrm{Ro} \ll 1$ and indicate strong rotational constraints.
This suggests that rotation is quite important for the bulk of these zones, but can be safely neglected in their outer regions.

The Ekman number, by contrast, is similar between the two boundaries.
Though the Ekman number is much smaller at the inner boundary than the outer one, it is tiny in both regions.
Hence throughout the zone we expect rotation to dominate over viscosity, except at very small length-scales.

\begin{figure*}
\centering
\begin{minipage}{0.433\textwidth}
\includegraphics[width=\textwidth]{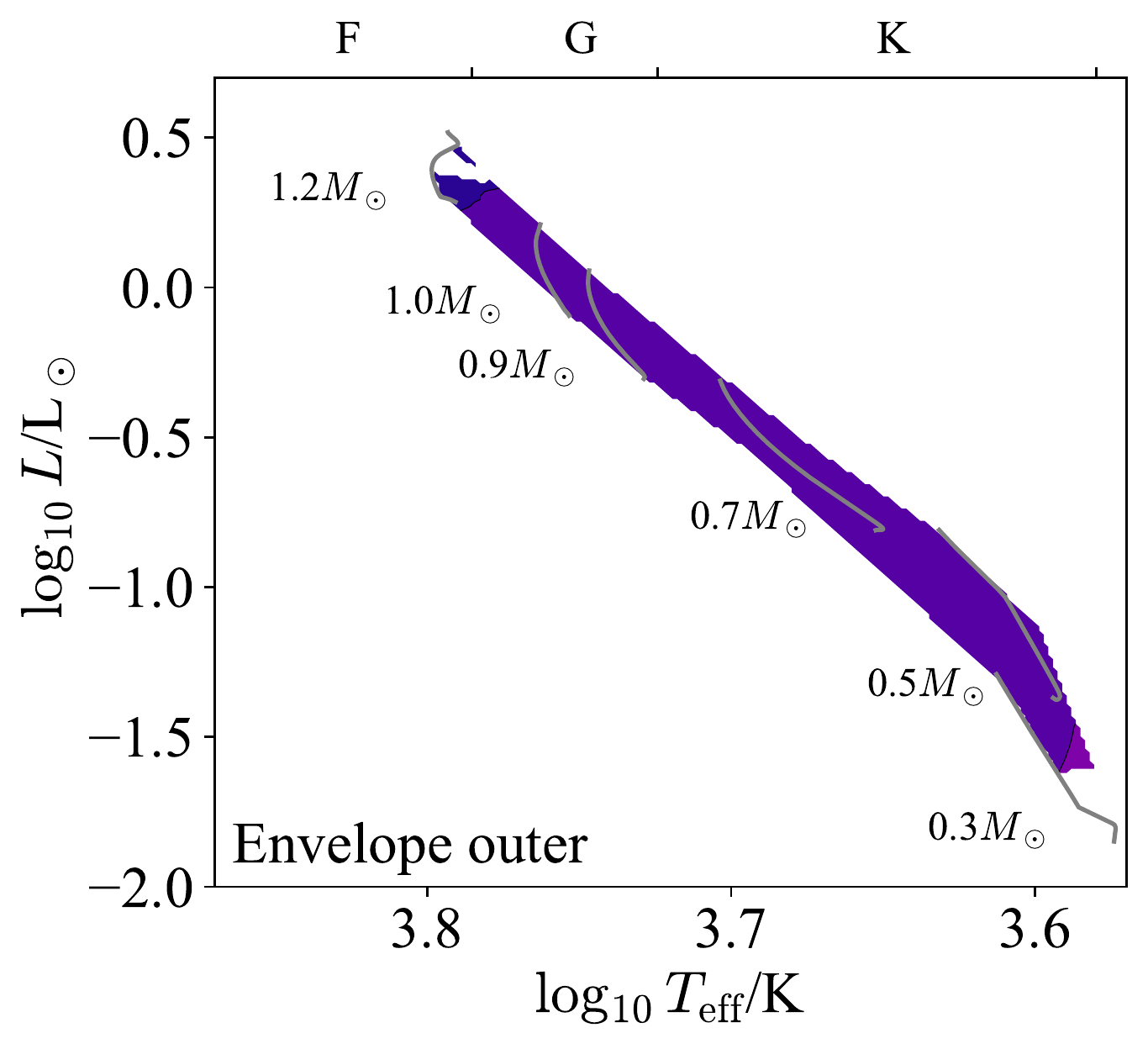}
\end{minipage}
\hfill
\begin{minipage}{0.537\textwidth}
\includegraphics[width=\textwidth]{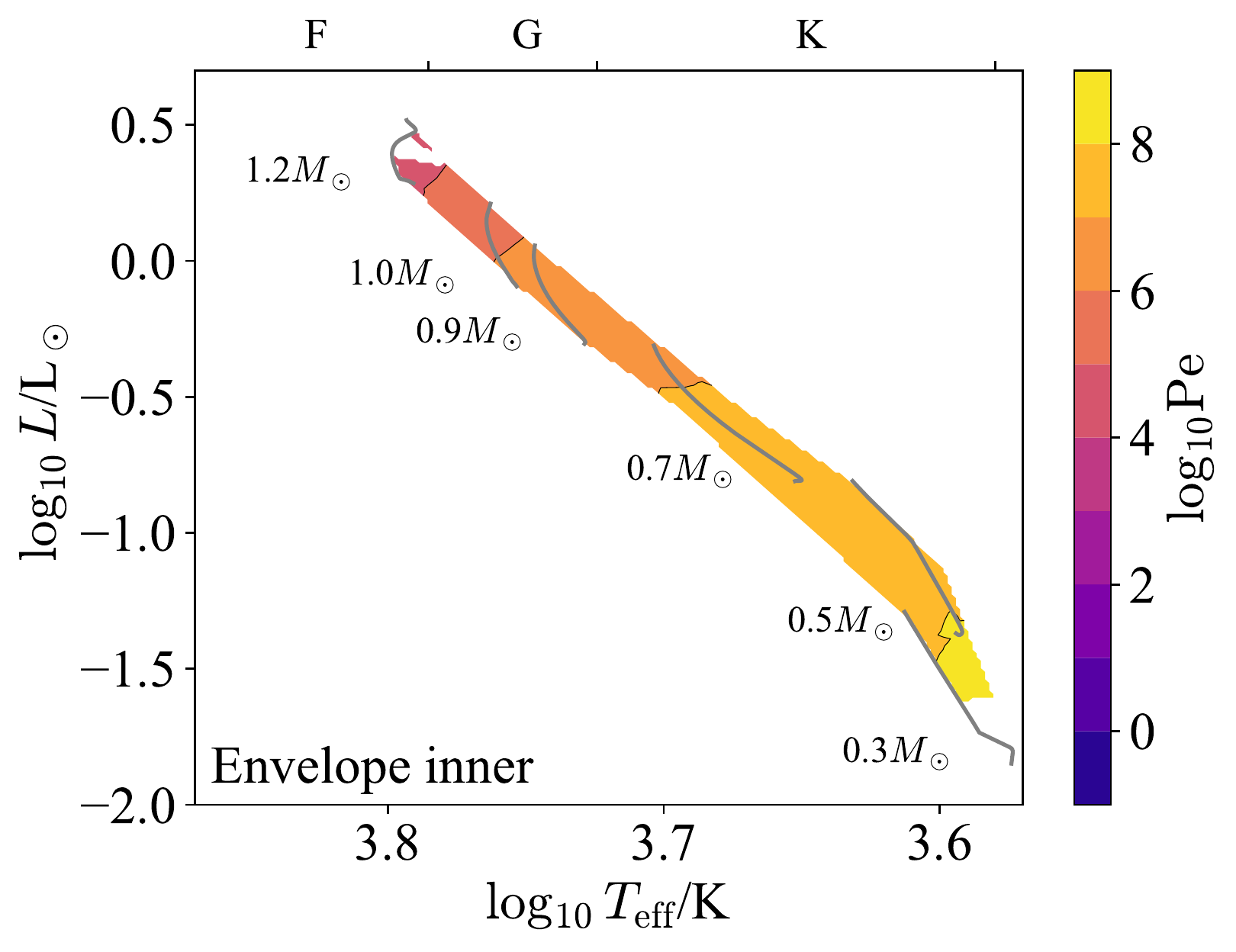}
\end{minipage}
\hfill
\begin{minipage}{0.433\textwidth}
\includegraphics[width=\textwidth]{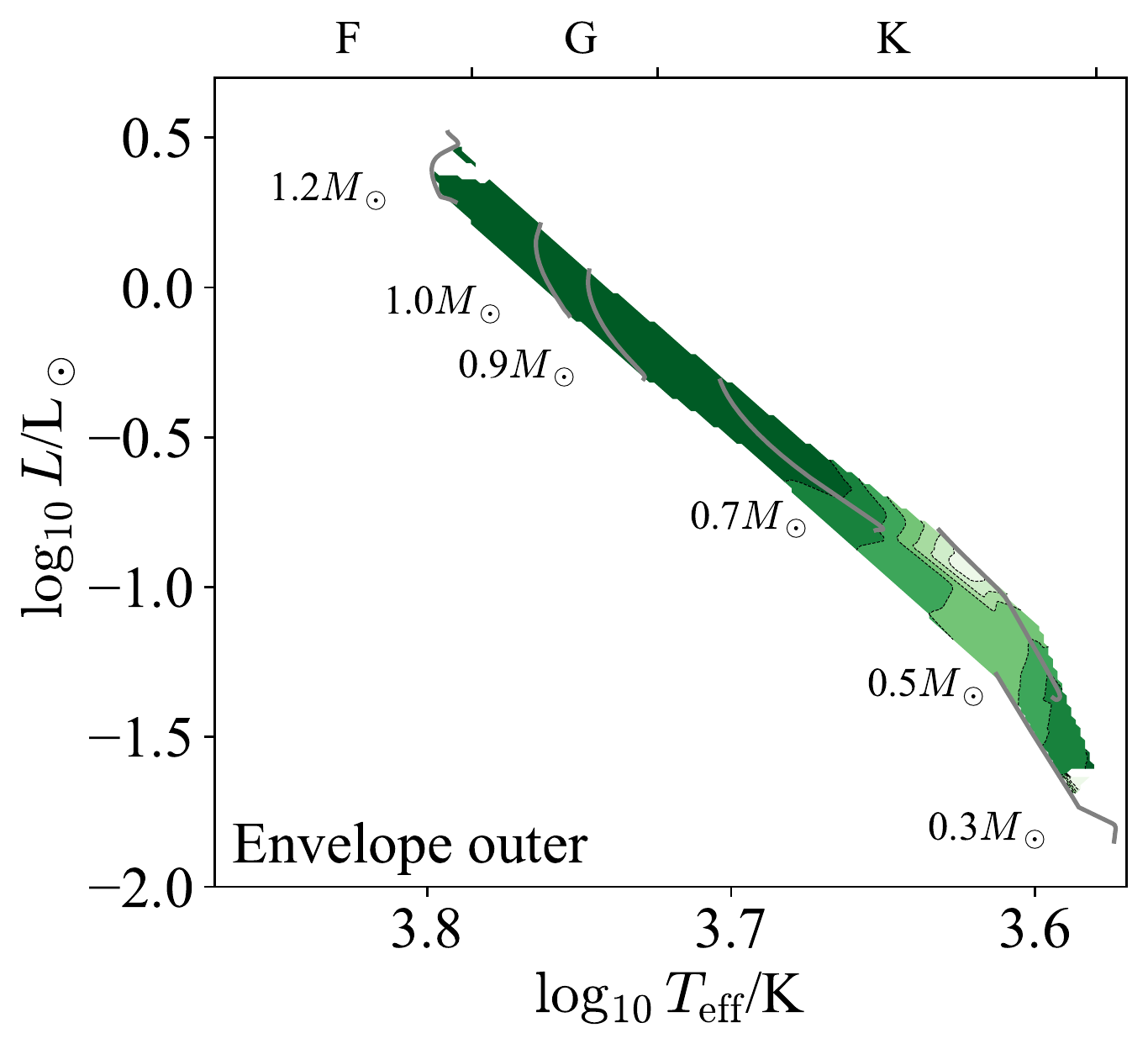}
\end{minipage}
\hfill
\begin{minipage}{0.537\textwidth}
\includegraphics[width=\textwidth]{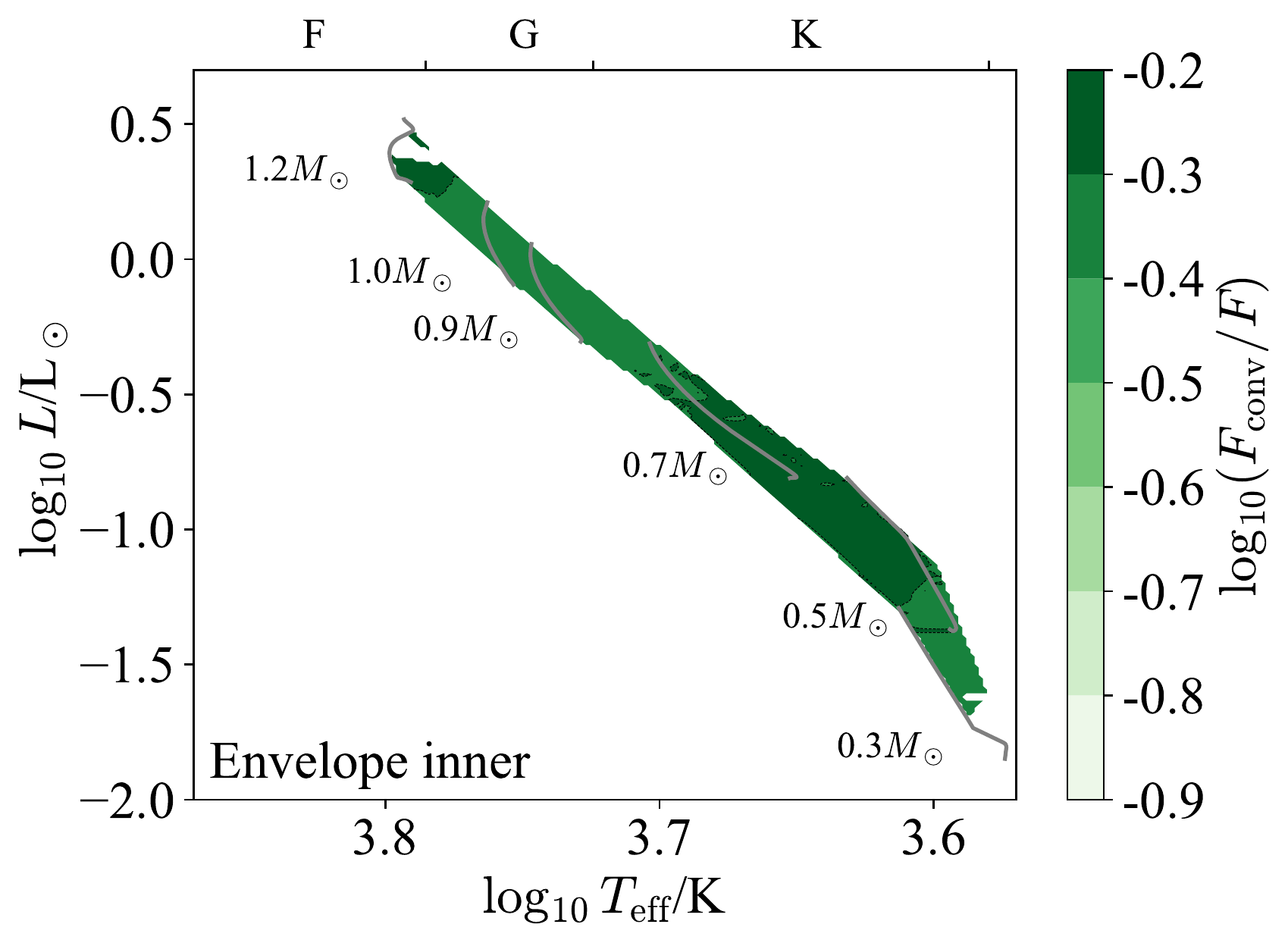}
\end{minipage}
\hfill

\caption{The P{\'e}clet number $\mathrm{Pe}$ (upper) and $F_{\rm conv}/F$ (lower) are shown averaged over the outermost pressure scale height (left) and innermost one (right) in terms of $\log T_{\rm eff}$/spectral type and $\log L$ for stellar models with deep envelope convection zones and Milky Way metallicity $Z=0.014$. Note that these quantities are calculated here using the pressure scale height at the relevant boundary rather than the full width of the convection zone $\delta r$.}
\label{fig:envelope_outputs_top_bottom_4}
\end{figure*}

Figure~\ref{fig:envelope_outputs_top_bottom_4} shows the P{\'e}clet number $\mathrm{Pe}$ (upper) and $F_{\rm conv}/F$ (lower).
The P{\'e}clet number is quite a bit smaller near the outer boundary than near the inner one, by factors of $10^{2}-10^{4}$.
Both quantities lie in the same qualitative regimes near both boundaries: the P{\'e}clet number indicates that advection dominates diffusion in the heat equation, and the flux ratio indicates that convection carries a substantial fraction of the flux.
The flux ratio is similar near the inner and outer boundaries of the convection zone, though it takes on a wider range of values near the outer boundary than the inner one.
Near both boundaries though it is smaller than in the bulk of the zone, which matches the intuition that convection ought to be more efficient in the bulk than near the boundaries.
The difference between the boundaries and bulk is as large as it is because the ratio $\nabla_{\rm rad} / \nabla_{\rm ad}$ is a gradual one near both boundaries.

\clearpage
\subsection{HI CZ}

We now examine the bulk structure of HI CZs, which occur in the subsurface layers of stars with masses $1.2M_\odot \la M_\star \la 3 M_\odot$.
Unlike in the case of Deep Envelope CZs, the HI CZs show large enough variation in all studied parameters to encompass many different regimes.
In this section the boundary between ``low'' and ``high'' masses is $\sim 1.5 M_\odot$.

Note that in some regions of the HR diagram this convection zone has a Rayleigh number below the $\sim 10^3$ critical value~\citep{1961hhs..book.....C}.
As a result while the region is superadiabatic, it is not unstable to convection.
We therefore neglect these stable regions in our analysis, and shade them in grey in our figures.

\begin{figure*}
\centering
\begin{minipage}{0.48\textwidth}
\includegraphics[width=\textwidth]{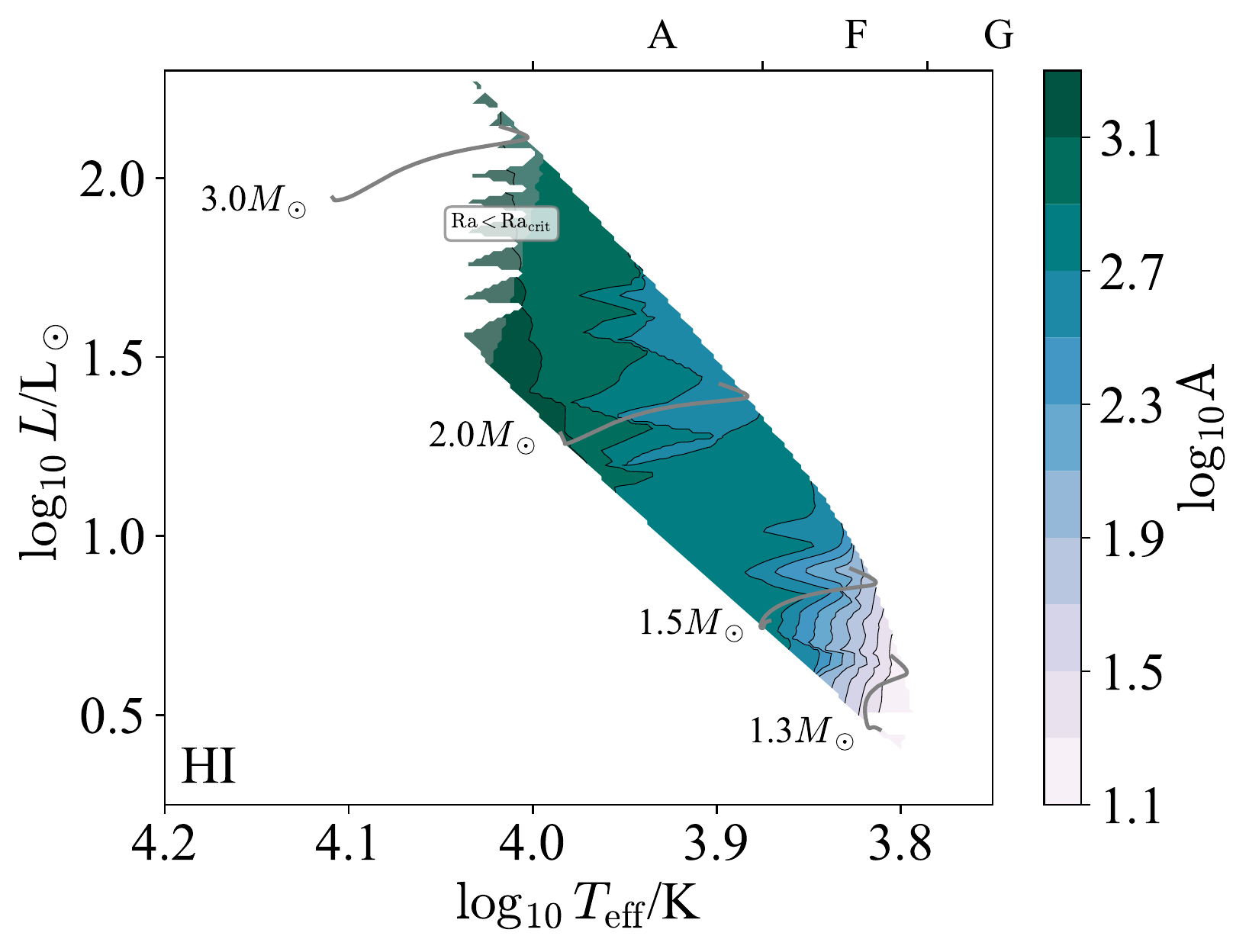}
\end{minipage}

\caption{The aspect ratio $\mathrm{A}$ is shown in terms of $\log T_{\rm eff}$/spectral type and $\log L$ for stellar models with HI~CZs and Milky Way metallicity $Z=0.014$. Note that the aspect ratio is an input parameter, and does not depend on a specific theory of convection. Regions with $\mathrm{Ra} < \mathrm{Ra}_{\rm crit}$ are stable to convection and shaded in grey.}
\label{fig:HI_structure}
\end{figure*}

Figure~\ref{fig:HI_structure} shows the aspect ratio $\mathrm{A}$, which ranges from $10-3000$.
These large aspect ratios suggest that local simulations are likely sufficient to capture their dynamics.

\begin{figure*}
\begin{minipage}{0.48\textwidth}
\includegraphics[width=\textwidth]{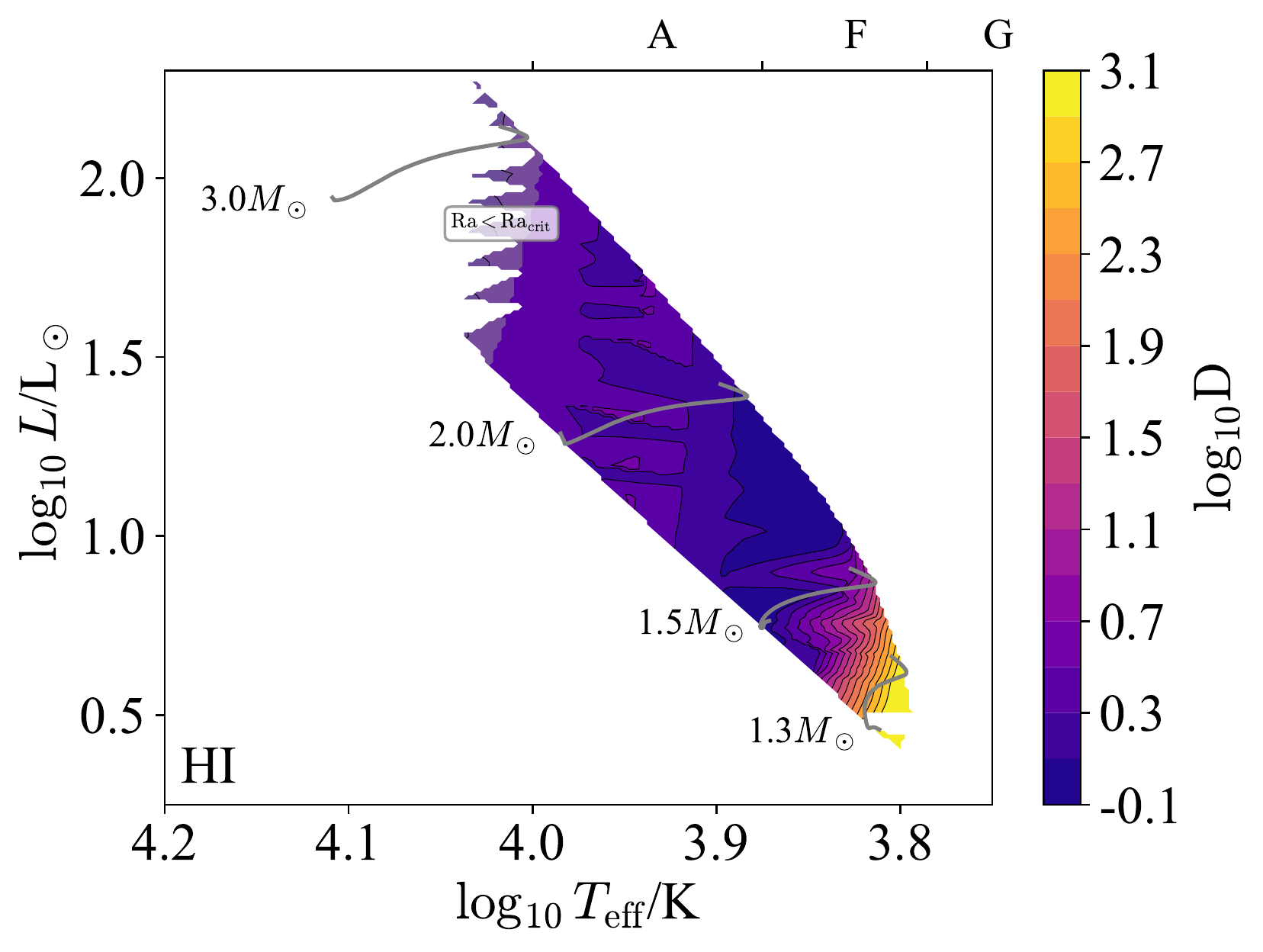}
\end{minipage}
\hfill
\begin{minipage}{0.48\textwidth}
\includegraphics[width=\textwidth]{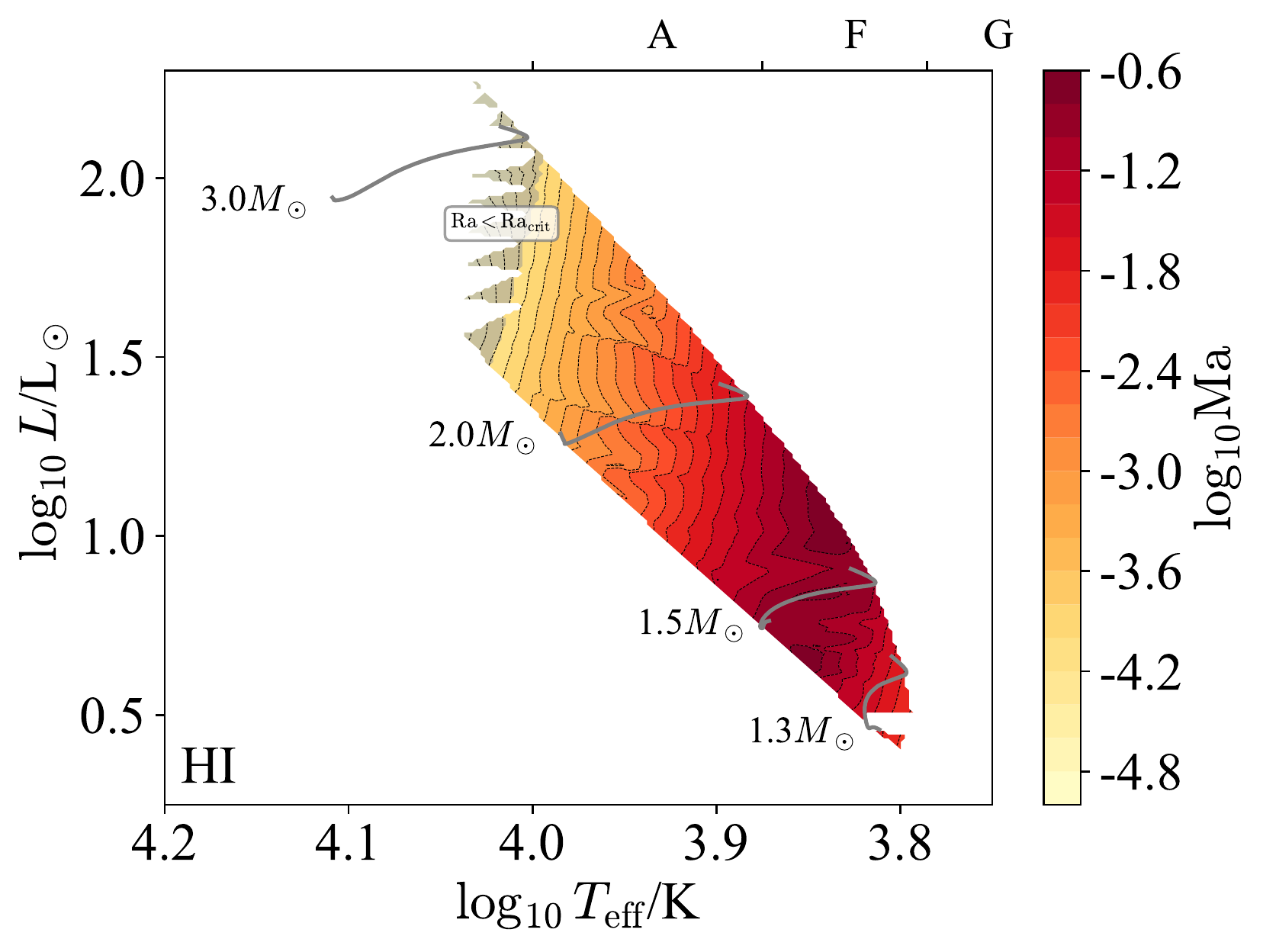}
\end{minipage}
\hfill

\caption{The density ratio $\mathrm{D}$ (left) and Mach number $\mathrm{Ma}$ (right) are shown in terms of $\log T_{\rm eff}$/spectral type and $\log L$ for stellar models with HI~CZs and Milky Way metallicity $Z=0.014$. Note that while the density ratio is an input parameter and does not depend on a specific theory of convection, the Mach number is an output of such a theory and so is model-dependent. Regions with $\mathrm{Ra} < \mathrm{Ra}_{\rm crit}$ are stable to convection and shaded in grey.}
\label{fig:HI_equations}
\end{figure*}

Next, the density ratio $\mathrm{D}$ (Figure~\ref{fig:HI_equations}, left) and Mach number $\mathrm{Ma}$ (Figure~\ref{fig:HI_equations}, right) inform which physics the fluid equations must include to model these zones.
At low masses the density ratio is large while at higher masses it is near-unity.
Likewise at low masses the Mach number is moderate ($\sim 0.3$) while at high masses it is small ($10^{-5}$).
This, along with the density ratio, suggests it is appropriate to use the Boussinesq approximation at high masses, while the fully compressible equations are necessary at low masses.

\begin{figure*}
\begin{minipage}{0.48\textwidth}
\includegraphics[width=\textwidth]{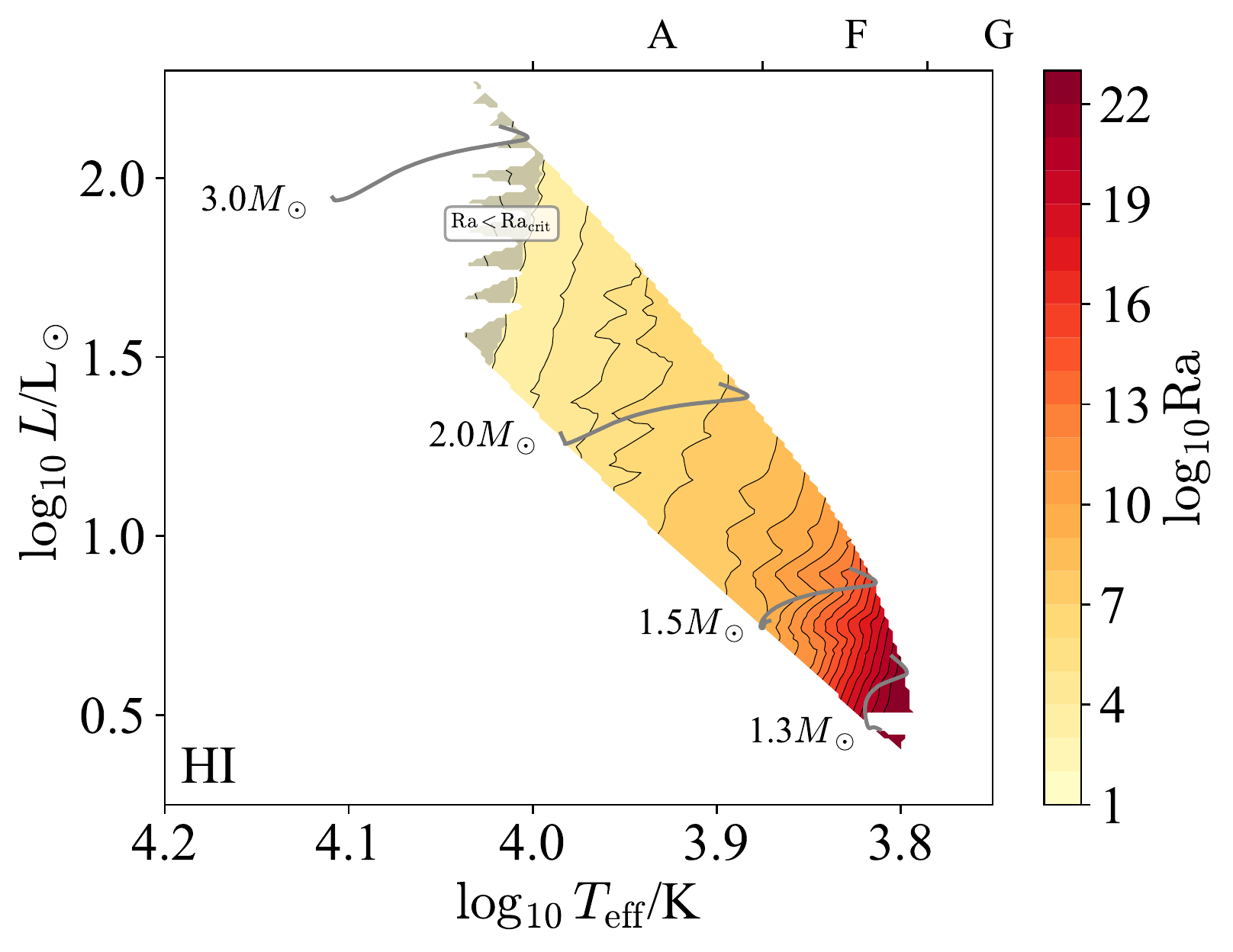}
\end{minipage}
\hfill
\begin{minipage}{0.48\textwidth}
\includegraphics[width=\textwidth]{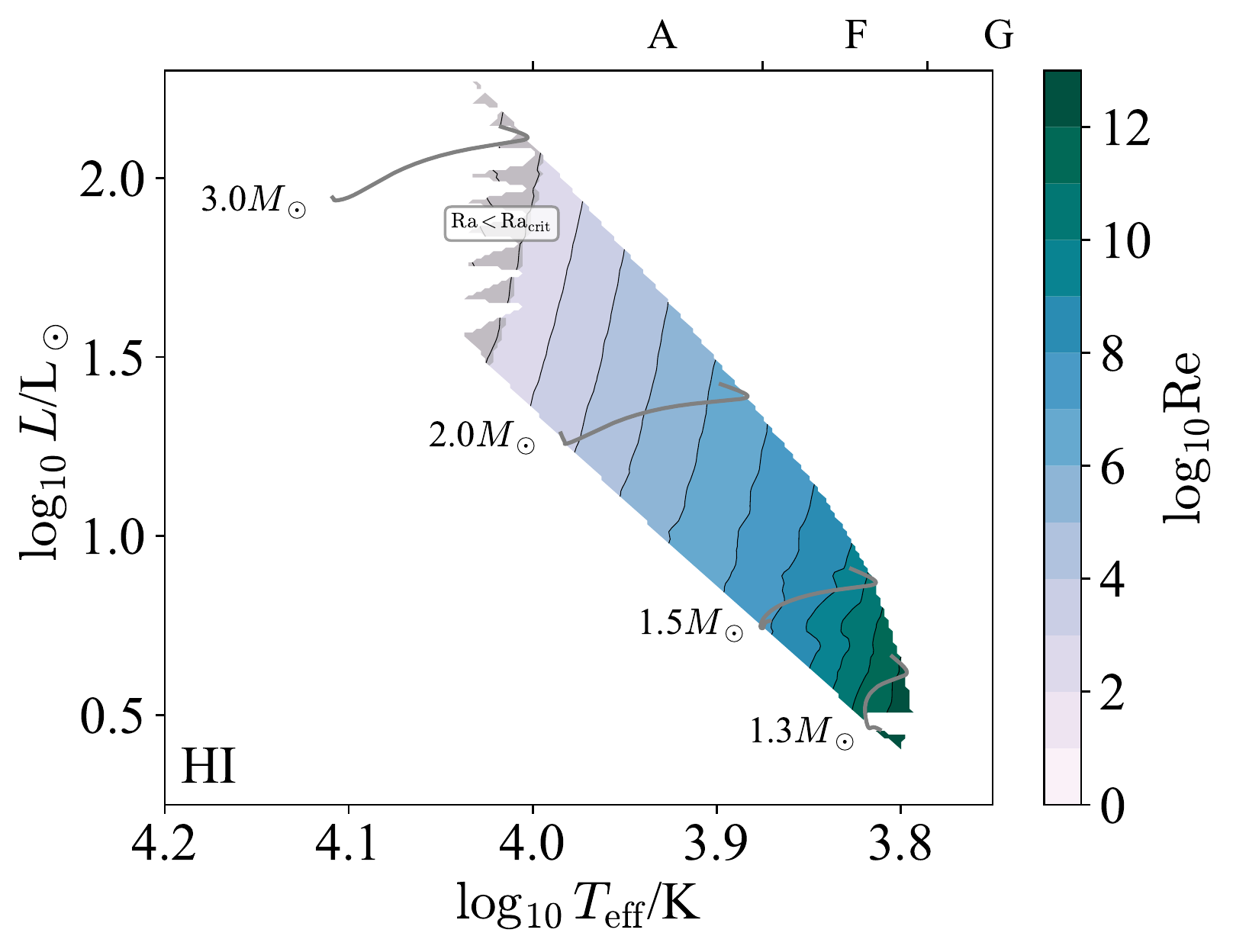}
\end{minipage}
\hfill

\caption{The Rayleigh number $\mathrm{Ra}$ (left) and Reynolds number $\mathrm{Re}$ (right) are shown in terms of $\log T_{\rm eff}$/spectral type and $\log L$ for stellar models with HI~CZs and Milky Way metallicity $Z=0.014$.  Note that while the Rayleigh number is an input parameter and does not depend on a specific theory of convection, the Reynolds number is an output of such a theory and so is model-dependent. Regions with $\mathrm{Ra} < \mathrm{Ra}_{\rm crit}$ are stable to convection and shaded in grey.}
\label{fig:HI_stability}
\end{figure*}

The Rayleigh number $\mathrm{Ra}$ (Figure~\ref{fig:HI_stability}, left) determines whether or not a putative convection zone is actually unstable to convection, and the Reynolds number $\mathrm{Re}$ determines how turbulent the zone is if instability sets in (Figure~\ref{fig:HI_stability}, right).
At low masses the Rayleigh number is large ($10^{25}$), at high masses it plummets and eventually becomes sub-critical, which we show in grey.
Likewise at low masses the Reynolds number is large ($10^{13}$) while at high masses it is quite small ($\sim 1)$.
These putative convection zones then span a wide range of properties, from being subcritical and \emph{stable}~\citep{1961hhs..book.....C} at high masses, to being marginally unstable and weakly turbulent at intermediate masses ($\sim 2 M_\odot$), to eventually being strongly unstable and having well-developed turbulence at low masses.

\begin{figure*}
\centering
\begin{minipage}{0.47\textwidth}
\includegraphics[width=\textwidth]{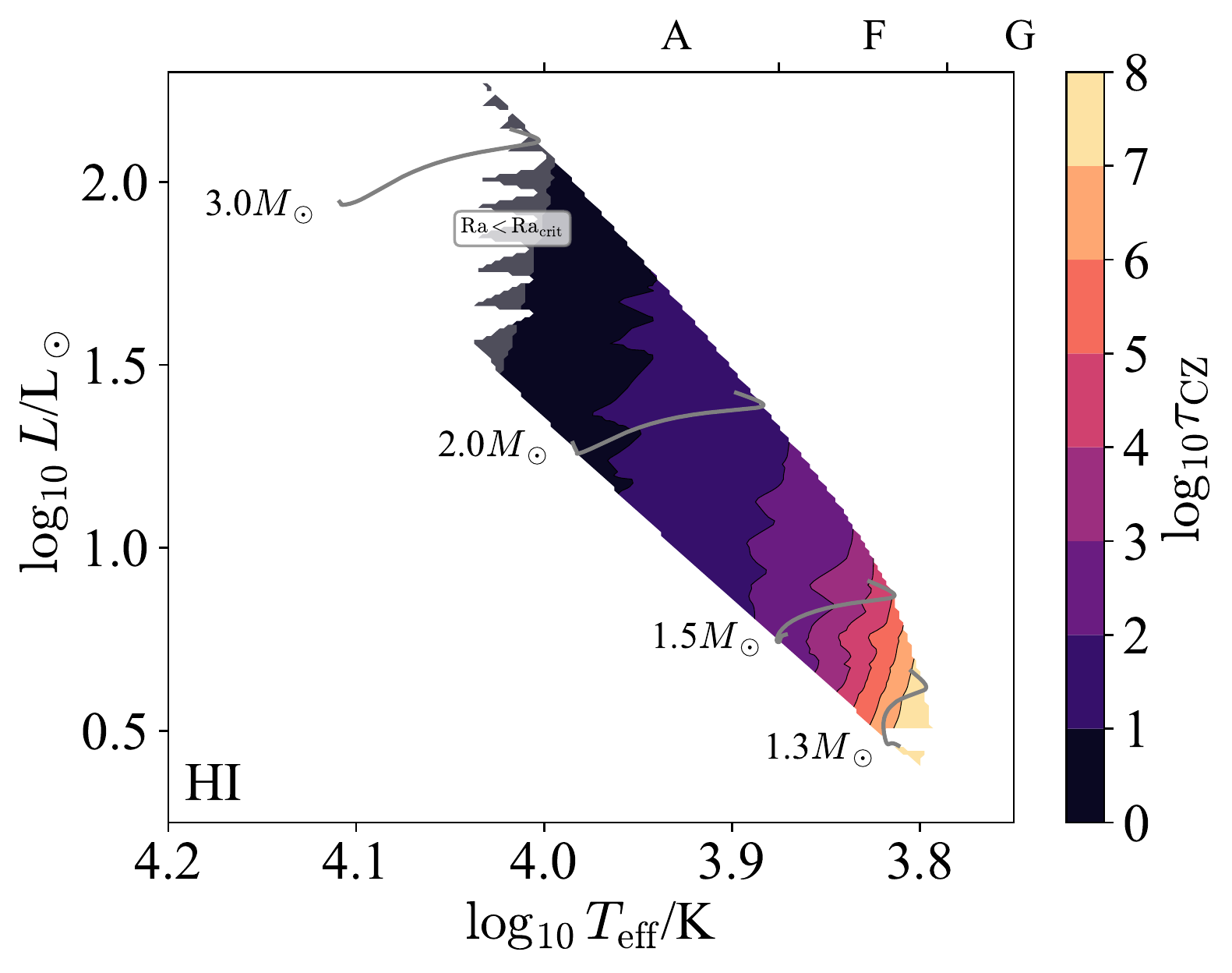}
\end{minipage}
\hfill
\begin{minipage}{0.49\textwidth}
\includegraphics[width=\textwidth]{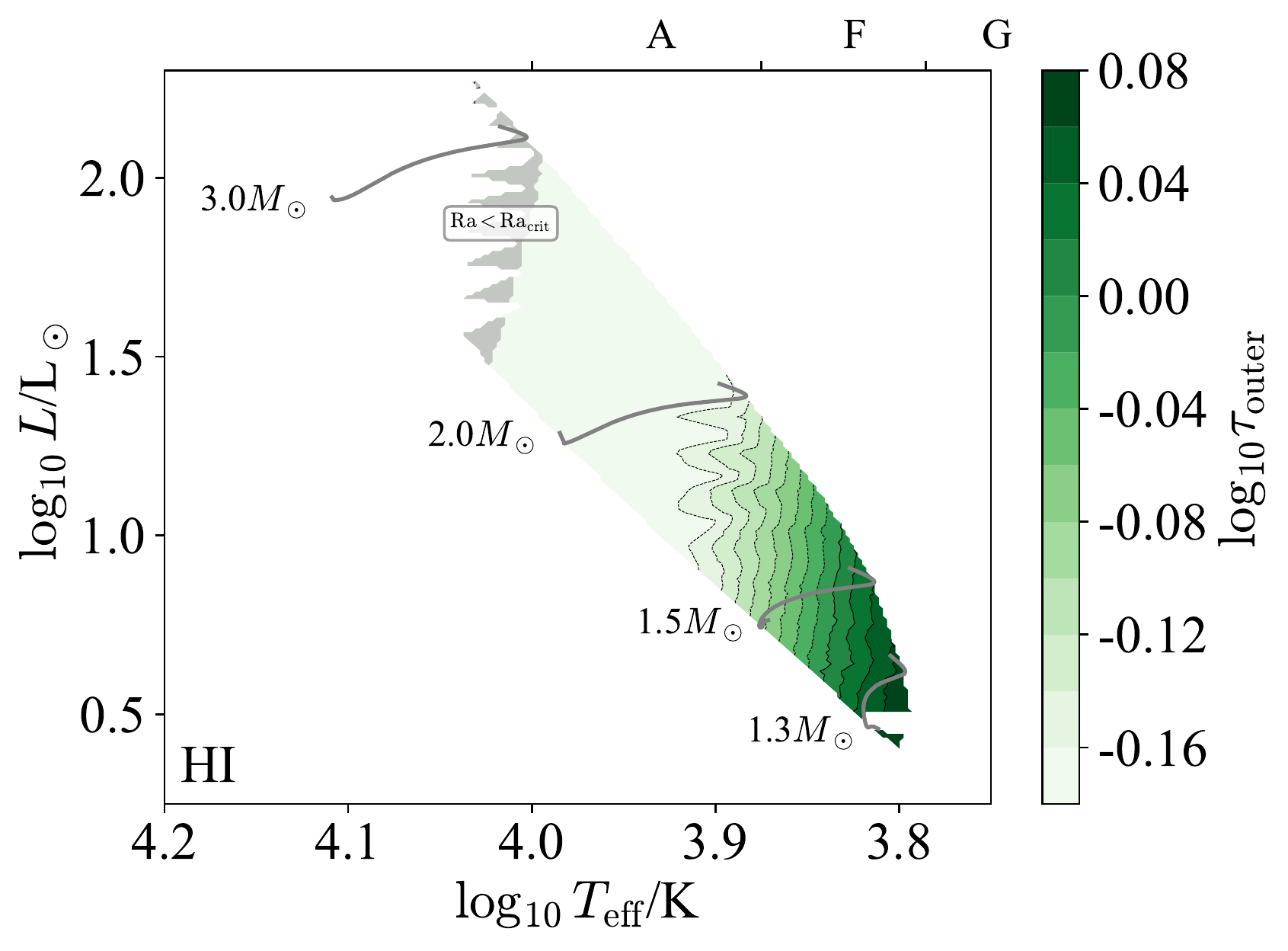}
\end{minipage}
\hfill
\caption{The convection optical depth $\tau_{\rm CZ}$ (left) and the optical depth to the surface $\tau_{\rm outer}$ (right) are shown in terms of $\log T_{\rm eff}$/spectral type and $\log L$ for stellar models with HI~CZs and Milky Way metallicity $Z=0.014$. Note that both of these are input parameters, and do not depend on a specific theory of convection. Regions with $\mathrm{Ra} < \mathrm{Ra}_{\rm crit}$ are stable to convection and shaded in grey.}
\label{fig:HI_optical}
\end{figure*}

The optical depth across a convection zone $\tau_{\rm CZ}$ (Figure~\ref{fig:HI_optical}, left) indicates whether or not radiation can be handled in the diffusive approximation, while the optical depth from the outer boundary to infinity $\tau_{\rm outer}$ (Figure~\ref{fig:HI_optical}, right) indicates the nature of radiative transfer and cooling in the outer regions of the convection zone.
The surface of the HI CZ is always at low optical depth ($\tau_{\rm outer} \sim 1$), meaning that radiation hydrodynamics is likely necessary near the outer boundary of this zone.
By contrast, the optical depth across the HI CZ is low at high masses ($\sim 1-10$) and large at low masses ($10^{3+}$).
This implies that radiation hydrodynamics is necessary to model the bulk of the HI CZ at high masses, but not at low masses.

\begin{figure*}
\centering
\begin{minipage}{0.48\textwidth}
\includegraphics[width=\textwidth]{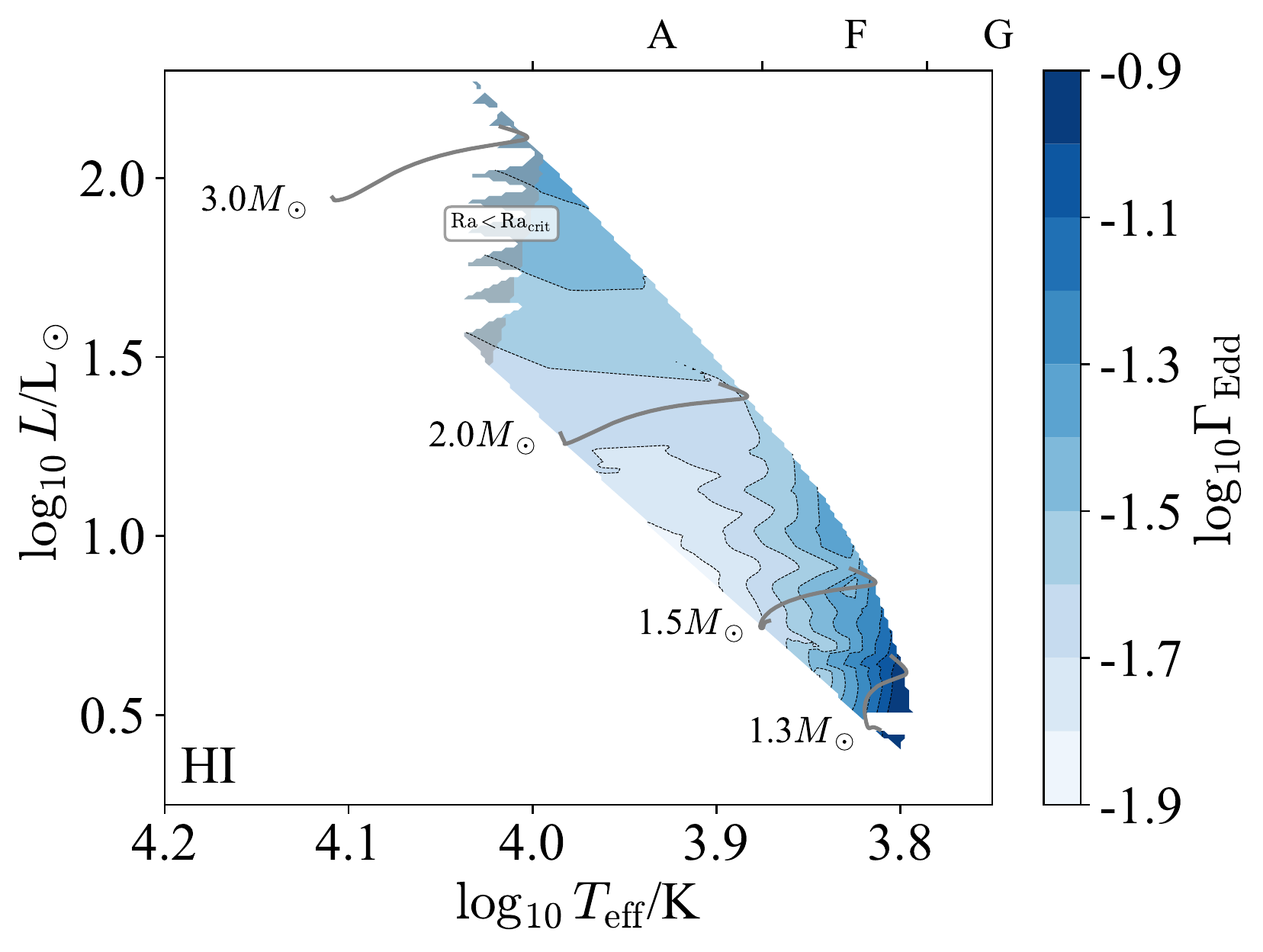}
\end{minipage}
\hfill
\begin{minipage}{0.48\textwidth}
\includegraphics[width=\textwidth]{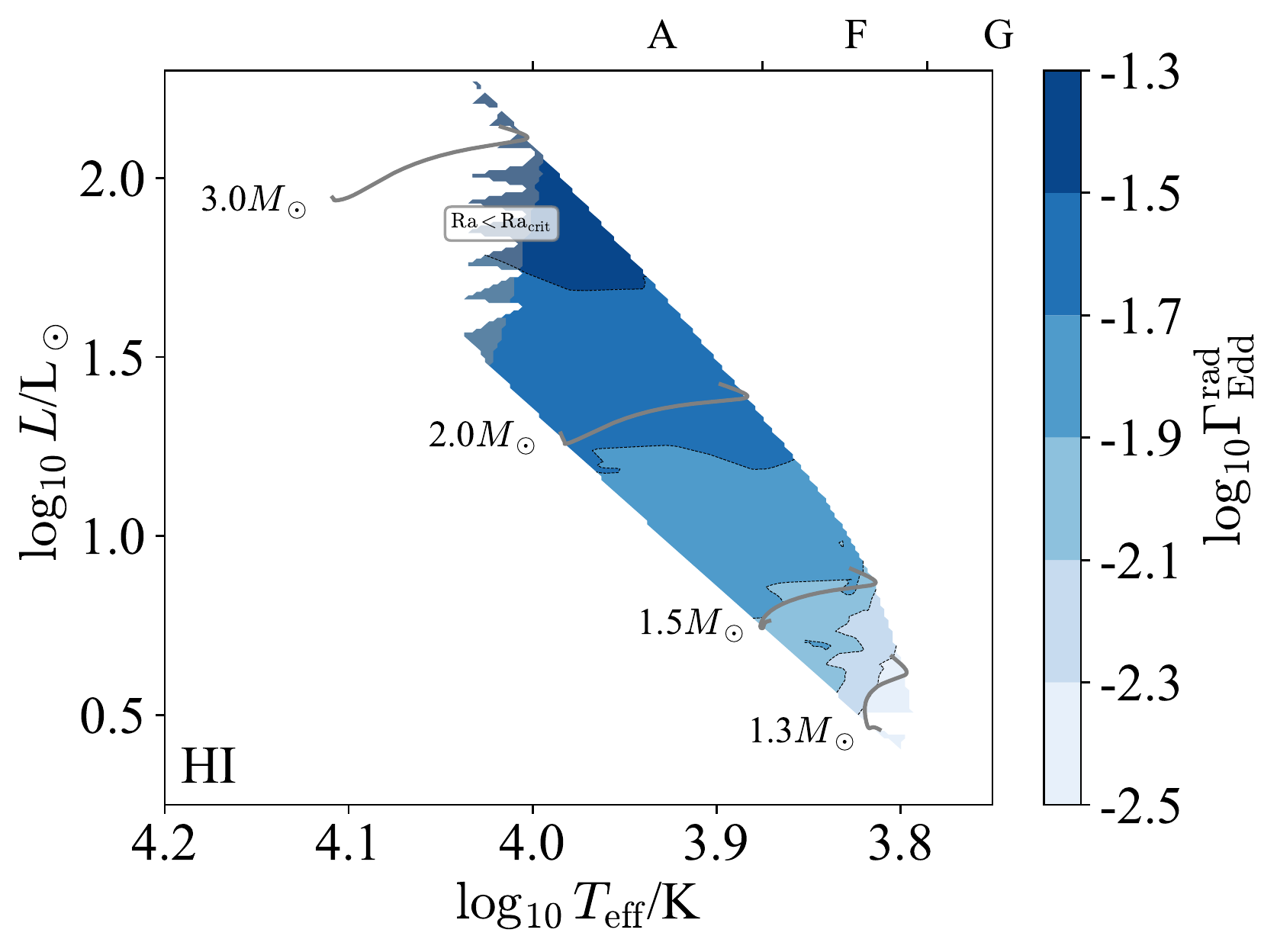}
\end{minipage}
\hfill
\caption{The Eddington ratio with the full luminosity $\Gamma_{\rm Edd}$ (left) and the radiative luminosity (right) are shown in terms of $\log T_{\rm eff}$/spectral type and $\log L$ for stellar models with HI~CZs and Milky Way metallicity $Z=0.014$. Note that while $\Gamma_{\rm Edd}$ is an input parameter and does not depend on a specific theory of convection, $\Gamma_{\rm Edd}^{\rm rad}$ is an output of such a theory and so is model-dependent. Regions with $\mathrm{Ra} < \mathrm{Ra}_{\rm crit}$ are stable to convection and shaded in grey.}
\label{fig:HI_eddington}
\end{figure*}

The Eddington ratio $\Gamma_{\rm Edd}$ (Figure~\ref{fig:HI_eddington}, left) indicates whether or not radiation hydrodynamic instabilities are important in the non-convecting state, and the radiative Eddington ratio $\Gamma_{\rm Edd}^{\rm rad}$ (Figure~\ref{fig:HI_eddington}, right) indicates the same in the developed convective state.
Both ratios are small in the HI CZ, so radiation hydrodynamic instabilities are unlikely to matter.

\begin{figure*}
\centering
\begin{minipage}{0.48\textwidth}
\includegraphics[width=\textwidth]{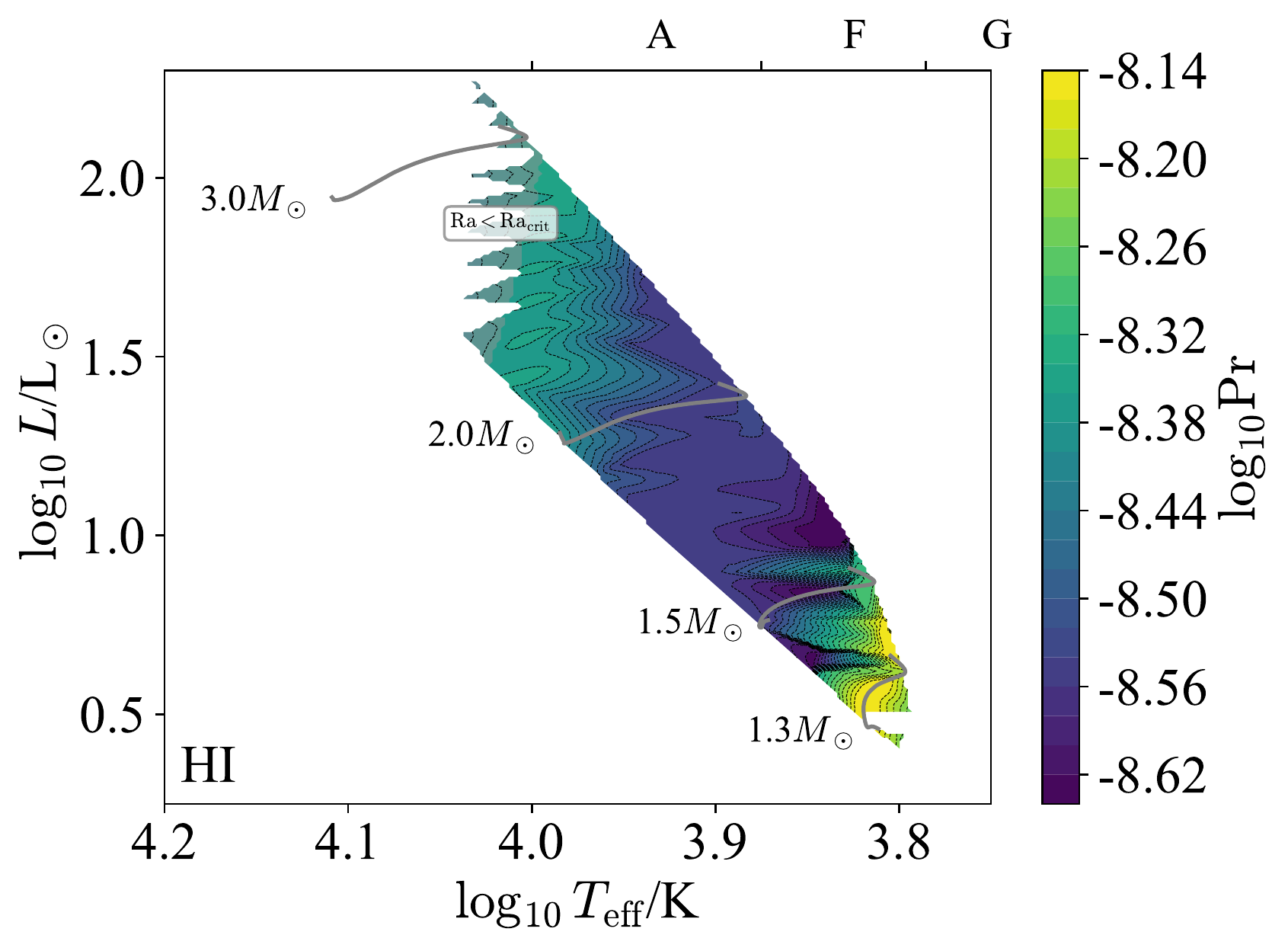}
\end{minipage}
\hfill
\begin{minipage}{0.48\textwidth}
\includegraphics[width=\textwidth]{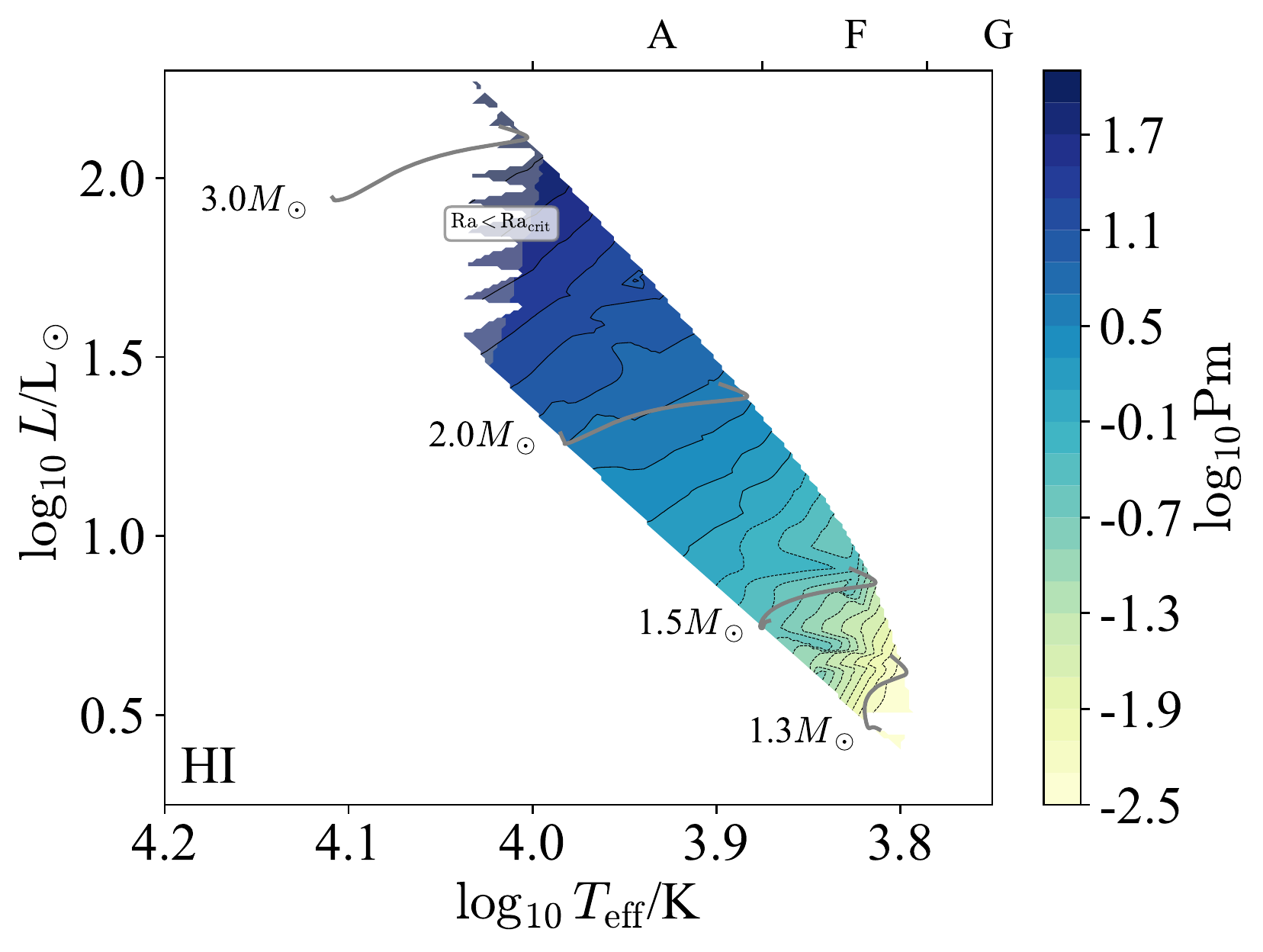}
\end{minipage}
\hfill

\caption{The Prandtl number $\mathrm{Pr}$ (left) and magnetic Prandtl number $\mathrm{Pm}$ (right) are shown in terms of $\log T_{\rm eff}$/spectral type and $\log L$ for stellar models with HI~CZs and Milky Way metallicity $Z=0.014$. Note that both $\mathrm{Pr}$ and $\mathrm{Pm}$ are input parameters, and so do not depend on a specific theory of convection. Regions with $\mathrm{Ra} < \mathrm{Ra}_{\rm crit}$ are stable to convection and shaded in grey.}
\label{fig:HI_diffusivities}
\end{figure*}

The Prandtl number $\mathrm{Pr}$ (Figure~\ref{fig:HI_diffusivities}, left) measures the relative importance of thermal diffusion and viscosity, and the magnetic Prandtl number $\mathrm{Pm}$ (Figure~\ref{fig:HI_diffusivities}, right) measures the same for magnetic diffusion and viscosity.
The Prandtl number is always small in these models, so the thermal diffusion length-scale is much larger than the viscous scale.
By contrast, the magnetic Prandtl number varies from small at low masses to large at high masses.

The fact that $\mathrm{Pm}$ is large at high masses is notable because the quasistatic approximation for magnetohydrodynamics has frequently been used to study magnetoconvection in minimal 3D MHD simulations of planetary and stellar interiors~\citep[e.g.][]{yan_calkins_maffei_julien_tobias_marti_2019} and assumes that $\mathrm{Rm} = \mathrm{Pm} \mathrm{Re} \rightarrow 0$; in doing so, this approximation assumes a global background magnetic field is dominant and neglects the nonlinear portion of the Lorentz force. This approximation breaks down in convection zones with $\mathrm{Pm} > 1$ and future numerical experiments should seek to understand how magnetoconvection operates in this regime.

\begin{figure*}
\centering
\begin{minipage}{0.48\textwidth}
\includegraphics[width=\textwidth]{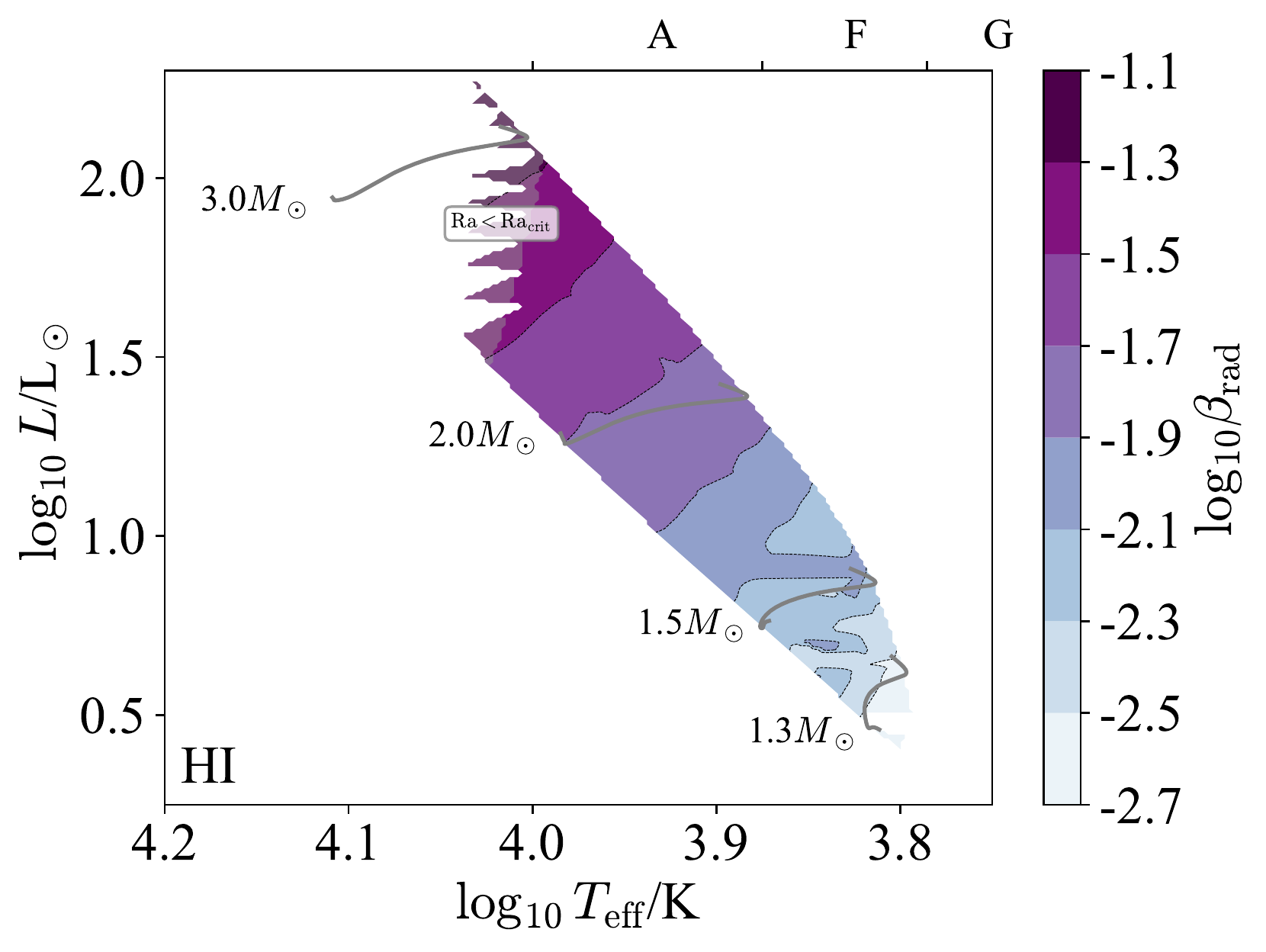}
\end{minipage}
\hfill

\caption{The radiation pressure ratio $\beta_{\rm rad}$ is shown in terms of $\log T_{\rm eff}$/spectral type and $\log L$ for stellar models with HI~CZs and Milky Way metallicity $Z=0.014$. Note that this ratio is an input parameter, and does not depend on a specific theory of convection. Regions with $\mathrm{Ra} < \mathrm{Ra}_{\rm crit}$ are stable to convection and shaded in grey.}
\label{fig:HI_beta}
\end{figure*}

The radiation pressure ratio $\beta_{\rm rad}$ (Figure~\ref{fig:HI_beta}) measures the importance of radiation in setting the thermodynamic properties of the fluid.
We see that this is uniformly small ($\la 0.1$) and so radiation pressure likely plays a sub-dominant role in these zones.

\begin{figure*}
\centering
\begin{minipage}{0.48\textwidth}
\includegraphics[width=\textwidth]{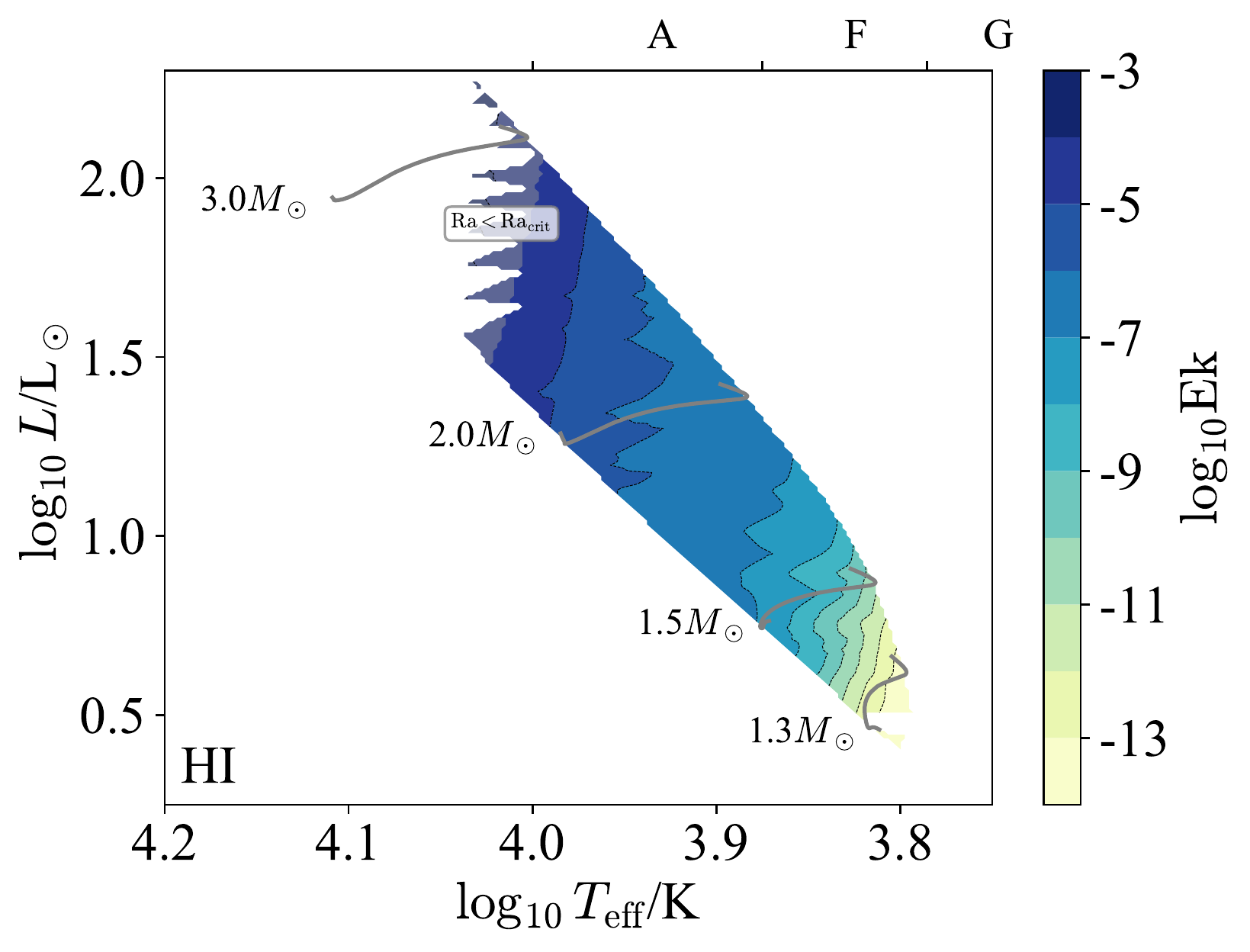}
\end{minipage}

\caption{The Ekman number $\mathrm{Ek}$ is shown in terms of $\log T_{\rm eff}$/spectral type and $\log L$ for stellar models with HI~CZs and Milky Way metallicity $Z=0.014$. Note that the Ekman number is an input parameter, and does not depend on a specific theory of convection. Regions with $\mathrm{Ra} < \mathrm{Ra}_{\rm crit}$ are stable to convection and shaded in grey.}
\label{fig:HI_ekman}
\end{figure*}

The Ekman number $\mathrm{Ek}$ (Figure~\ref{fig:HI_ekman}) indicates the relative importance of viscosity and rotation.
This is tiny across the HRD~\footnote{Note that, because the Prandtl number is also very small, this does not significantly alter the critical Rayleigh number~(see Ch3 of~\cite{1961hhs..book.....C} and appendix D of~\cite{2022arXiv220110567J}).}, so we expect rotation to dominate over viscosity, except at very small length-scales.

\begin{figure*}
\centering
\begin{minipage}{0.48\textwidth}
\includegraphics[width=\textwidth]{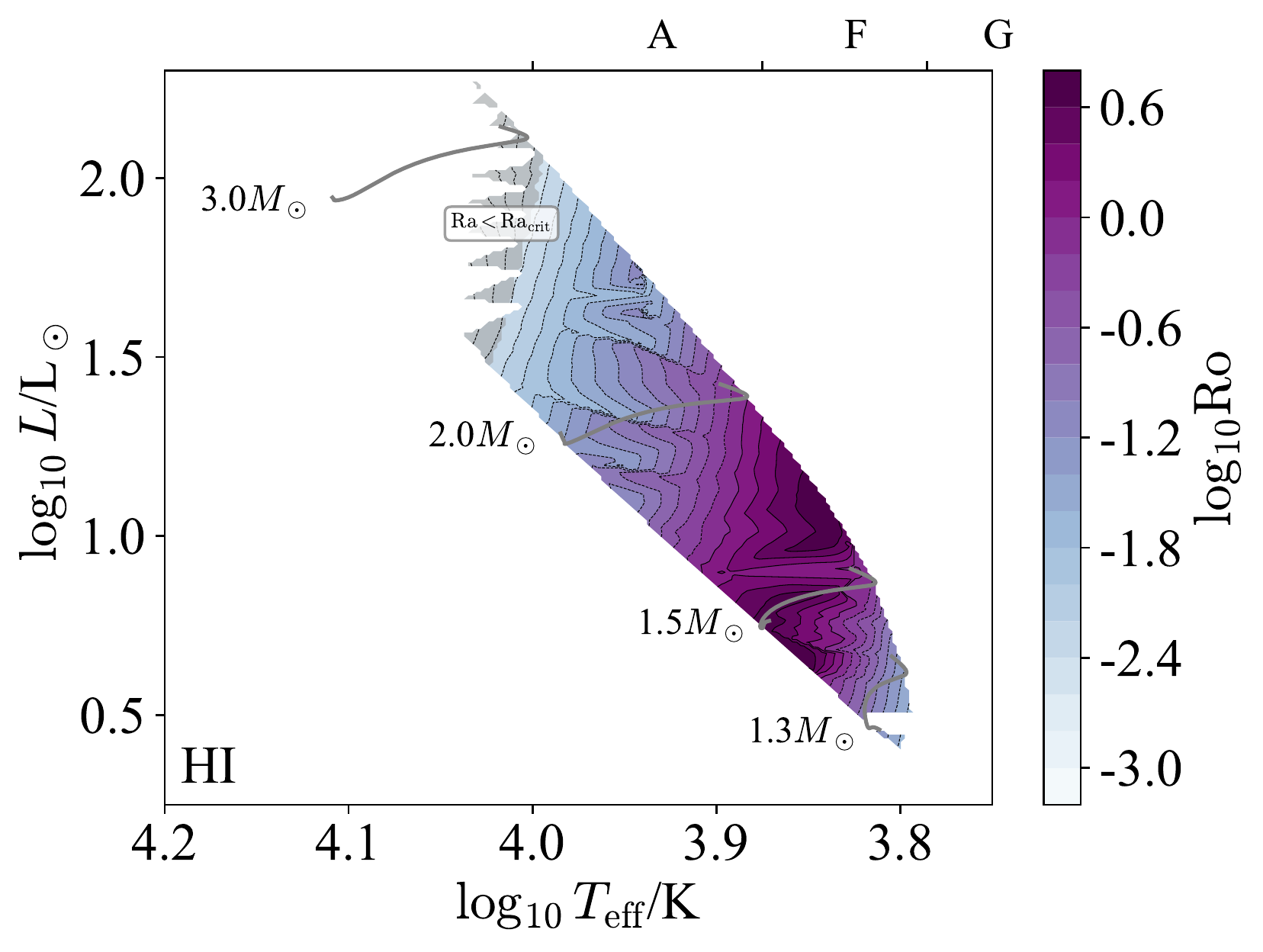}
\end{minipage}
\hfill
\begin{minipage}{0.48\textwidth}
\includegraphics[width=\textwidth]{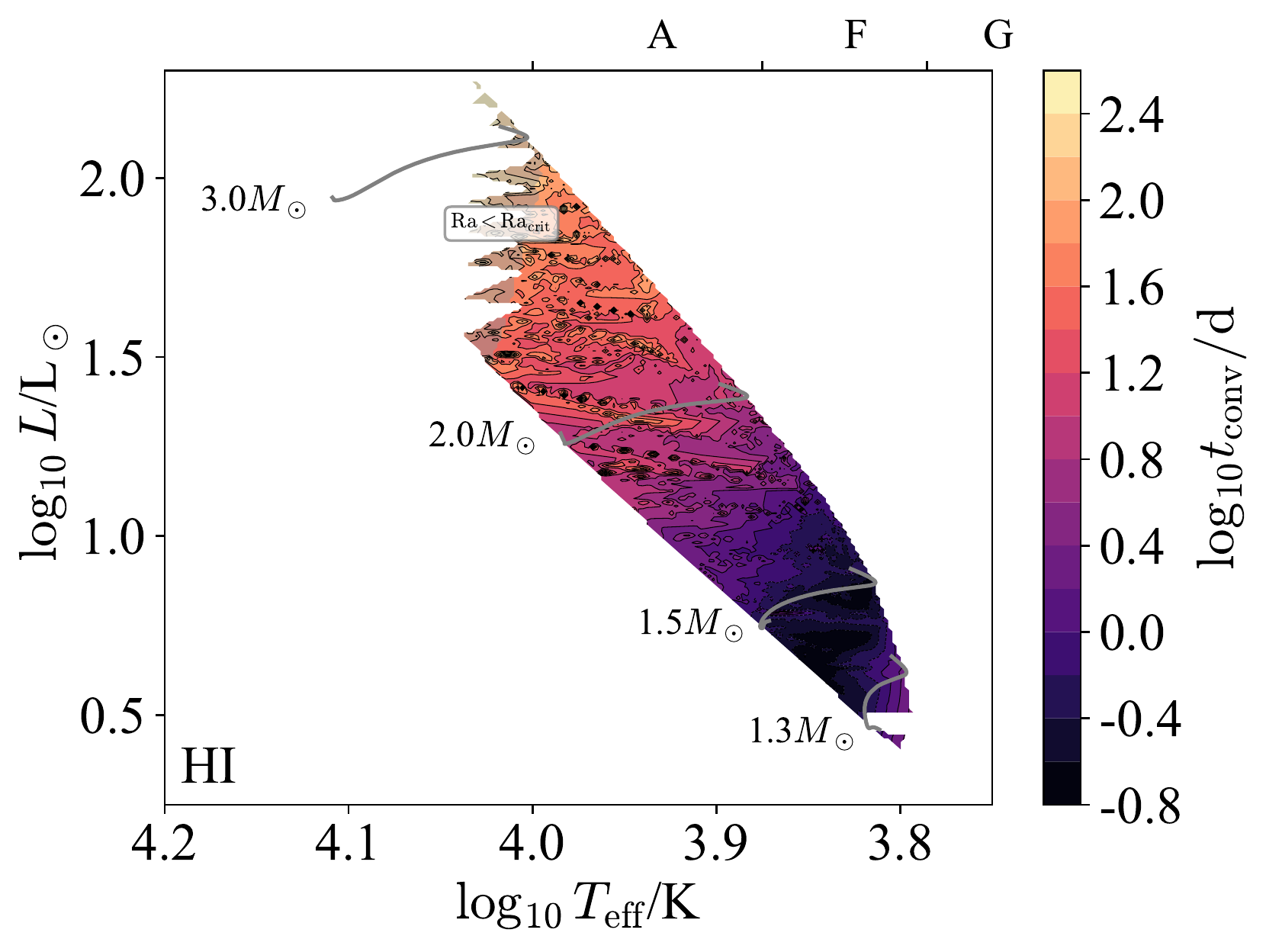}
\end{minipage}
\hfill

\caption{The Rossby number $\mathrm{Ro}$ (left) and turnover time $t_{\rm conv}$ (right) are shown in terms of $\log T_{\rm eff}$/spectral type and $\log L$ for stellar models with HI~CZs and Milky Way metallicity $Z=0.014$. Note that both $\mathrm{Ro}$ and $t_{\rm conv}$ are outputs of a theory of convection and so are model-dependent. Regions with $\mathrm{Ra} < \mathrm{Ra}_{\rm crit}$ are stable to convection and shaded in grey.}
\label{fig:HI_rotation}
\end{figure*}

The Rossby number $\mathrm{Ro}$ (Figure~\ref{fig:HI_rotation}, left) measures the relative importance of rotation and inertia.
This is small at high masses but greater than unity at low masses, meaning that the HI CZ is rotationally constrained at high masses but not at low masses for typical rotation rates~\citep{2013A&A...557L..10N}.

We have assumed a fiducial rotation law to calculate $\mathrm{Ro}$.
Stars exhibit a variety of different rotation rates, so we also show the convective turnover time $t_{\rm conv}$ (Figure~\ref{fig:HI_rotation}, right) which may be used to estimate the Rossby number for different rotation periods.

\begin{figure*}
\centering
\begin{minipage}{0.48\textwidth}
\includegraphics[width=\textwidth]{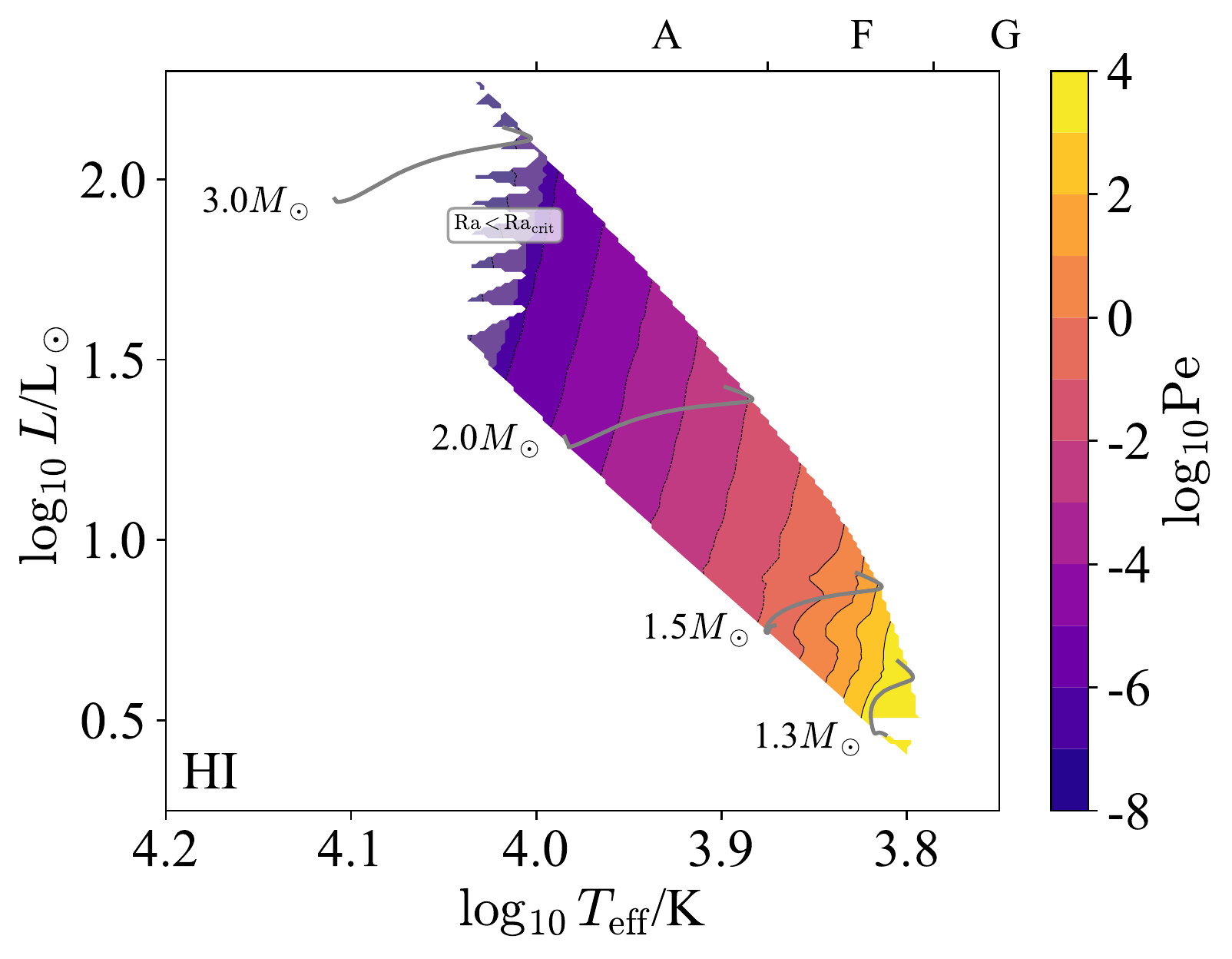}
\end{minipage}
\hfill
\begin{minipage}{0.48\textwidth}
\includegraphics[width=\textwidth]{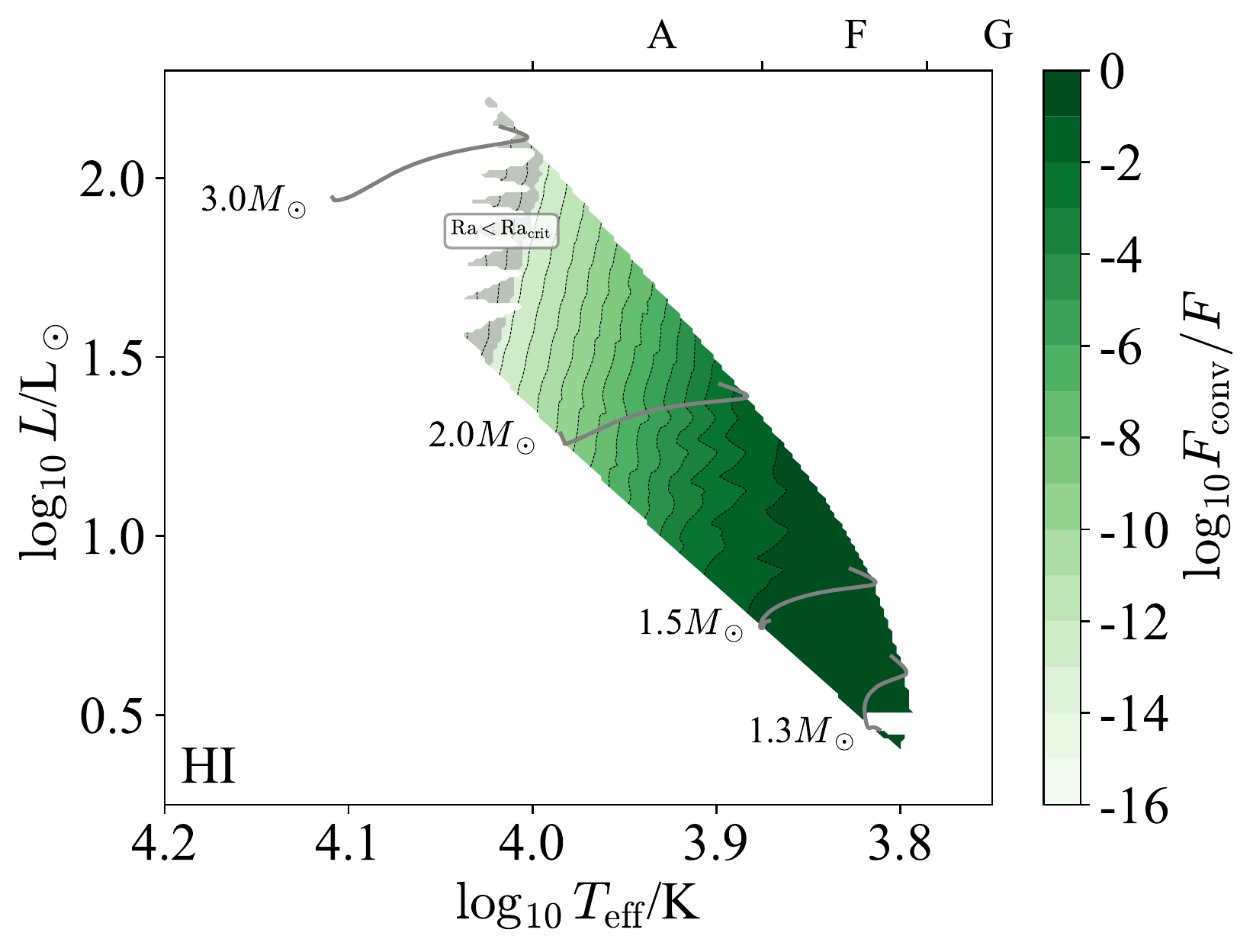}
\end{minipage}

\caption{The P{\'e}clet number $\mathrm{Pe}$ (left) and $F_{\rm conv}/F$ (right) are shown in terms of $\log T_{\rm eff}$/spectral type and $\log L$ for stellar models with HI~CZs and Milky Way metallicity $Z=0.014$. Note that both $\mathrm{Pe}$ and $F_{\rm conv}/F$ are outputs of a theory of convection and so are model-dependent. Regions with $\mathrm{Ra} < \mathrm{Ra}_{\rm crit}$ are stable to convection and shaded in grey.}
\label{fig:HI_efficiency}
\end{figure*}

The P{\'e}clet number $\mathrm{Pe}$ (Figure~\ref{fig:HI_efficiency}, left) measures the relative importance of advection and diffusion in transporting heat, and the flux ratio $F_{\rm conv}/F$ (Figure~\ref{fig:HI_efficiency}, right) reports the fraction of the energy flux which is advected.
Both exhibit substantial variation with mass.
The P{\'e}clet number varies from large ($10^4$) at low masses to very small at high masses ($10^{-9}$), and the flux ratio similarly varies from near-unity at low masses to tiny ($10^{-16}$) at high masses.
That is, there is a large gradient in convective efficiency with mass, with efficient convection at low masses and very inefficient convection at high masses.

Of particular interest at intermediate masses ($1.5 M_\odot \la M \la 1.7 M_\odot$) are stars for which the Reynolds number is still large ($\mathrm{Re} > 10^4$) but the P{\'e}clet number is small ($\mathrm{Pe} < 1$).
In these stars the HI CZ should exhibit turbulent velocity fields but very laminar thermodynamic fields, which could be quite interesting to study numerically.

We further note that at the low mass end the Mach number is moderate ($\sim 0.3$) high $F_{\rm conv}/F$ is near-unity, so convection likely produces a significant luminosity in internal gravity waves~\citep{1990ApJ...363..694G,2013MNRAS.430.2363L}.

\begin{figure*}
\centering
\begin{minipage}{0.48\textwidth}
\includegraphics[width=\textwidth]{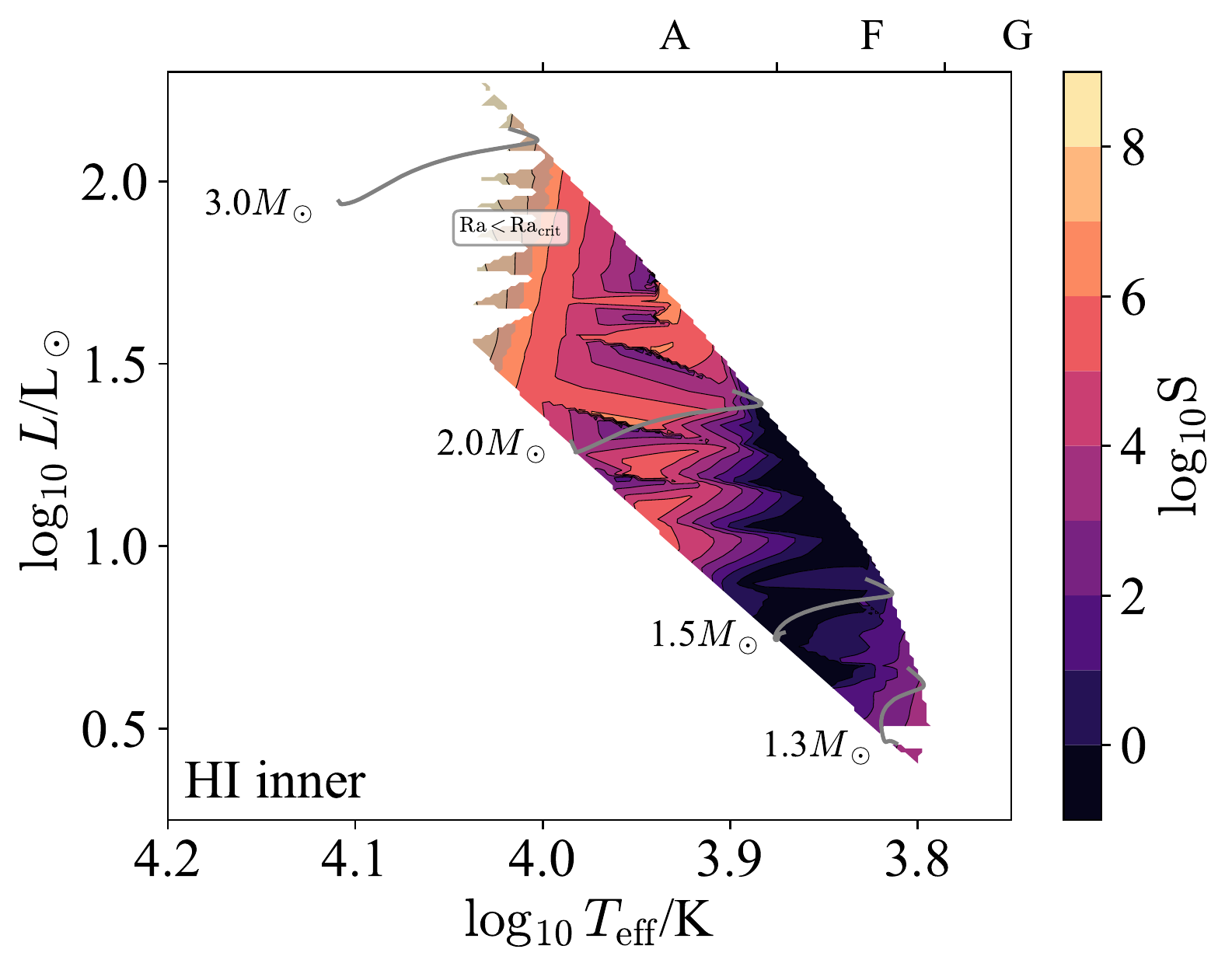}
\end{minipage}
\hfill
\begin{minipage}{0.48\textwidth}
\includegraphics[width=\textwidth]{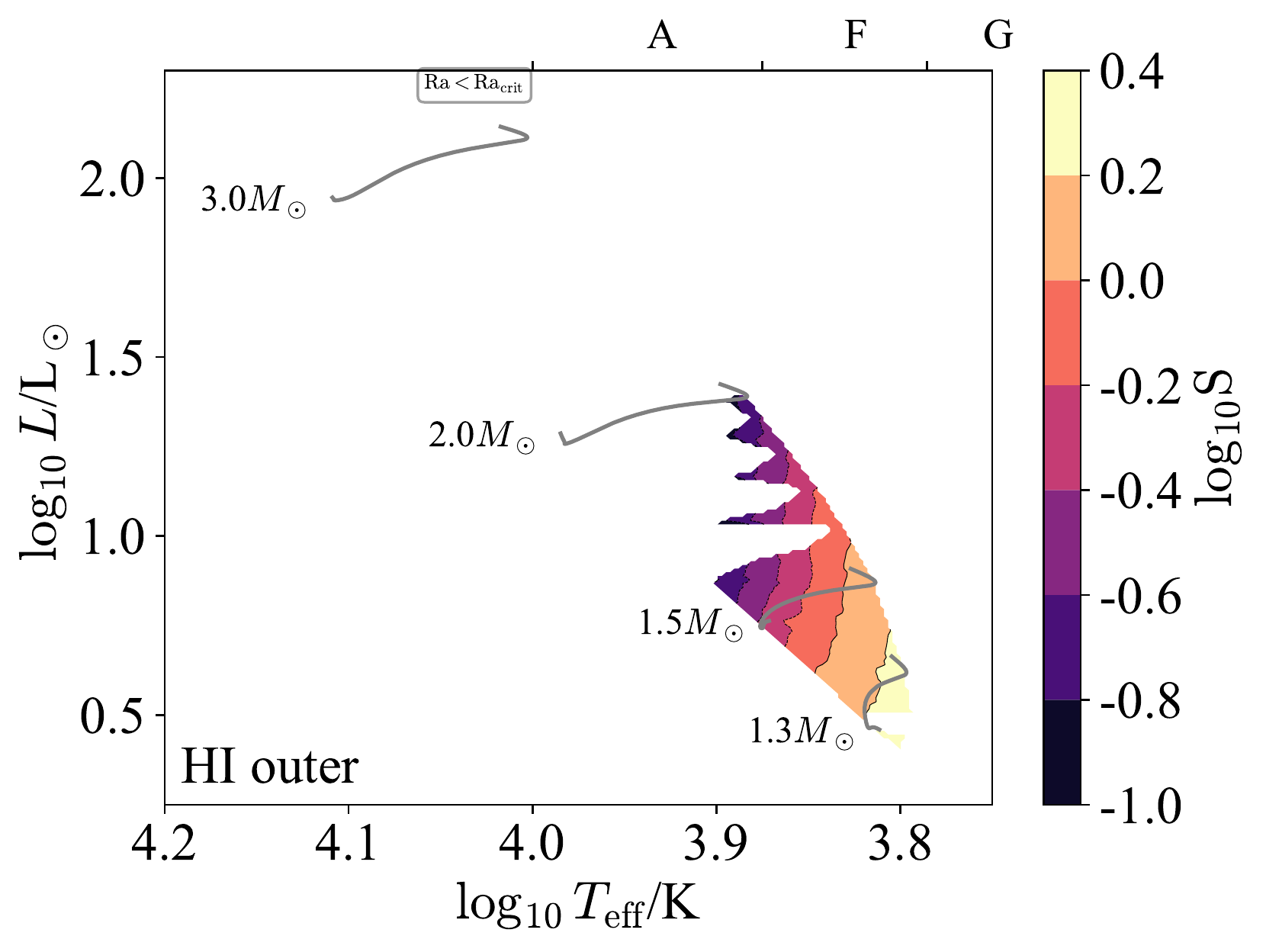}
\end{minipage}
\hfill

\caption{The stiffness of the inner (left) and outer (right) convective boundaries are shown in terms of $\log T_{\rm eff}$ and $\log L$ for stellar models with HI~CZs and Milky Way metallicity $Z=0.014$. Note that the stiffness is an output of a theory of convection and so is model-dependent. Regions with $\mathrm{Ra} < \mathrm{Ra}_{\rm crit}$ are stable against convection and shaded in grey.}
\label{fig:HI_stiff}
\end{figure*}

Finally, Figure~\ref{fig:HI_stiff} shows the stiffness of both the inner and outer boundaries of the HI CZ.
Both range from very stiff ($S \sim 10^{4-8}$) to very weak ($S \sim 1$), with decreasing stiffness towards decreasing mass.
For instance, for masses $M \ga 1.5 M_\odot$ we do not expect much mechanical overshooting, whereas for $M \la 1.5 M_\odot$ both boundaries should show substantial overshooting, because their low stiffness causes convective flows to decelerate over large length scales.

Note that at the inner boundary the stiffness shows sharp changes along evolutionary tracks.
This is because the emergence of the HeI CZ just below the HI CZ makes the typical radiative $N^2$ near the lower boundary of the HI CZ much smaller, thereby reducing the stiffness.

\clearpage
\subsection{HeI CZ}

We now examine the bulk structure of HeI CZs, which occur in the subsurface layers of stars with masses $2M_\odot \la M_\star \la 5 M_\odot$.
Note that in some regions of the HR diagram this convection zone has a Rayleigh number below the $\sim 10^3$ critical value~\citep{1961hhs..book.....C}.
As a result while the region is superadiabatic, it is not unstable to convection.
We therefore neglect these stable regions in our analysis, and shade them in grey in our figures.

\begin{figure*}
\centering
\begin{minipage}{0.48\textwidth}
\includegraphics[width=\textwidth]{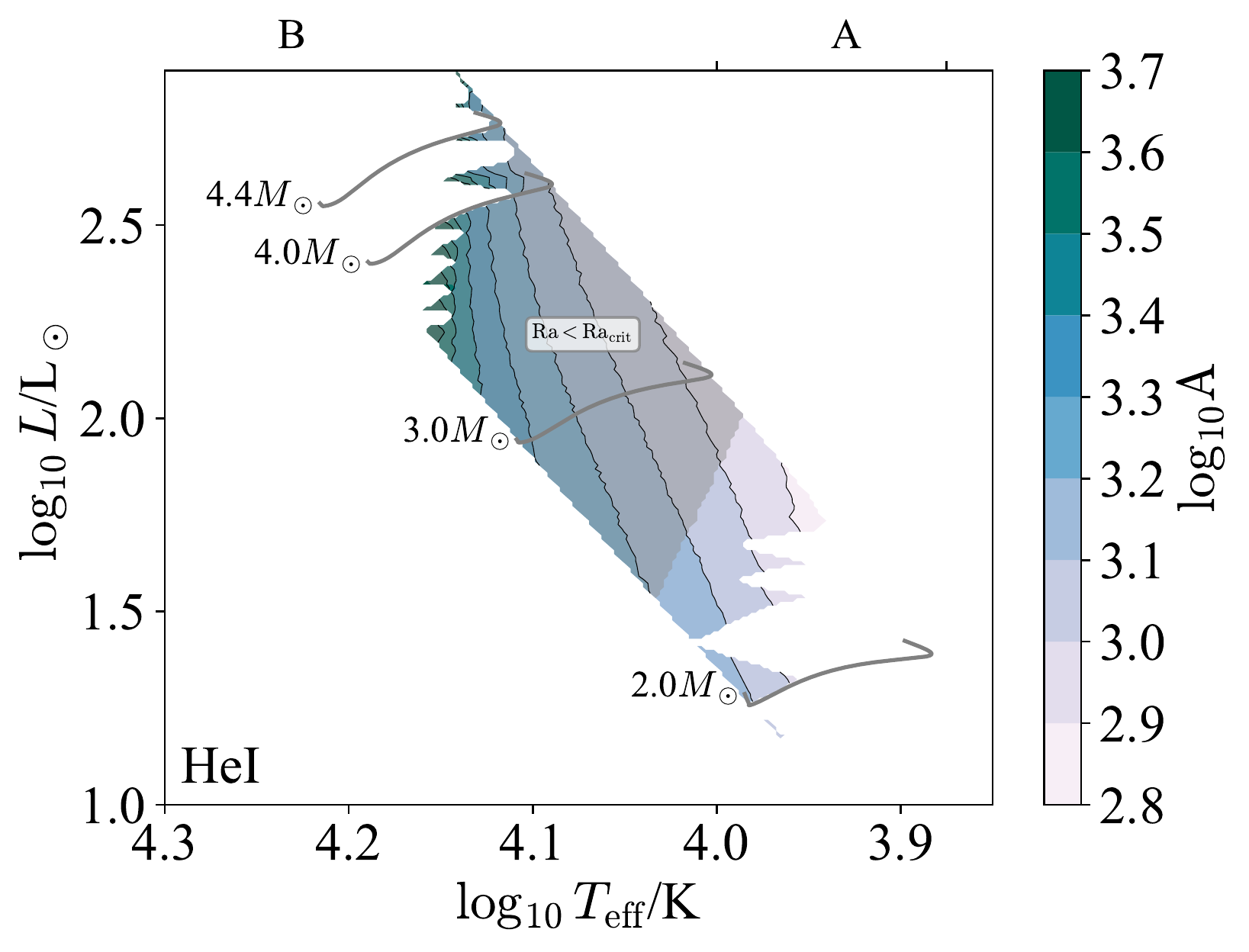}
\end{minipage}

\caption{The aspect ratio $\mathrm{A}$ is shown in terms of $\log T_{\rm eff}$/spectral type and $\log L$ for stellar models with HeI~CZs and Milky Way metallicity $Z=0.014$. Note that the aspect ratio is an input parameter, and does not depend on a specific theory of convection. Regions with $\mathrm{Ra} < \mathrm{Ra}_{\rm crit}$ are stable to convection and shaded in grey.}
\label{fig:HeI_structure}
\end{figure*}

Figure~\ref{fig:HeI_structure} shows the aspect ratio $\mathrm{A}$, which is of order $10^{3}$.
These large aspect ratios suggest that local simulations spanning the full depth of the CZ and only a fraction of $4\pi$ angularly can capture the convective dynamics.

\begin{figure*}
\begin{minipage}{0.48\textwidth}
\includegraphics[width=\textwidth]{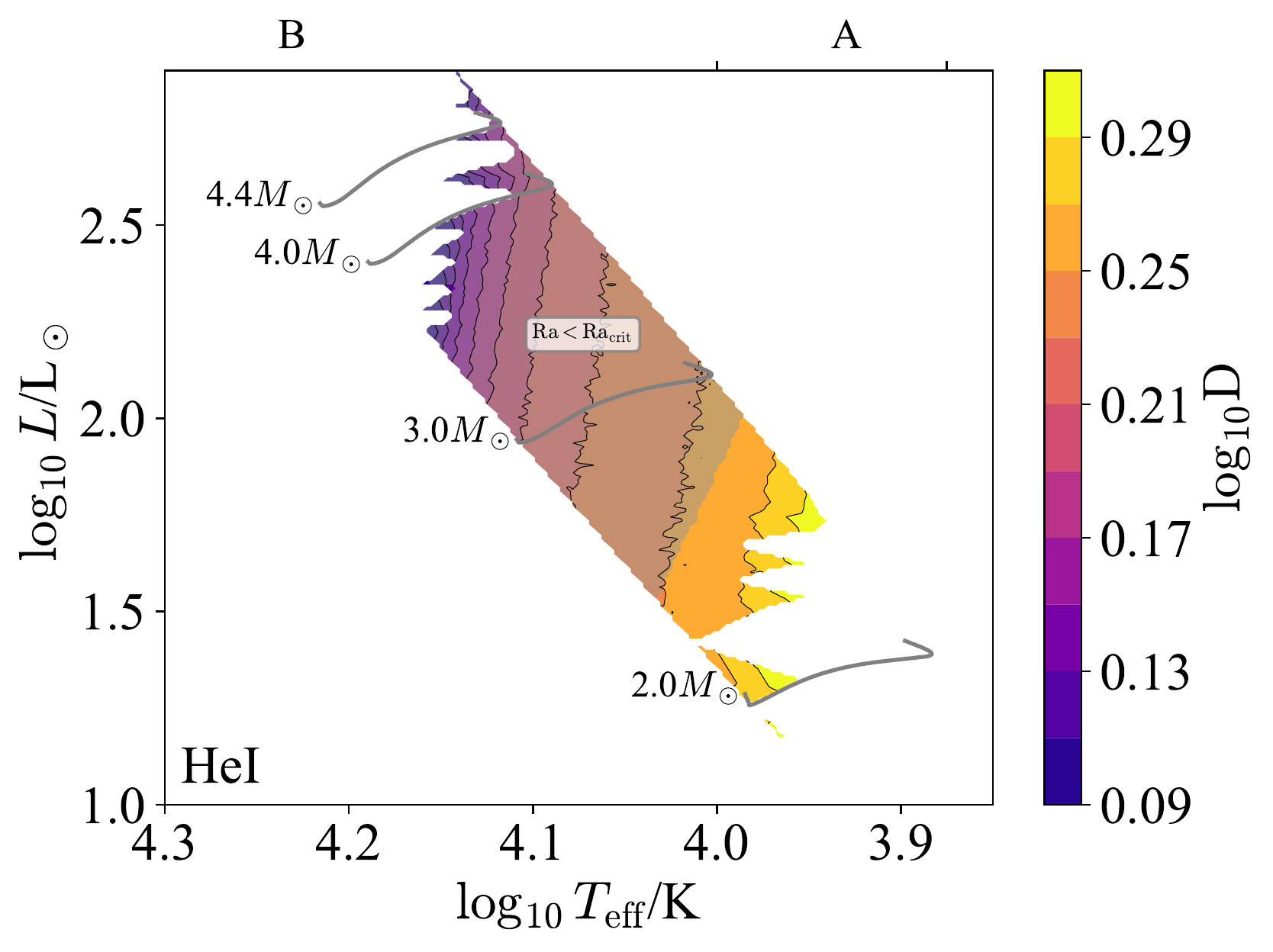}
\end{minipage}
\hfill
\begin{minipage}{0.48\textwidth}
\includegraphics[width=\textwidth]{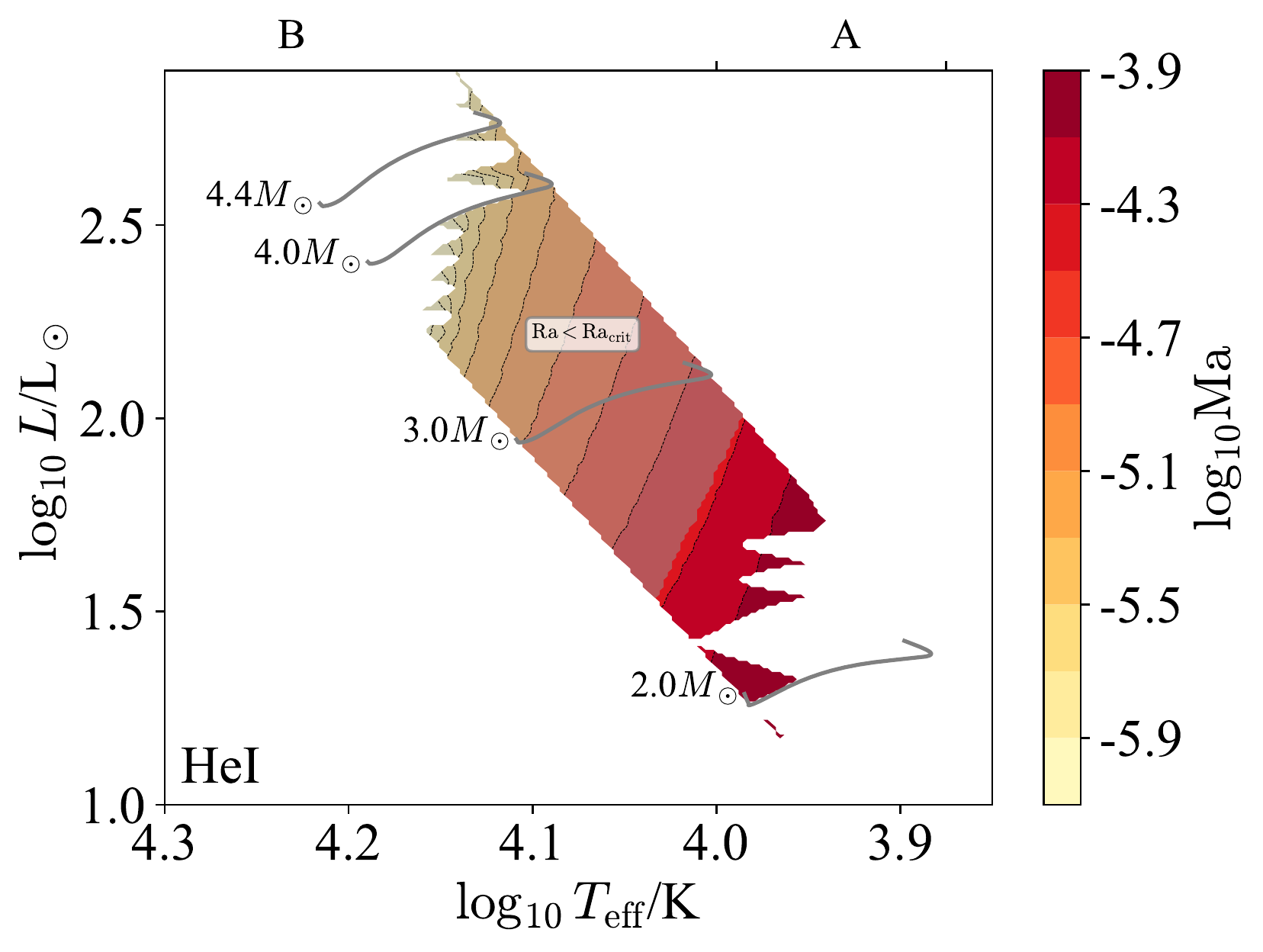}
\end{minipage}
\hfill

\caption{The density ratio $\mathrm{D}$ (left) and Mach number $\mathrm{Ma}$ (right) are shown in terms of $\log T_{\rm eff}$/spectral type and $\log L$ for stellar models with HeI~CZs and Milky Way metallicity $Z=0.014$. Note that while the density ratio is an input parameter and does not depend on a specific theory of convection, the Mach number is an output of such a theory and so is model-dependent. Regions with $\mathrm{Ra} < \mathrm{Ra}_{\rm crit}$ are stable to convection and shaded in grey.}
\label{fig:HeI_equations}
\end{figure*}

Next, the density ratio $\mathrm{D}$ (Figure~\ref{fig:HeI_equations}, left) and Mach number $\mathrm{Ma}$ (Figure~\ref{fig:HeI_equations}, right) inform which physics the fluid equations must include to model these zones.
The density ratio is always of order unity, and the Mach number is always small ($\la 10^{-4}$).
This suggests it is always appropriate to use the Boussinesq approximation in modelling these convection zones.

\begin{figure*}
\begin{minipage}{0.48\textwidth}
\includegraphics[width=\textwidth]{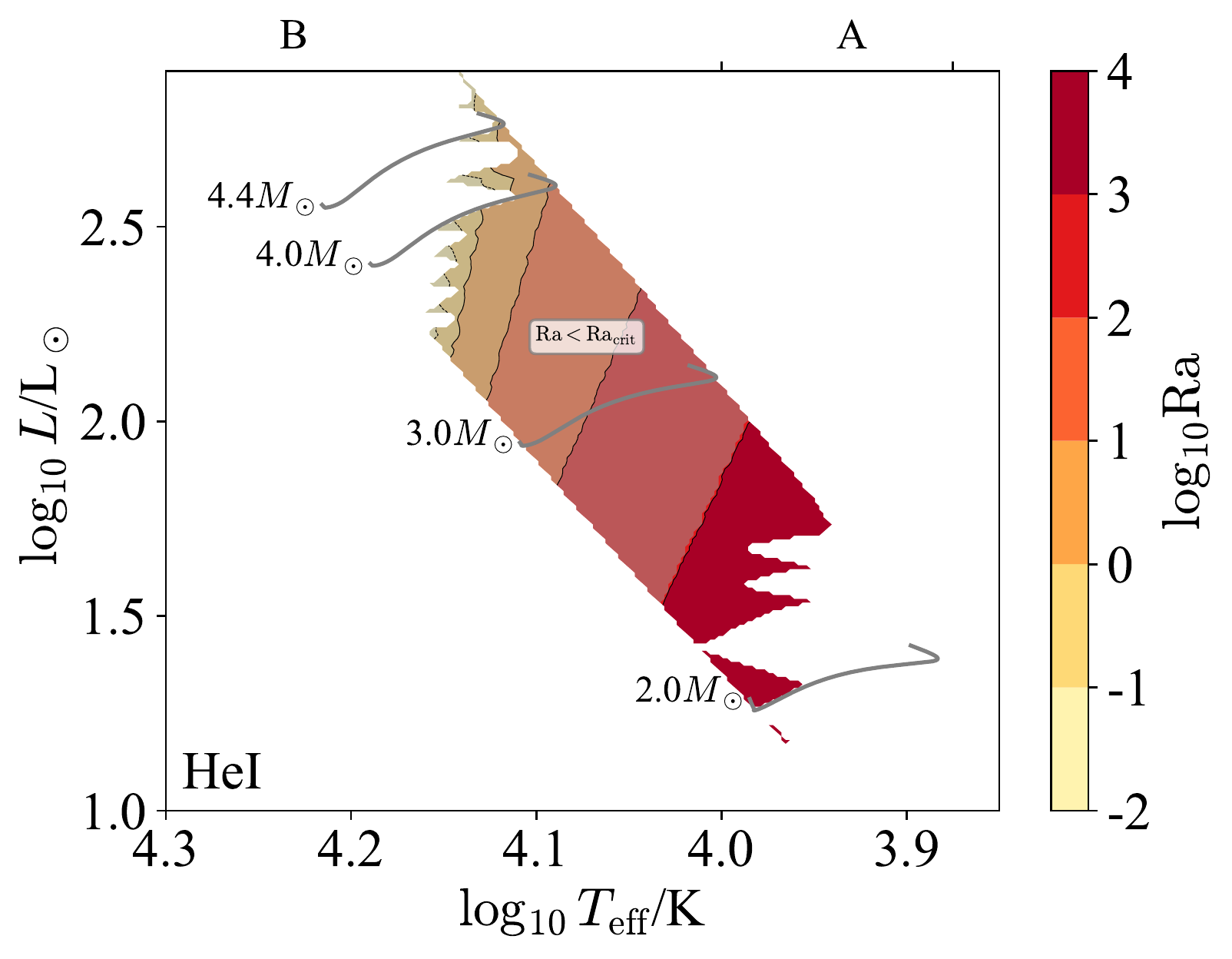}
\end{minipage}
\hfill
\begin{minipage}{0.48\textwidth}
\includegraphics[width=\textwidth]{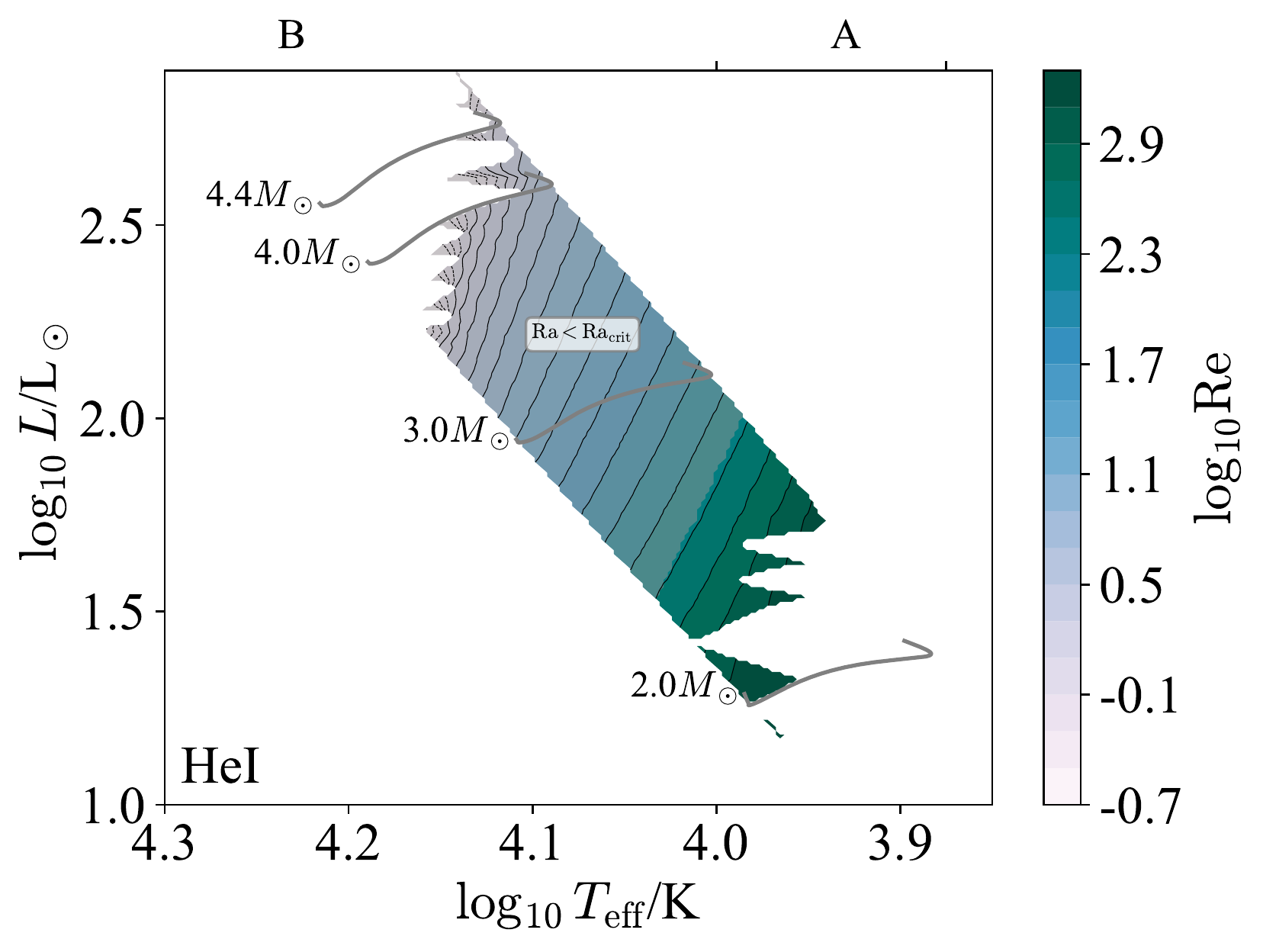}
\end{minipage}
\hfill

\caption{The Rayleigh number $\mathrm{Ra}$ (left) and Reynolds number $\mathrm{Re}$ (right) are shown in terms of $\log T_{\rm eff}$/spectral type and $\log L$ for stellar models with HeI~CZs and Milky Way metallicity $Z=0.014$.  Note that while the Rayleigh number is an input parameter and does not depend on a specific theory of convection, the Reynolds number is an output of such a theory and so is model-dependent. Regions with $\mathrm{Ra} < \mathrm{Ra}_{\rm crit}$ are stable to convection and shaded in grey.}
\label{fig:HeI_stability}
\end{figure*}

The Rayleigh number $\mathrm{Ra}$ (Figure~\ref{fig:HeI_stability}, left) determines whether or not a putative convection zone is actually unstable to convection, and the Reynolds number $\mathrm{Re}$ determines how turbulent the zone is if convection sets in (Figure~\ref{fig:HeI_stability}, right).
At low masses the Rayleigh number is slightly super-critical ($10^4-10^5$), at high masses it plummets and eventually becomes sub-critical, which we show in grey.
Likewise at low masses the Reynolds number is around the threshold for turbulence to develop ($\sim 10^3$) while at high masses it is quite small ($\sim 1)$.
These putative convection zones then span a wide range of properties, from being subcritical and \emph{stable} at high masses, to being weakly unstable and weakly turbulent at low masses ($\sim 2 M_\odot$).

\begin{figure*}
\centering
\begin{minipage}{0.48\textwidth}
\includegraphics[width=\textwidth]{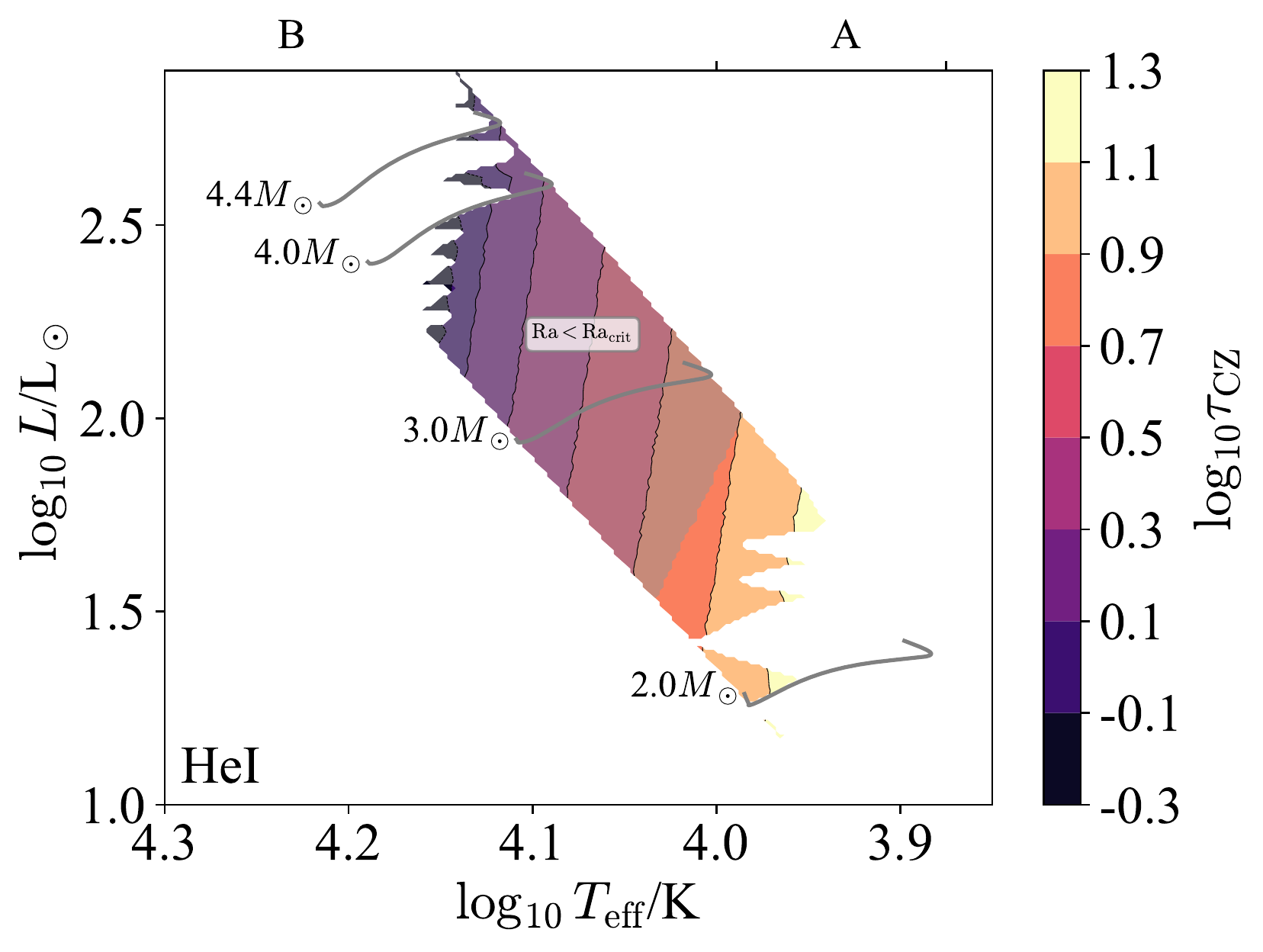}
\end{minipage}
\hfill
\begin{minipage}{0.48\textwidth}
\includegraphics[width=\textwidth]{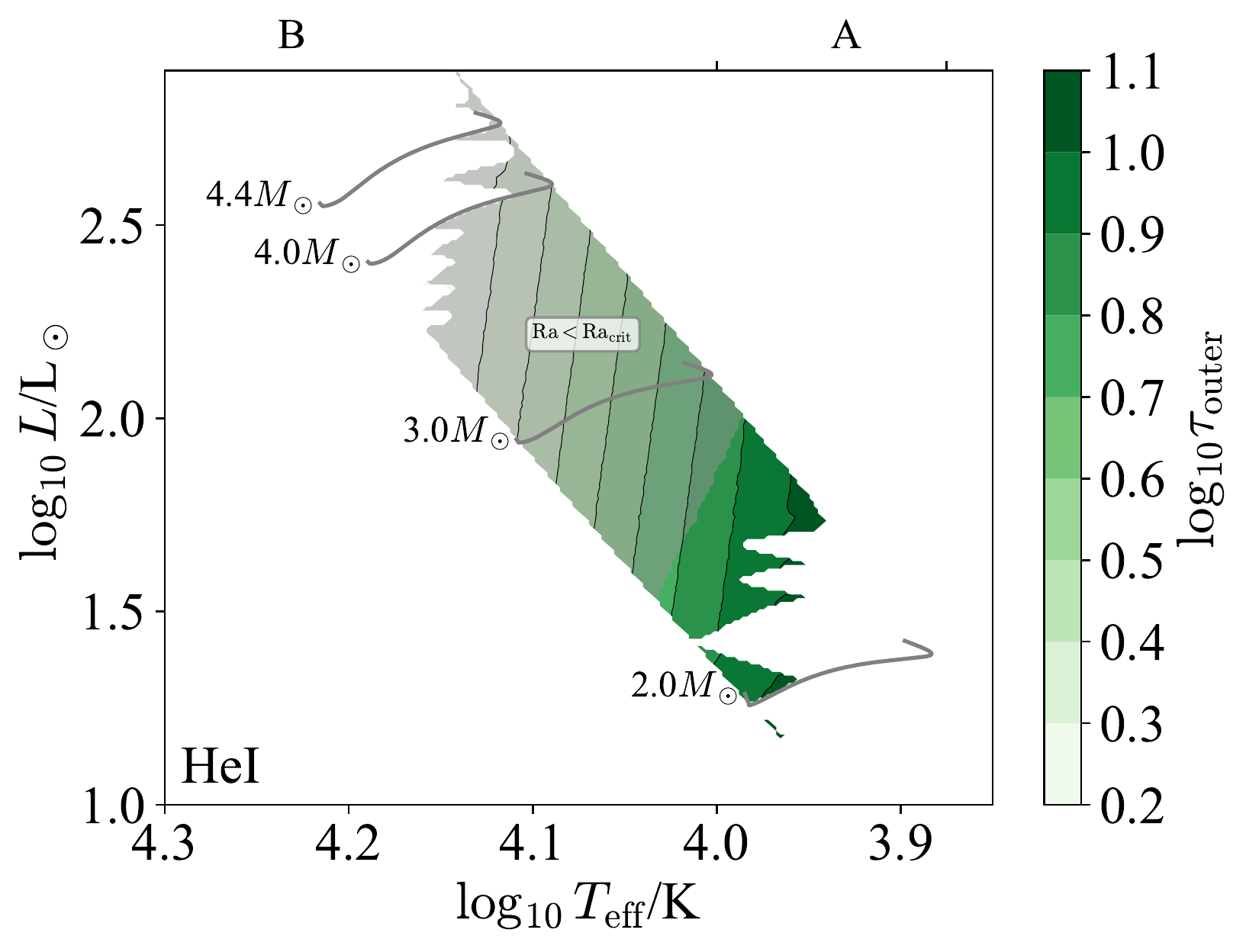}
\end{minipage}
\hfill
\caption{The convection optical depth $\tau_{\rm CZ}$ (left) and the optical depth to the surface $\tau_{\rm outer}$ (right) are shown in terms of $\log T_{\rm eff}$/spectral type and $\log L$ for stellar models with HeI~CZs and Milky Way metallicity $Z=0.014$. Note that both of these are input parameters, and do not depend on a specific theory of convection. Regions with $\mathrm{Ra} < \mathrm{Ra}_{\rm crit}$ are stable to convection and shaded in grey.}
\label{fig:HeI_optical}
\end{figure*}

The optical depth across a convection zone $\tau_{\rm CZ}$ (Figure~\ref{fig:HeI_optical}, left) indicates whether or not radiation can be handled in the diffusive approximation, while the optical depth from the outer boundary to infinity $\tau_{\rm outer}$ (Figure~\ref{fig:HeI_optical}, right) indicates the nature of radiative transfer and cooling in the outer regions of the convection zone.
The surface of the HeI CZ is always at moderate low optical depth in the unstable region ($\tau_{\rm outer} \sim 10$), and the optical depth across the HeI CZ is of the same order.
This means that both the bulk and outer boundary of the HeI CZ can be treated within the diffusive approximation for radiation.

\begin{figure*}
\centering
\begin{minipage}{0.48\textwidth}
\includegraphics[width=\textwidth]{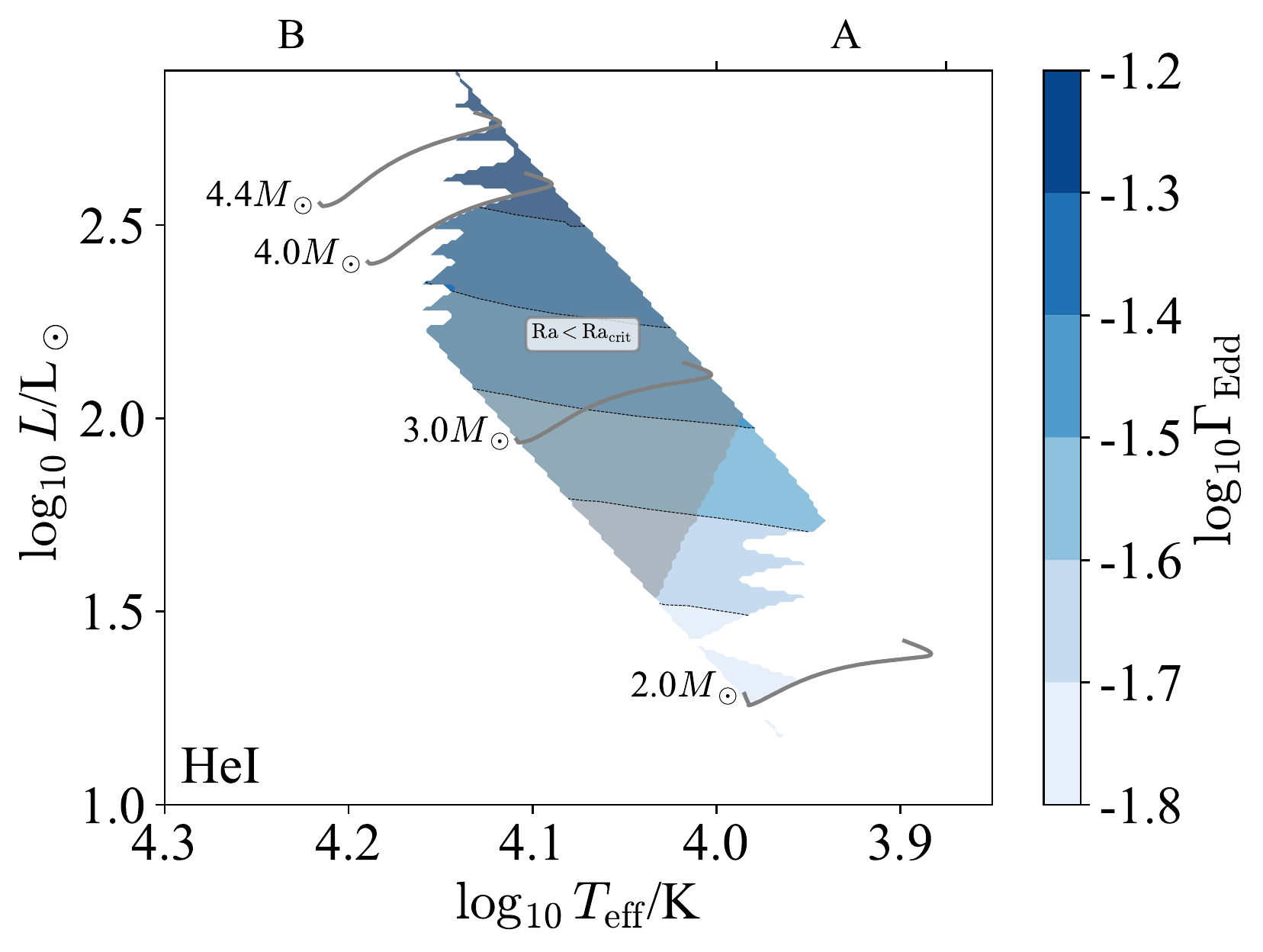}
\end{minipage}
\hfill
\begin{minipage}{0.48\textwidth}
\includegraphics[width=\textwidth]{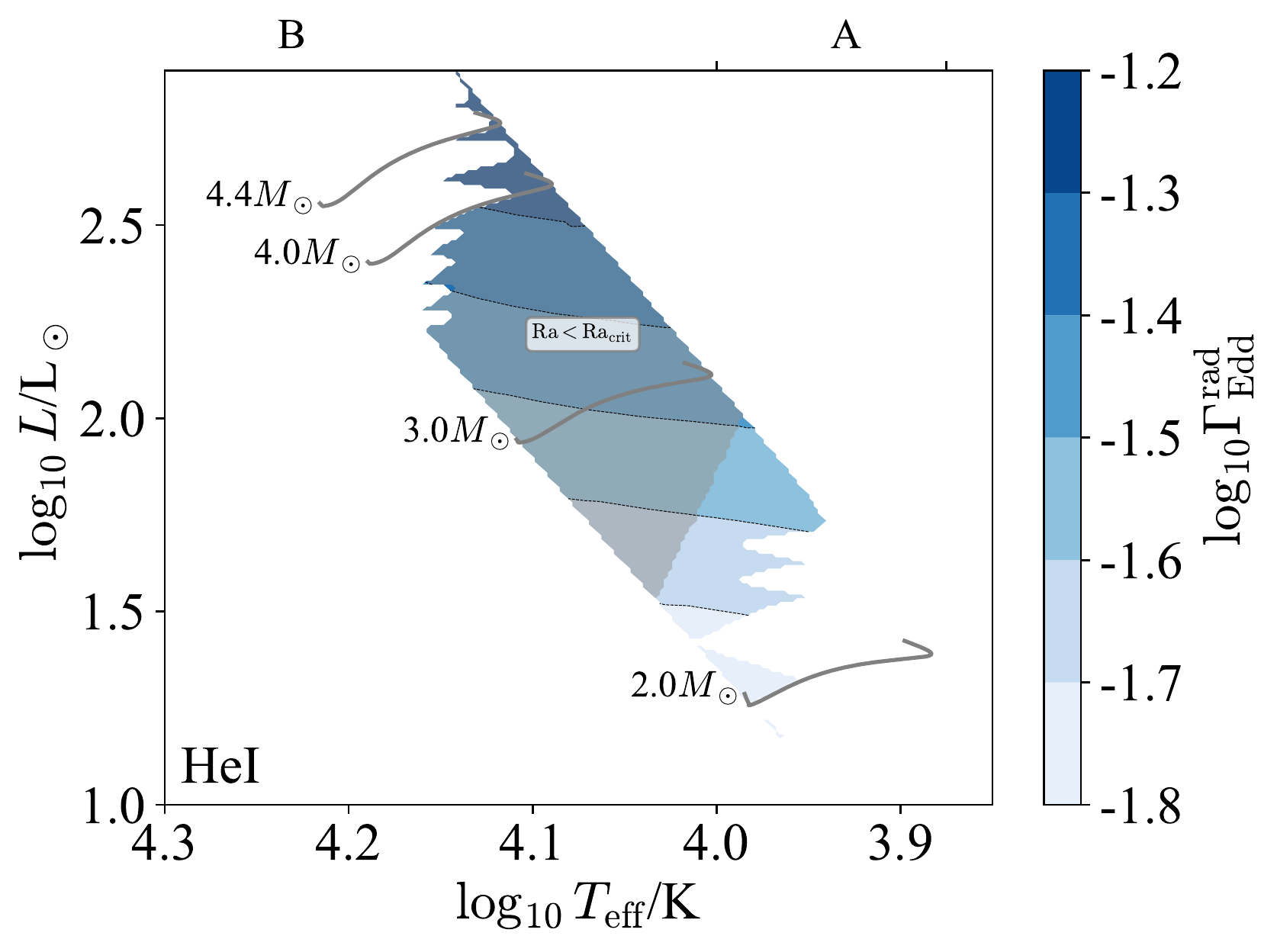}
\end{minipage}
\hfill
\caption{The Eddington ratio with the full luminosity $\Gamma_{\rm Edd}$ (left) and the radiative luminosity (right) are shown in terms of $\log T_{\rm eff}$/spectral type and $\log L$ for stellar models with HeI~CZs and Milky Way metallicity $Z=0.014$. Note that while $\Gamma_{\rm Edd}$ is an input parameter and does not depend on a specific theory of convection, $\Gamma_{\rm Edd}^{\rm rad}$ is an output of such a theory and so is model-dependent. Regions with $\mathrm{Ra} < \mathrm{Ra}_{\rm crit}$ are stable to convection and shaded in grey. Note that the two panels appear very similar because HeI~CZs are so inefficient that they transport almost no flux, and so the radiative temperature gradient is very similar to the realized temperature gradient in MLT.}
\label{fig:HeI_eddington}
\end{figure*}

The Eddington ratio $\Gamma_{\rm Edd}$ (Figure~\ref{fig:HeI_eddington}, left) indicates whether or not radiation hydrodynamic instabilities are important in the non-convecting state, and the radiative Eddington ratio $\Gamma_{\rm Edd}^{\rm rad}$ (Figure~\ref{fig:HeI_eddington}, right) indicates the same in the developed convective state.
Both ratios are small in the HeI CZ, so radiation hydrodynamic instabilities are unlikely to matter.

\begin{figure*}
\centering
\begin{minipage}{0.48\textwidth}
\includegraphics[width=\textwidth]{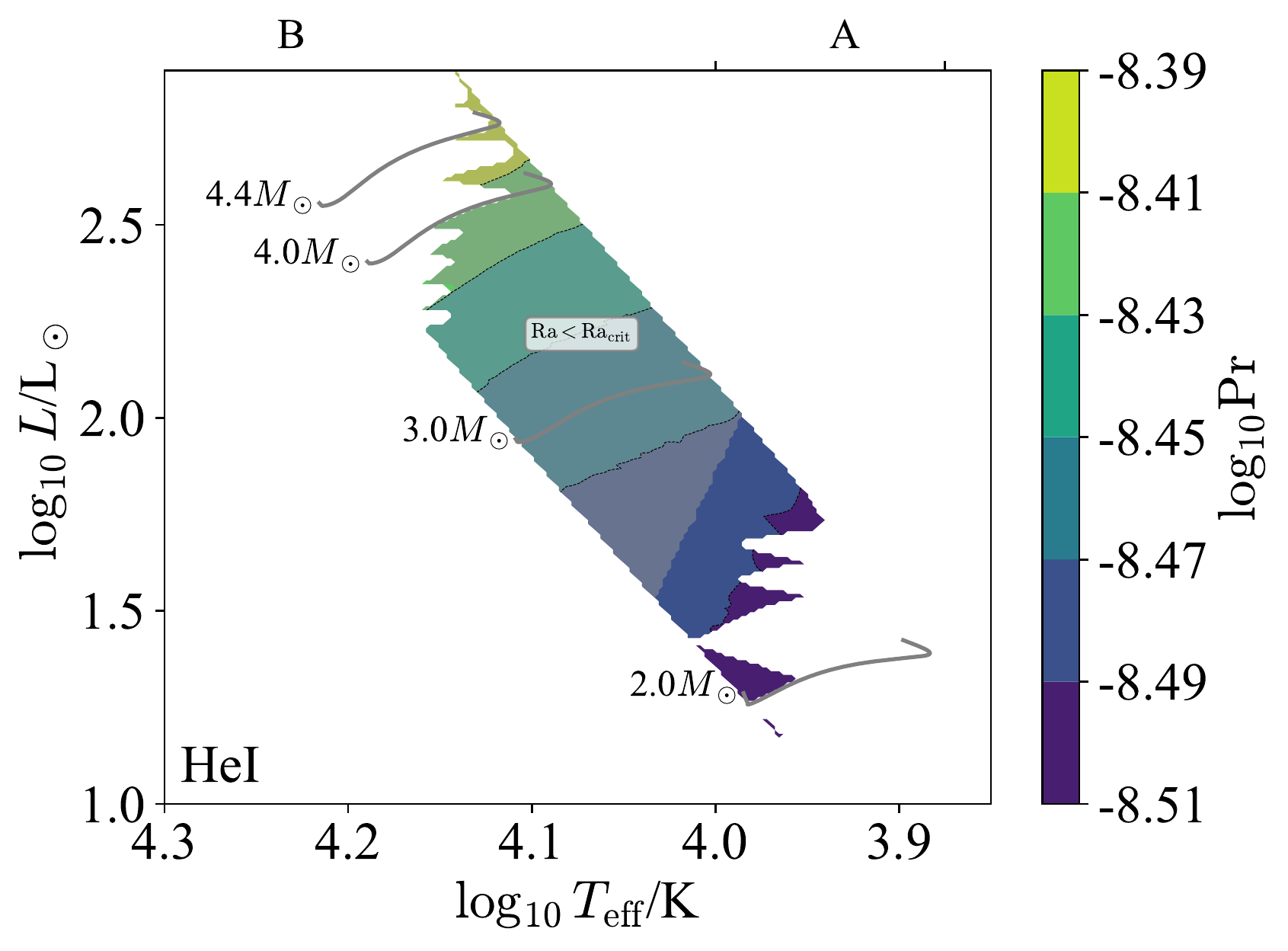}
\end{minipage}
\hfill
\begin{minipage}{0.48\textwidth}
\includegraphics[width=\textwidth]{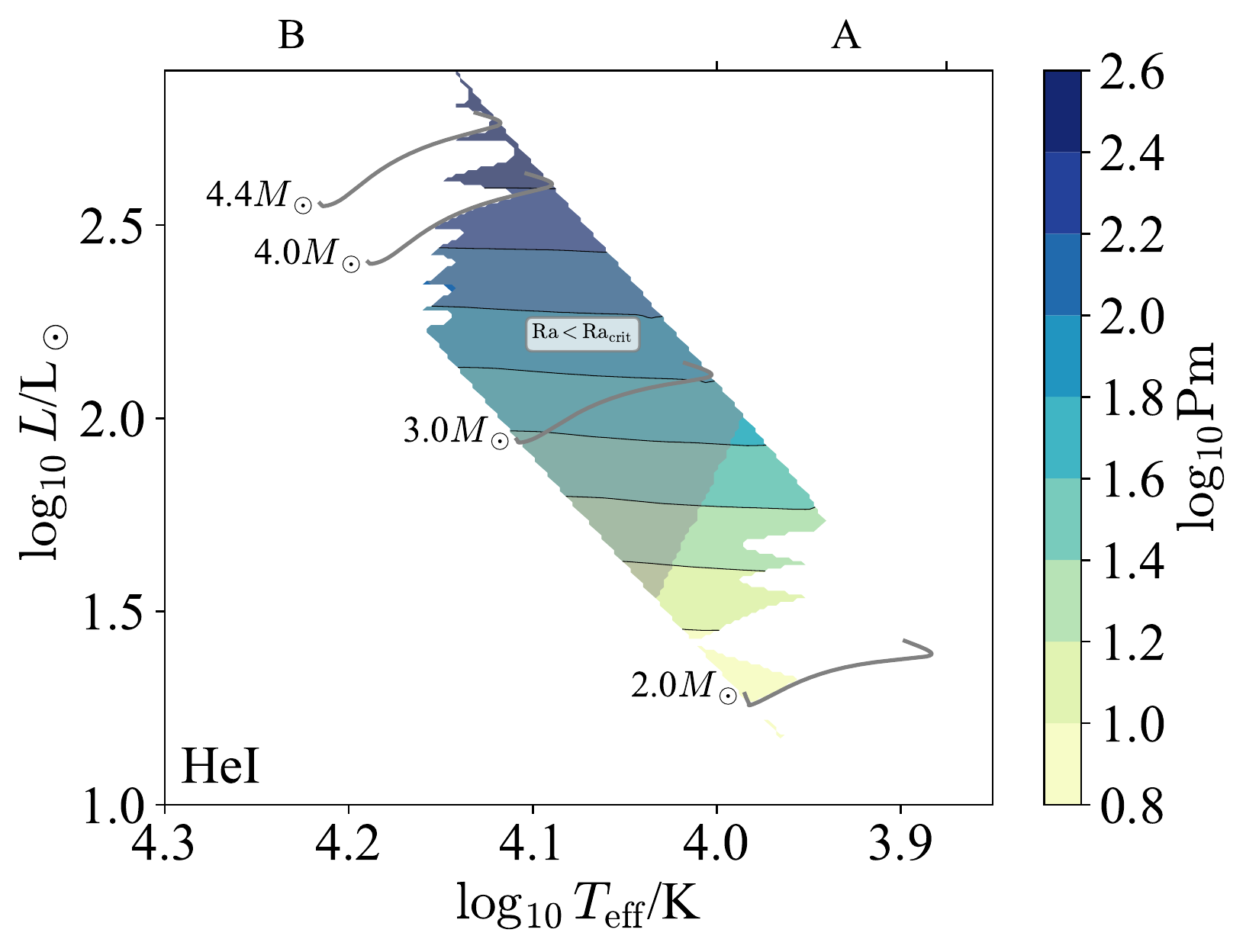}
\end{minipage}
\hfill

\caption{The Prandtl number $\mathrm{Pr}$ (left) and magnetic Prandtl number $\mathrm{Pm}$ (right) are shown in terms of $\log T_{\rm eff}$/spectral type and $\log L$ for stellar models with HeI~CZs and Milky Way metallicity $Z=0.014$. Note that both $\mathrm{Pr}$ and $\mathrm{Pm}$ are input parameters, and so do not depend on a specific theory of convection. Regions with $\mathrm{Ra} < \mathrm{Ra}_{\rm crit}$ are stable to convection and shaded in grey.}
\label{fig:HeI_diffusivities}
\end{figure*}

The Prandtl number $\mathrm{Pr}$ (Figure~\ref{fig:HeI_diffusivities}, left) measures the relative importance of thermal diffusion and viscosity, and the magnetic Prandtl number $\mathrm{Pm}$ (Figure~\ref{fig:HeI_diffusivities}, right) measures the same for magnetic diffusion and viscosity.
The Prandtl number is always small in these models, so the thermal diffusion length-scale is much larger than the viscous scale.
By contrast, the magnetic Prandtl number is always greater than unity, and reaches nearly $10^2$ in the unstable regions.

The fact that $\mathrm{Pm}$ is large is notable because the quasistatic approximation for magnetohydrodynamics has frequently been used to study magnetoconvection in minimal 3D MHD simulations of planetary and stellar interiors~\citep[e.g.][]{yan_calkins_maffei_julien_tobias_marti_2019} and assumes that $\mathrm{Rm} = \mathrm{Pm} \mathrm{Re} \rightarrow 0$; in doing so, this approximation assumes a global background magnetic field is dominant and neglects the nonlinear portion of the Lorentz force. This approximation breaks down in convection zones with $\mathrm{Pm} > 1$ and future numerical experiments should seek to understand how magnetoconvection operates in this regime.

\begin{figure*}
\centering
\begin{minipage}{0.48\textwidth}
\includegraphics[width=\textwidth]{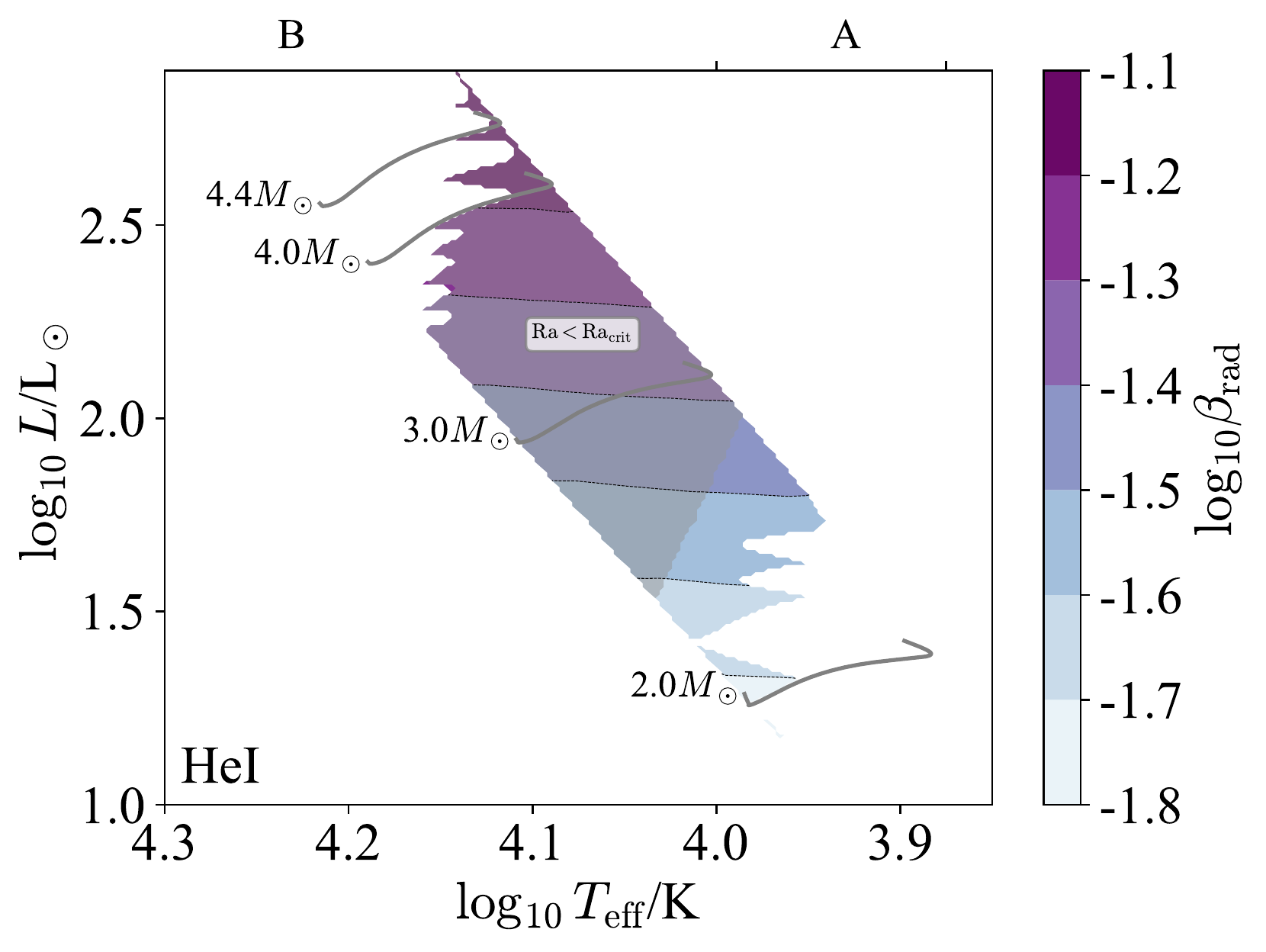}
\end{minipage}
\hfill

\caption{The radiation pressure ratio $\beta_{\rm rad}$ is shown in terms of $\log T_{\rm eff}$/spectral type and $\log L$ for stellar models with HeI~CZs and Milky Way metallicity $Z=0.014$. Note that this ratio is an input parameter, and does not depend on a specific theory of convection. Regions with $\mathrm{Ra} < \mathrm{Ra}_{\rm crit}$ are stable to convection and shaded in grey.}
\label{fig:HeI_beta}
\end{figure*}

The radiation pressure ratio $\beta_{\rm rad}$ (Figure~\ref{fig:HeI_beta}) measures the importance of radiation in setting the thermodynamic properties of the fluid.
We see that this is uniformly small ($\la 0.1$) and so radiation pressure likely plays a sub-dominant role in these zones.

\begin{figure*}
\centering
\begin{minipage}{0.48\textwidth}
\includegraphics[width=\textwidth]{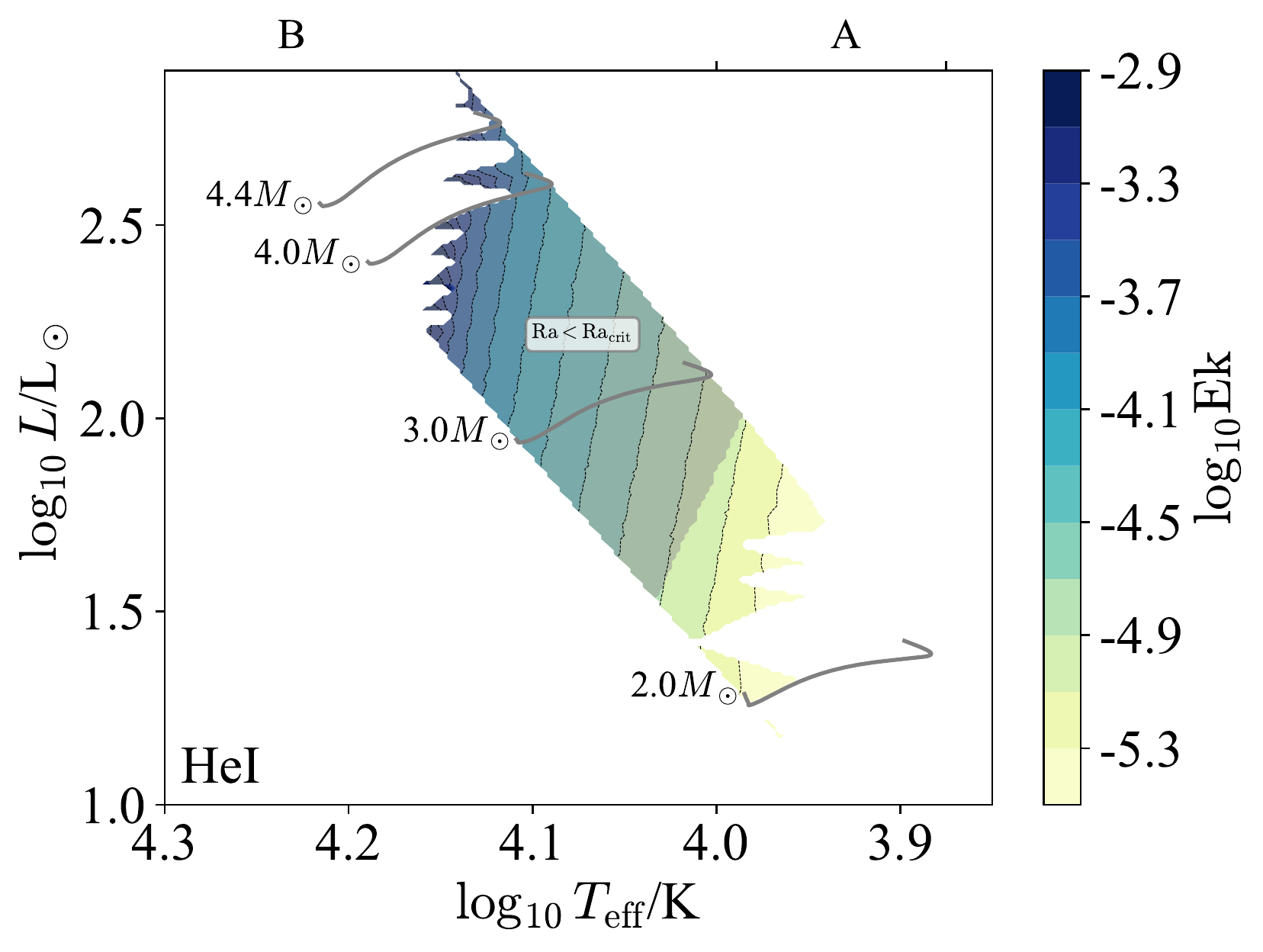}
\end{minipage}

\caption{The Ekman number $\mathrm{Ek}$ is shown in terms of $\log T_{\rm eff}$/spectral type and $\log L$ for stellar models with HeI~CZs and Milky Way metallicity $Z=0.014$. Note that the Ekman number is an input parameter, and does not depend on a specific theory of convection. Regions with $\mathrm{Ra} < \mathrm{Ra}_{\rm crit}$ are stable to convection and shaded in grey.}
\label{fig:HeI_ekman}
\end{figure*}

The Ekman number $\mathrm{Ek}$ (Figure~\ref{fig:HeI_ekman}) indicates the relative importance of viscosity and rotation.
This is tiny across the HRD~\footnote{Note that, because the Prandtl number is also very small, this does not significantly alter the critical Rayleigh number~(see Ch3 of~\cite{1961hhs..book.....C} and appendix D of~\cite{2022arXiv220110567J}).}, so we expect rotation to dominate over viscosity, except at very small length-scales.

\begin{figure*}
\centering
\begin{minipage}{0.48\textwidth}
\includegraphics[width=\textwidth]{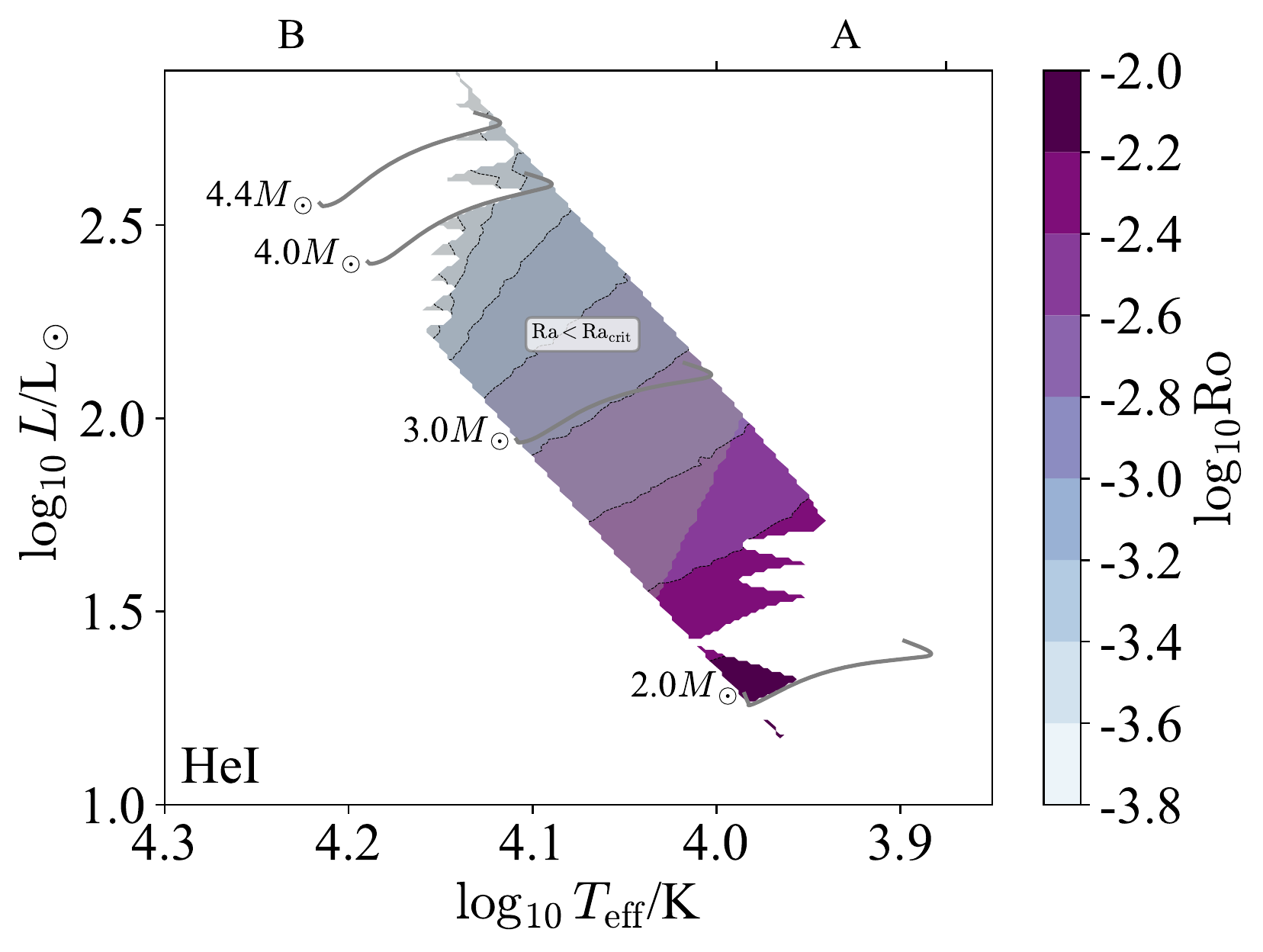}
\end{minipage}
\hfill
\begin{minipage}{0.48\textwidth}
\includegraphics[width=\textwidth]{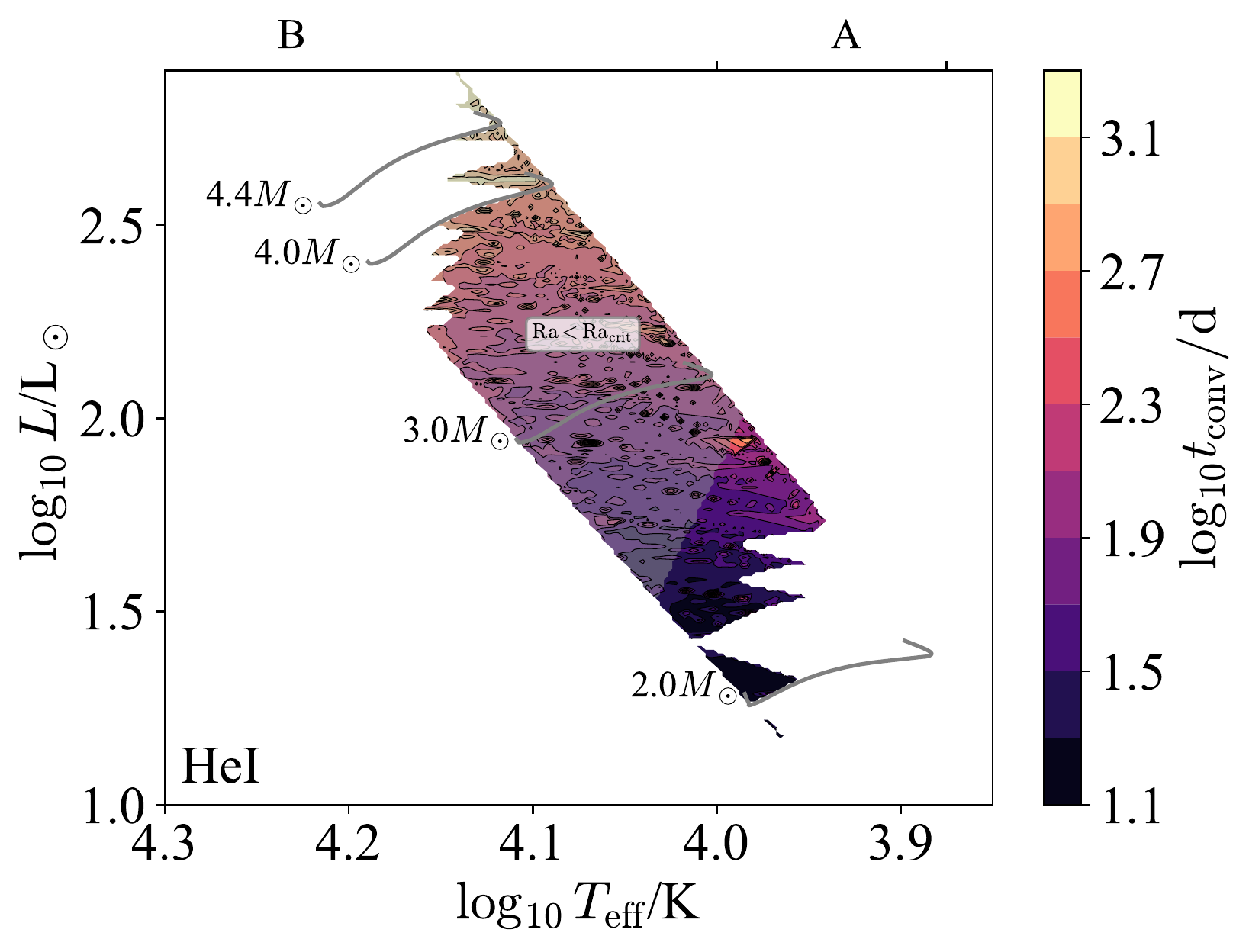}
\end{minipage}
\hfill

\caption{The Rossby number $\mathrm{Ro}$ (left) and turnover time $t_{\rm conv}$ (right) are shown in terms of $\log T_{\rm eff}$/spectral type and $\log L$ for stellar models with HeI~CZs and Milky Way metallicity $Z=0.014$. Note that both $\mathrm{Ro}$ and $t_{\rm conv}$ are outputs of a theory of convection and so are model-dependent. Regions with $\mathrm{Ra} < \mathrm{Ra}_{\rm crit}$ are stable to convection and shaded in grey. Note that the turnover time exhibits numerical noise related to the model mesh resolution because the integrand $1/v_c$ diverges towards the convective boundaries.}
\label{fig:HeI_rotation}
\end{figure*}

The Rossby number $\mathrm{Ro}$ (Figure~\ref{fig:HeI_rotation}, left) measures the relative importance of rotation and inertia.
This is uniformly small, meaning that the HeI CZ is rotationally constrained for typical rotation rates~\citep{2013A&A...557L..10N}.

We have assumed a fiducial rotation law to calculate $\mathrm{Ro}$.
Stars exhibit a variety of different rotation rates, so we also show the convective turnover time $t_{\rm conv}$ (Figure~\ref{fig:HeI_rotation}, right) which may be used to estimate the Rossby number for different rotation periods.

\begin{figure*}
\centering
\begin{minipage}{0.48\textwidth}
\includegraphics[width=\textwidth]{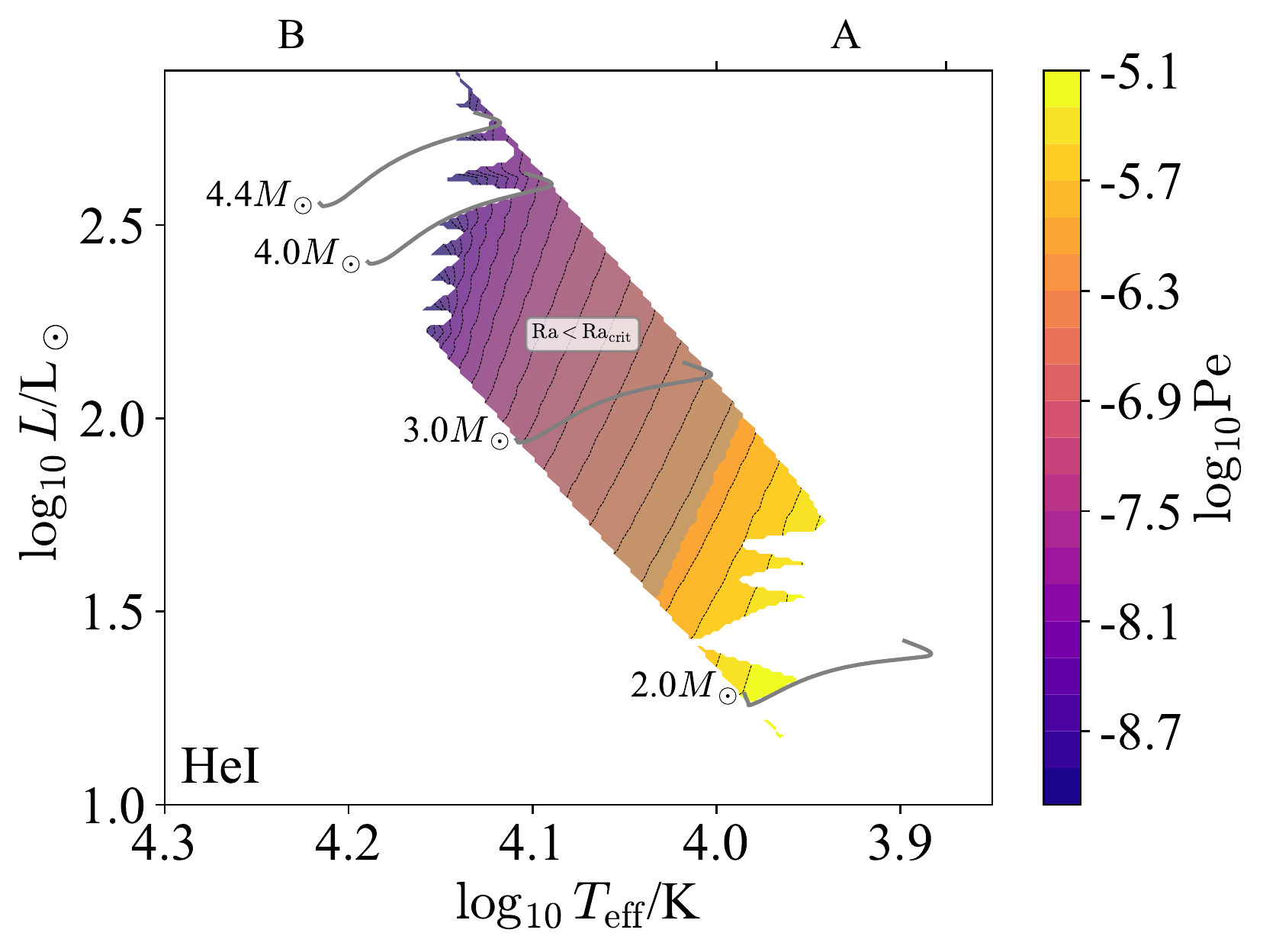}
\end{minipage}
\hfill
\begin{minipage}{0.48\textwidth}
\includegraphics[width=\textwidth]{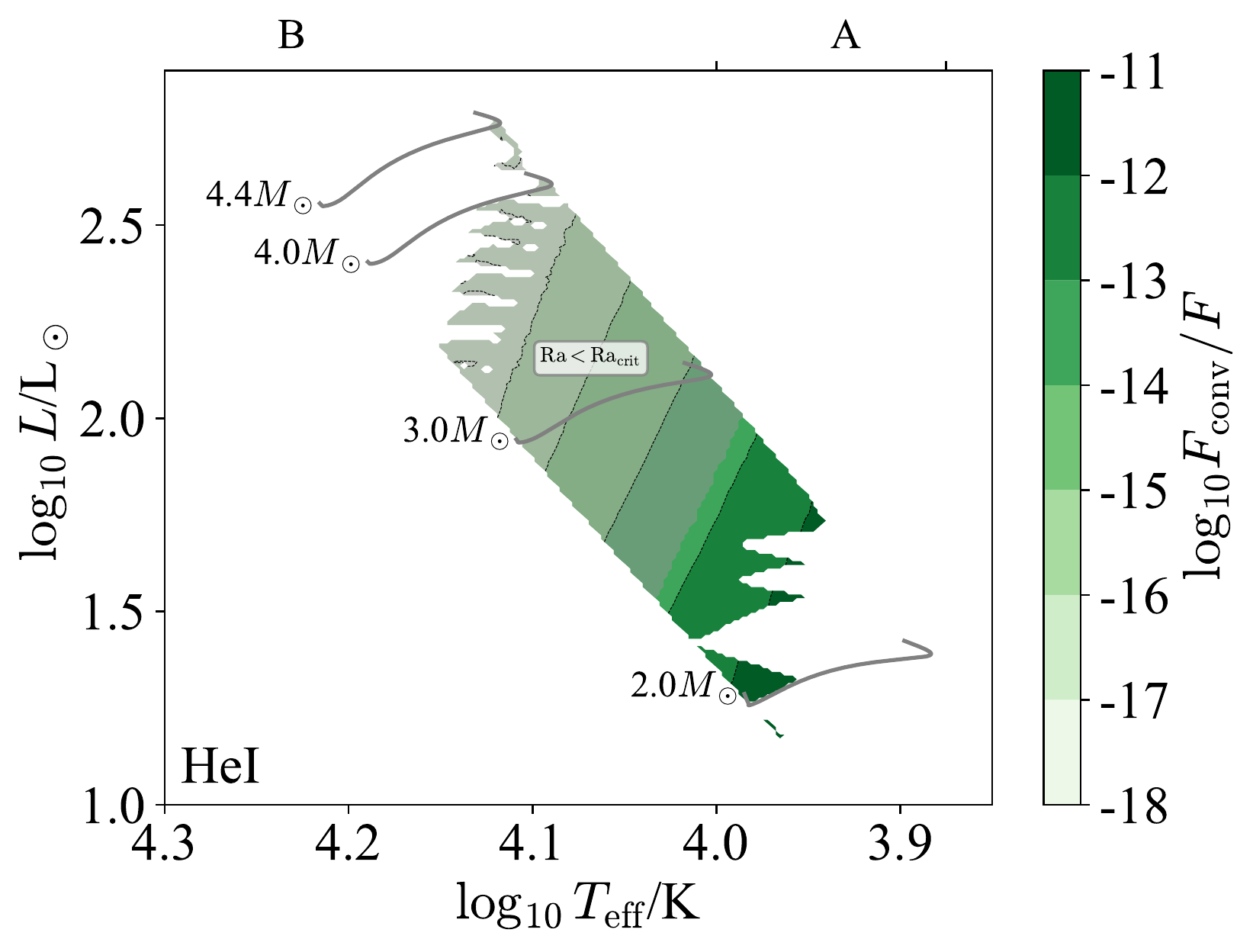}
\end{minipage}

\caption{The P{\'e}clet number $\mathrm{Pe}$ (left) and $F_{\rm conv}/F$ (right) are shown in terms of $\log T_{\rm eff}$/spectral type and $\log L$ for stellar models with HeI~CZs and Milky Way metallicity $Z=0.014$. Note that both $\mathrm{Pe}$ and $F_{\rm conv}/F$ are outputs of a theory of convection and so are model-dependent. Regions with $\mathrm{Ra} < \mathrm{Ra}_{\rm crit}$ are stable to convection and shaded in grey.}
\label{fig:HeI_efficiency}
\end{figure*}

The P{\'e}clet number $\mathrm{Pe}$ (Figure~\ref{fig:HeI_efficiency}, left) measures the relative importance of advection and diffusion in transporting heat, and the flux ratio $F_{\rm conv}/F$ (Figure~\ref{fig:HeI_efficiency}, right) reports the fraction of the energy flux which is advected.
Both are extremely small, meaning that these convection zones are very inefficient.

\begin{figure*}
\centering
\begin{minipage}{0.48\textwidth}
\includegraphics[width=\textwidth]{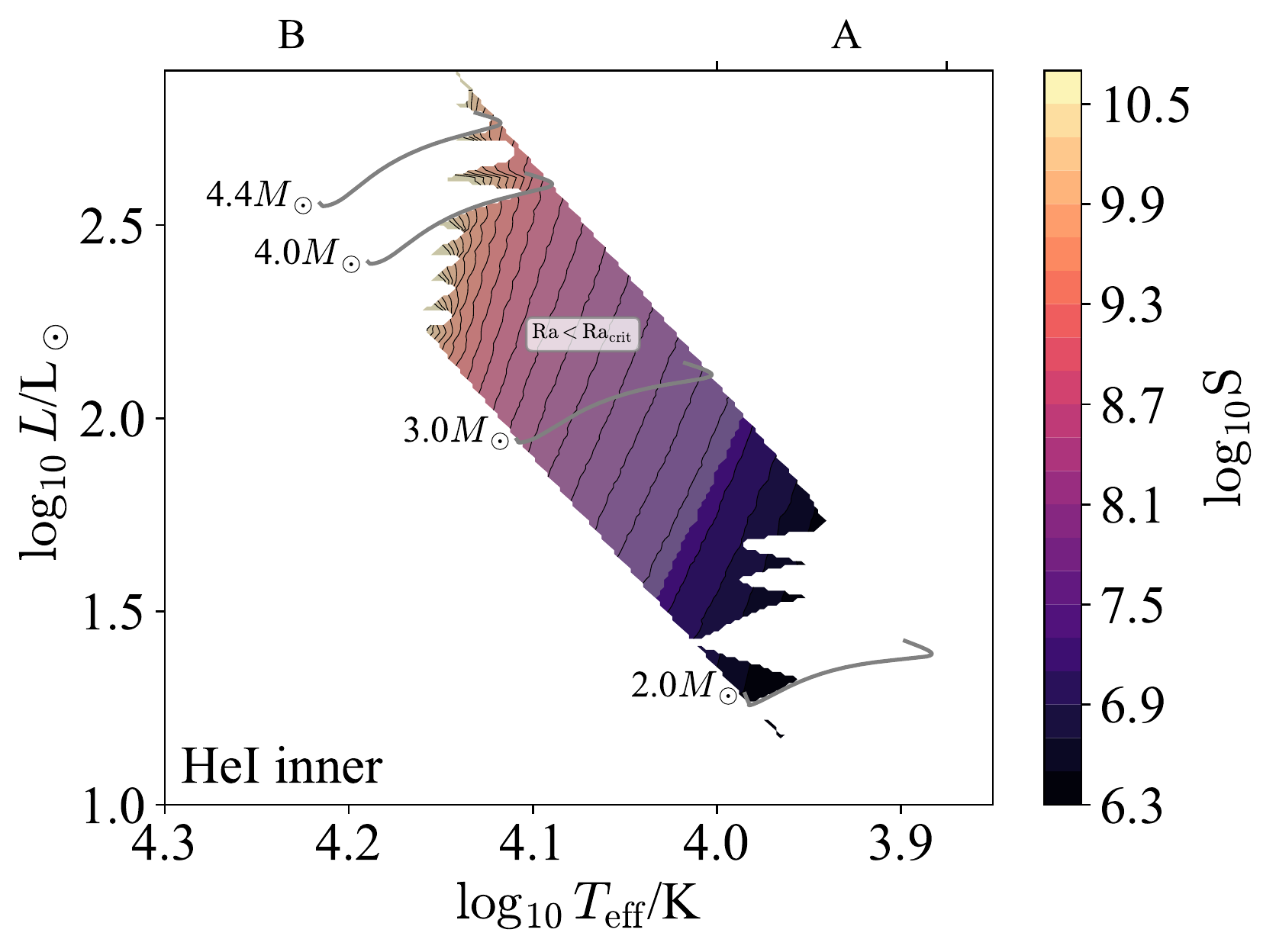}
\end{minipage}
\hfill
\begin{minipage}{0.48\textwidth}
\includegraphics[width=\textwidth]{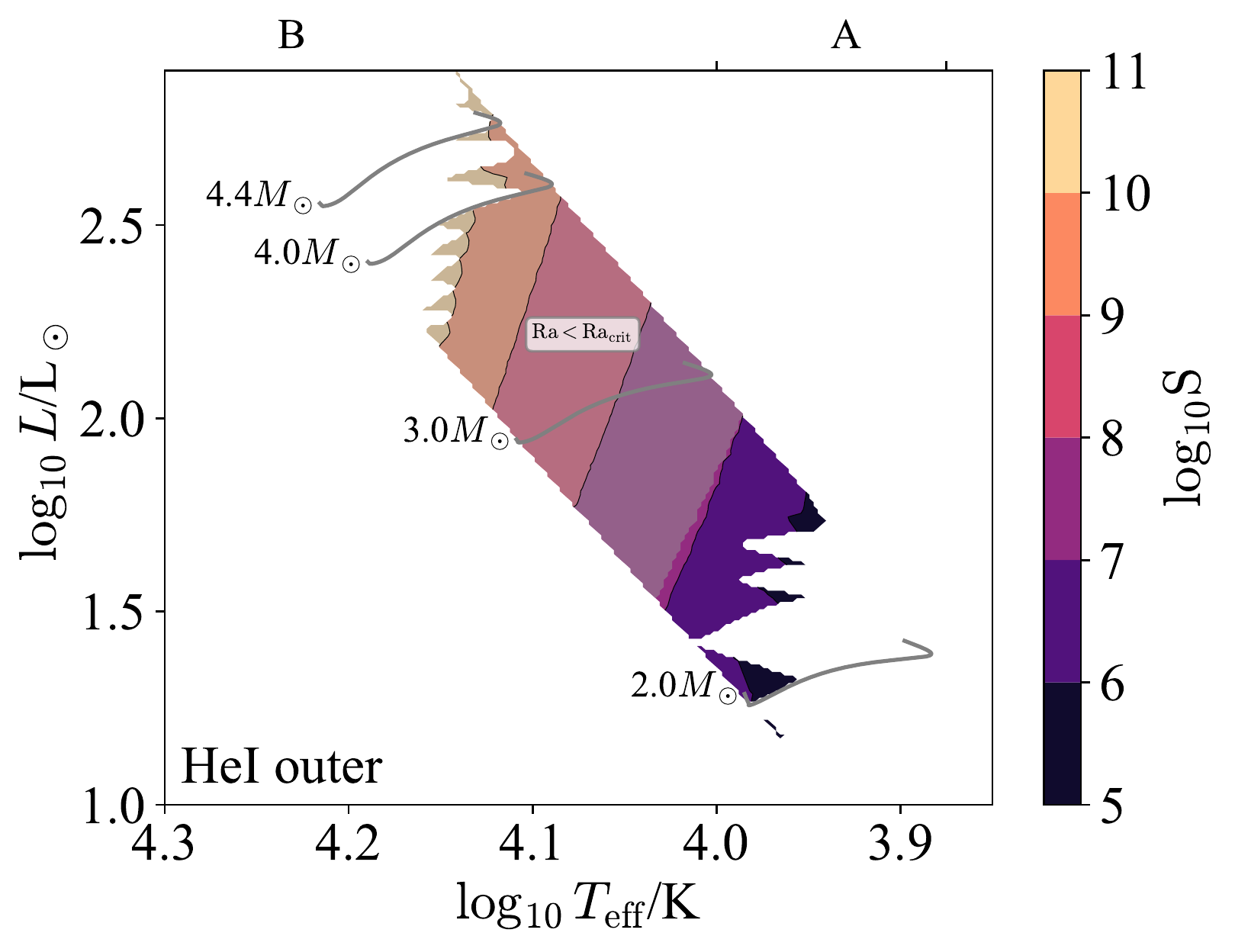}
\end{minipage}
\hfill

\caption{The stiffness of the inner (left) and outer (right) convective boundaries are shown in terms of $\log T_{\rm eff}$ and $\log L$ for stellar models with HeI~CZs and Milky Way metallicity $Z=0.014$. Note that the stiffness is an output of a theory of convection and so is model-dependent. Regions with $\mathrm{Ra} < \mathrm{Ra}_{\rm crit}$ are stable to convection and shaded in grey.}
\label{fig:HeI_stiff}
\end{figure*}

Finally, Figure~\ref{fig:HeI_stiff} shows the stiffness of both the inner and outer boundaries of the HeI CZ.
Both are very stiff at all masses ($S \sim 10^{5-10}$), so we do not expect much mechanical overshooting, though there could still well be convective penetration~\citep{2021arXiv211011356A}.

\clearpage
\subsection{HeII CZ}

We now examine the bulk structure of HeII CZs, which occur in the subsurface layers of stars with masses $1.3M_\odot \la M_\star \la 60 M_\odot$
Note that in some regions of the HR diagram this convection zone has a Rayleigh number below the $\sim 10^3$ critical value~\citep{1961hhs..book.....C}.
As a result while the region is superadiabatic, it is not unstable to convection.
We therefore neglect these stable regions in our analysis, and shade them in grey in our figures.

\begin{figure*}
\centering
\begin{minipage}{0.48\textwidth}
\includegraphics[width=\textwidth]{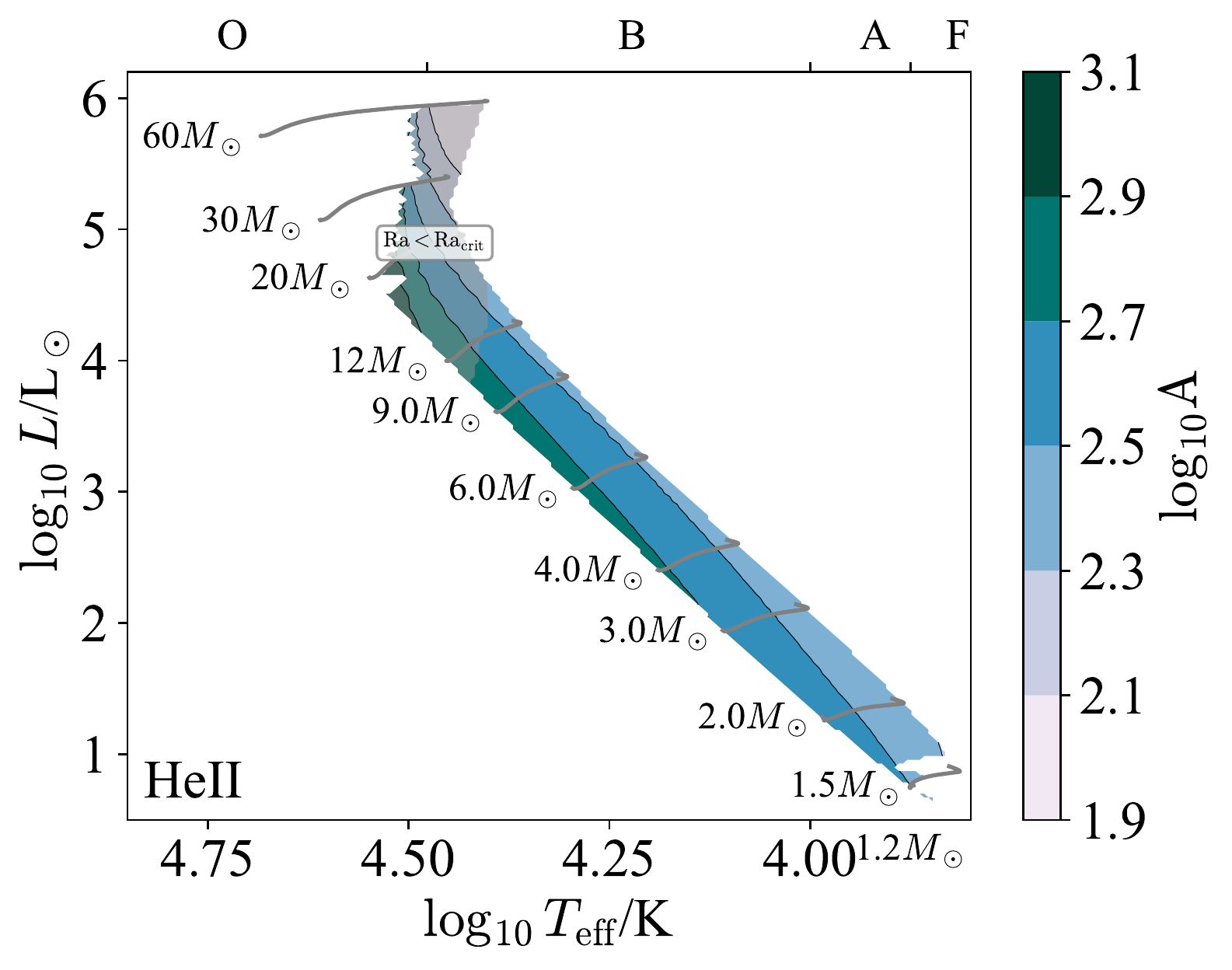}
\end{minipage}

\caption{The aspect ratio $\mathrm{A}$ is shown in terms of $\log T_{\rm eff}$/spectral type and $\log L$ for stellar models with HeII~CZs and Milky Way metallicity $Z=0.014$. Note that the aspect ratio is an input parameter, and does not depend on a specific theory of convection. Regions with $\mathrm{Ra} < \mathrm{Ra}_{\rm crit}$ are stable to convection and shaded in grey.}
\label{fig:HeII_structure}
\end{figure*}

Figure~\ref{fig:HeII_structure} shows the aspect ratio $\mathrm{A}$, which ranges from $10^2-10^3$.
These large aspect ratios suggest that local simulations are likely sufficient to capture their dynamics.

\begin{figure*}
\begin{minipage}{0.48\textwidth}
\includegraphics[width=\textwidth]{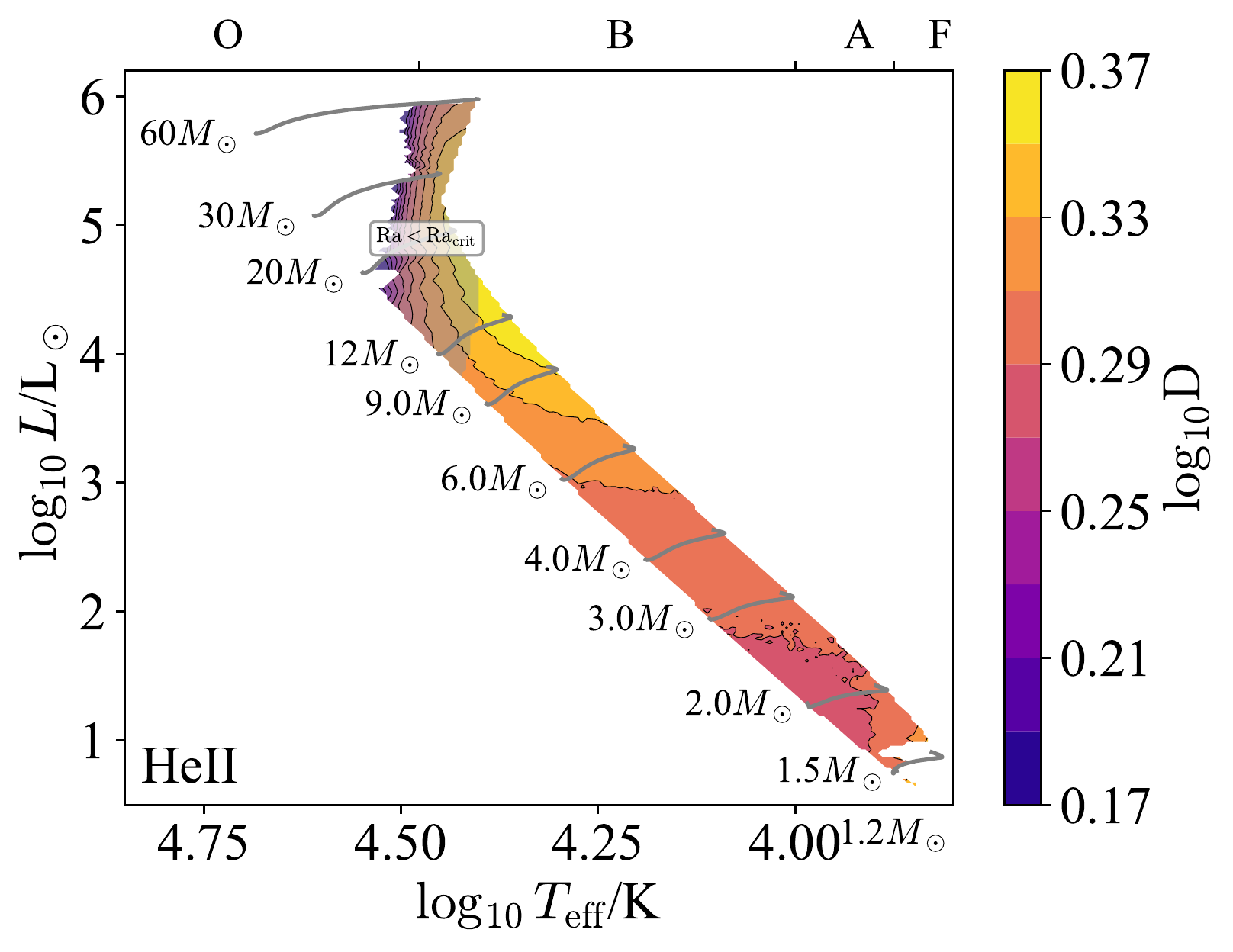}
\end{minipage}
\hfill
\begin{minipage}{0.48\textwidth}
\includegraphics[width=\textwidth]{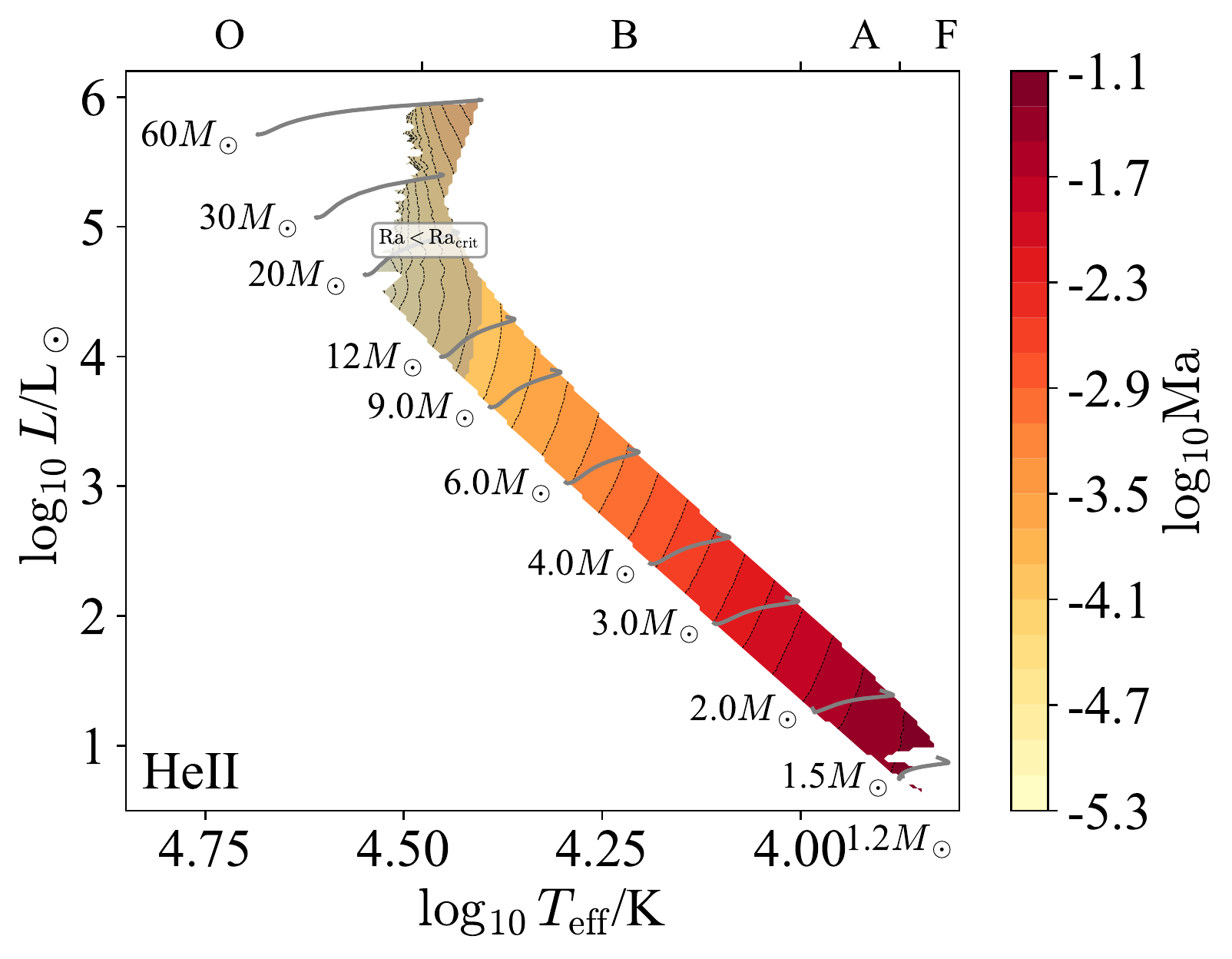}
\end{minipage}
\hfill

\caption{The density ratio $\mathrm{D}$ (left) and Mach number $\mathrm{Ma}$ (right) are shown in terms of $\log T_{\rm eff}$/spectral type and $\log L$ for stellar models with HeII~CZs and Milky Way metallicity $Z=0.014$. Note that while the density ratio is an input parameter and does not depend on a specific theory of convection, the Mach number is an output of such a theory and so is model-dependent. Regions with $\mathrm{Ra} < \mathrm{Ra}_{\rm crit}$ are stable to convection and shaded in grey.}
\label{fig:HeII_equations}
\end{figure*}

Next, the density ratio $\mathrm{D}$ (Figure~\ref{fig:HeII_equations}, left) and Mach number $\mathrm{Ma}$ (Figure~\ref{fig:HeII_equations}, right) inform which physics the fluid equations must include to model these zones.
The density ratio is typically small, of order $2-3$, and the Mach number ranges from $\sim 0.1$ at $M \la 2 M_\odot$ down to $10^{-4}$ at $M \approx 9 M_\odot$.
This suggests that above $\approx 2 M_\odot$ the Boussinesq approximation is valid, whereas below this the fully compressible equations may be needed to capture the dynamics at moderate Mach numbers.

\begin{figure*}
\begin{minipage}{0.48\textwidth}
\includegraphics[width=\textwidth]{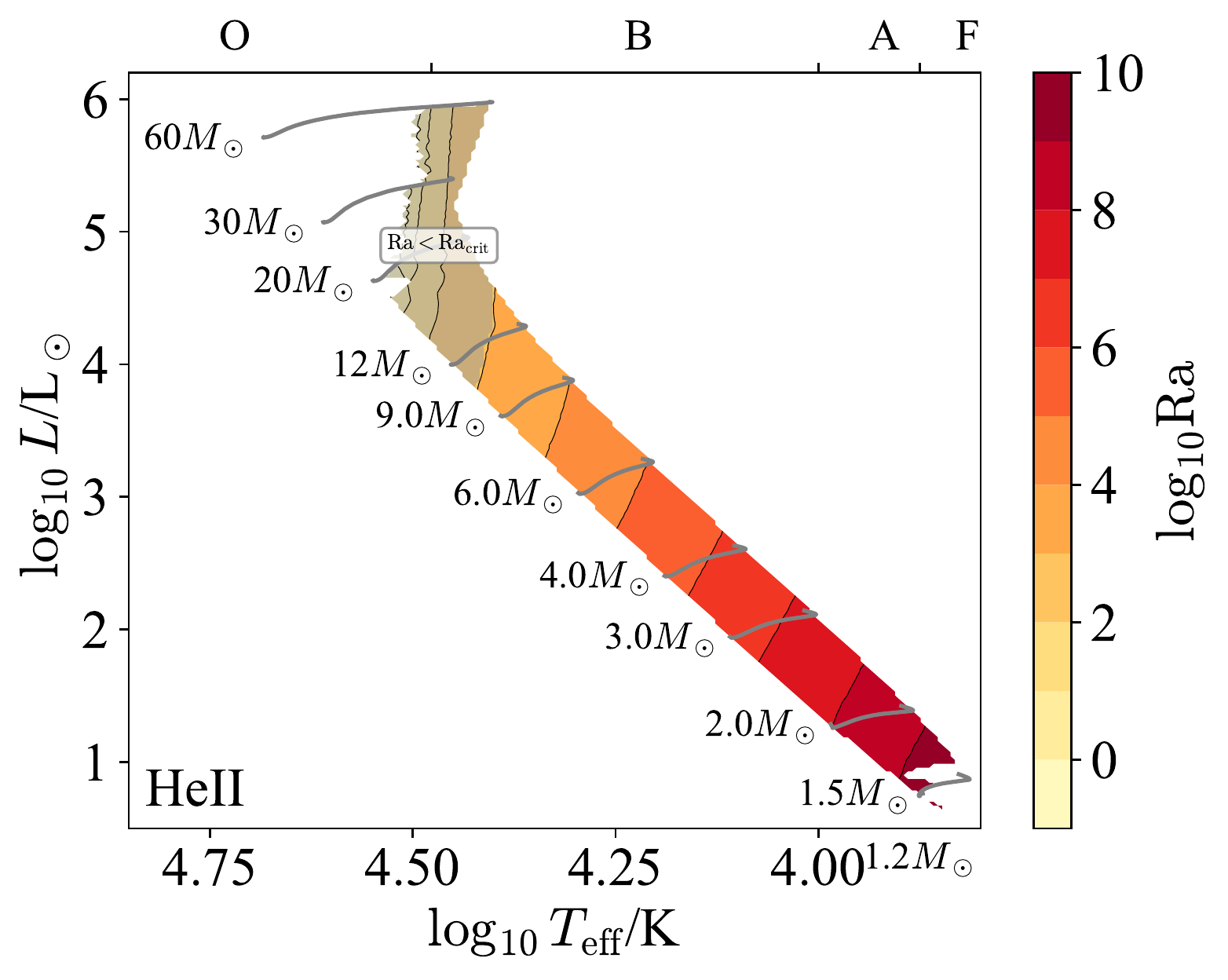}
\end{minipage}
\hfill
\begin{minipage}{0.48\textwidth}
\includegraphics[width=\textwidth]{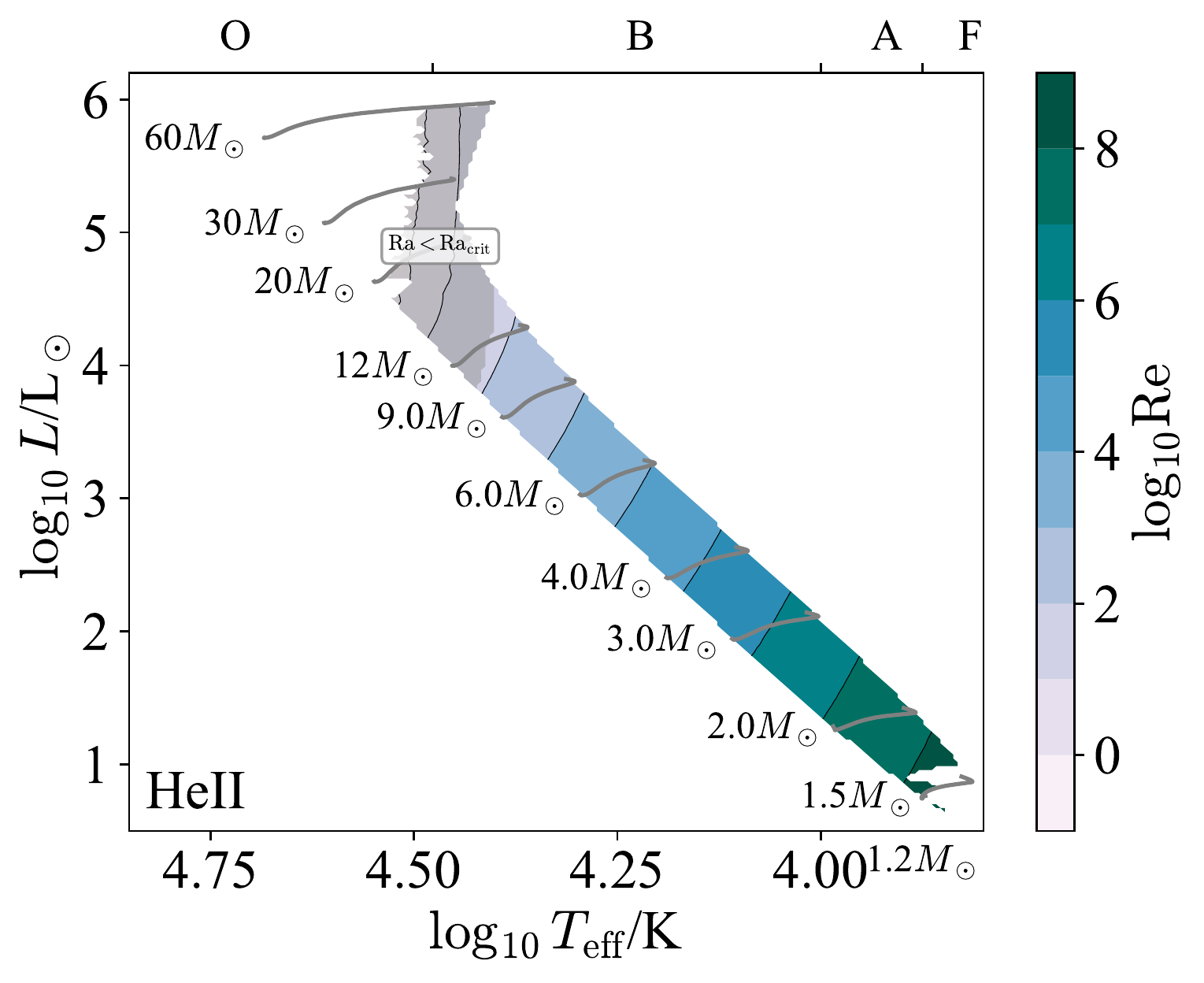}
\end{minipage}
\hfill

\caption{The Rayleigh number $\mathrm{Ra}$ (left) and Reynolds number $\mathrm{Re}$ (right) are shown in terms of $\log T_{\rm eff}$/spectral type and $\log L$ for stellar models with HeII~CZs and Milky Way metallicity $Z=0.014$.  Note that while the Rayleigh number is an input parameter and does not depend on a specific theory of convection, the Reynolds number is an output of such a theory and so is model-dependent. Regions with $\mathrm{Ra} < \mathrm{Ra}_{\rm crit}$ are stable to convection and shaded in grey.}
\label{fig:HeII_stability}
\end{figure*}

The Rayleigh number $\mathrm{Ra}$ (Figure~\ref{fig:HeII_stability}, left) determines whether or not a putative convection zone is actually unstable to convection, and the Reynolds number $\mathrm{Re}$ determines how turbulent the zone is if instability sets in (Figure~\ref{fig:HeII_stability}, right).
At low masses the Rayleigh number is large ($10^{10}$), at high masses it plummets and eventually becomes sub-critical, which we show in grey.
Likewise at low masses the Reynolds number is large ($10^8$) while at high masses it is quite small ($\sim 10^2)$.
These putative convection zones then span a wide range of properties, from being subcritical and \emph{stable}~\citep{1961hhs..book.....C} at high masses, to being marginally unstable and weakly turbulent at intermediate masses ($\sim 9 M_\odot$), to eventually being strongly unstable and having well-developed turbulence at low masses ($\sim 2 M_\odot$).

\begin{figure*}
\centering
\begin{minipage}{0.48\textwidth}
\includegraphics[width=\textwidth]{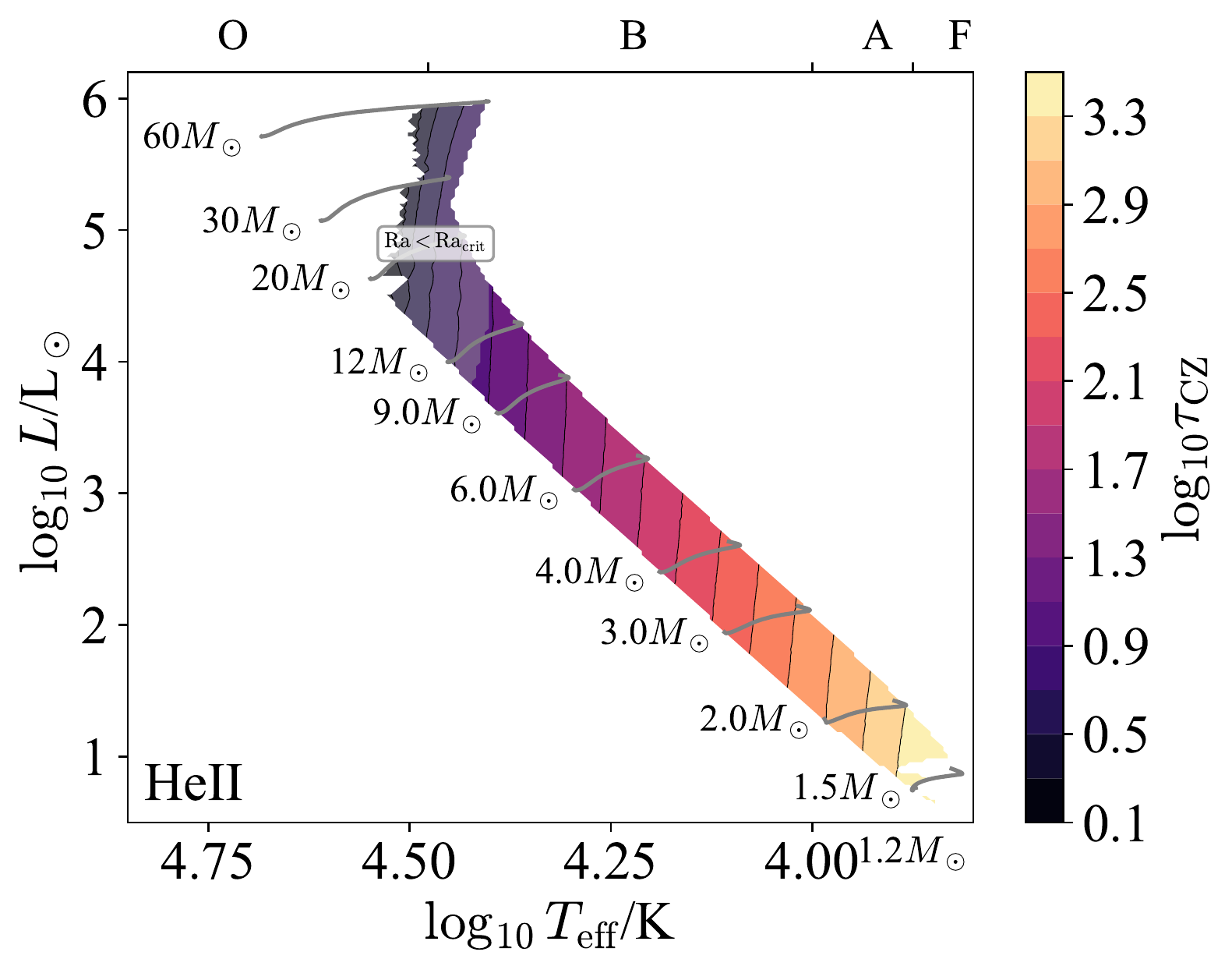}
\end{minipage}
\hfill
\begin{minipage}{0.48\textwidth}
\includegraphics[width=\textwidth]{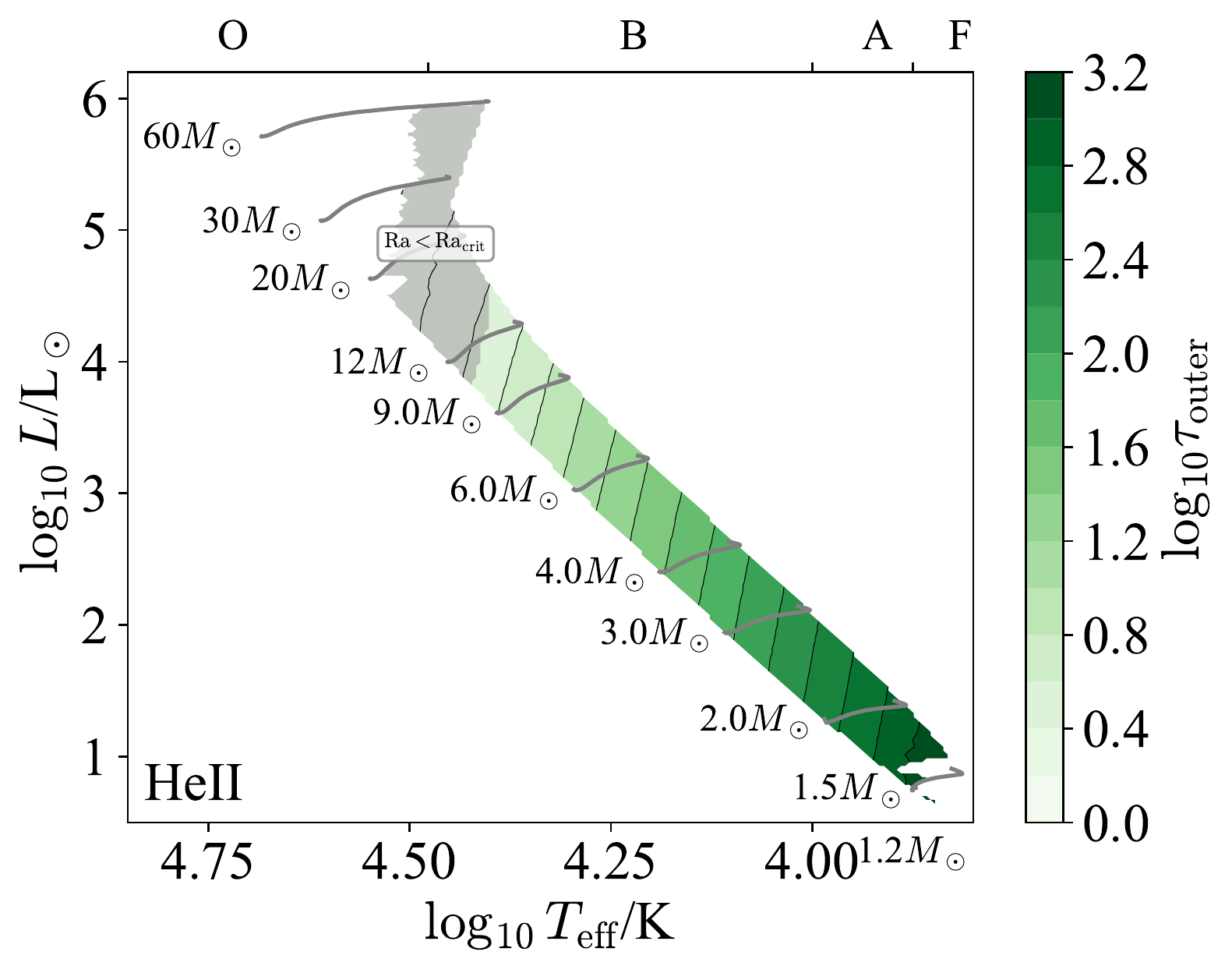}
\end{minipage}
\hfill
\caption{The convection optical depth $\tau_{\rm CZ}$ (left) and the optical depth to the surface $\tau_{\rm outer}$ (right) are shown in terms of $\log T_{\rm eff}$/spectral type and $\log L$ for stellar models with HeII~CZs and Milky Way metallicity $Z=0.014$. Note that both of these are input parameters, and do not depend on a specific theory of convection. Regions with $\mathrm{Ra} < \mathrm{Ra}_{\rm crit}$ are stable to convection and shaded in grey.}
\label{fig:HeII_optical}
\end{figure*}

The optical depth across a convection zone $\tau_{\rm CZ}$ (Figure~\ref{fig:HeII_optical}, left) indicates whether or not radiation can be handled in the diffusive approximation, while the optical depth from the outer boundary to infinity $\tau_{\rm outer}$ (Figure~\ref{fig:HeII_optical}, right) indicates the nature of radiative transfer and cooling in the outer regions of the convection zone.
At high masses ($M \ga 9 M_\odot$) the surface of the HeII CZ is at low optical depth ($\tau_{\rm outer} \sim 1-3$), while at lower masses the optical depth quickly becomes large.
Similarly, the optical depth across the HeII CZ is moderate at high masses ($\sim 10$) and becomes large towards lower masses.
Overall, then, the bulk of the HeII CZ can likely be treated in the diffusive approximation, as can the outer boundary for $M \la 9 M_\odot$), while the outer boundary at higher masses likely requires a treatment with radiation hydrodynamics.

\begin{figure*}
\centering
\begin{minipage}{0.48\textwidth}
\includegraphics[width=\textwidth]{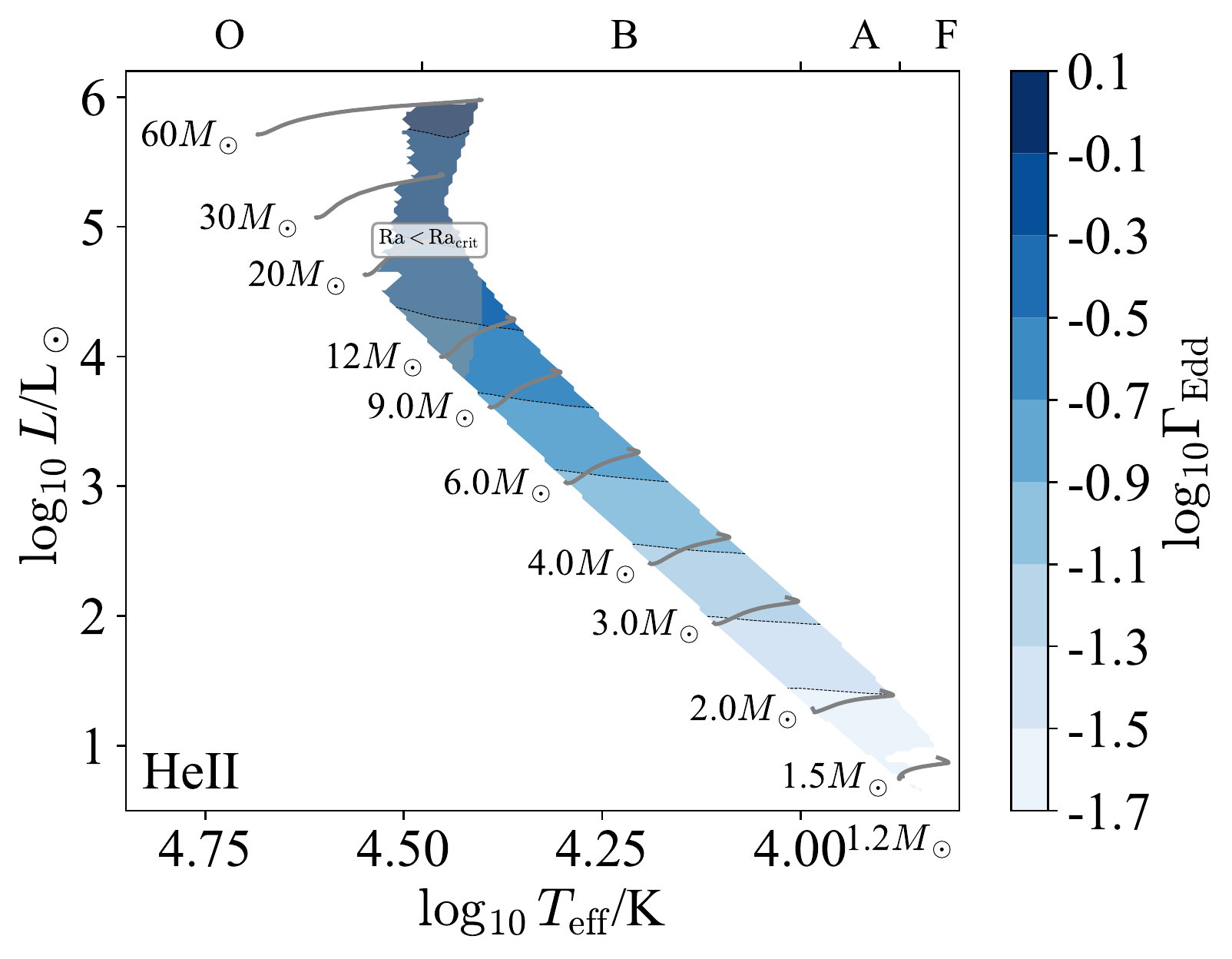}
\end{minipage}
\hfill
\begin{minipage}{0.48\textwidth}
\includegraphics[width=\textwidth]{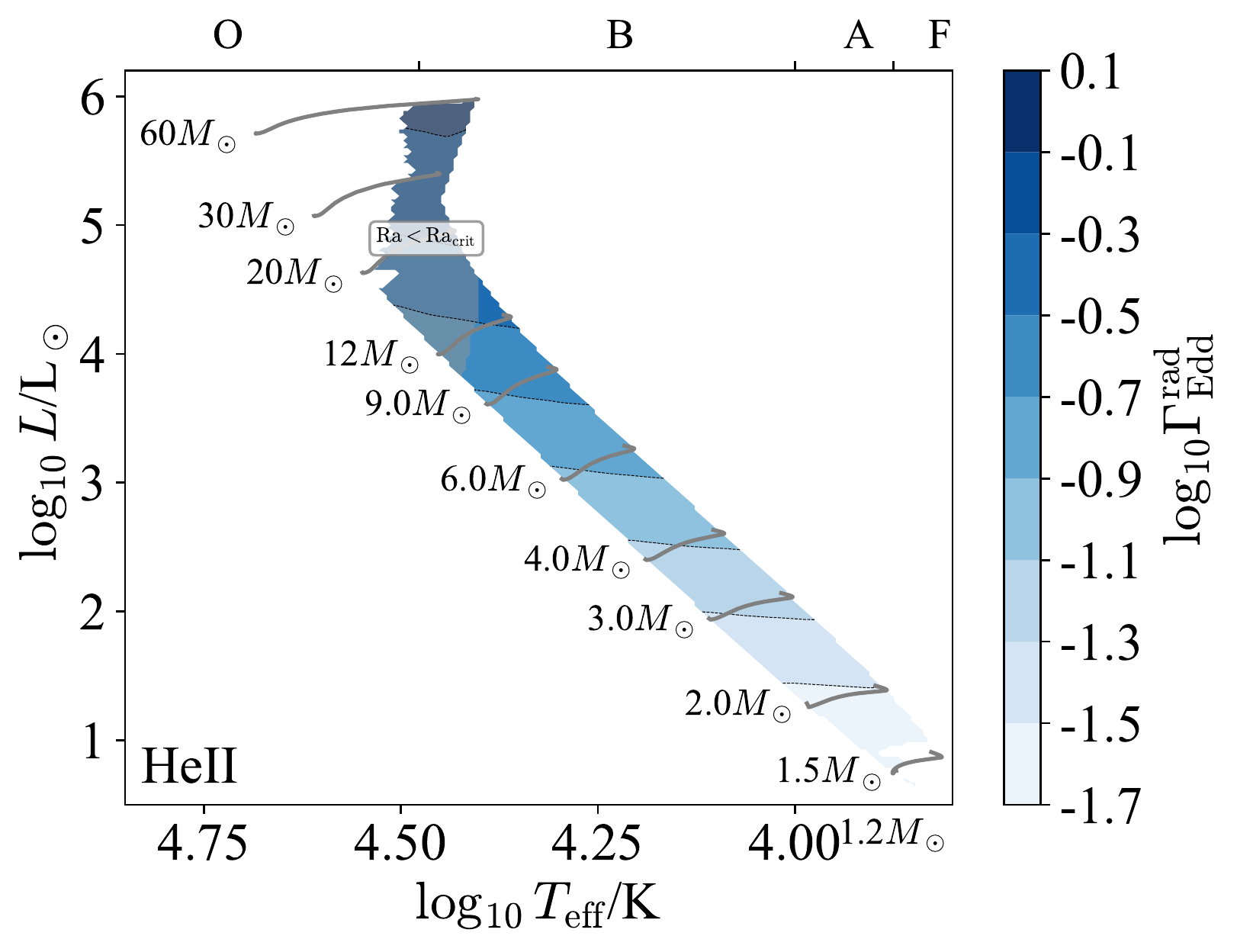}
\end{minipage}
\hfill
\caption{The Eddington ratio with the full luminosity $\Gamma_{\rm Edd}$ (left) and the radiative luminosity (right) are shown in terms of $\log T_{\rm eff}$/spectral type and $\log L$ for stellar models with HeII~CZs and Milky Way metallicity $Z=0.014$. Note that while $\Gamma_{\rm Edd}$ is an input parameter and does not depend on a specific theory of convection, $\Gamma_{\rm Edd}^{\rm rad}$ is an output of such a theory and so is model-dependent. Regions with $\mathrm{Ra} < \mathrm{Ra}_{\rm crit}$ are stable to convection and shaded in grey.}
\label{fig:HeII_eddington}
\end{figure*}

The Eddington ratio $\Gamma_{\rm Edd}$ (Figure~\ref{fig:HeII_eddington}, left) indicates whether or not radiation hydrodynamic instabilities are important in the non-convecting state, and the radiative Eddington ratio $\Gamma_{\rm Edd}^{\rm rad}$ (Figure~\ref{fig:HeII_eddington}, right) indicates the same in the developed convective state.
Both ratios are moderate at high masses ($\Gamma \sim 0.3$ at $M \sim 10 M_\odot$), and radiation hydrodynamic instabilities could be important in this regime.
By contrast at lower masses ($M \la 6 M_\odot$) these ratios are both small, and radiation hydrodynamic instabilities are unlikely to matter.

\begin{figure*}
\centering
\begin{minipage}{0.48\textwidth}
\includegraphics[width=\textwidth]{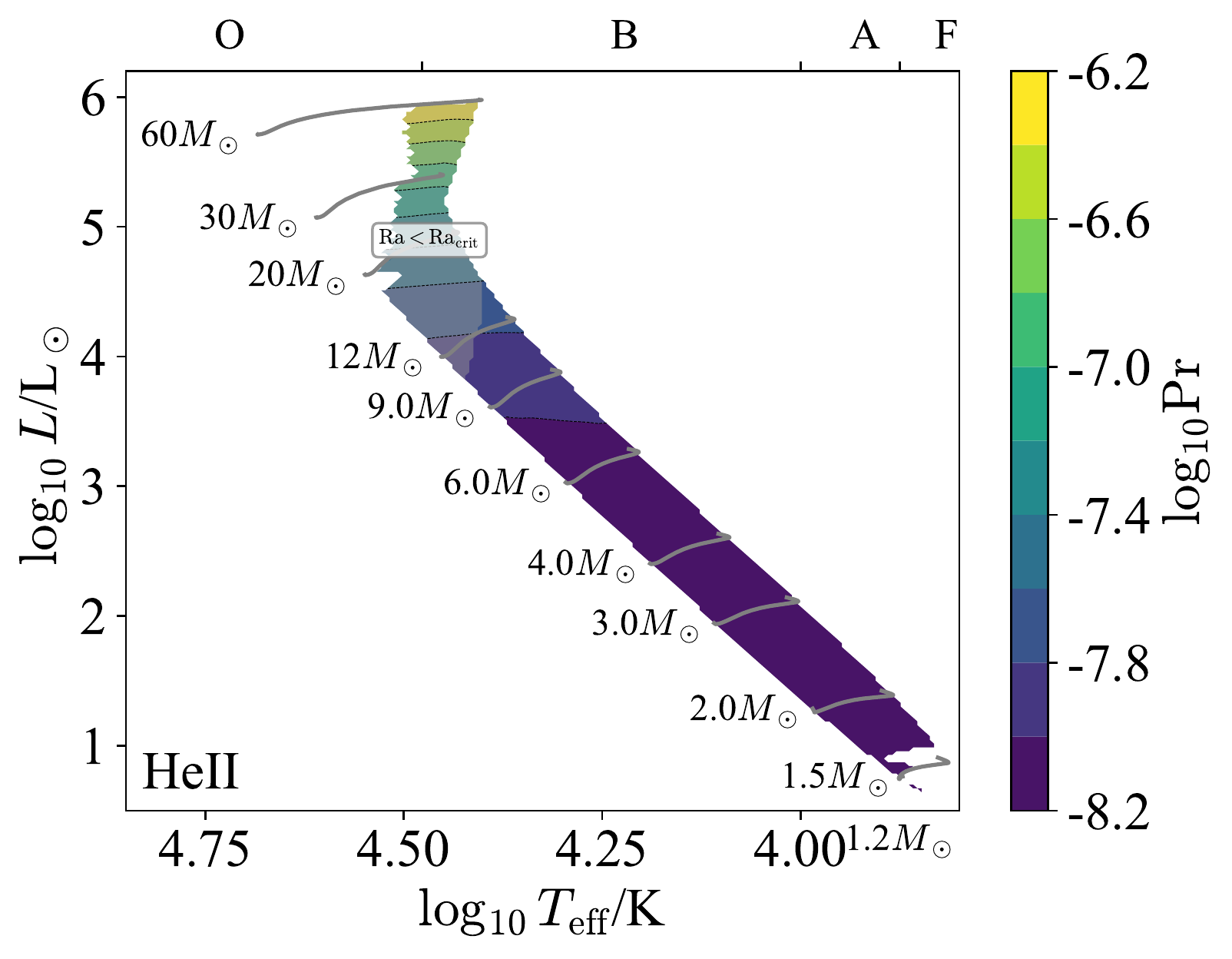}
\end{minipage}
\hfill
\begin{minipage}{0.48\textwidth}
\includegraphics[width=\textwidth]{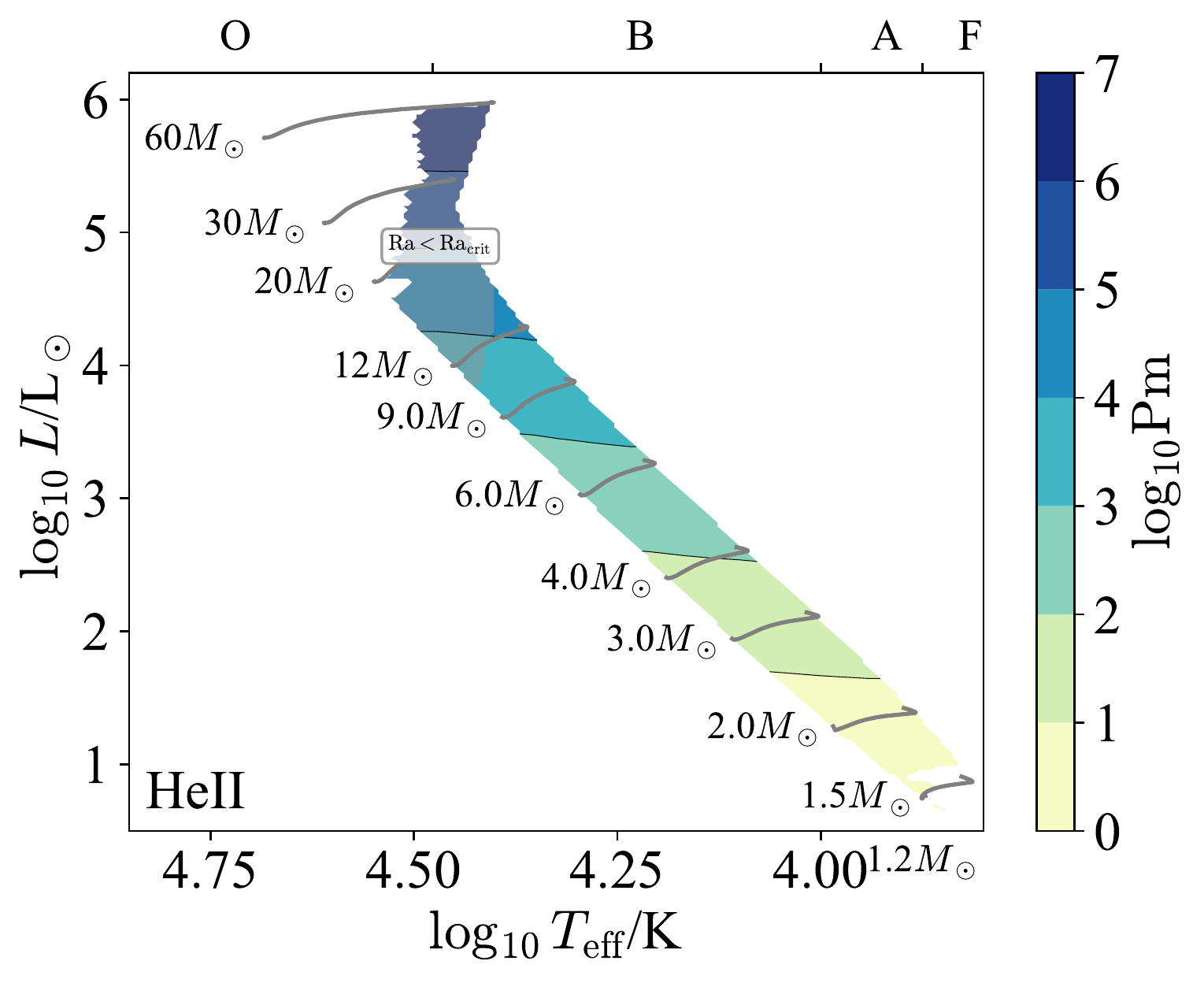}
\end{minipage}
\hfill

\caption{The Prandtl number $\mathrm{Pr}$ (left) and magnetic Prandtl number $\mathrm{Pm}$ (right) are shown in terms of $\log T_{\rm eff}$/spectral type and $\log L$ for stellar models with HeII~CZs and Milky Way metallicity $Z=0.014$. Note that both $\mathrm{Pr}$ and $\mathrm{Pm}$ are input parameters, and so do not depend on a specific theory of convection. Regions with $\mathrm{Ra} < \mathrm{Ra}_{\rm crit}$ are stable to convection and shaded in grey.}
\label{fig:HeII_diffusivities}
\end{figure*}

The Prandtl number $\mathrm{Pr}$ (Figure~\ref{fig:HeII_diffusivities}, left) measures the relative importance of thermal diffusion and viscosity, and the magnetic Prandtl number $\mathrm{Pm}$ (Figure~\ref{fig:HeII_diffusivities}, right) measures the same for magnetic diffusion and viscosity.
The Prandtl number is always small in these models, so the thermal diffusion length-scale is much larger than the viscous scale.
By contrast, the magnetic Prandtl number varies from order-unity at low masses to large ($10^4$) at high masses.

The fact that $\mathrm{Pm}$ is large at high masses is notable because the quasistatic approximation for magnetohydrodynamics has frequently been used to study magnetoconvection in minimal 3D MHD simulations of planetary and stellar interiors~\citep[e.g.][]{yan_calkins_maffei_julien_tobias_marti_2019} and assumes that $\mathrm{Rm} = \mathrm{Pm} \mathrm{Re} \rightarrow 0$; in doing so, this approximation assumes a global background magnetic field is dominant and neglects the nonlinear portion of the Lorentz force. This approximation breaks down in convection zones with $\mathrm{Pm} > 1$ and future numerical experiments should seek to understand how magnetoconvection operates in this regime.

\begin{figure*}
\centering
\begin{minipage}{0.48\textwidth}
\includegraphics[width=\textwidth]{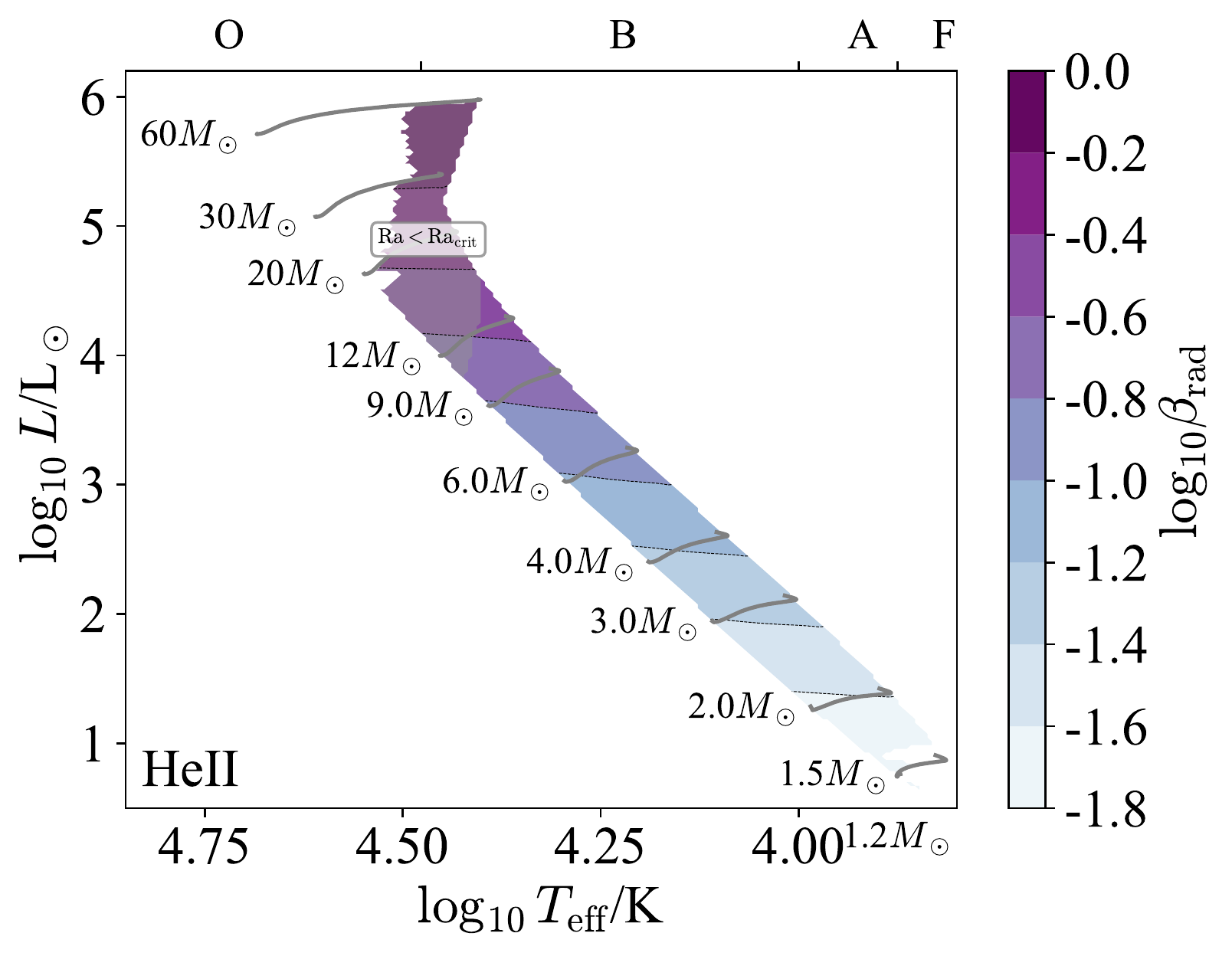}
\end{minipage}
\hfill

\caption{The radiation pressure ratio $\beta_{\rm rad}$ is shown in terms of $\log T_{\rm eff}$/spectral type and $\log L$ for stellar models with HeII~CZs and Milky Way metallicity $Z=0.014$. Note that this ratio is an input parameter, and does not depend on a specific theory of convection. Regions with $\mathrm{Ra} < \mathrm{Ra}_{\rm crit}$ are stable to convection and shaded in grey.}
\label{fig:HeII_beta}
\end{figure*}

The radiation pressure ratio $\beta_{\rm rad}$ (Figure~\ref{fig:HeII_beta}) measures the importance of radiation in setting the thermodynamic properties of the fluid.
We see that this is uniformly small ($\la 0.1$) and so radiation pressure likely plays a sub-dominant role in these zones.

\begin{figure*}
\centering
\begin{minipage}{0.48\textwidth}
\includegraphics[width=\textwidth]{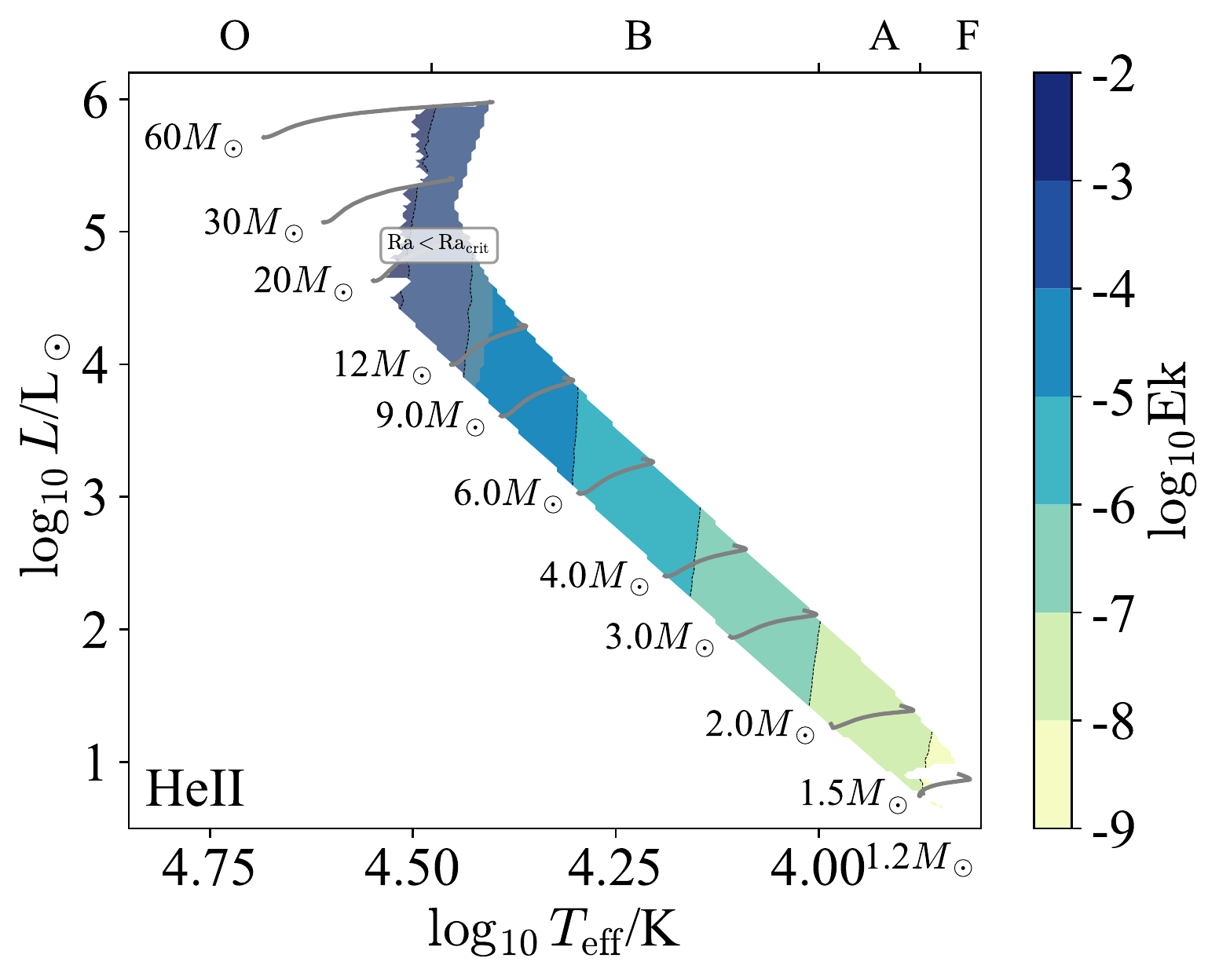}
\end{minipage}

\caption{The Ekman number $\mathrm{Ek}$ is shown in terms of $\log T_{\rm eff}$/spectral type and $\log L$ for stellar models with HeII~CZs and Milky Way metallicity $Z=0.014$. Note that the Ekman number is an input parameter, and does not depend on a specific theory of convection. Regions with $\mathrm{Ra} < \mathrm{Ra}_{\rm crit}$ are stable to convection and shaded in grey.}
\label{fig:HeII_ekman}
\end{figure*}

The Ekman number $\mathrm{Ek}$ (Figure~\ref{fig:HeII_ekman}) indicates the relative importance of viscosity and rotation.
This is tiny across the HRD~\footnote{Note that, because the Prandtl number is also very small, this does not significantly alter the critical Rayleigh number~(see Ch3 of~\cite{1961hhs..book.....C} and appendix D of~\cite{2022arXiv220110567J}).}, so we expect rotation to dominate over viscosity, except at very small length-scales.

\begin{figure*}
\centering
\begin{minipage}{0.48\textwidth}
\includegraphics[width=\textwidth]{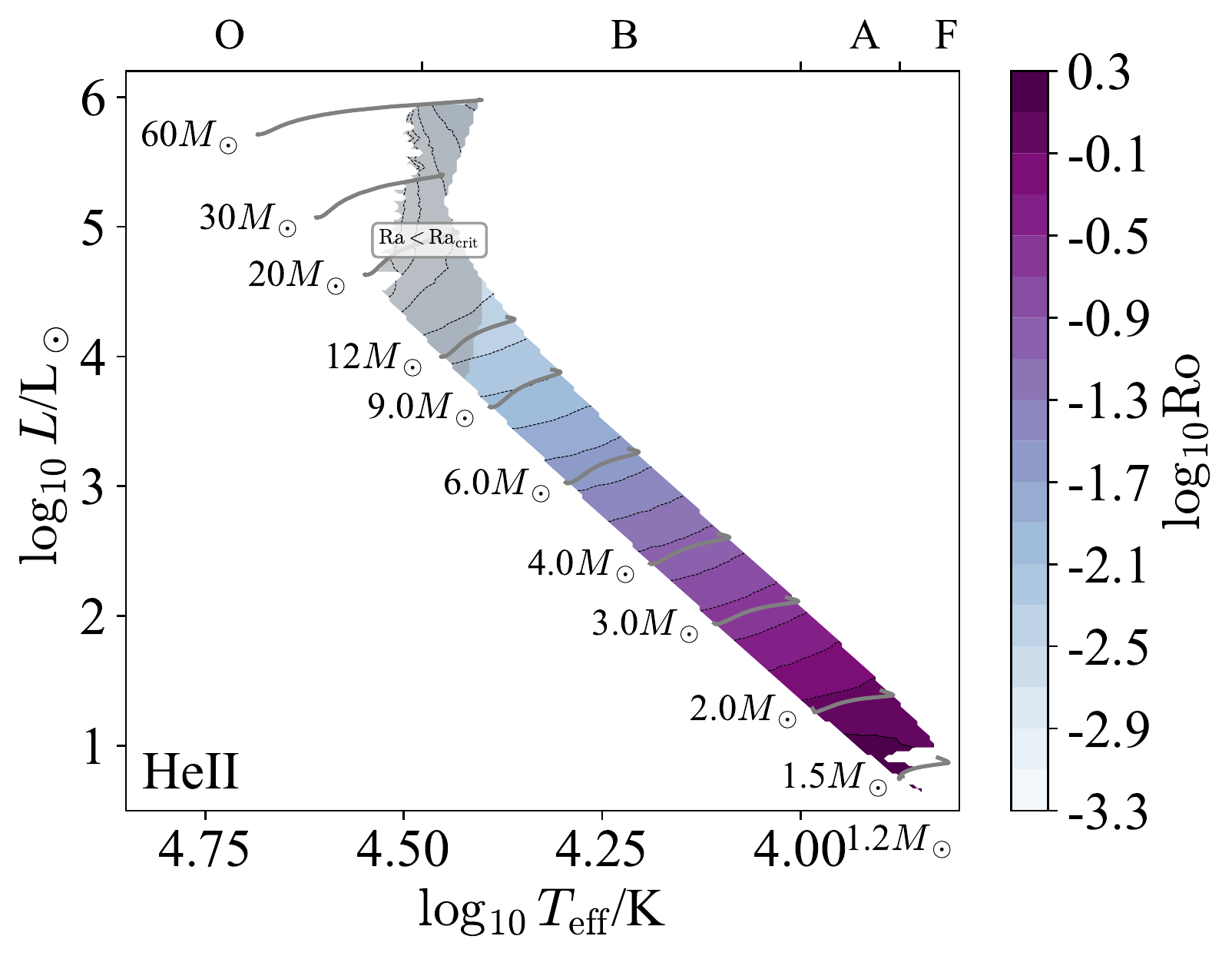}
\end{minipage}
\hfill
\begin{minipage}{0.48\textwidth}
\includegraphics[width=\textwidth]{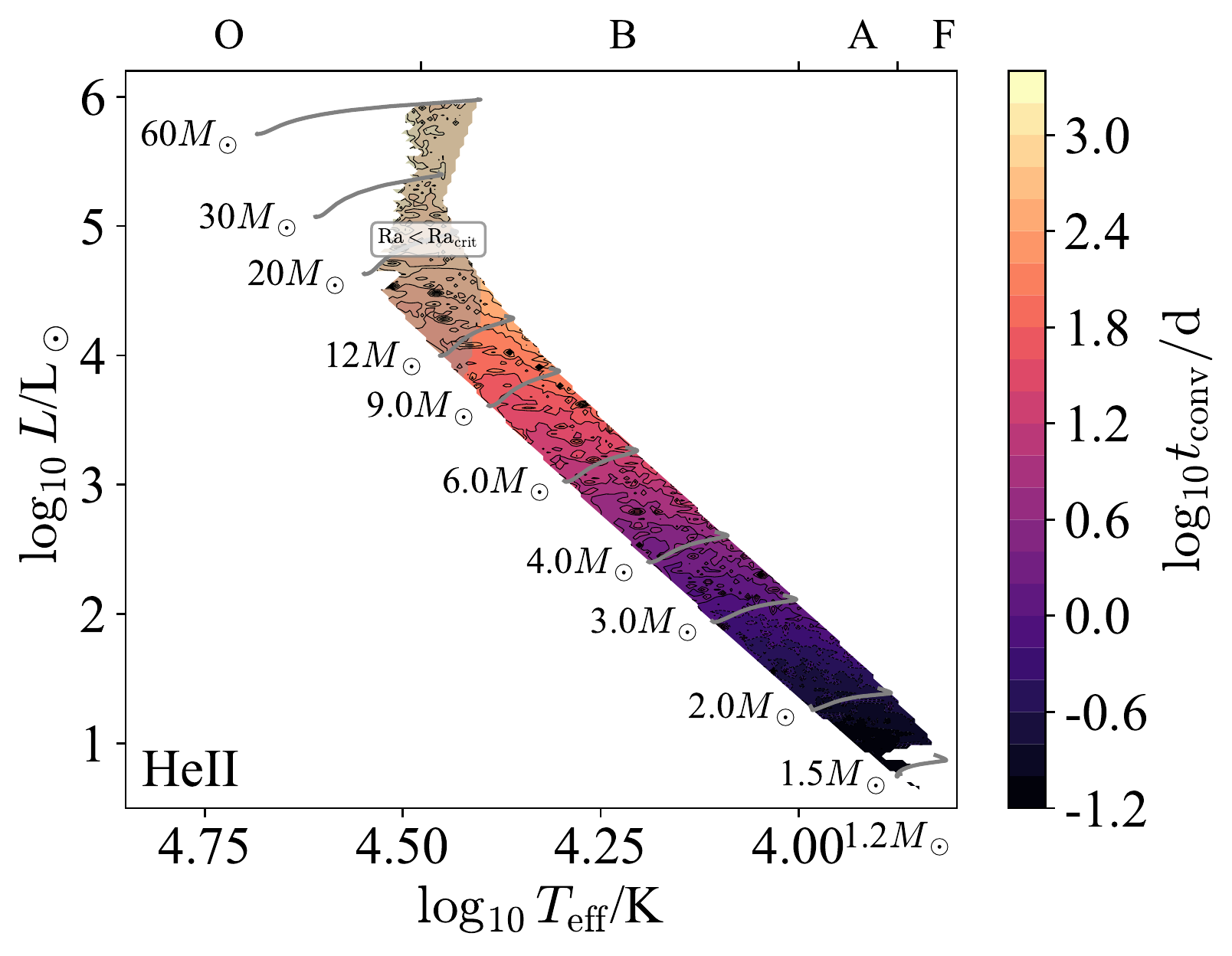}
\end{minipage}
\hfill

\caption{The Rossby number $\mathrm{Ro}$ (left) and turnover time $t_{\rm conv}$ (right) are shown in terms of $\log T_{\rm eff}$/spectral type and $\log L$ for stellar models with HeII~CZs and Milky Way metallicity $Z=0.014$. Note that both $\mathrm{Ro}$ and $t_{\rm conv}$ are outputs of a theory of convection and so are model-dependent. Regions with $\mathrm{Ra} < \mathrm{Ra}_{\rm crit}$ are stable to convection and shaded in grey. Note that the turnover time exhibits numerical noise related to the model mesh resolution because the integrand $1/v_c$ diverges towards the convective boundaries.}
\label{fig:HeII_rotation}
\end{figure*}

The Rossby number $\mathrm{Ro}$ (Figure~\ref{fig:HeII_rotation}, left) measures the relative importance of rotation and inertia.
This is small ($10^{-3}$) at high masses but greater than unity at low masses, meaning that the HeII CZ is rotationally constrained at high masses but not at low masses for typical rotation rates~\citep{2013A&A...557L..10N}.

We have assumed a fiducial rotation law to calculate $\mathrm{Ro}$.
Stars exhibit a variety of different rotation rates, so we also show the convective turnover time $t_{\rm conv}$ (Figure~\ref{fig:HeII_rotation}, right) which may be used to estimate the Rossby number for different rotation periods.

\begin{figure*}
\centering
\begin{minipage}{0.48\textwidth}
\includegraphics[width=\textwidth]{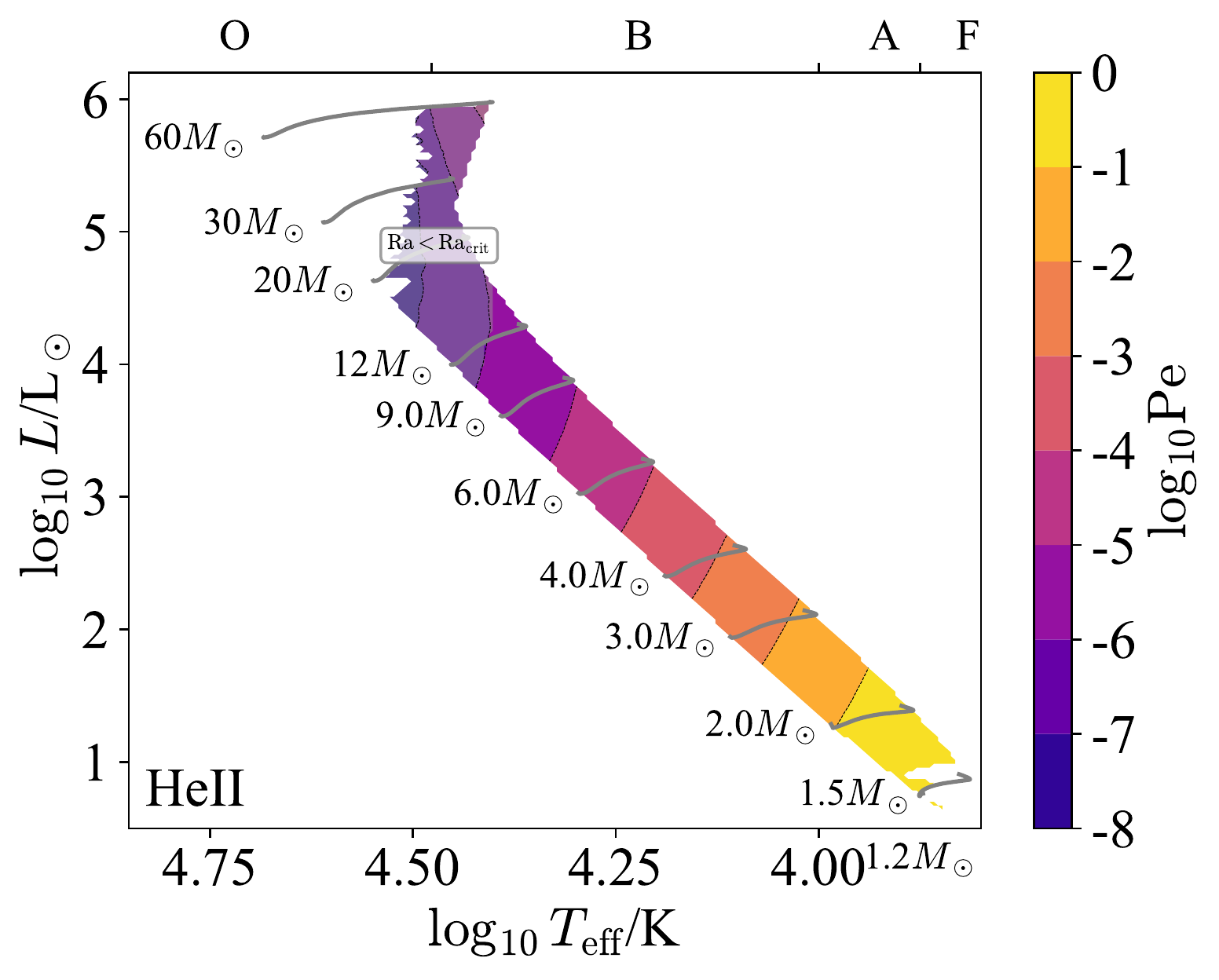}
\end{minipage}
\hfill
\begin{minipage}{0.48\textwidth}
\includegraphics[width=\textwidth]{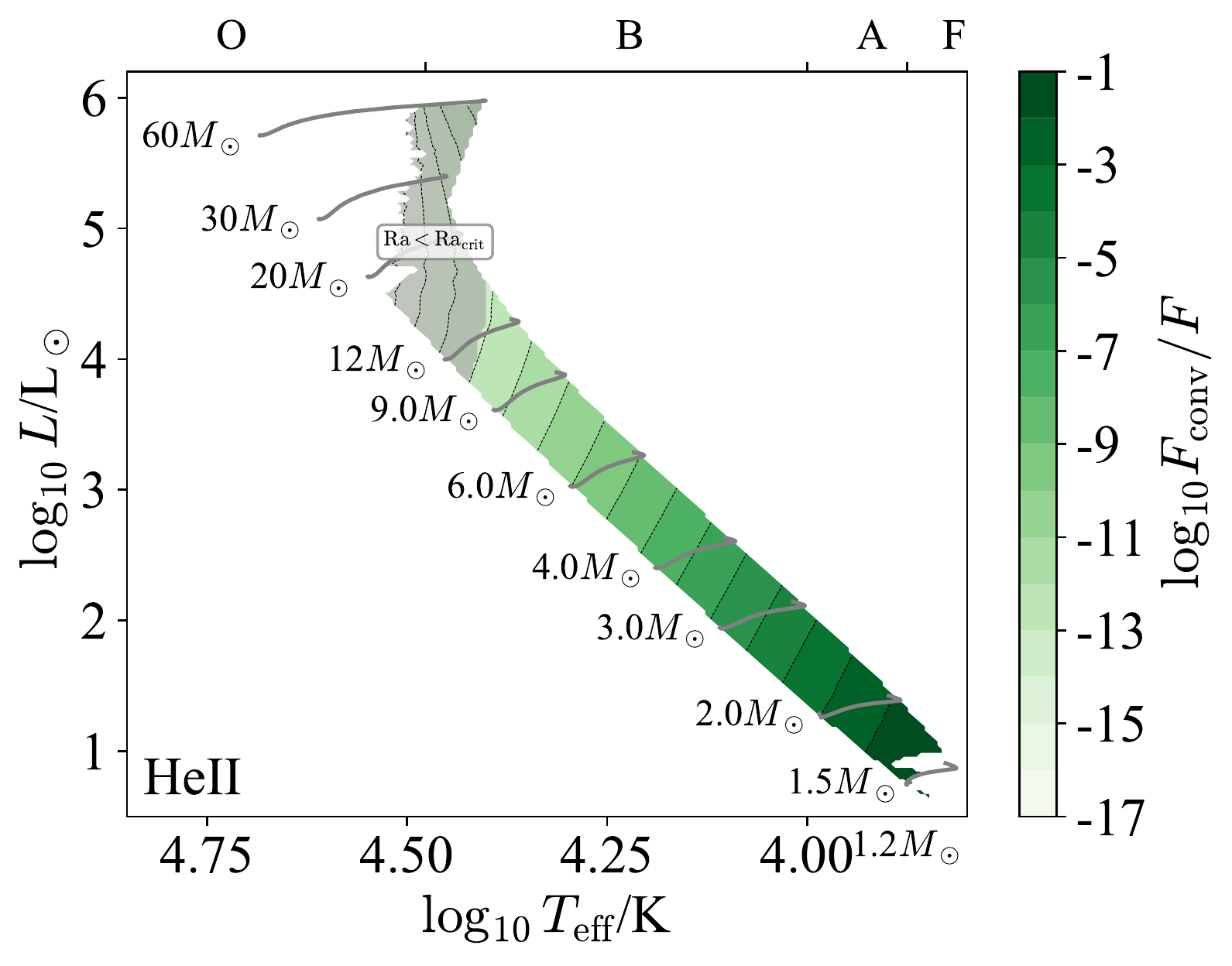}
\end{minipage}

\caption{The P{\'e}clet number $\mathrm{Pe}$ (left) and $F_{\rm conv}/F$ (right) are shown in terms of $\log T_{\rm eff}$/spectral type and $\log L$ for stellar models with HeII~CZs and Milky Way metallicity $Z=0.014$. Note that both $\mathrm{Pe}$ and $F_{\rm conv}/F$ are outputs of a theory of convection and so are model-dependent. Regions with $\mathrm{Ra} < \mathrm{Ra}_{\rm crit}$ are stable to convection and shaded in grey.}
\label{fig:HeII_efficiency}
\end{figure*}

The P{\'e}clet number $\mathrm{Pe}$ (Figure~\ref{fig:HeII_efficiency}, left) measures the relative importance of advection and diffusion in transporting heat, and the flux ratio $F_{\rm conv}/F$ (Figure~\ref{fig:HeII_efficiency}, right) reports the fraction of the energy flux which is advected.
Both exhibit substantial variation with mass.
The P{\'e}clet number varies from order unity at low masses to very small at high masses ($10^{-5}$), and the flux ratio similarly varies from near-unity at low masses to tiny ($10^{-14}$) at high masses.
That is, there is a large gradient in convective efficiency with mass, with efficient convection at low masses and very inefficient convection at high masses.

\begin{figure*}
\centering
\begin{minipage}{0.48\textwidth}
\includegraphics[width=\textwidth]{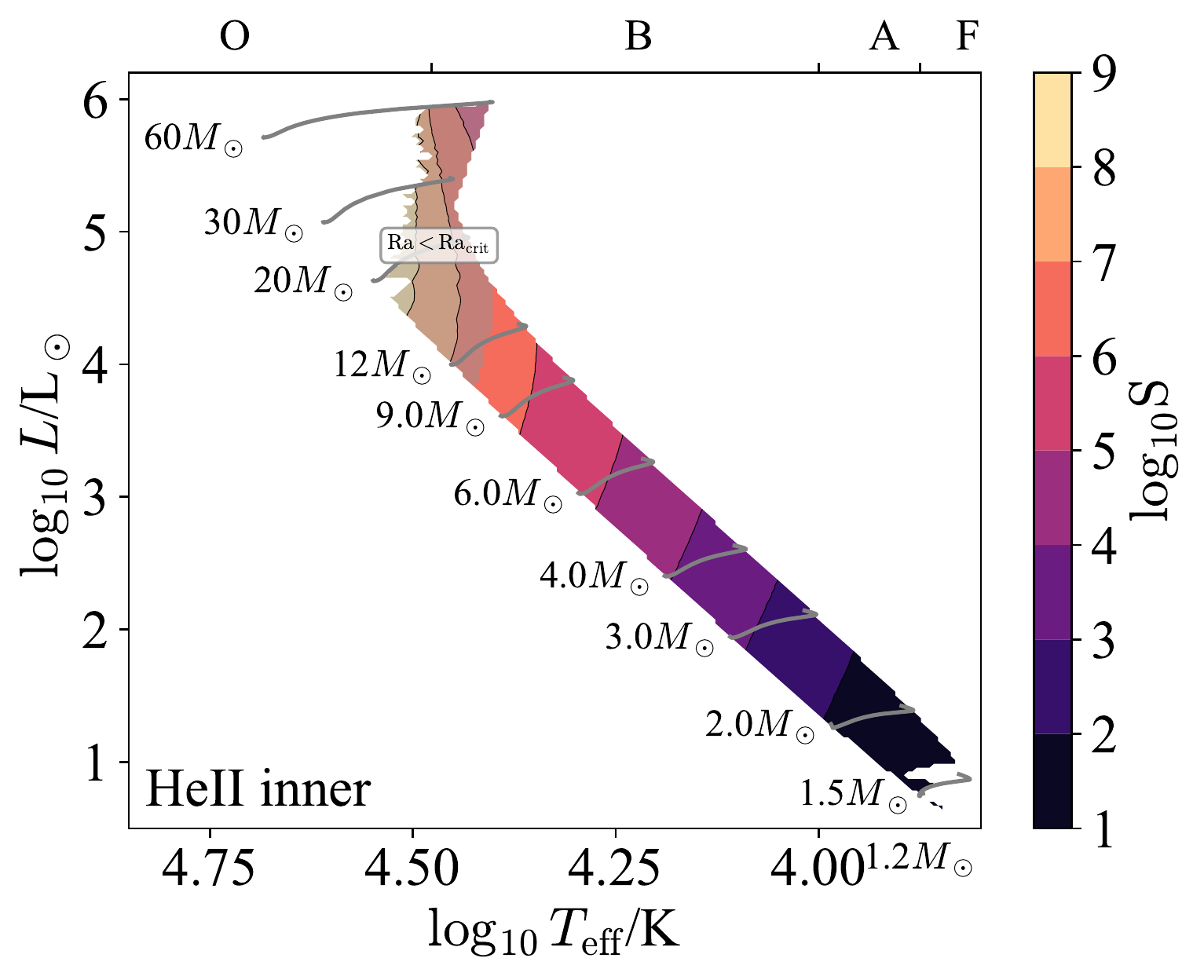}
\end{minipage}
\hfill
\begin{minipage}{0.48\textwidth}
\includegraphics[width=\textwidth]{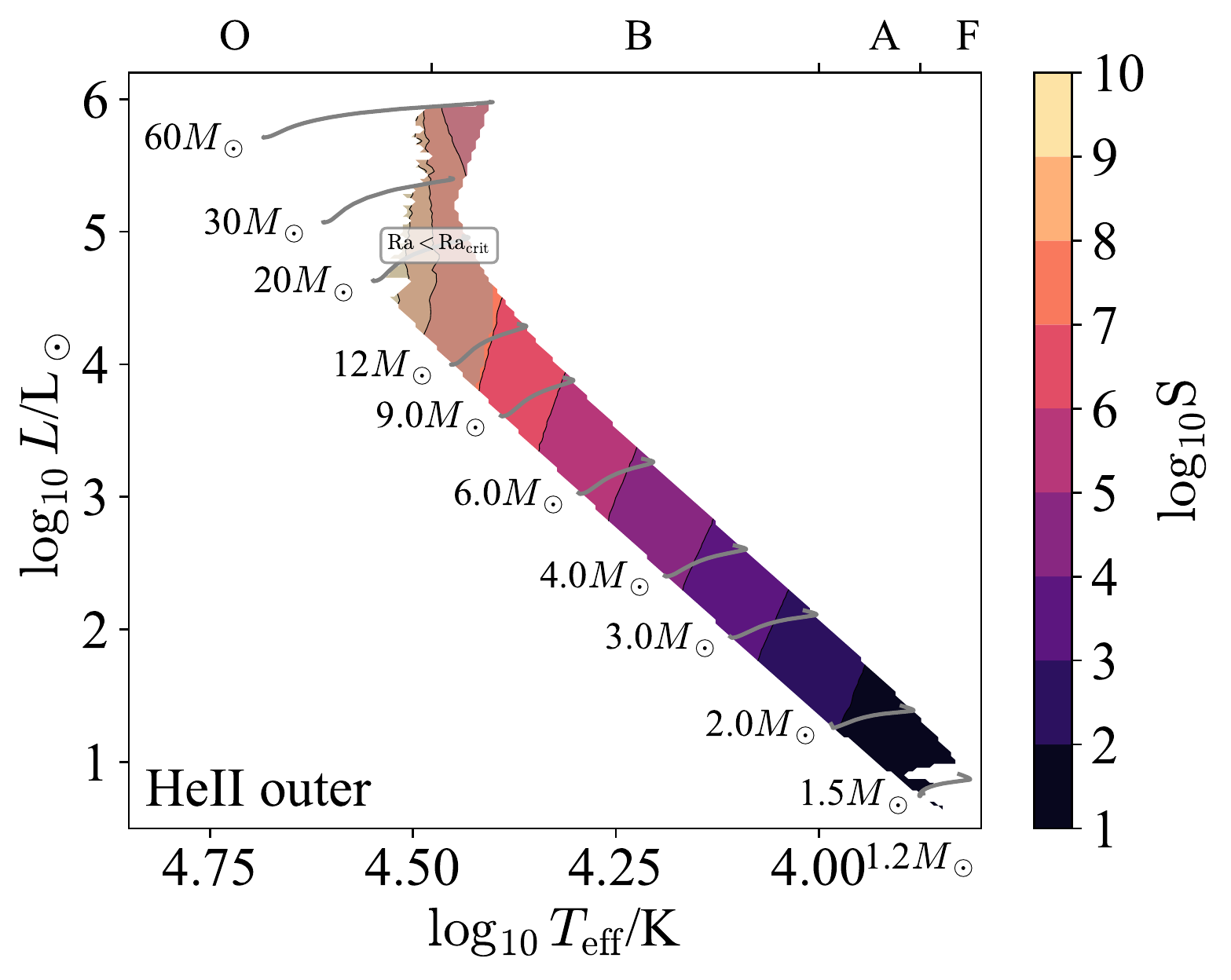}
\end{minipage}
\hfill

\caption{The stiffness of the inner (left) and outer (right) convective boundaries are shown in terms of $\log T_{\rm eff}$ and $\log L$ for stellar models with HeII~CZs and Milky Way metallicity $Z=0.014$. Note that the stiffness is an output of a theory of convection and so is model-dependent. Regions with $\mathrm{Ra} < \mathrm{Ra}_{\rm crit}$ are stable to convection and shaded in grey.}
\label{fig:HeII_stiff}
\end{figure*}

Finally, Figure~\ref{fig:HeII_stiff} shows the stiffness of both the inner and outer boundaries of the HeII CZ.
Both range from very stiff ($S \sim 10^{4-8}$) to very weak ($S \sim 1$), with decreasing stiffness towards decreasing mass.
So for instance for masses $M \ga 3 M_\odot$ we do not expect much mechanical overshooting, whereas for $M \la 3 M_\odot$ both boundaries should show substantial overshooting, because their low stiffness causes convective flows to decelerate over large length scales.

\clearpage
\subsection{FeCZ}

We round out the ionization-driven convection zones by turning to the Fe CZs, which occur in the subsurface layers of solar metallicity stars with masses $M_\star \ga 7 M_\odot$, and we first turn to our input parameters.

\begin{figure*}
\centering
\begin{minipage}{0.48\textwidth}
\includegraphics[width=\textwidth]{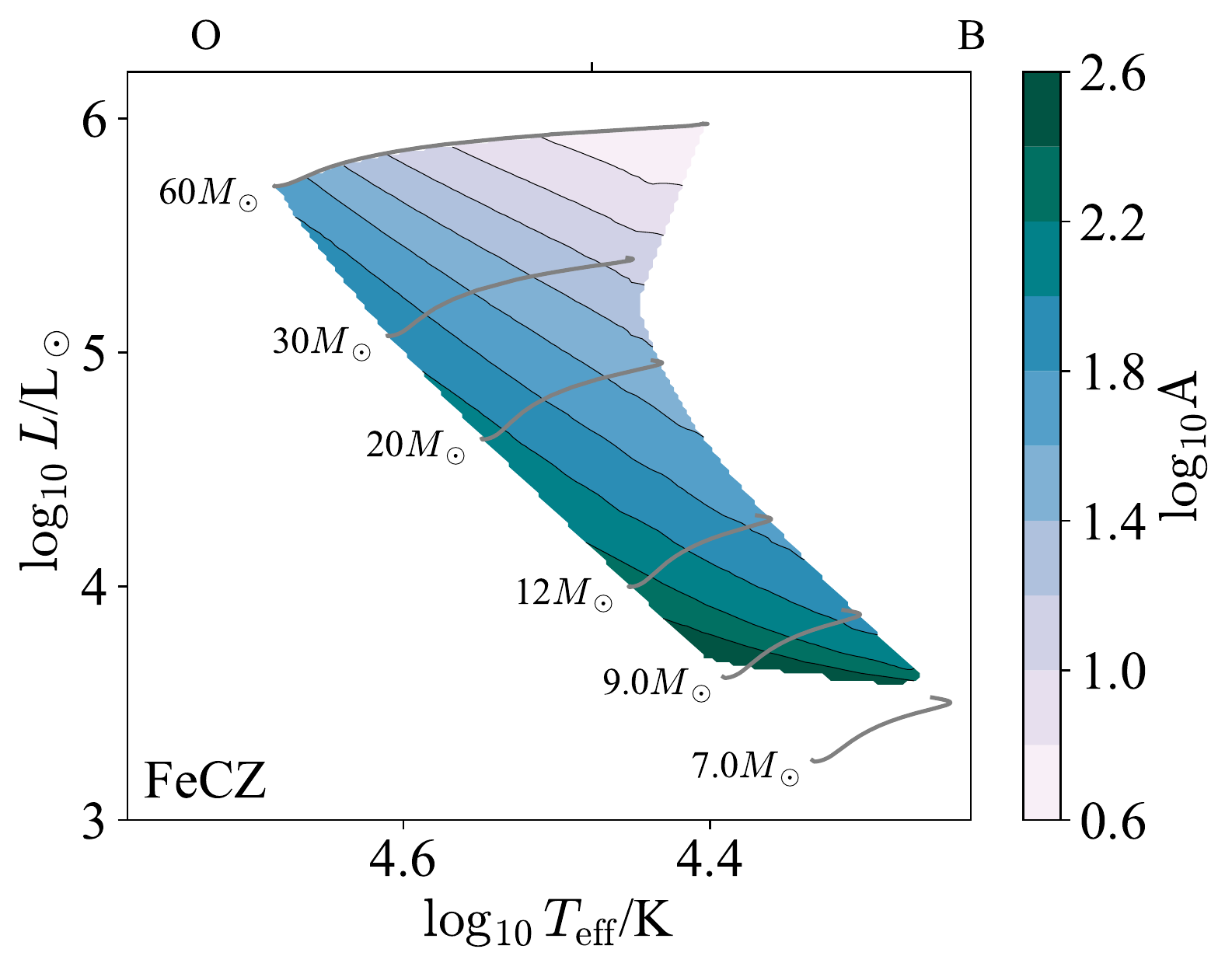}
\end{minipage}

\caption{The aspect ratio $\mathrm{A}$ is shown in terms of $\log T_{\rm eff}$/spectral type and $\log L$ for stellar models with FeCZs and Milky Way metallicity $Z=0.014$. Note that the aspect ratio is an input parameter, and does not depend on a specific theory of convection. Regions with $\mathrm{Ra} < \mathrm{Ra}_{\rm crit}$ are stable to convection and shaded in grey.}
\label{fig:FeCZ_structure}
\end{figure*}

Figure~\ref{fig:FeCZ_structure} shows the aspect ratio $\mathrm{A}$.
The aspect ratios are typically large ($10-1000$) except for very massive ($M_\star \sim 60 M_\odot$) stars on the Terminal Age Main Sequence (TAMS\footnote{The TAMS is defined by hydrogen exhaustion in the core.}) so local simulations are likely sufficient to capture their dynamics.
At high masses on the TAMS the aspect ratio is high enough that global (spherical shell) geometry could be important.

\begin{figure*}
\begin{minipage}{0.48\textwidth}
\includegraphics[width=\textwidth]{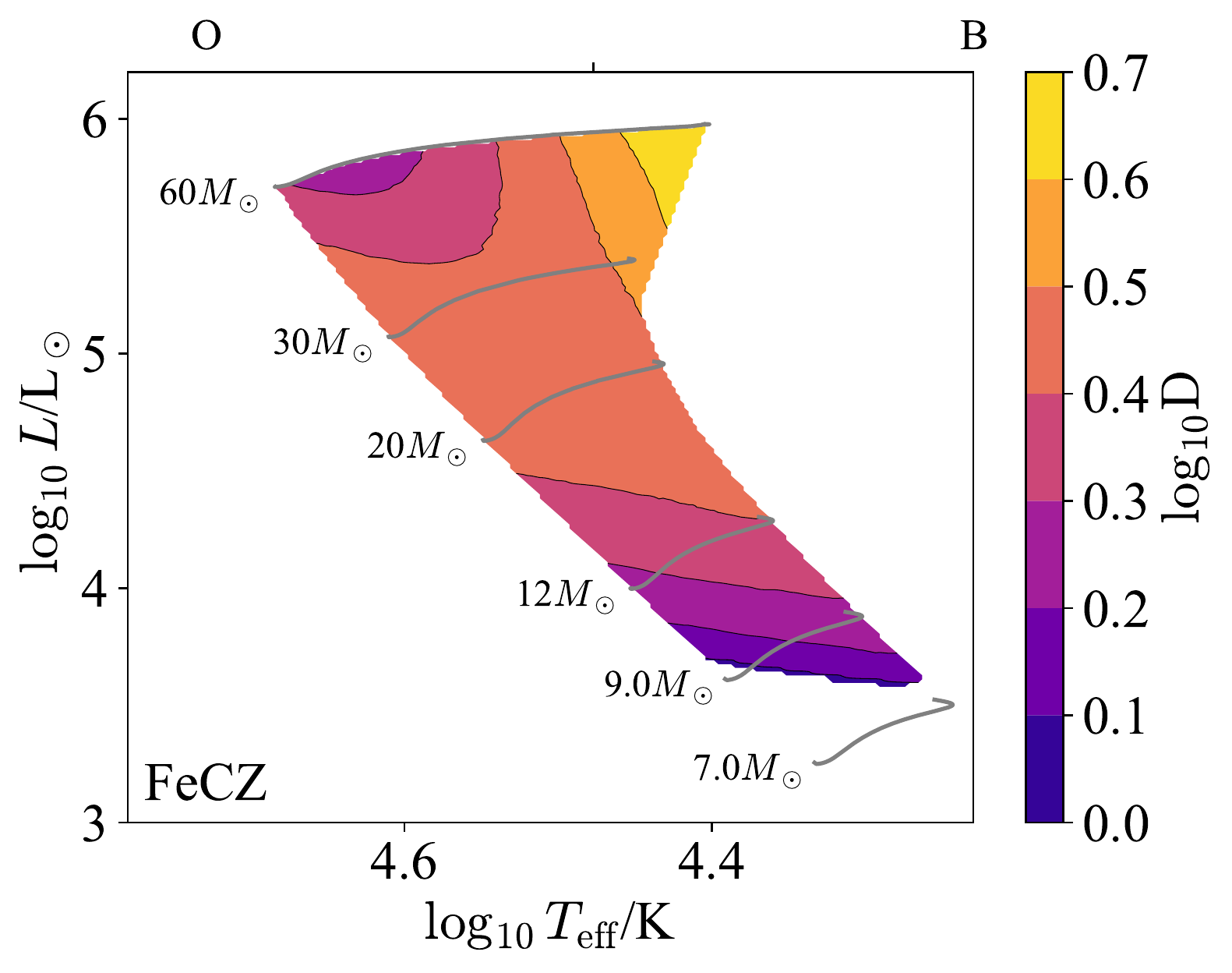}
\end{minipage}
\hfill
\begin{minipage}{0.48\textwidth}
\includegraphics[width=\textwidth]{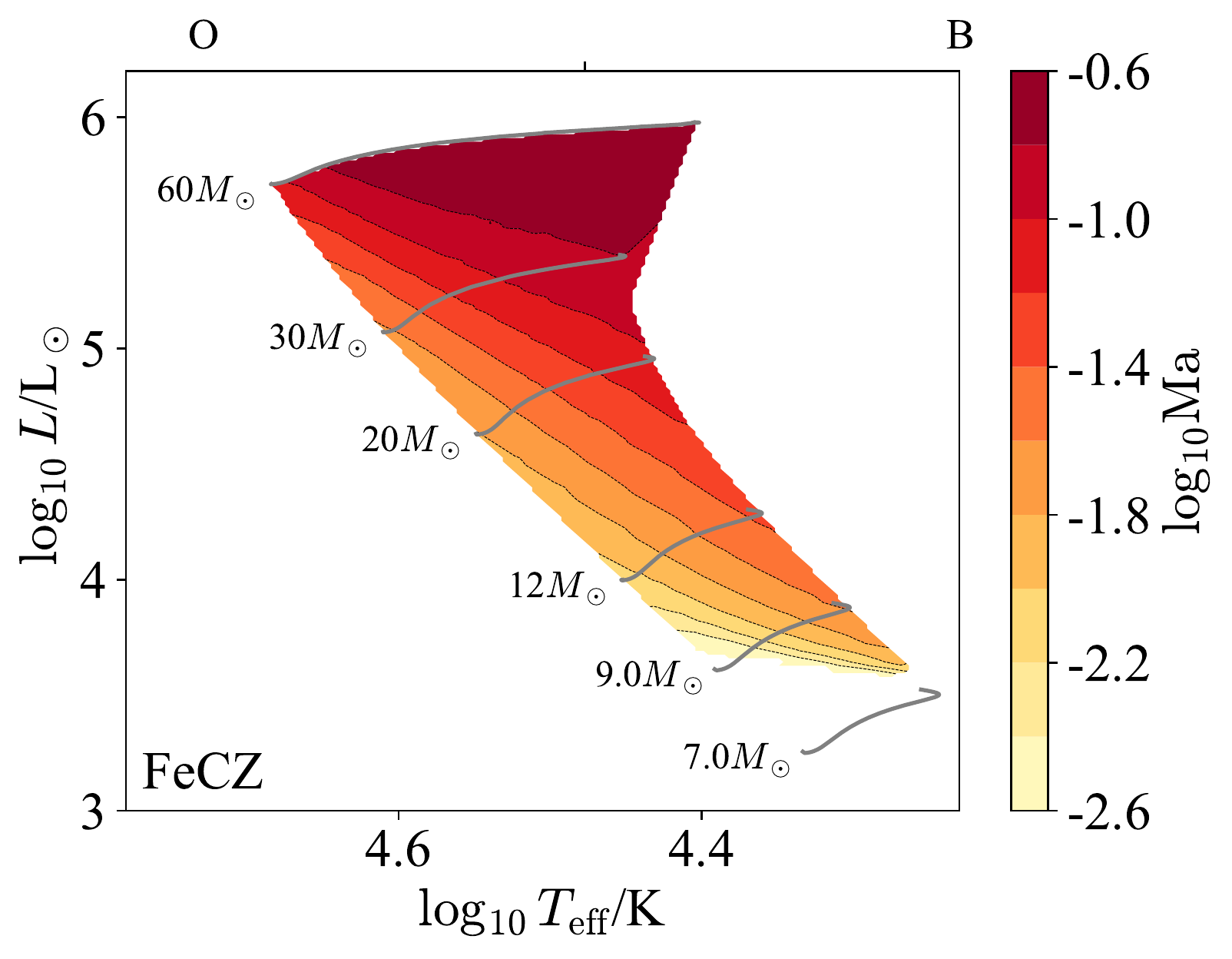}
\end{minipage}
\hfill

\caption{The density ratio $\mathrm{D}$ (left) and Mach number $\mathrm{Ma}$ (right) are shown in terms of $\log T_{\rm eff}$/spectral type and $\log L$ for stellar models with FeCZs and Milky Way metallicity $Z=0.014$. Note that while the density ratio is an input parameter and does not depend on a specific theory of convection, the Mach number is an output of such a theory and so is model-dependent. Regions with $\mathrm{Ra} < \mathrm{Ra}_{\rm crit}$ are stable to convection and shaded in grey.}
\label{fig:FeCZ_equations}
\end{figure*}

Next, the density ratio $\mathrm{D}$ (Figure~\ref{fig:FeCZ_equations}, left) and Mach number $\mathrm{Ma}$ (Figure~\ref{fig:FeCZ_equations}, right) inform which physics the fluid equations must include to model these zones.
The density ratio is typically small, of order $2-3$, and the Mach number ranges from $\la 0.1$ at $M \la 20 M_\odot$ up to $0.3$ at $M \approx 50-60 M_\odot$.
This suggests that below $\approx 20 M_\odot$ the Boussinesq approximation is valid, whereas above this the fully compressible equations may be needed to capture the dynamics at moderate Mach numbers.

\begin{figure}
\centering
\begin{minipage}{0.48\textwidth}
\includegraphics[width=\textwidth]{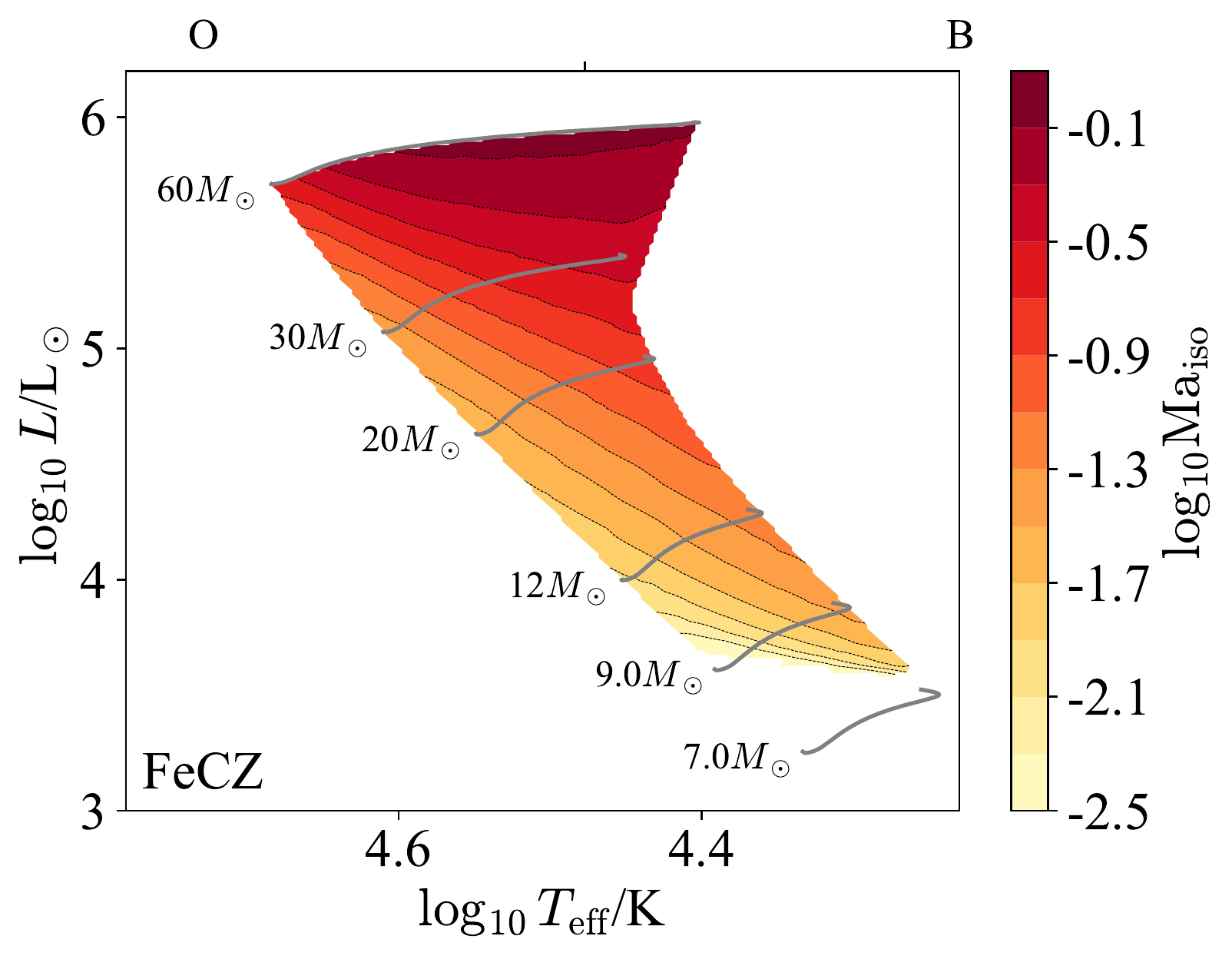}
\end{minipage}

\caption{The Mach number computed with the \emph{isothermal} sound speed is shown in terms of $\log T_{\rm eff}$ and $\log L$ for stellar models with Fe convection zones and Milky Way metallicity $Z=0.014$.}
\label{fig:FeCZ_MachIso}
\end{figure}

In fact the Mach number in Figure~\ref{fig:FeCZ_equations} is an underestimate of the importance of density fluctuations because at high masses the zone is radiation pressure dominated ($\beta_{\rm rad} \sim 1$) and has a moderate P{\'e}clet number ($\mathrm{Pe} \sim 10$), so fluctuations occur isothermally and we should really be comparing the convection speed with the isothermal sound speed rather than the adiabatic one.
Figure~\ref{fig:FeCZ_MachIso} shows this comparison, which reveals even larger Mach numbers ($\mathrm{Ma}_{\rm iso} \sim 1$) at high masses.
Taking this into account, we suggest using the fully compressible equations down to $\approx 12 M_\odot$ to ensure that density fluctuations are correctly accounted for.

\begin{figure*}
\begin{minipage}{0.48\textwidth}
\includegraphics[width=\textwidth]{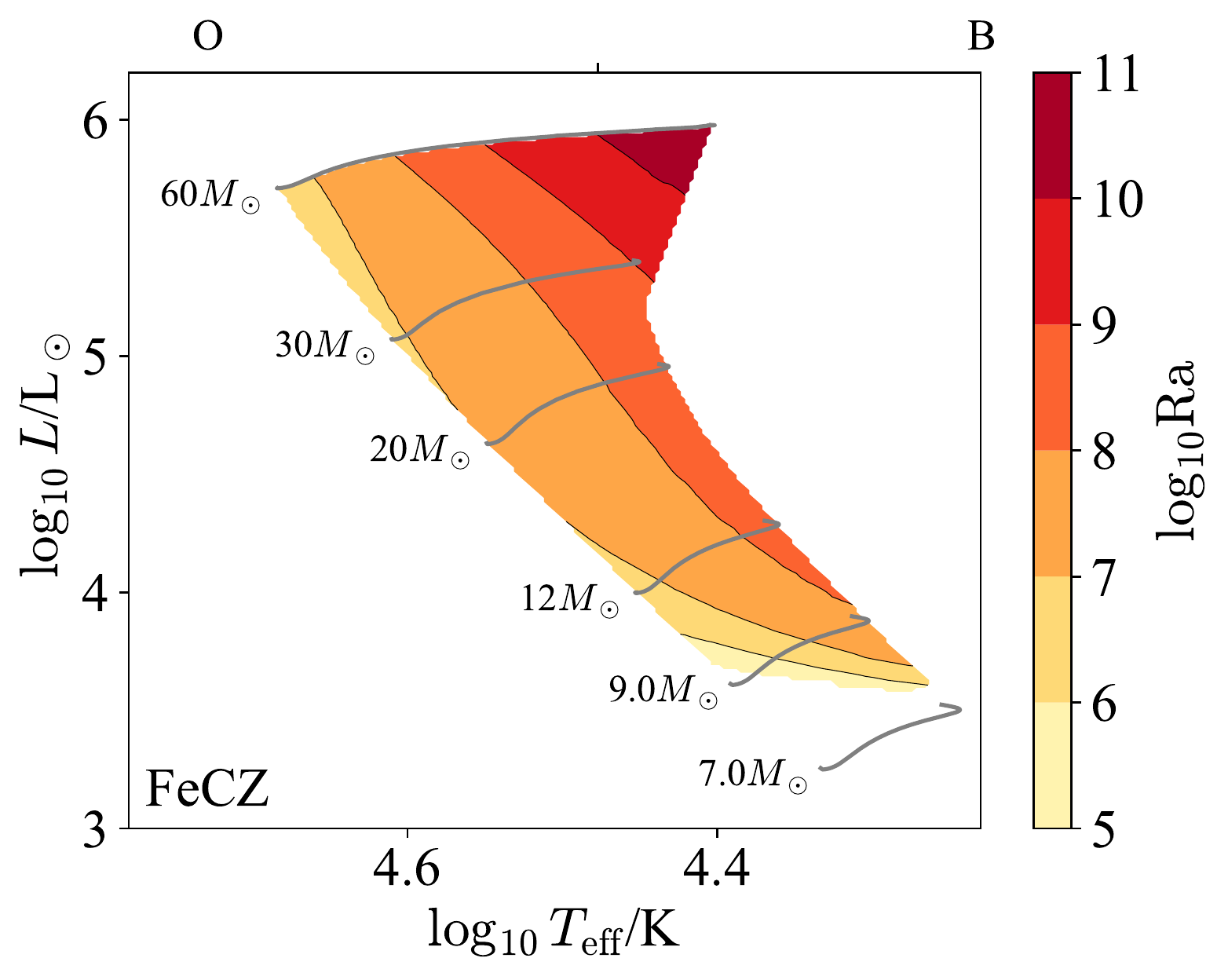}
\end{minipage}
\hfill
\begin{minipage}{0.48\textwidth}
\includegraphics[width=\textwidth]{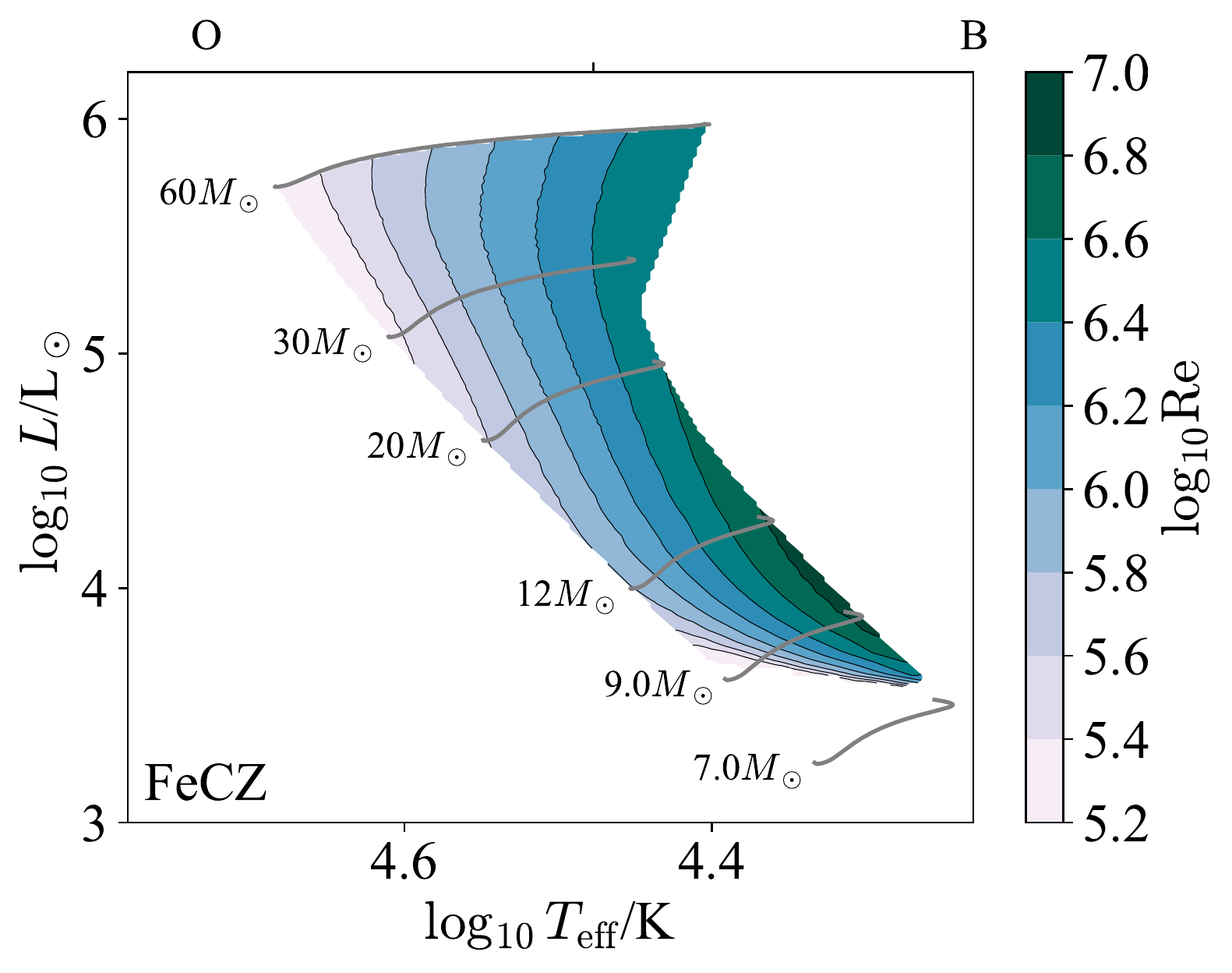}
\end{minipage}
\hfill

\caption{The Rayleigh number $\mathrm{Ra}$ (left) and Reynolds number $\mathrm{Re}$ (right) are shown in terms of $\log T_{\rm eff}$/spectral type and $\log L$ for stellar models with FeCZs and Milky Way metallicity $Z=0.014$.  Note that while the Rayleigh number is an input parameter and does not depend on a specific theory of convection, the Reynolds number is an output of such a theory and so is model-dependent. Regions with $\mathrm{Ra} < \mathrm{Ra}_{\rm crit}$ are stable to convection and shaded in grey.}
\label{fig:FeCZ_stability}
\end{figure*}

The Rayleigh number $\mathrm{Ra}$ (Figure~\ref{fig:FeCZ_stability}, left) determines whether or not a putative convection zone is actually unstable to convection, and the Reynolds number $\mathrm{Re}$ determines how turbulent the zone is if instability sets in (Figure~\ref{fig:FeCZ_stability}, right).
The Rayleigh number is generally large ($10^{5}-10^{11}$), as is the Reynolds number ($10^{5}-10^{7}$), suggesting that the FeCZ is strongly unstable to convection and exhibits well-developed turbulence.

\begin{figure*}
\centering
\begin{minipage}{0.48\textwidth}
\includegraphics[width=\textwidth]{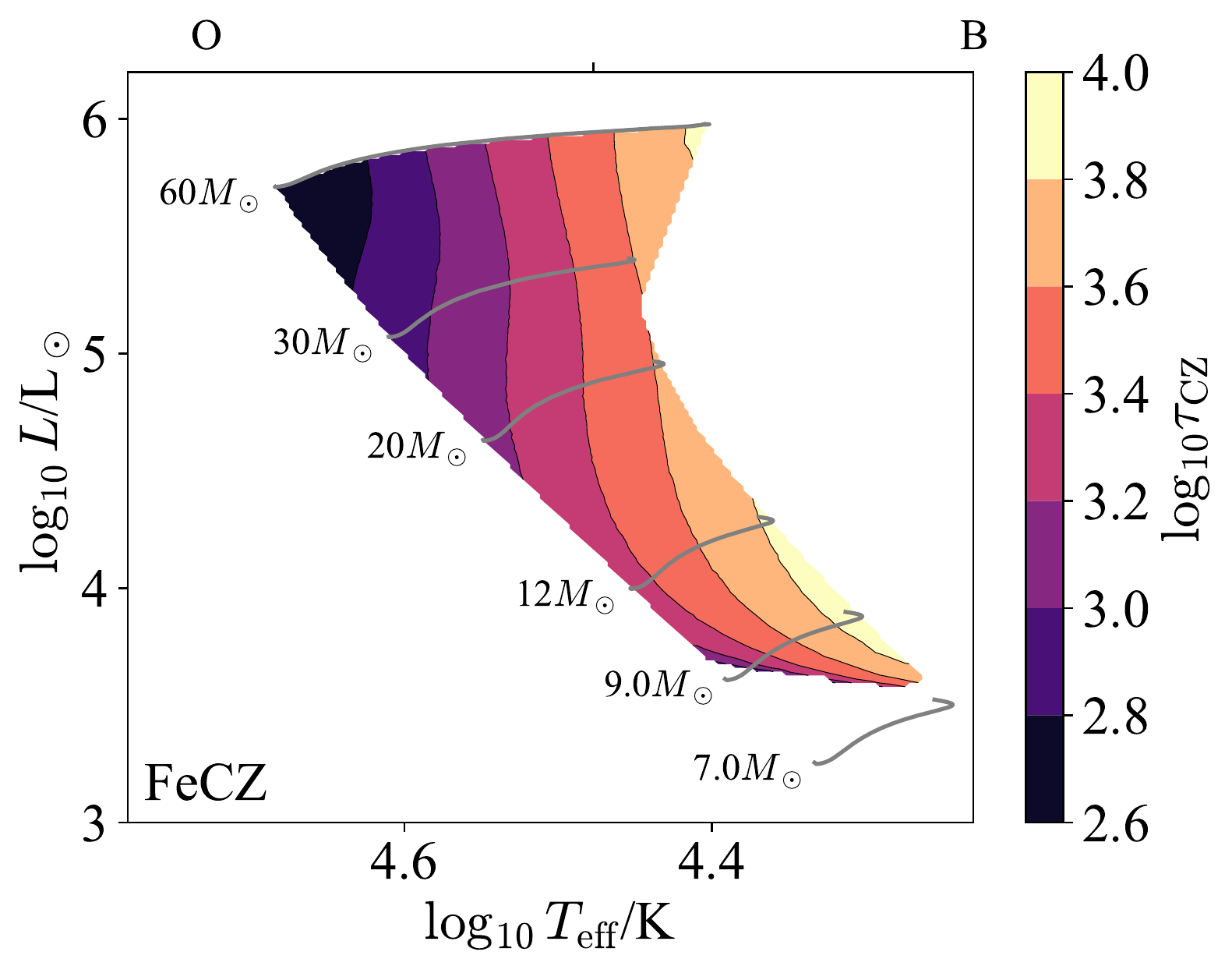}
\end{minipage}
\hfill
\begin{minipage}{0.48\textwidth}
\includegraphics[width=\textwidth]{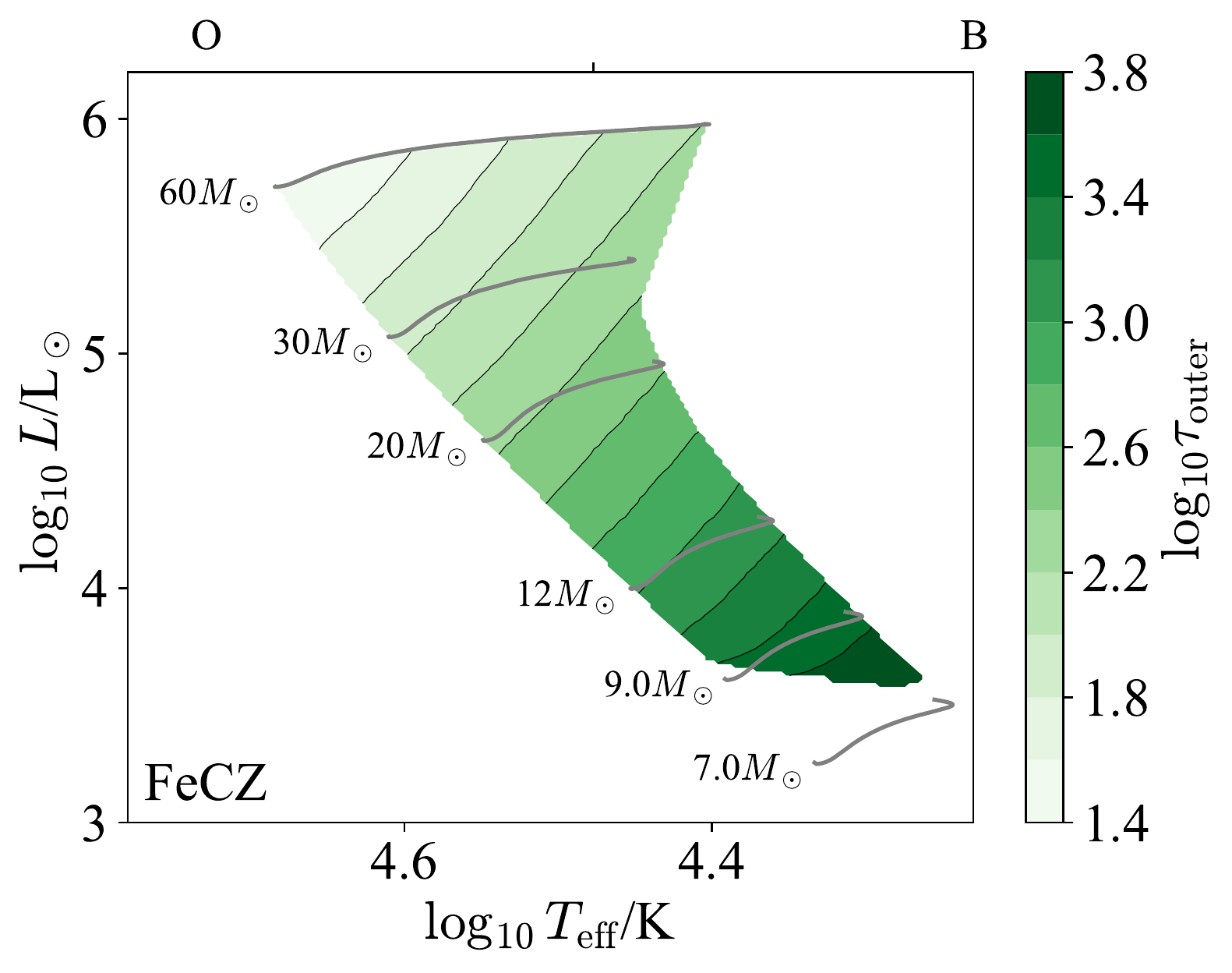}
\end{minipage}
\hfill
\caption{The convection optical depth $\tau_{\rm CZ}$ (left) and the optical depth to the surface $\tau_{\rm outer}$ (right) are shown in terms of $\log T_{\rm eff}$/spectral type and $\log L$ for stellar models with FeCZs and Milky Way metallicity $Z=0.014$. Note that both of these are input parameters, and do not depend on a specific theory of convection. Regions with $\mathrm{Ra} < \mathrm{Ra}_{\rm crit}$ are stable to convection and shaded in grey.}
\label{fig:FeCZ_optical}
\end{figure*}

The optical depth across a convection zone $\tau_{\rm CZ}$ (Figure~\ref{fig:FeCZ_optical}, left) indicates whether or not radiation can be handled in the diffusive approximation, while the optical depth from the outer boundary to infinity $\tau_{\rm outer}$ (Figure~\ref{fig:FeCZ_optical}, right) indicates the nature of radiative transfer and cooling in the outer regions of the convection zone.
We see that the optical depth across these zones is large ($\tau_{\rm CZ} \sim 10^{3}$), as is that from the outer boundary to infinity ($\tau_{\rm outer} \ga 10$).
This suggests that radiation can be treated in the diffusive approximation.

Our only reservation with this conclusion is that both the Mach number and Eddington ratios can be large in the FeCZ, so density fluctuations can be large and can open up low-density optically thin ``tunnels'' through the FeCZ.
This is what is seen in 3D radiation hydrodynamics simulations of the FeCZ~\citep{2020ApJ...902...67S}, which show strong correlations between the radiative flux and the attenuation length $(\kappa \rho)^{-1}$.
Thus radiation hydrodynamics seems to be essential for modelling the FeCZ, at least at the higher masses ($M \ga 15 M_\odot$) which host near-unity Eddington ratios and moderate Mach numbers.

\begin{figure*}
\centering
\begin{minipage}{0.48\textwidth}
\includegraphics[width=\textwidth]{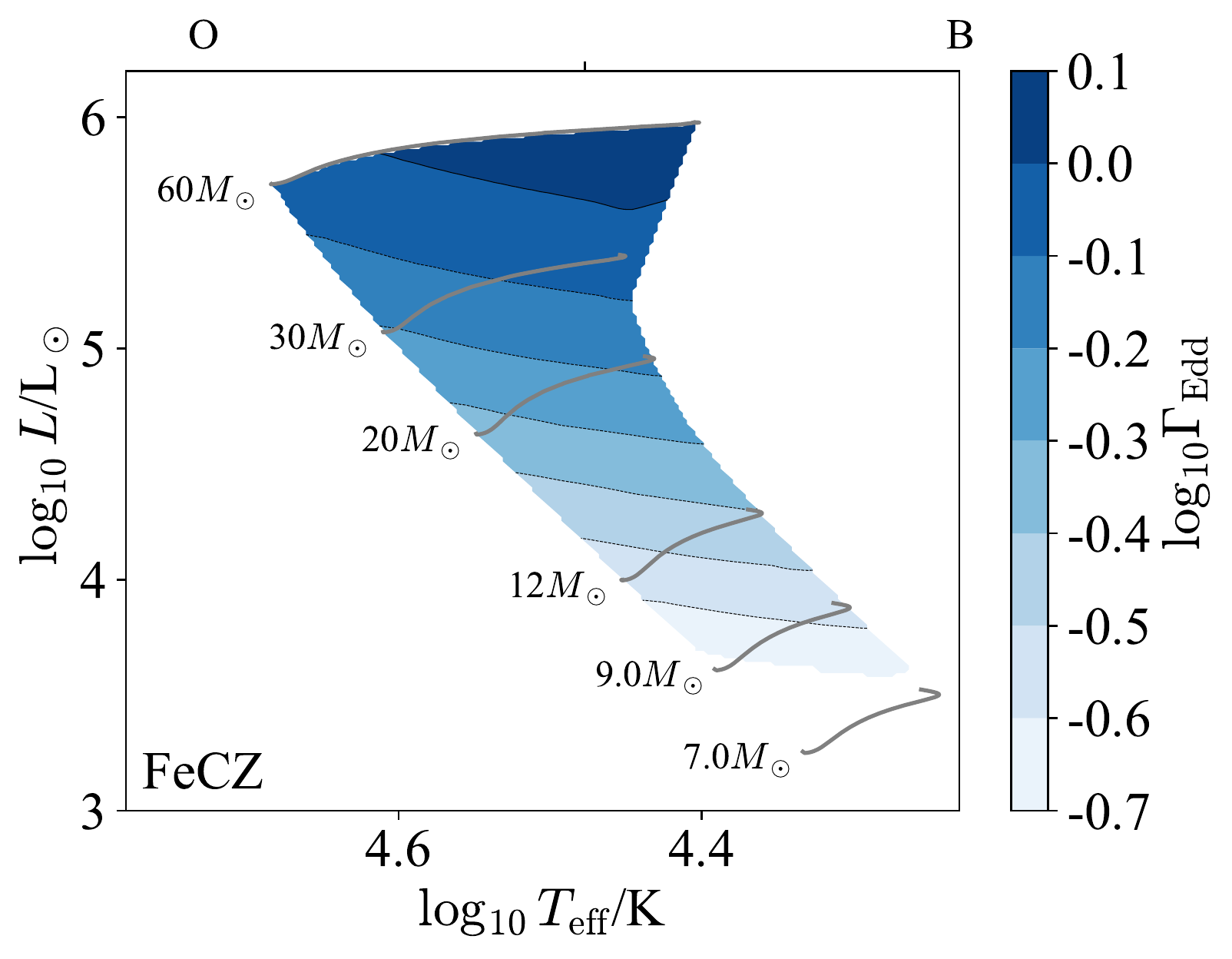}
\end{minipage}
\hfill
\begin{minipage}{0.48\textwidth}
\includegraphics[width=\textwidth]{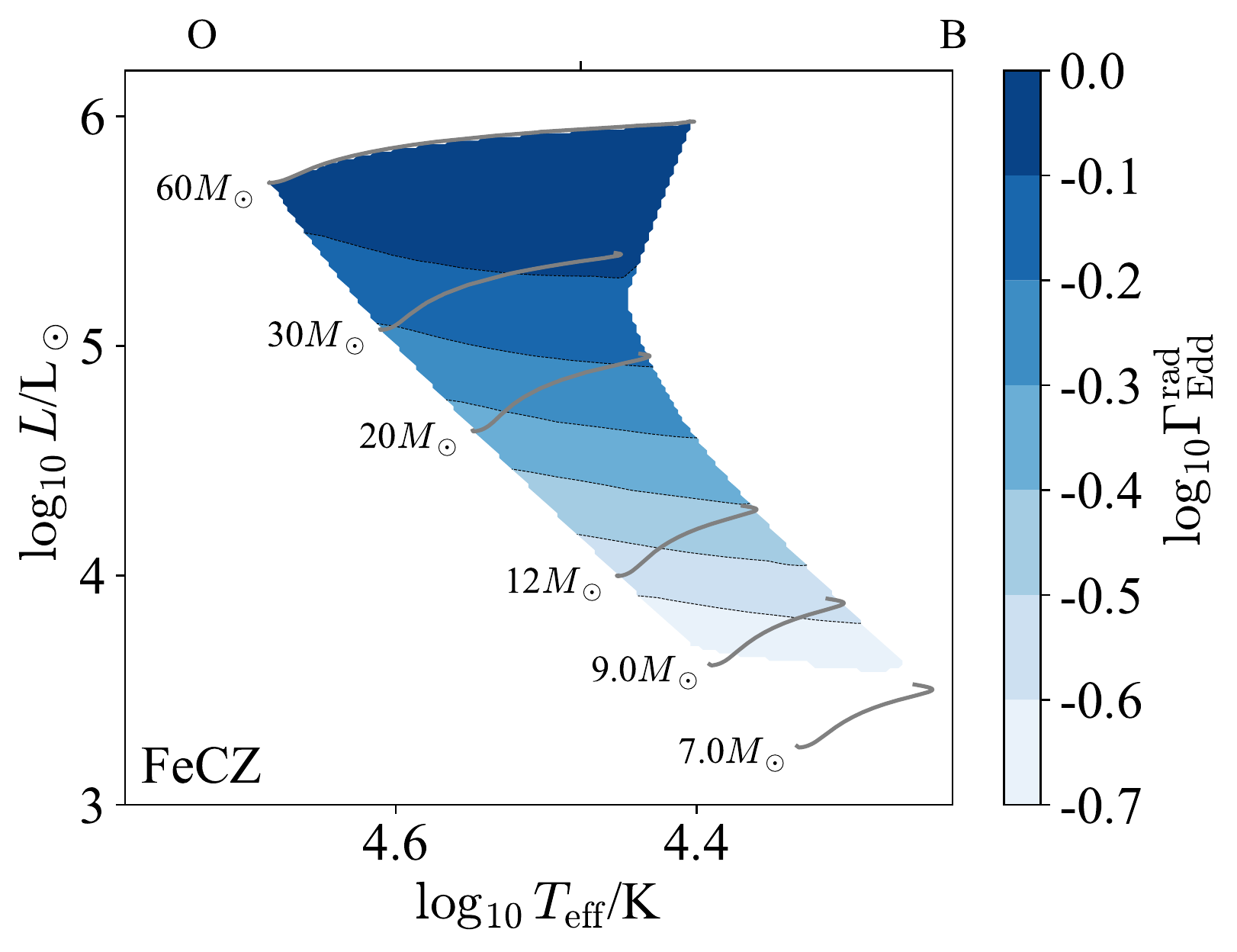}
\end{minipage}
\hfill
\caption{The Eddington ratio with the full luminosity $\Gamma_{\rm Edd}$ (left) and the radiative luminosity (right) are shown in terms of $\log T_{\rm eff}$/spectral type and $\log L$ for stellar models with FeCZs and Milky Way metallicity $Z=0.014$. Note that while $\Gamma_{\rm Edd}$ is an input parameter and does not depend on a specific theory of convection, $\Gamma_{\rm Edd}^{\rm rad}$ is an output of such a theory and so is model-dependent. Regions with $\mathrm{Ra} < \mathrm{Ra}_{\rm crit}$ are stable to convection and shaded in grey.}
\label{fig:FeCZ_eddington}
\end{figure*}

The Eddington ratio $\Gamma_{\rm Edd}$ (Figure~\ref{fig:FeCZ_eddington}, left) indicates whether or not radiation hydrodynamic instabilities are important in the non-convecting state, and the radiative Eddington ratio $\Gamma_{\rm Edd}^{\rm rad}$ (Figure~\ref{fig:FeCZ_eddington}, right) indicates the same in the developed convective state.
Both ratios are moderate at low masses ($\Gamma \sim 0.3$ at $M \sim 10 M_\odot$) and reach unity at high masses ($M \ga 25 M_\odot$), so radiation hydrodynamic instabilities are almost certainly important in the FeCZ.

\begin{figure*}
\centering
\begin{minipage}{0.48\textwidth}
\includegraphics[width=\textwidth]{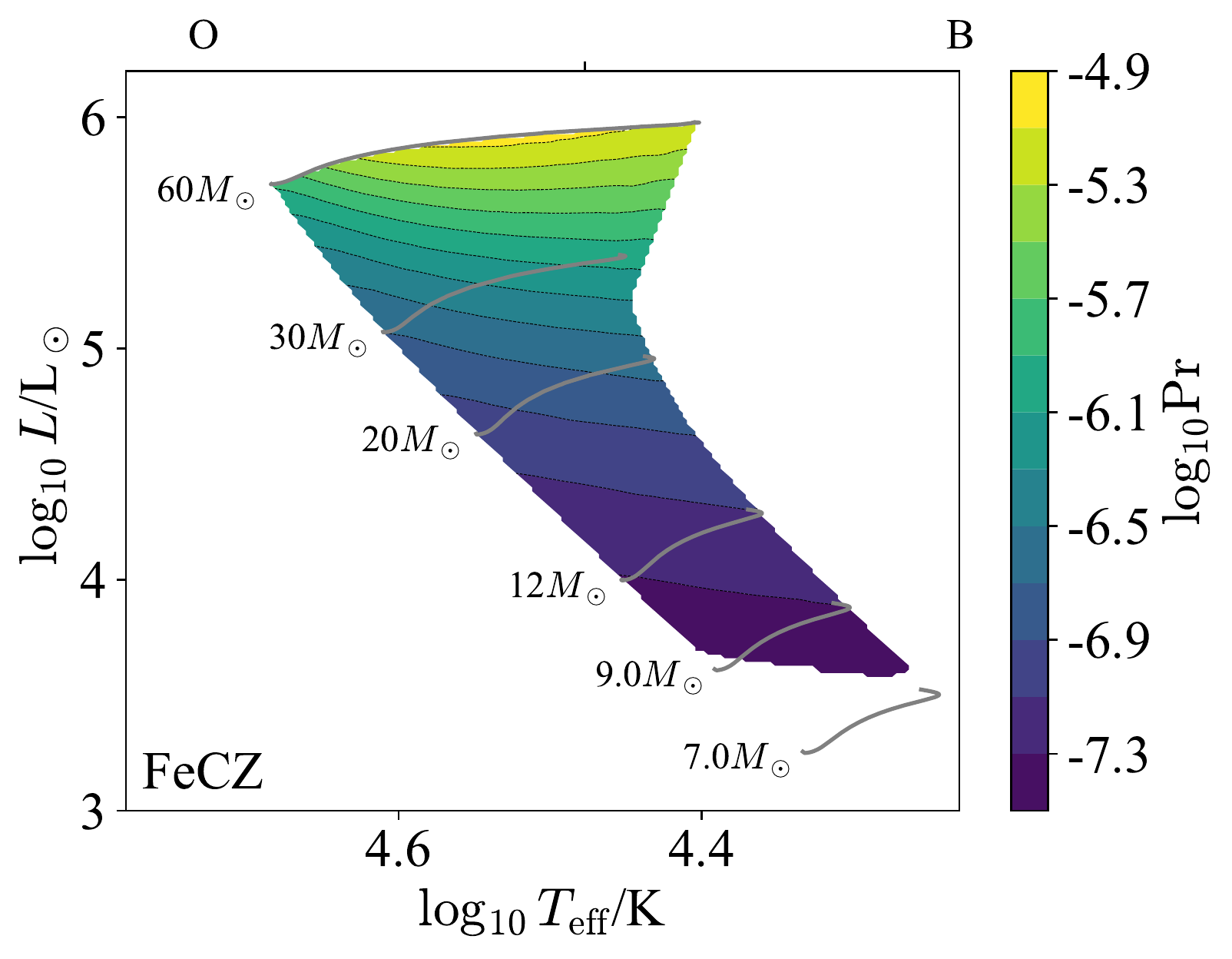}
\end{minipage}
\hfill
\begin{minipage}{0.48\textwidth}
\includegraphics[width=\textwidth]{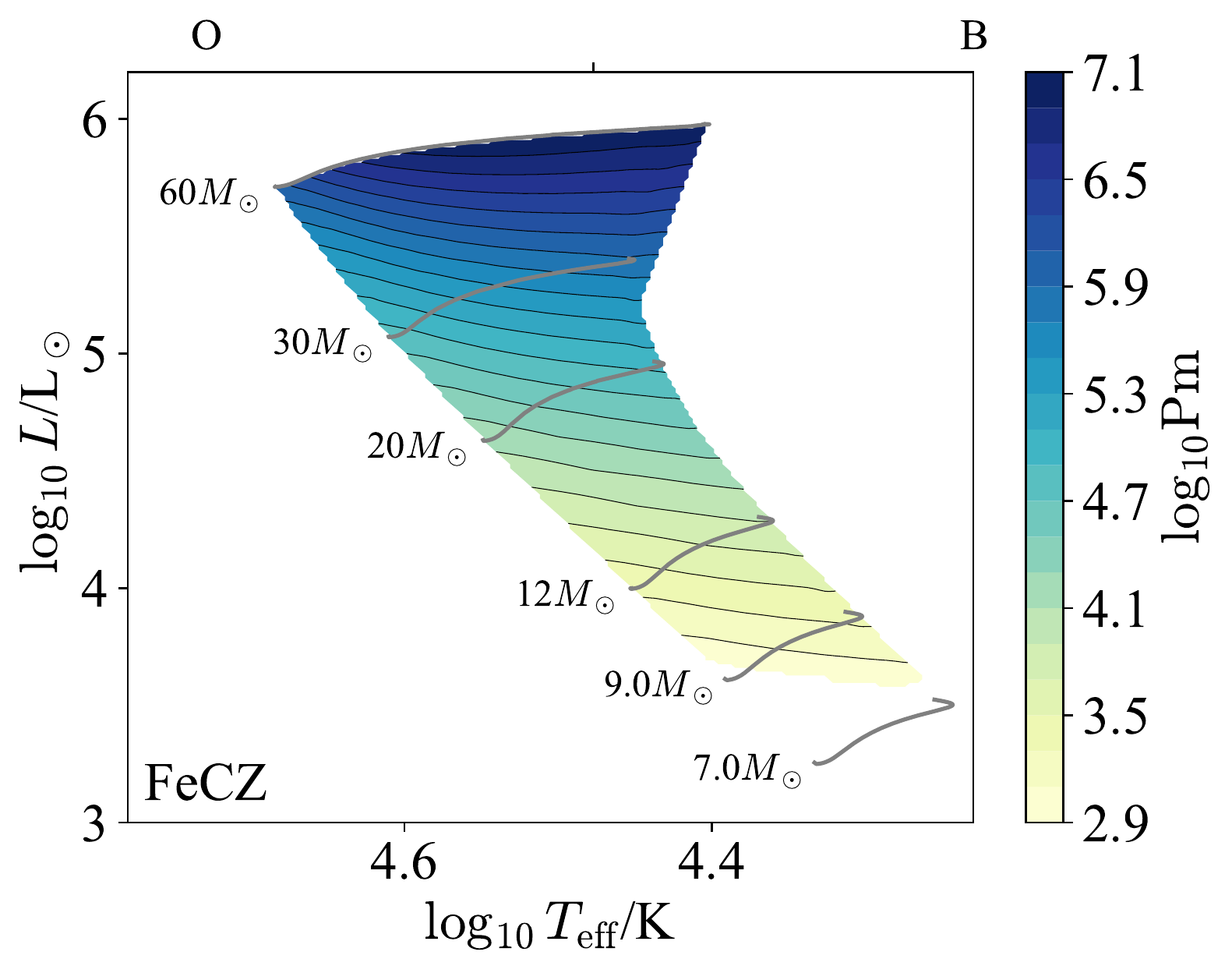}
\end{minipage}
\hfill

\caption{The Prandtl number $\mathrm{Pr}$ (left) and magnetic Prandtl number $\mathrm{Pm}$ (right) are shown in terms of $\log T_{\rm eff}$/spectral type and $\log L$ for stellar models with FeCZs and Milky Way metallicity $Z=0.014$. Note that both $\mathrm{Pr}$ and $\mathrm{Pm}$ are input parameters, and so do not depend on a specific theory of convection. Regions with $\mathrm{Ra} < \mathrm{Ra}_{\rm crit}$ are stable to convection and shaded in grey.}
\label{fig:FeCZ_diffusivities}
\end{figure*}

The Prandtl number $\mathrm{Pr}$ (Figure~\ref{fig:FeCZ_diffusivities}, left) measures the relative importance of thermal diffusion and viscosity, and the magnetic Prandtl number $\mathrm{Pm}$ (Figure~\ref{fig:FeCZ_diffusivities}, right) measures the same for magnetic diffusion and viscosity.
The Prandtl number is always small in these models, so the thermal diffusion length-scale is much larger than the viscous scale.
By contrast, the magnetic Prandtl number is very large ($10^{2}-10^{8}$), so the viscous scale is much larger than the magnetic diffusion scale.

The fact that $\mathrm{Pm}$ is large at high masses is notable because the quasistatic approximation for magnetohydrodynamics has frequently been used to study magnetoconvection in minimal 3D MHD simulations of planetary and stellar interiors~\citep[e.g.][]{yan_calkins_maffei_julien_tobias_marti_2019} and assumes that $\mathrm{Rm} = \mathrm{Pm} \mathrm{Re} \rightarrow 0$; in doing so, this approximation assumes a global background magnetic field is dominant and neglects the nonlinear portion of the Lorentz force. This approximation breaks down in convection zones with $\mathrm{Pm} > 1$ and future numerical experiments should seek to understand how magnetoconvection operates in this regime.

\begin{figure*}
\centering
\begin{minipage}{0.48\textwidth}
\includegraphics[width=\textwidth]{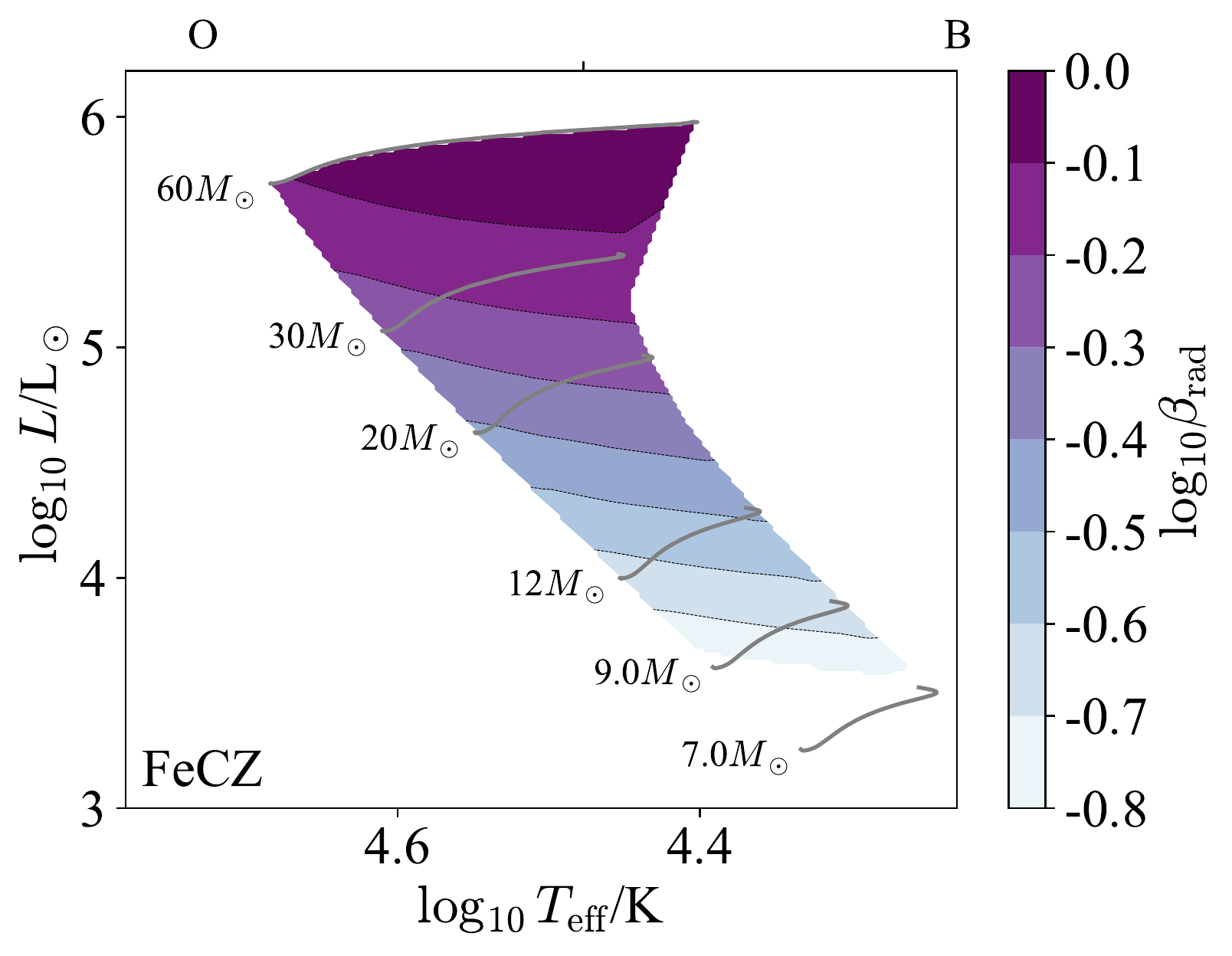}
\end{minipage}
\hfill

\caption{The radiation pressure ratio $\beta_{\rm rad}$ is shown in terms of $\log T_{\rm eff}$/spectral type and $\log L$ for stellar models with FeCZs and Milky Way metallicity $Z=0.014$. Note that this ratio is an input parameter, and does not depend on a specific theory of convection. Regions with $\mathrm{Ra} < \mathrm{Ra}_{\rm crit}$ are stable to convection and shaded in grey.}
\label{fig:FeCZ_beta}
\end{figure*}

The radiation pressure ratio $\beta_{\rm rad}$ (Figure~\ref{fig:FeCZ_beta}) measures the importance of radiation in setting the thermodynamic properties of the fluid.
This is a 30-100\% correction and so is very important to capture in modelling the FeCZ.

\begin{figure*}
\centering
\begin{minipage}{0.48\textwidth}
\includegraphics[width=\textwidth]{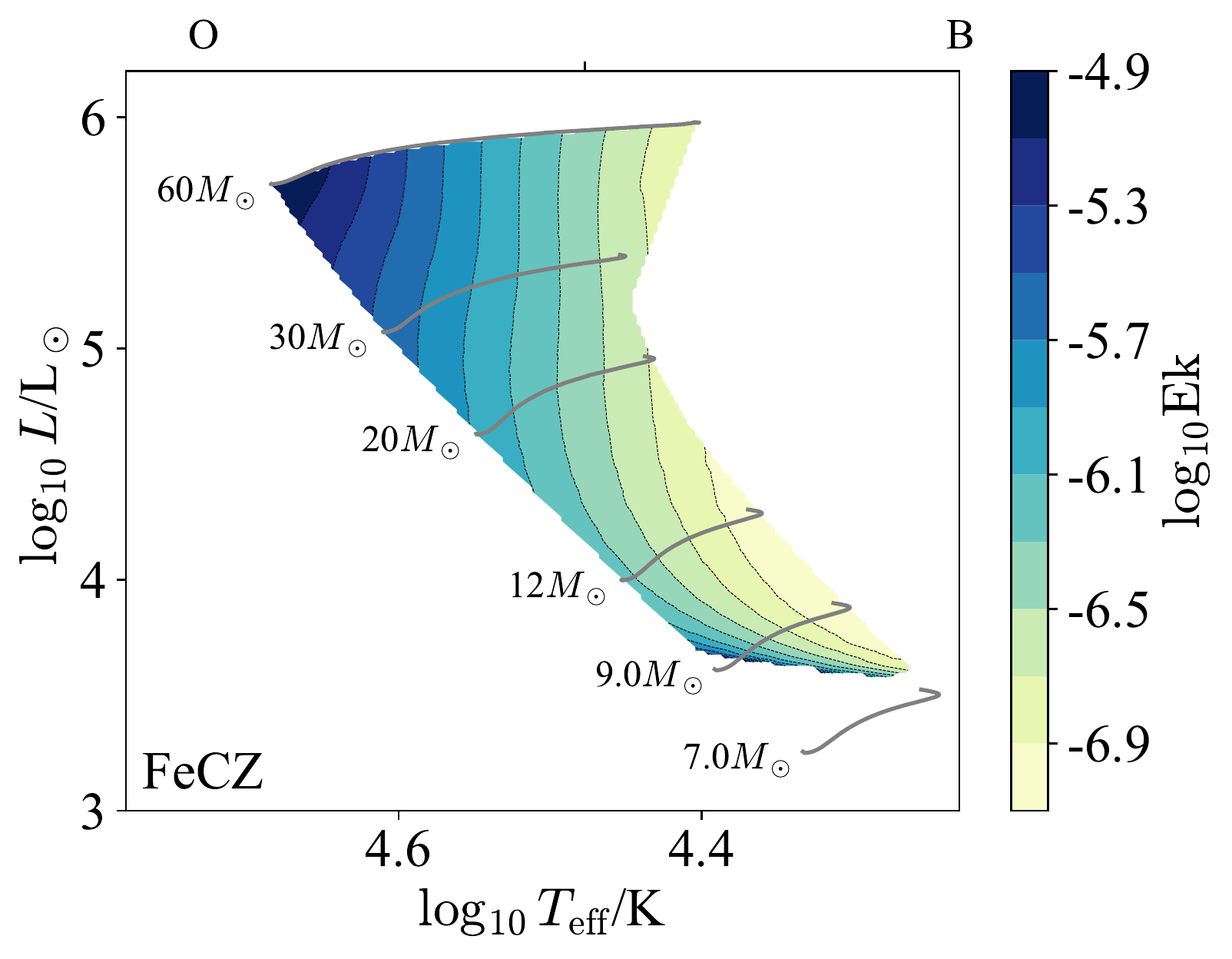}
\end{minipage}

\caption{The Ekman number $\mathrm{Ek}$ is shown in terms of $\log T_{\rm eff}$/spectral type and $\log L$ for stellar models with FeCZs and Milky Way metallicity $Z=0.014$. Note that the Ekman number is an input parameter, and does not depend on a specific theory of convection. Regions with $\mathrm{Ra} < \mathrm{Ra}_{\rm crit}$ are stable to convection and shaded in grey.}
\label{fig:FeCZ_ekman}
\end{figure*}

The Ekman number $\mathrm{Ek}$ (Figure~\ref{fig:FeCZ_ekman}) indicates the relative importance of viscosity and rotation.
This is tiny across the HRD~\footnote{Note that, because the Prandtl number is also very small, this does not significantly alter the critical Rayleigh number~(see Ch3 of~\cite{1961hhs..book.....C} and appendix D of~\cite{2022arXiv220110567J}).}, so we expect rotation to dominate over viscosity, except at very small length-scales.

\begin{figure*}
\centering
\begin{minipage}{0.48\textwidth}
\includegraphics[width=\textwidth]{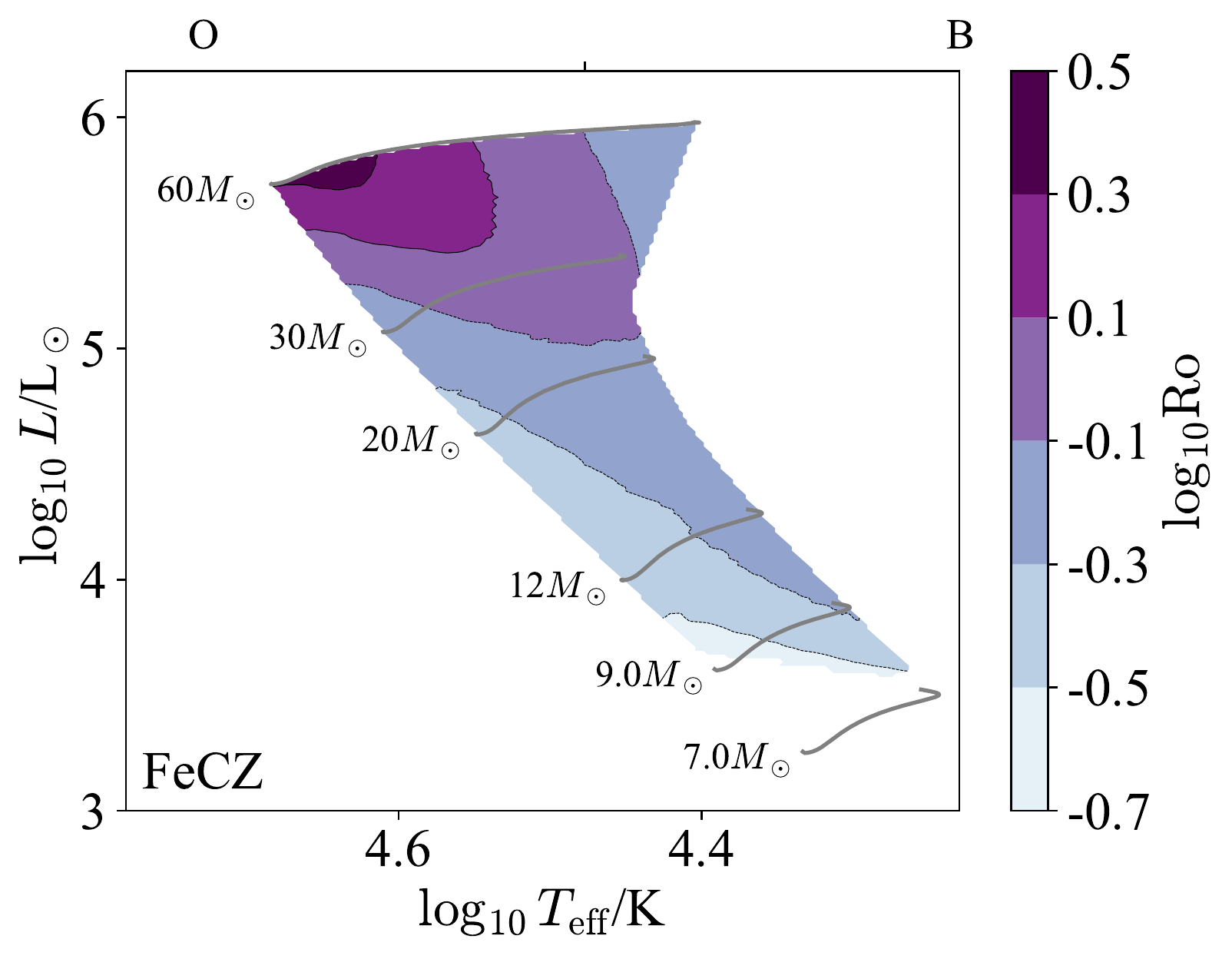}
\end{minipage}
\hfill
\begin{minipage}{0.48\textwidth}
\includegraphics[width=\textwidth]{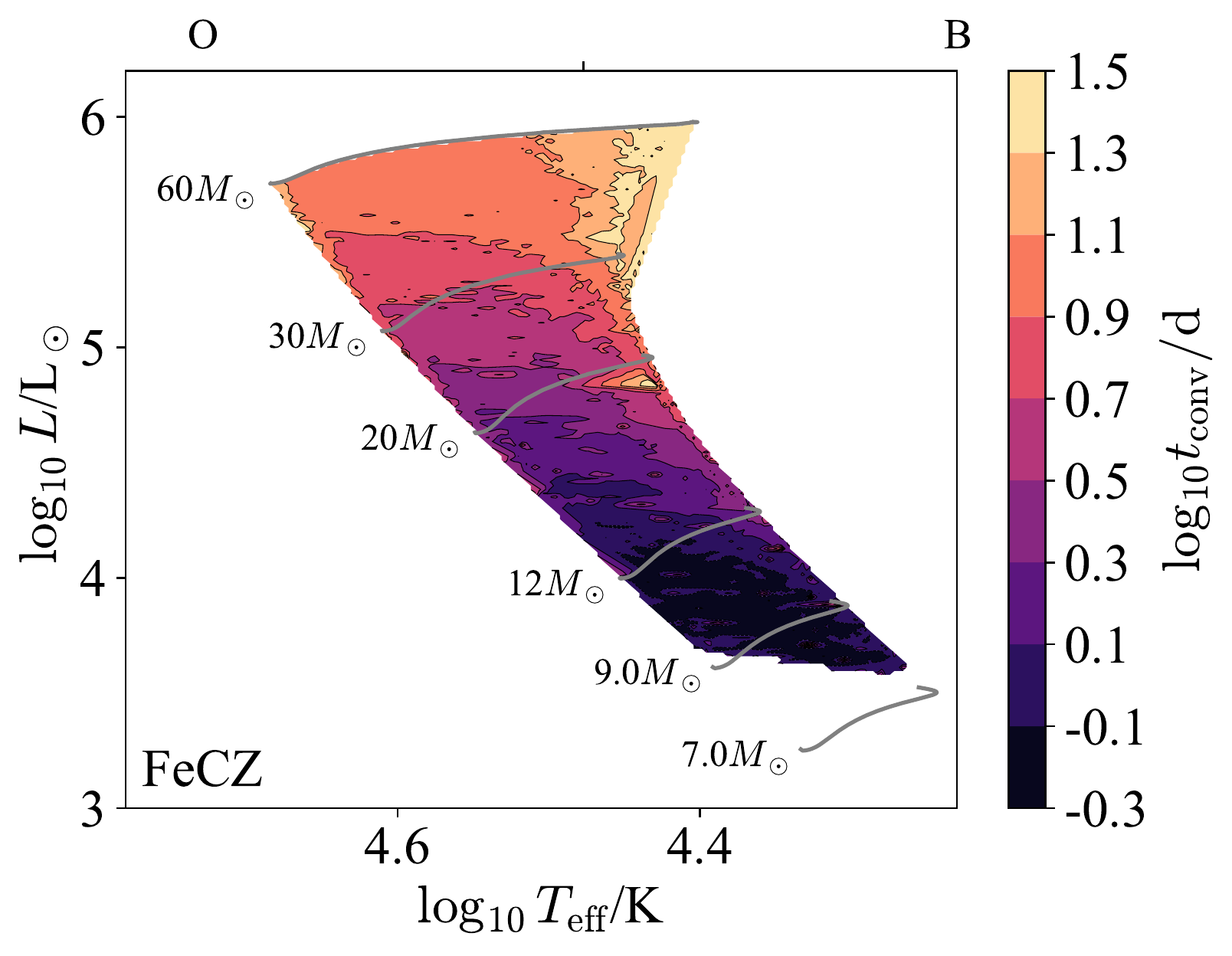}
\end{minipage}
\hfill

\caption{The Rossby number $\mathrm{Ro}$ (left) and turnover time $t_{\rm conv}$ (right) are shown in terms of $\log T_{\rm eff}$/spectral type and $\log L$ for stellar models with FeCZs and Milky Way metallicity $Z=0.014$. Note that both $\mathrm{Ro}$ and $t_{\rm conv}$ are outputs of a theory of convection and so are model-dependent. Regions with $\mathrm{Ra} < \mathrm{Ra}_{\rm crit}$ are stable to convection and shaded in grey.}
\label{fig:FeCZ_rotation}
\end{figure*}

The Rossby number $\mathrm{Ro}$ (Figure~\ref{fig:FeCZ_rotation}, left) measures the relative importance of rotation and inertia.
This is of order unity, with a gradient from moderately smaller ($\sim 0.1$) to moderately larger ($\sim 3$) running from low to high mass.
We conclude then that for typical rotation rates~\citep{2013A&A...557L..10N} the FeCZ is rotationally constrained at low masses ($M \la 12 M_\odot$), weakly so at intermediate masses ($12 M_\odot \la M \la 30 M_\odot$) and not constrained at high masses ($M \ga 30 M_\odot$).

We have assumed a fiducial rotation law to calculate $\mathrm{Ro}$.
Stars exhibit a variety of different rotation rates, so we also show the convective turnover time $t_{\rm conv}$ (Figure~\ref{fig:FeCZ_rotation}, right) which may be used to estimate the Rossby number for different rotation periods.

\begin{figure*}
\centering
\begin{minipage}{0.48\textwidth}
\includegraphics[width=\textwidth]{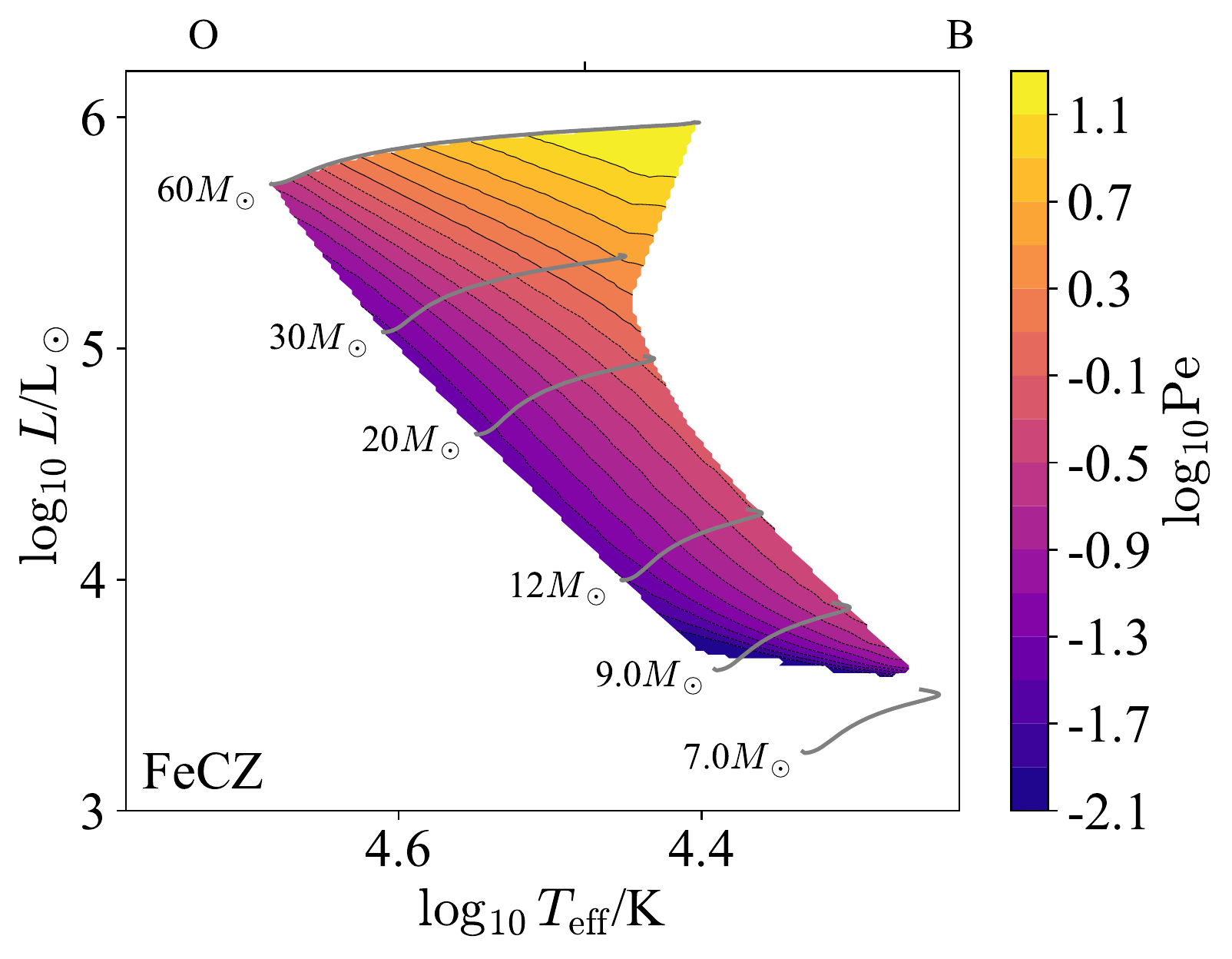}
\end{minipage}
\hfill
\begin{minipage}{0.48\textwidth}
\includegraphics[width=\textwidth]{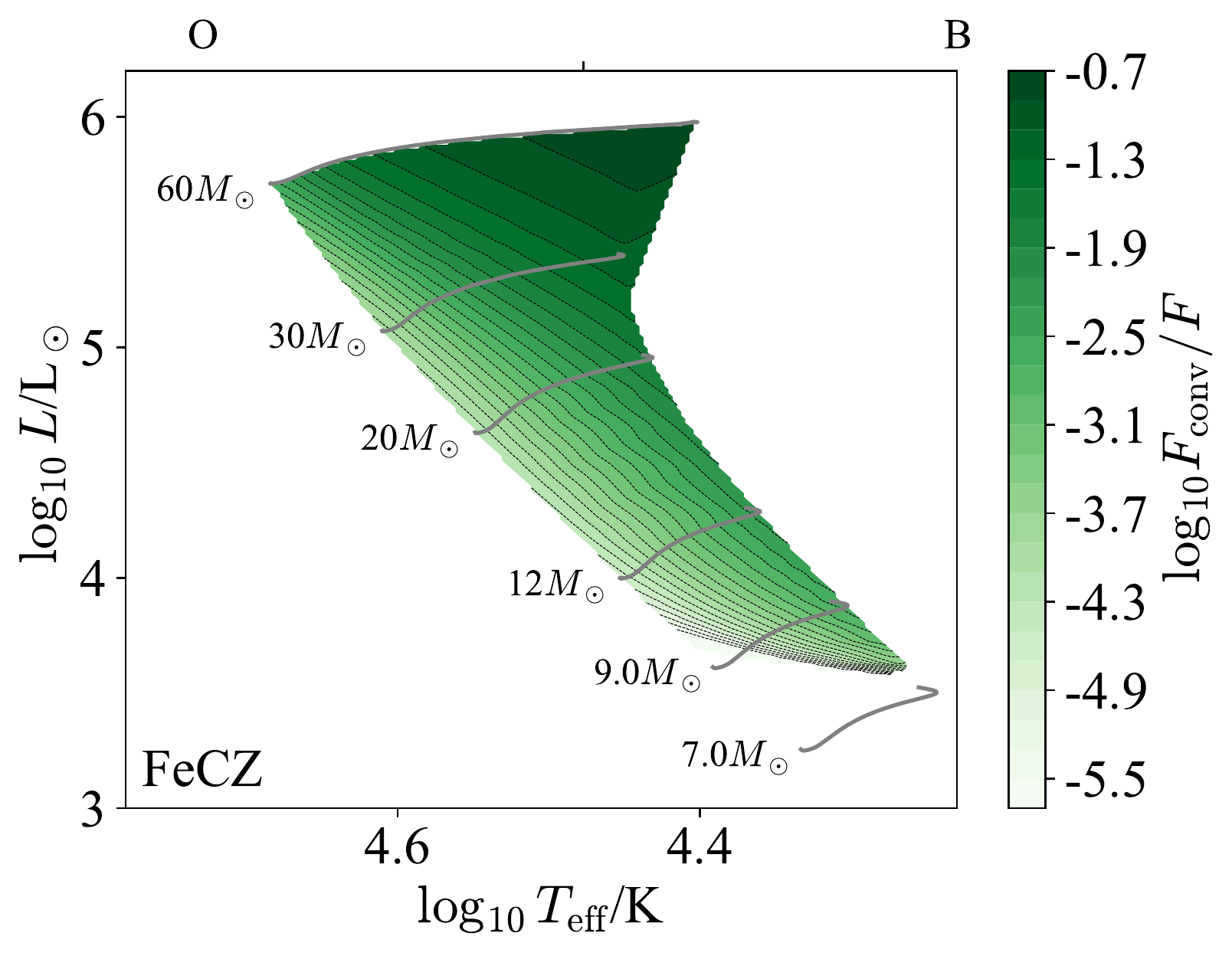}
\end{minipage}

\caption{The P{\'e}clet number $\mathrm{Pe}$ (left) and $F_{\rm conv}/F$ (right) are shown in terms of $\log T_{\rm eff}$/spectral type and $\log L$ for stellar models with FeCZs and Milky Way metallicity $Z=0.014$. Note that both $\mathrm{Pe}$ and $F_{\rm conv}/F$ are outputs of a theory of convection and so are model-dependent. Regions with $\mathrm{Ra} < \mathrm{Ra}_{\rm crit}$ are stable to convection and shaded in grey.}
\label{fig:FeCZ_efficiency}
\end{figure*}

The P{\'e}clet number $\mathrm{Pe}$ (Figure~\ref{fig:FeCZ_efficiency}, left) measures the relative importance of advection and diffusion in transporting heat, and the flux ratio $F_{\rm conv}/F$ (Figure~\ref{fig:FeCZ_efficiency}, right) reports the fraction of the energy flux which is advected.
Both exhibit substantial variation with mass.
The P{\'e}clet number varies from order $30$ at high masses on the TAMS to very small at low masses ($10^{-2}$), and the flux ratio similarly varies from $\sim 0.3$ at high masses on the TAMS to tiny ($10^{-6}$) at low masses.
That is, there is a large gradient in convective efficiency with mass, with moderately efficient convection at high masses and very inefficient convection at low masses.

\begin{figure*}
\centering
\begin{minipage}{0.48\textwidth}
\includegraphics[width=\textwidth]{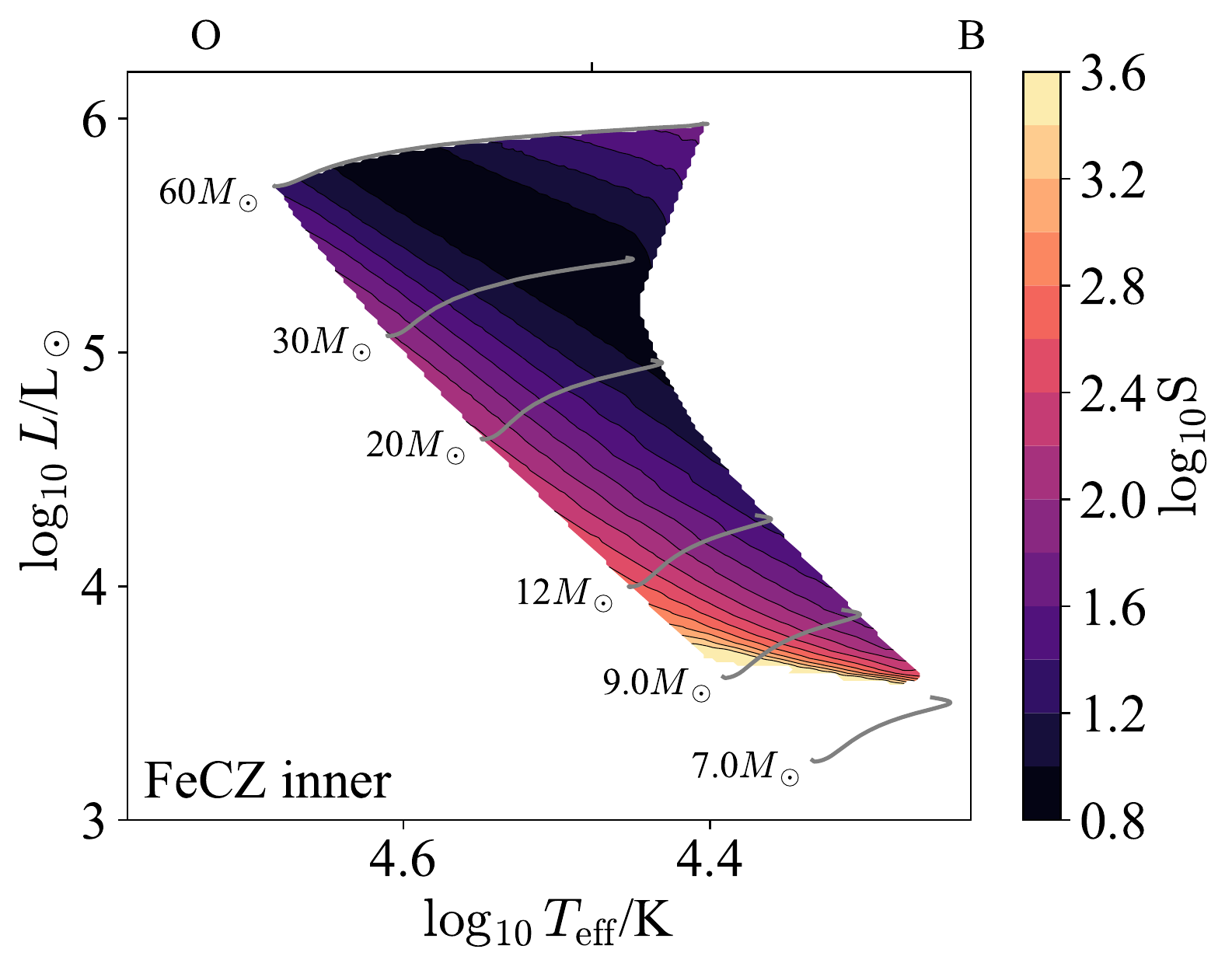}
\end{minipage}
\hfill
\begin{minipage}{0.48\textwidth}
\includegraphics[width=\textwidth]{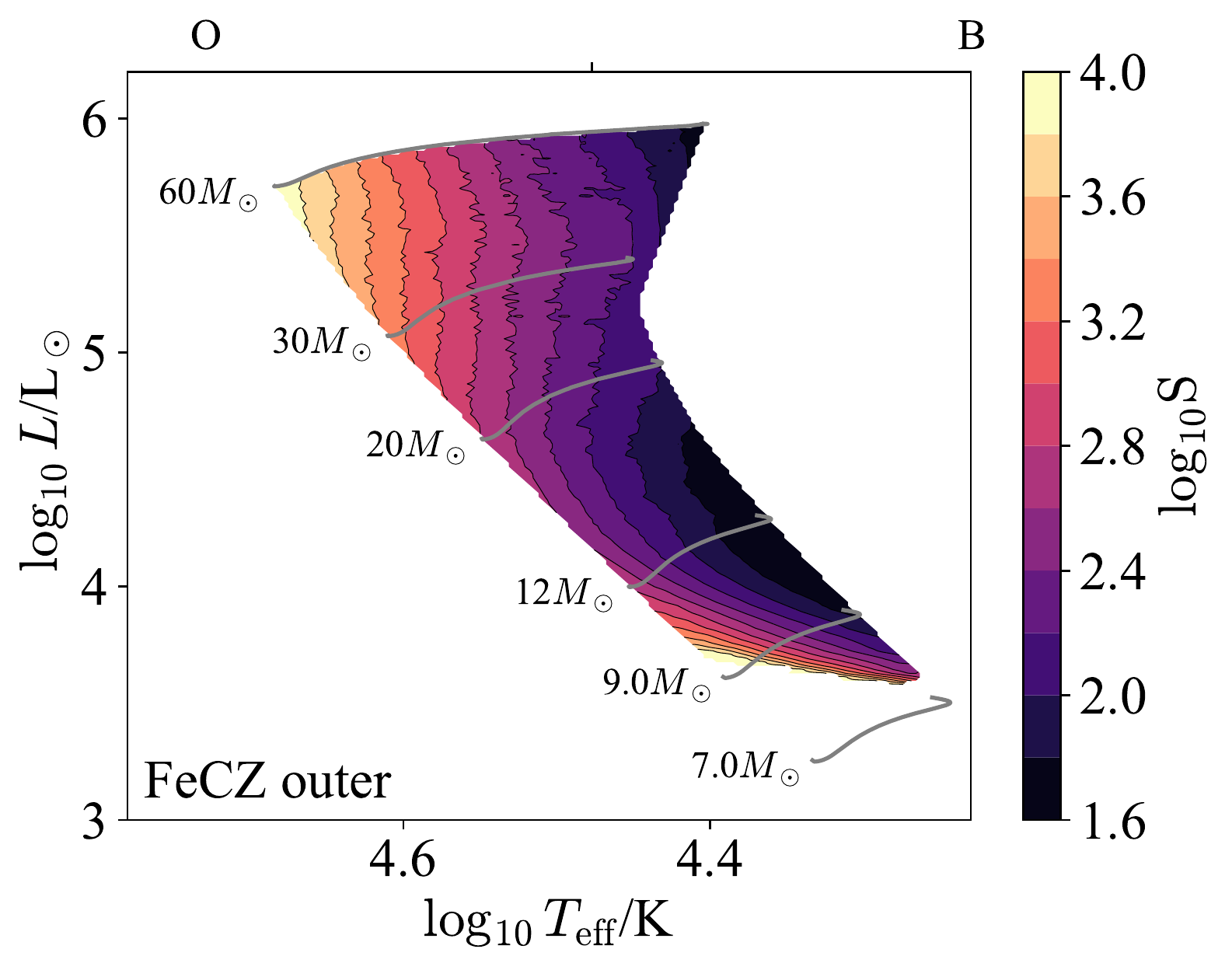}
\end{minipage}
\hfill

\caption{The stiffness of the inner (left) and outer (right) convective boundaries are shown in terms of $\log T_{\rm eff}$ and $\log L$ for stellar models with FeCZs and Milky Way metallicity $Z=0.014$. Note that the stiffness is an output of a theory of convection and so is model-dependent. Regions with $\mathrm{Ra} < \mathrm{Ra}_{\rm crit}$ are stable to convection and shaded in grey.}
\label{fig:FeCZ_stiff}
\end{figure*}

Finally, Figure~\ref{fig:FeCZ_stiff} shows the stiffness of both the inner and outer boundaries of the FeCZ.
The outer boundary is considerably more stiff ($S \sim 10^{2-6}$) than the inner boundary ($S \sim 10^{0-3}$).
Over most of the mass range and much of the main-sequence we expect significant overshooting past the relatively weak inner boundary.
Naively we would expect the opposite for the outer boundary, but there is numerical evidence of significant motion past the convective boundary~\citep{2020ApJ...902...67S}.
We suspect this is due to the moderate-to-small P{\'e}clet numbers in and near the FeCZ, which mean that motions approach isothermal and so see a reduced entropy gradient.
This lowers the effective stiffness of the outer boundary.

\clearpage
\subsection{Core CZ}

We now examine the bulk structure of Core CZs, which occur in stars with masses $M_\star \ga 1.1 M_\odot$.
The aspect ratios are unity, by definition, so the global (spherical) geometry is important.

\begin{figure*}
\begin{minipage}{0.48\textwidth}
\includegraphics[width=\textwidth]{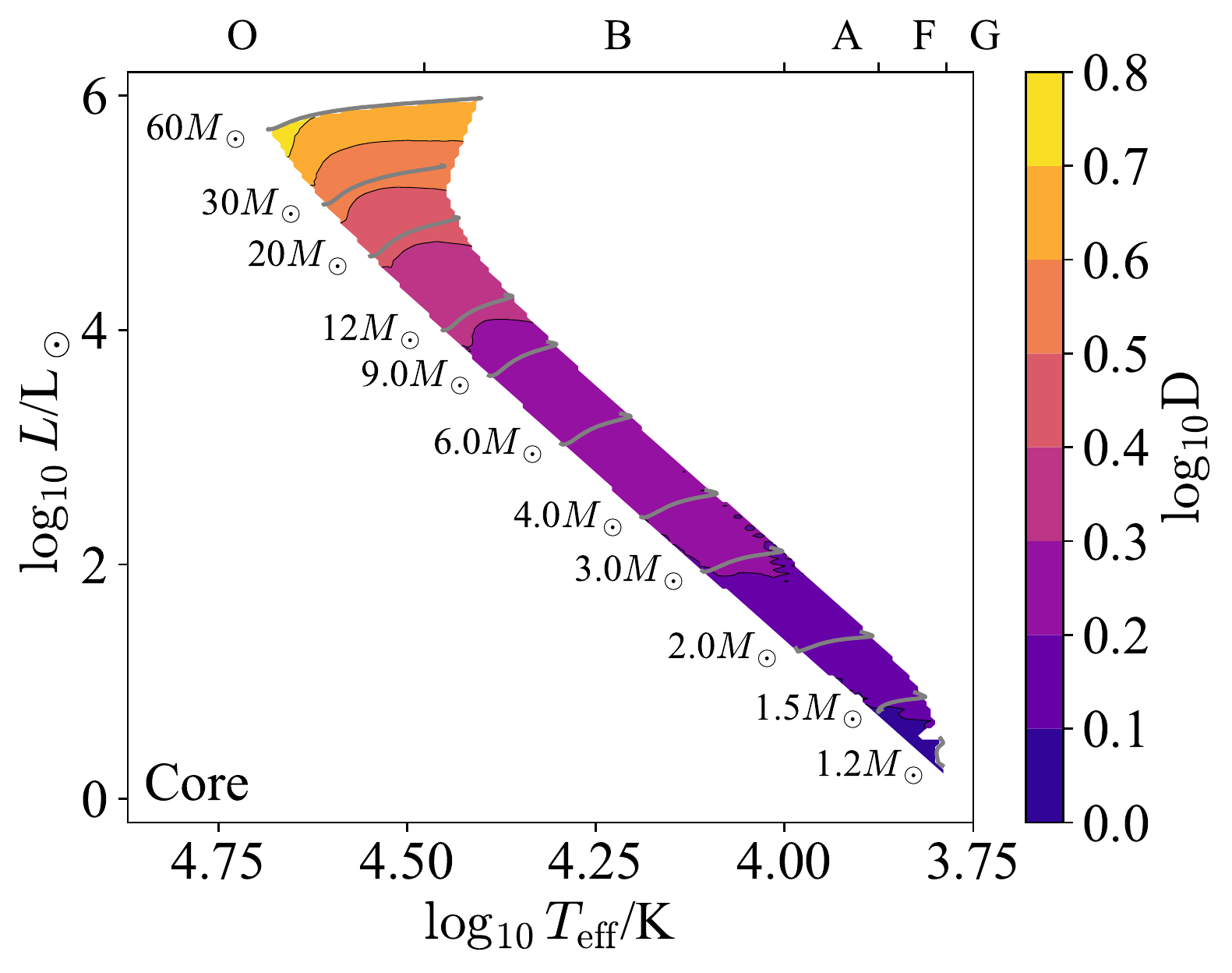}
\end{minipage}
\hfill
\begin{minipage}{0.48\textwidth}
\includegraphics[width=\textwidth]{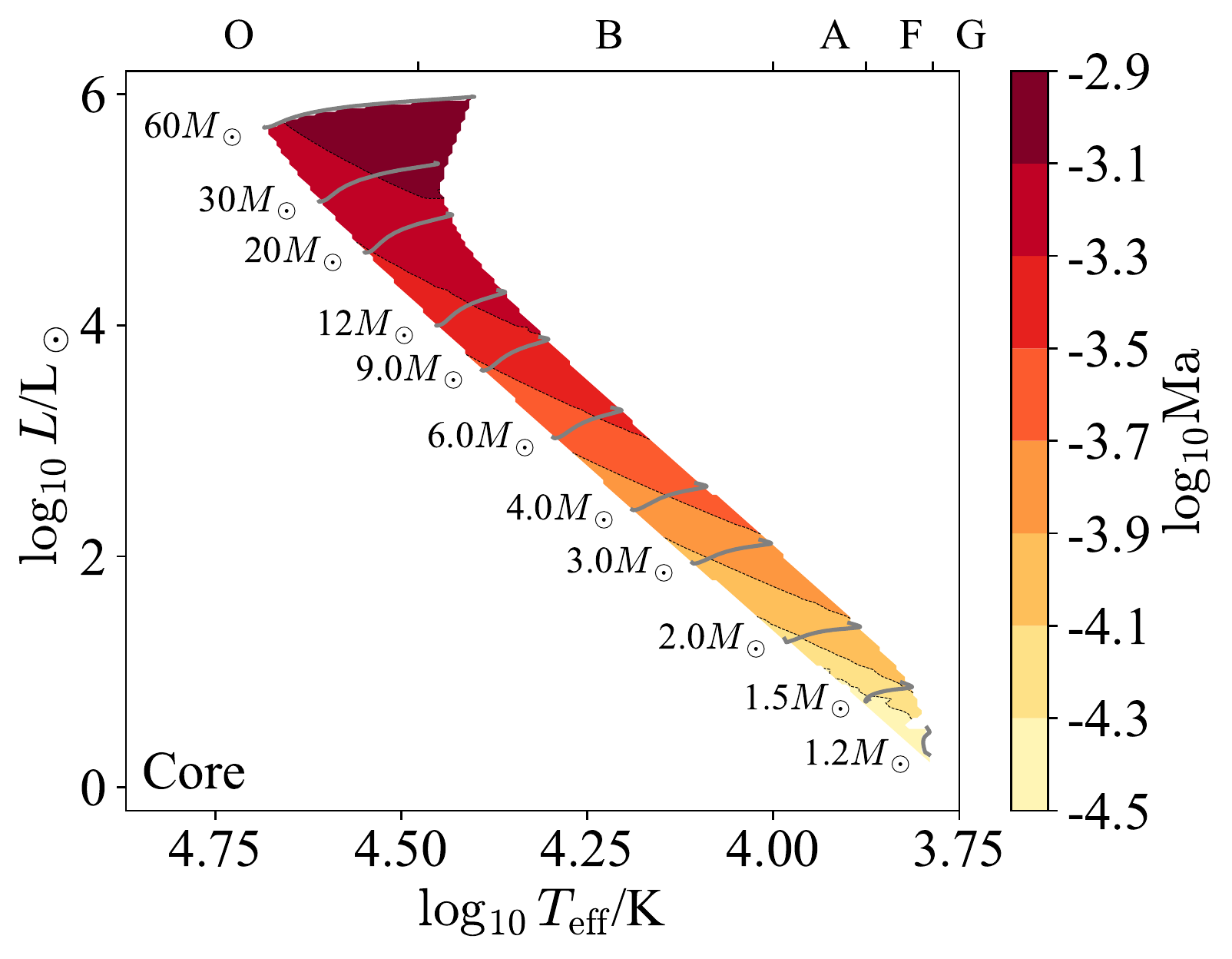}
\end{minipage}
\hfill

\caption{The density ratio $\mathrm{D}$ (left) and Mach number $\mathrm{Ma}$ (right) are shown in terms of $\log T_{\rm eff}$/spectral type and $\log L$ for stellar models with Core CZs and Milky Way metallicity $Z=0.014$. Note that while the density ratio is an input parameter and does not depend on a specific theory of convection, the Mach number is an output of such a theory and so is model-dependent. Regions with $\mathrm{Ra} < \mathrm{Ra}_{\rm crit}$ are stable to convection and shaded in grey.}
\label{fig:Core_equations}
\end{figure*}

The density ratio $\mathrm{D}$ (Figure~\ref{fig:Core_equations}, left) and Mach number $\mathrm{Ma}$ (Figure~\ref{fig:Core_equations}, right) inform which physics the fluid equations must include to model these zones.
The density ratio is typically small, of order $2-3$, and the Mach number ranges from $\sim 0.1$ at $M \la 2 M_\odot$ down to $10^{-4}$ at $M \approx 9 M_\odot$.
This suggests that above $\approx 2 M_\odot$ the Boussinesq approximation is valid, whereas below this the fully compressible equations may be needed to capture the dynamics at moderate Mach numbers.

\begin{figure*}
\begin{minipage}{0.48\textwidth}
\includegraphics[width=\textwidth]{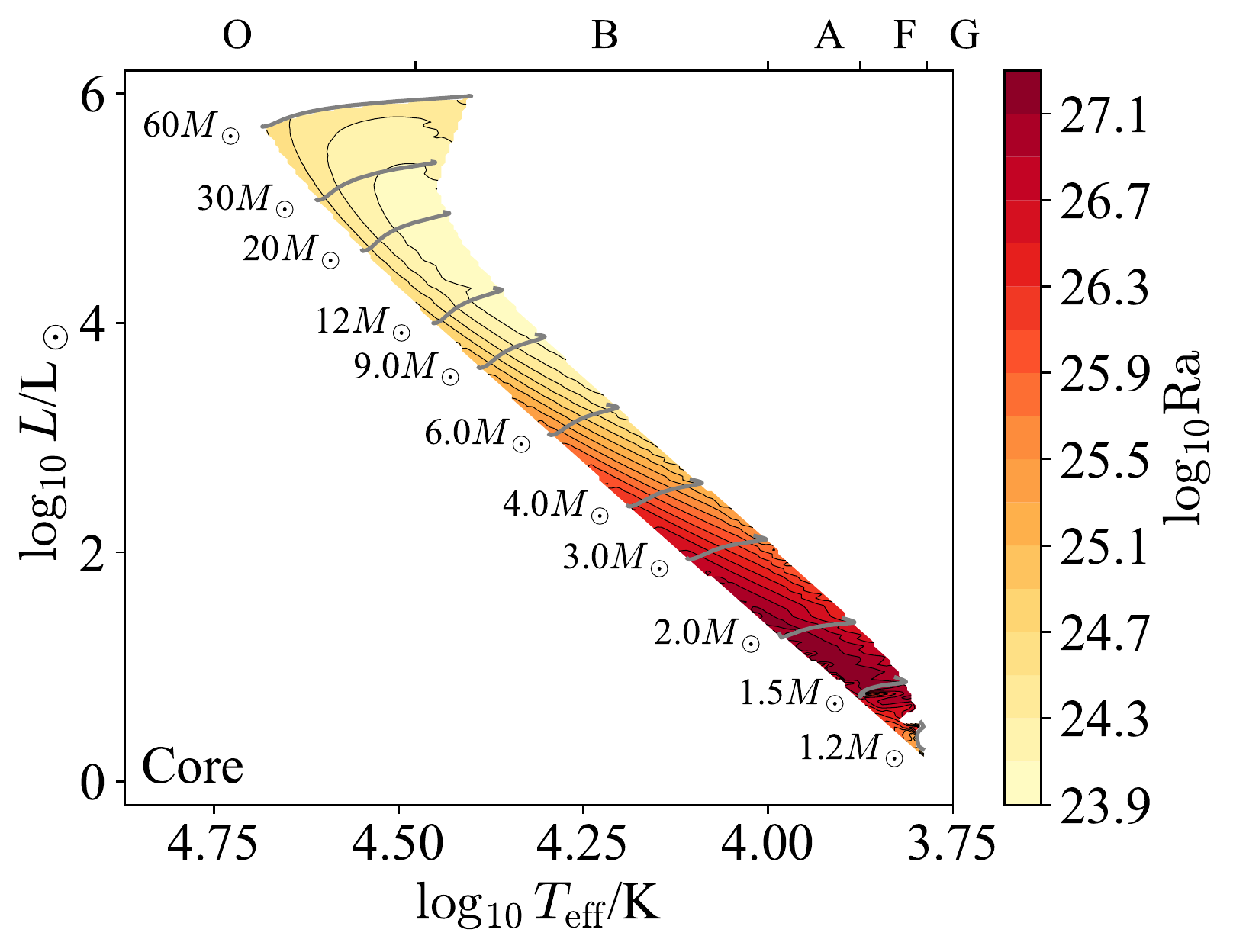}
\end{minipage}
\hfill
\begin{minipage}{0.48\textwidth}
\includegraphics[width=\textwidth]{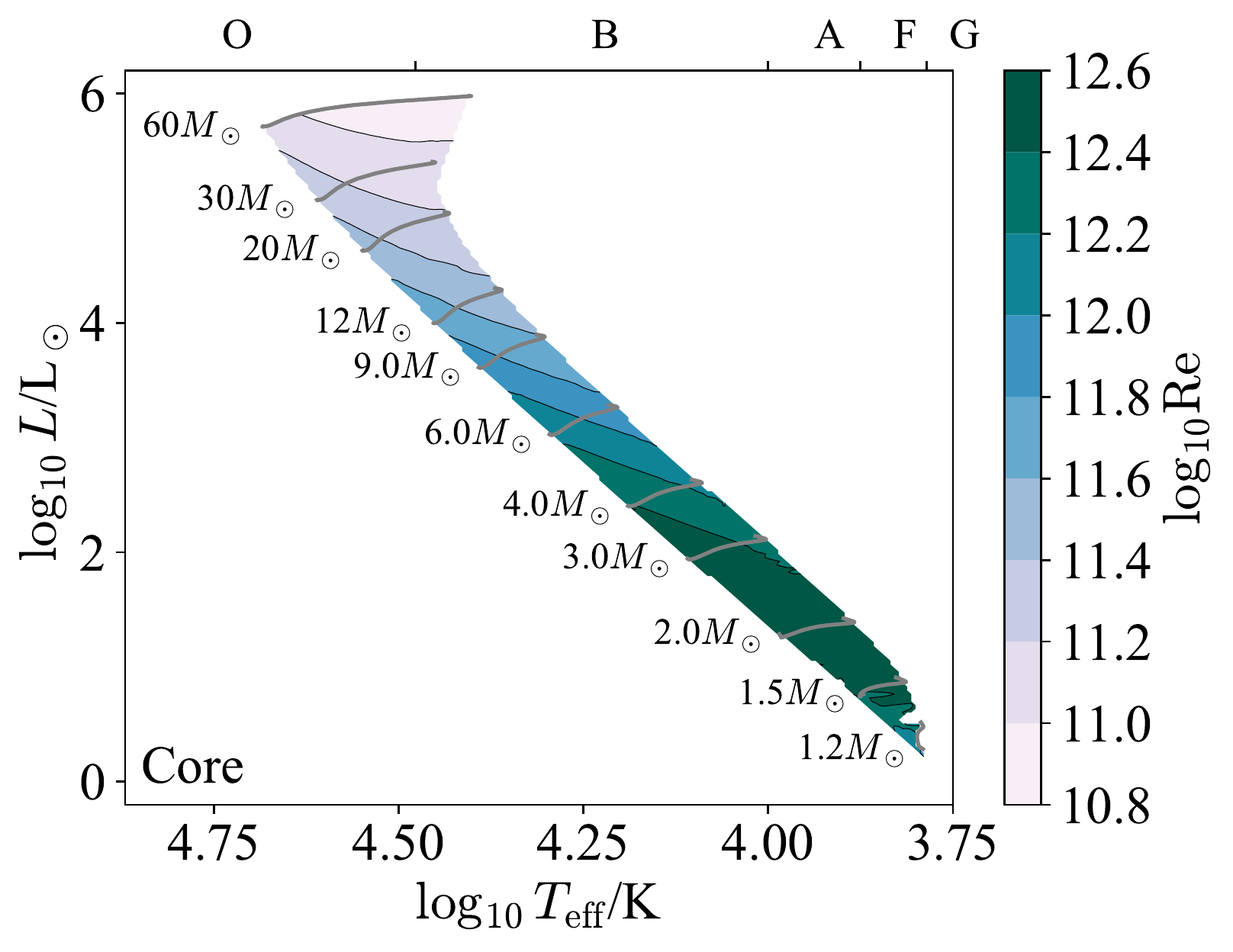}
\end{minipage}
\hfill

\caption{The Rayleigh number $\mathrm{Ra}$ (left) and Reynolds number $\mathrm{Re}$ (right) are shown in terms of $\log T_{\rm eff}$/spectral type and $\log L$ for stellar models with Core CZs and Milky Way metallicity $Z=0.014$.  Note that while the Rayleigh number is an input parameter and does not depend on a specific theory of convection, the Reynolds number is an output of such a theory and so is model-dependent. Regions with $\mathrm{Ra} < \mathrm{Ra}_{\rm crit}$ are stable to convection and shaded in grey.}
\label{fig:Core_stability}
\end{figure*}

The Rayleigh number $\mathrm{Ra}$ (Figure~\ref{fig:Core_stability}, left) determines whether or not a putative convection zone is actually unstable to convection, and the Reynolds number $\mathrm{Re}$ determines how turbulent the zone is if instability sets in (Figure~\ref{fig:Core_stability}, right).
In these zones both numbers are enormous, so we should expect convective instability to result in highly turbulent flows.

\begin{figure*}
\centering
\begin{minipage}{0.48\textwidth}
\includegraphics[width=\textwidth]{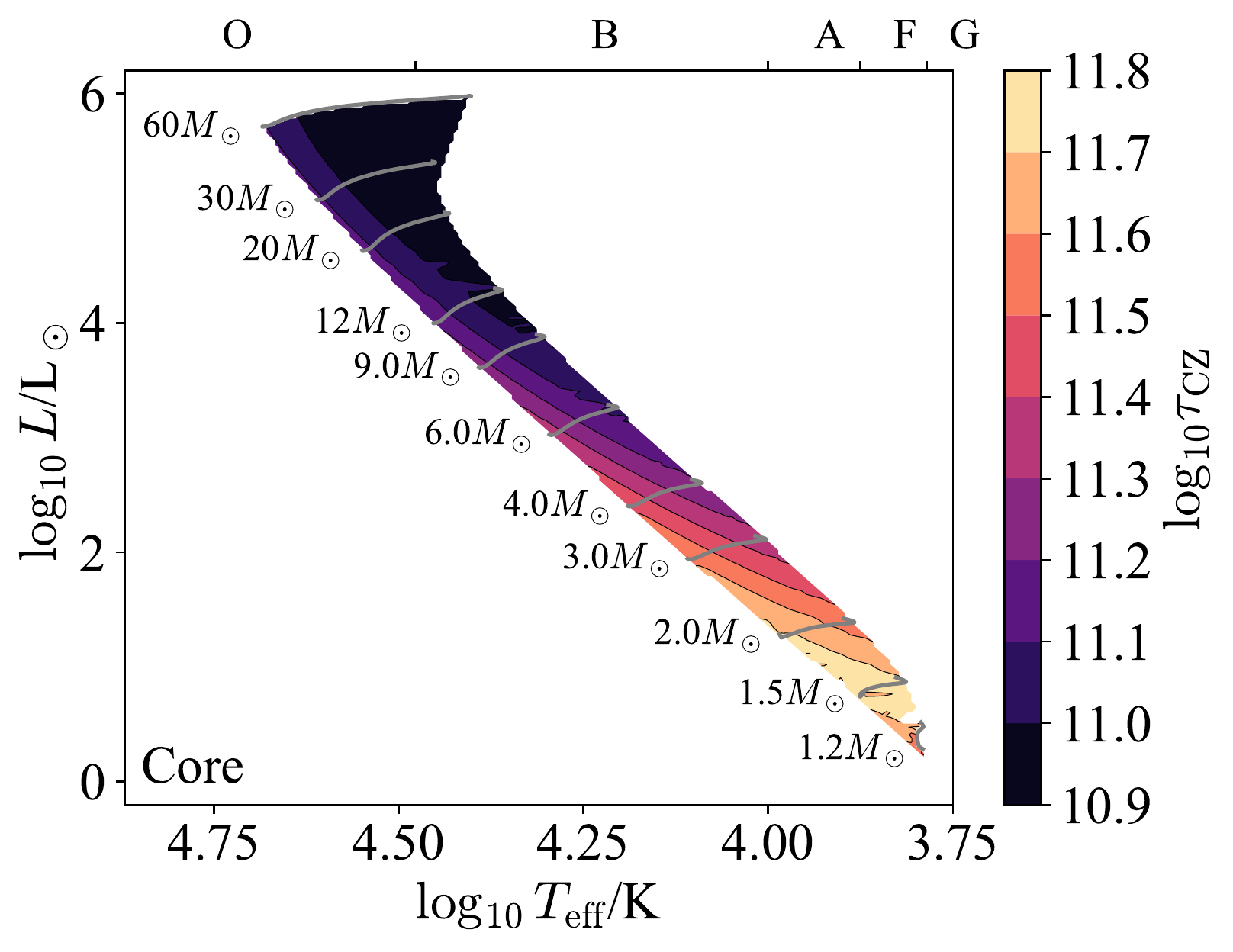}
\end{minipage}
\hfill
\begin{minipage}{0.48\textwidth}
\includegraphics[width=\textwidth]{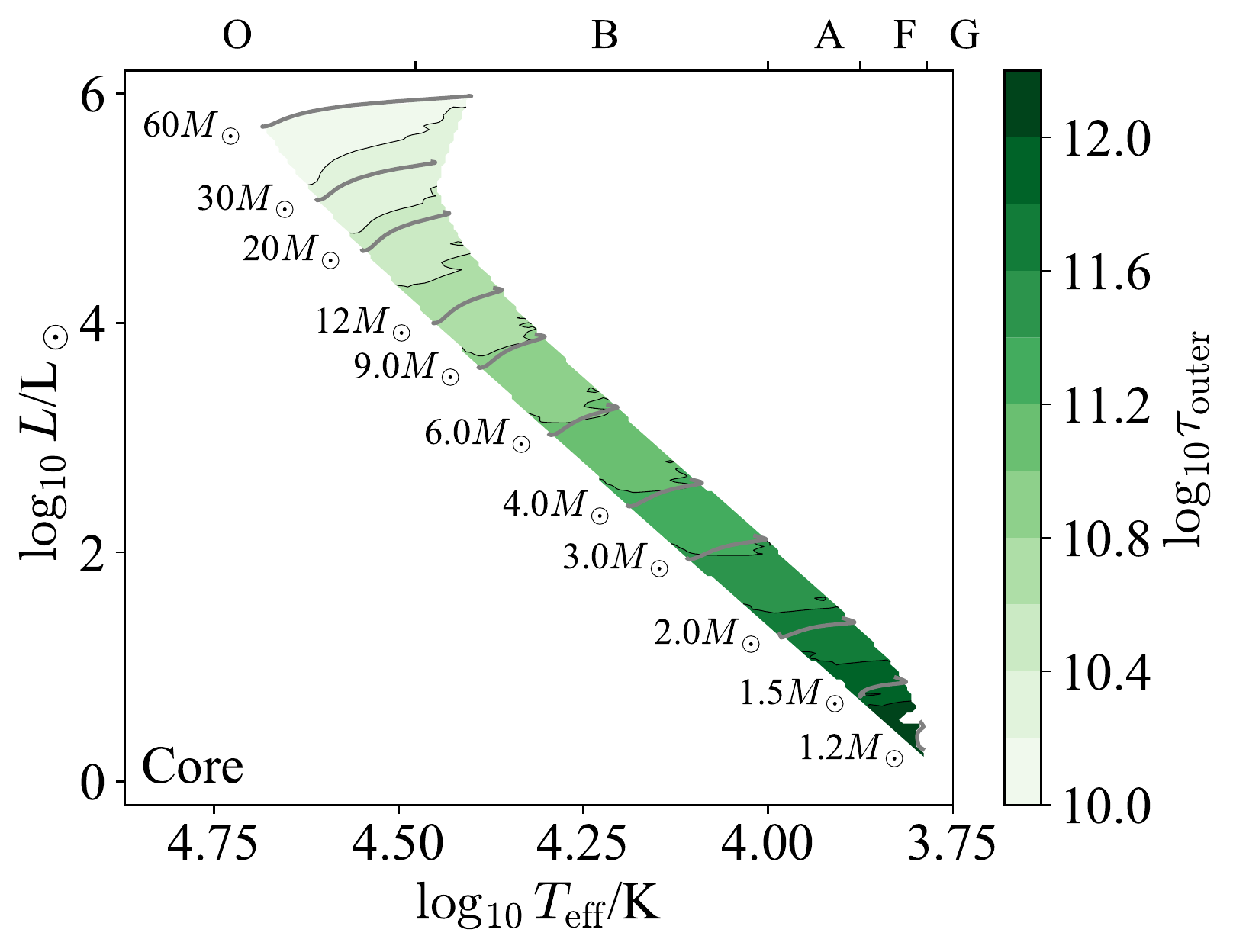}
\end{minipage}
\hfill
\caption{The convection optical depth $\tau_{\rm CZ}$ (left) and the optical depth to the surface $\tau_{\rm outer}$ (right) are shown in terms of $\log T_{\rm eff}$/spectral type and $\log L$ for stellar models with Core CZs and Milky Way metallicity $Z=0.014$. Note that both of these are input parameters, and do not depend on a specific theory of convection. Regions with $\mathrm{Ra} < \mathrm{Ra}_{\rm crit}$ are stable to convection and shaded in grey.}
\label{fig:Core_optical}
\end{figure*}

The optical depth across a convection zone $\tau_{\rm CZ}$ (Figure~\ref{fig:Core_optical}, left) indicates whether or not radiation can be handled in the diffusive approximation, while the optical depth from the outer boundary to infinity $\tau_{\rm outer}$ (Figure~\ref{fig:Core_optical}, right) indicates the nature of radiative transfer and cooling in the outer regions of the convection zone.
We see that the optical depth across these zones is enormous ($\tau_{\rm CZ} \sim 10^{11}$) and their outer boundaries lie at very large optical depths ($\tau_{\rm outer} \ga 10^{10}$). This means that both the bulk and the outer boundary of the Core CZ can likely be modeled in the limit of radiative diffusion.

\begin{figure*}
\centering
\begin{minipage}{0.48\textwidth}
\includegraphics[width=\textwidth]{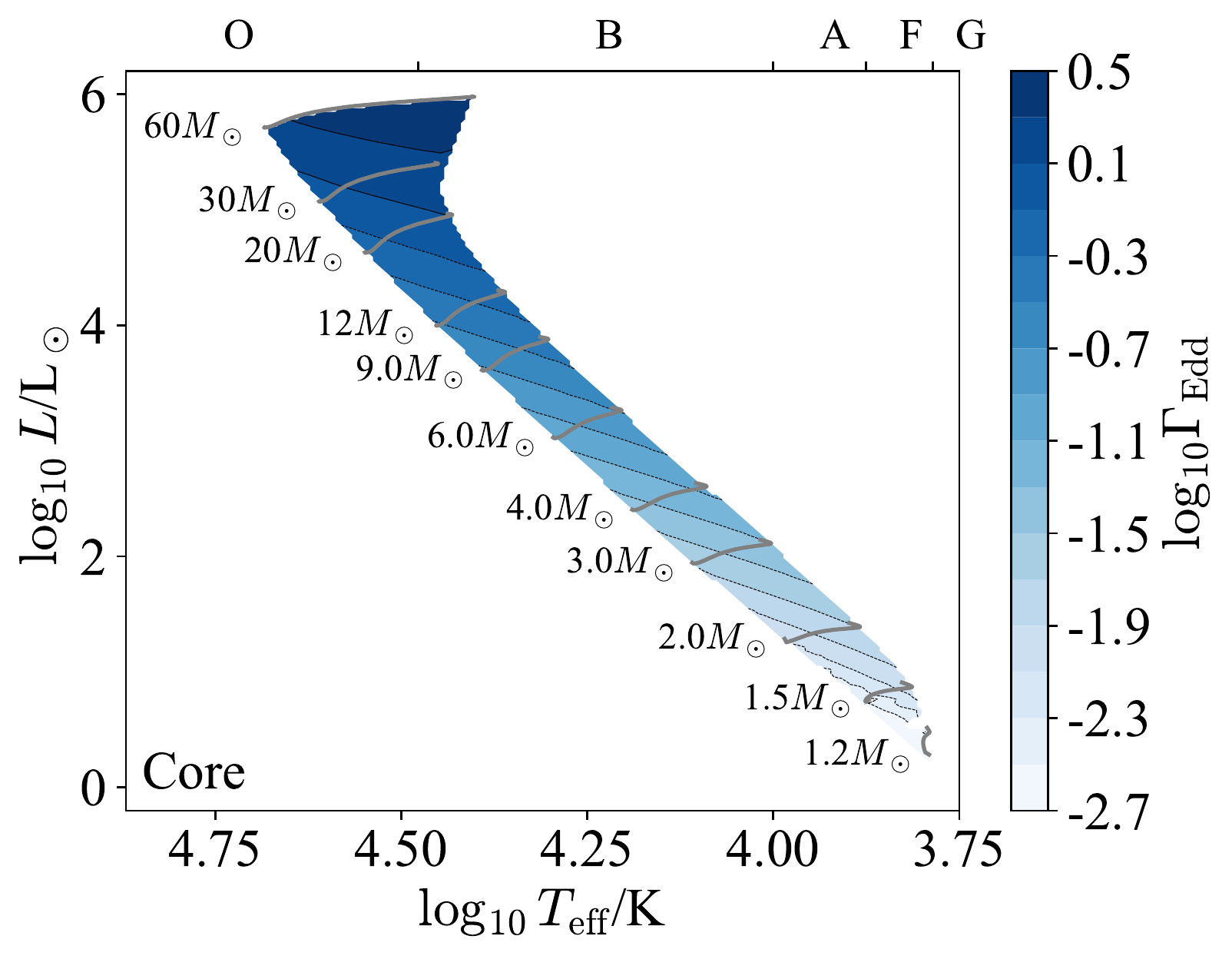}
\end{minipage}
\hfill
\begin{minipage}{0.48\textwidth}
\includegraphics[width=\textwidth]{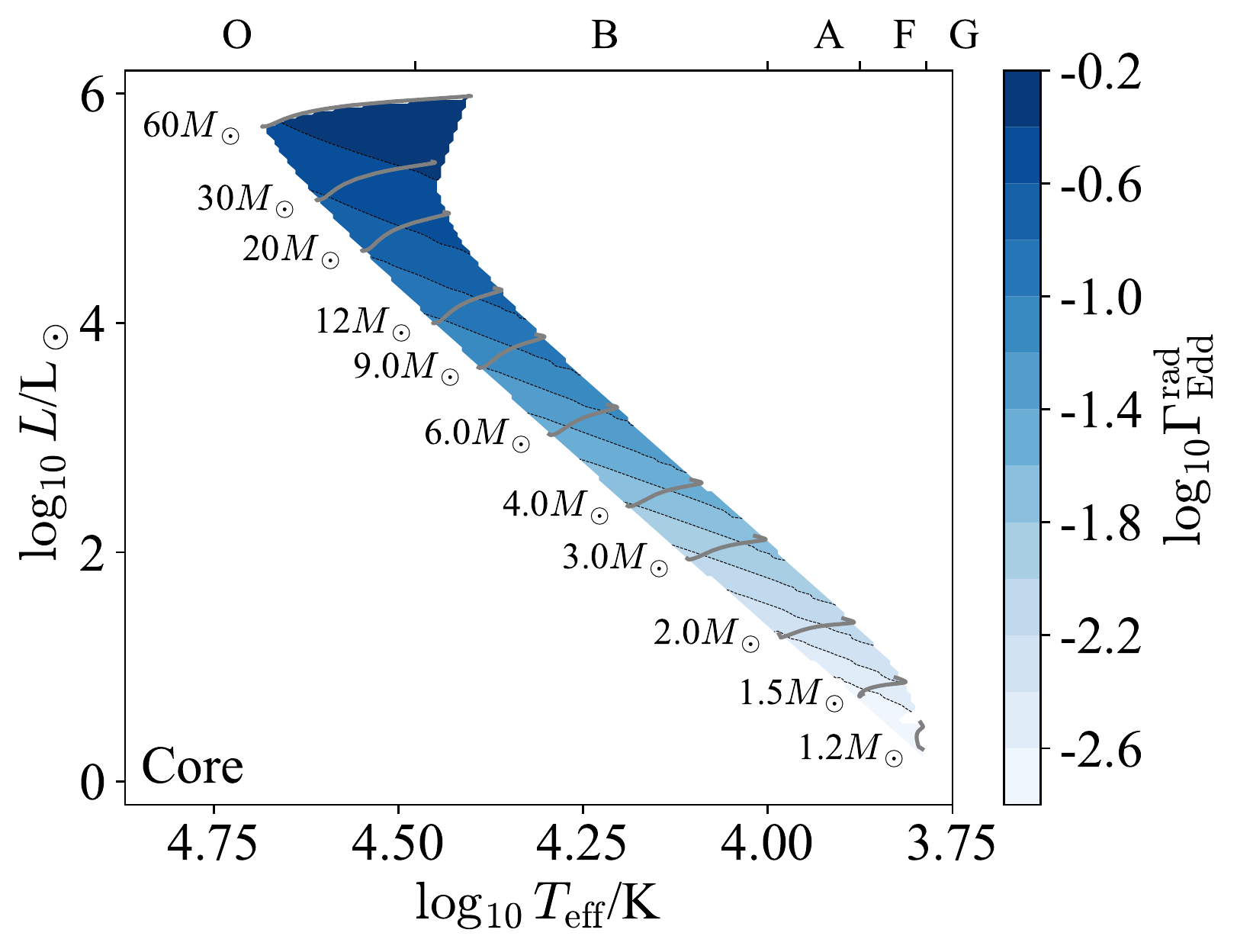}
\end{minipage}
\hfill
\caption{The Eddington ratio with the full luminosity $\Gamma_{\rm Edd}$ (left) and the radiative luminosity (right) are shown in terms of $\log T_{\rm eff}$/spectral type and $\log L$ for stellar models with Core CZs and Milky Way metallicity $Z=0.014$. Note that while $\Gamma_{\rm Edd}$ is an input parameter and does not depend on a specific theory of convection, $\Gamma_{\rm Edd}^{\rm rad}$ is an output of such a theory and so is model-dependent. Regions with $\mathrm{Ra} < \mathrm{Ra}_{\rm crit}$ are stable to convection and shaded in grey.}
\label{fig:Core_eddington}
\end{figure*}

The Eddington ratio $\Gamma_{\rm Edd}$ (Figure~\ref{fig:Core_eddington}, left) indicates whether or not radiation hydrodynamic instabilities are important in the non-convecting state, and the radiative Eddington ratio $\Gamma_{\rm Edd}^{\rm rad}$ (Figure~\ref{fig:Core_eddington}, right) indicates the same in the developed convective state.
The Eddington ratio with the full luminosity ($\Gamma_{\rm Edd}$) approaches unity around masses of $20 M_\odot$ and continues to rise with increasing mass from there, resulting in a nominally super-Eddington system.
Convection is able to carry much of this luminosity, however, and so the radiative Eddington ratio $\Gamma_{\rm Edd}^{\rm rad}$ approaches but does not exceed unity.

These near-unity Eddington ratios suggest that radiation hydrodynamic instabilities are important.
We have two reservations with this conclusion.
First, the Mach numbers in the Core CZ are tiny, so the velocity field does not drive large density fluctuations and hence we do not expect large opacity fluctuations like those reported in the FeCZ~\citep{2020ApJ...902...67S}.
Secondly, the optical depth across the Core CZ is enormous, so even if radiation hydrodynamics instabilities arise they should remain limited to scales which are small compared to the size of the convection zone.

\begin{figure*}
\centering
\begin{minipage}{0.48\textwidth}
\includegraphics[width=\textwidth]{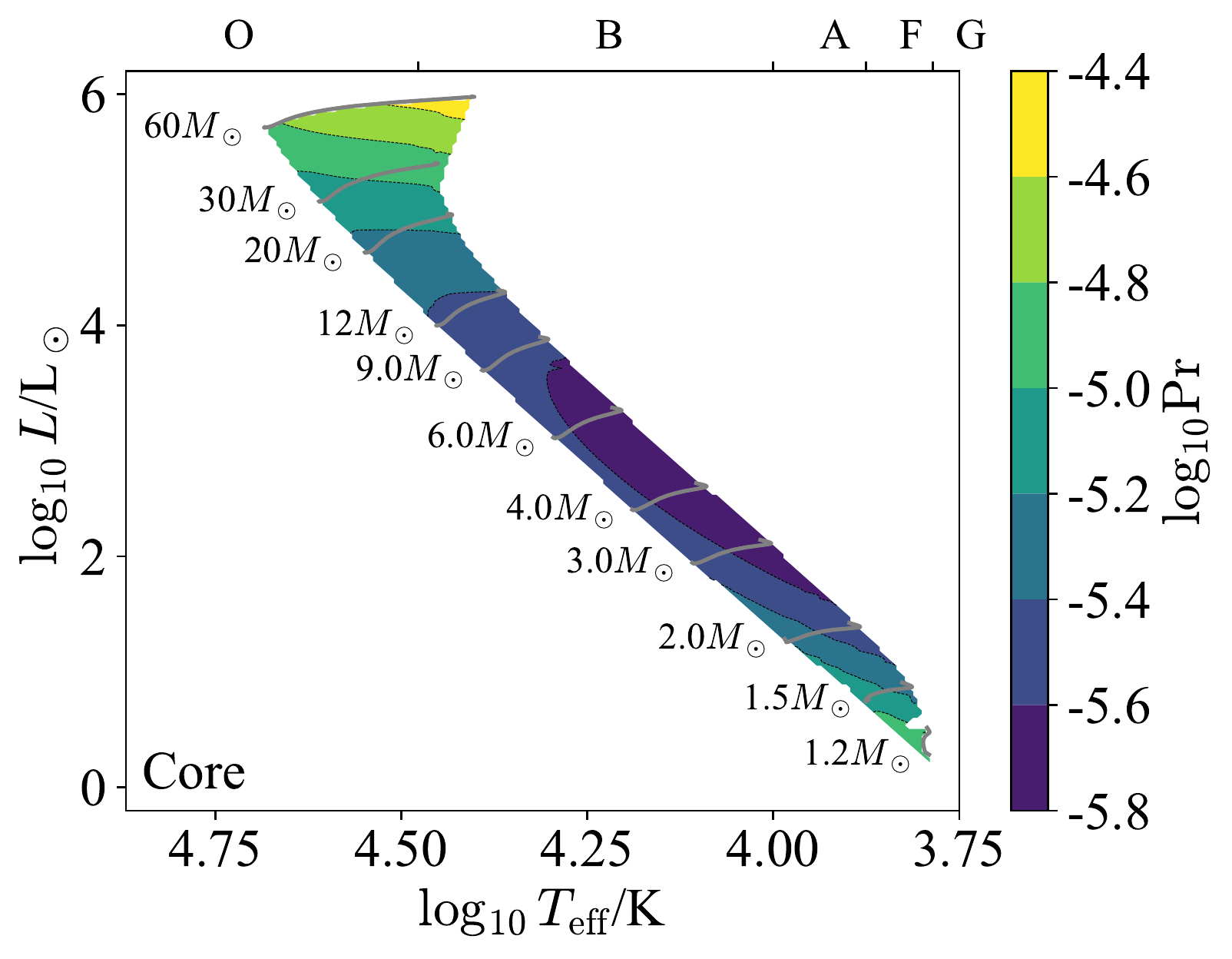}
\end{minipage}
\hfill
\begin{minipage}{0.48\textwidth}
\includegraphics[width=\textwidth]{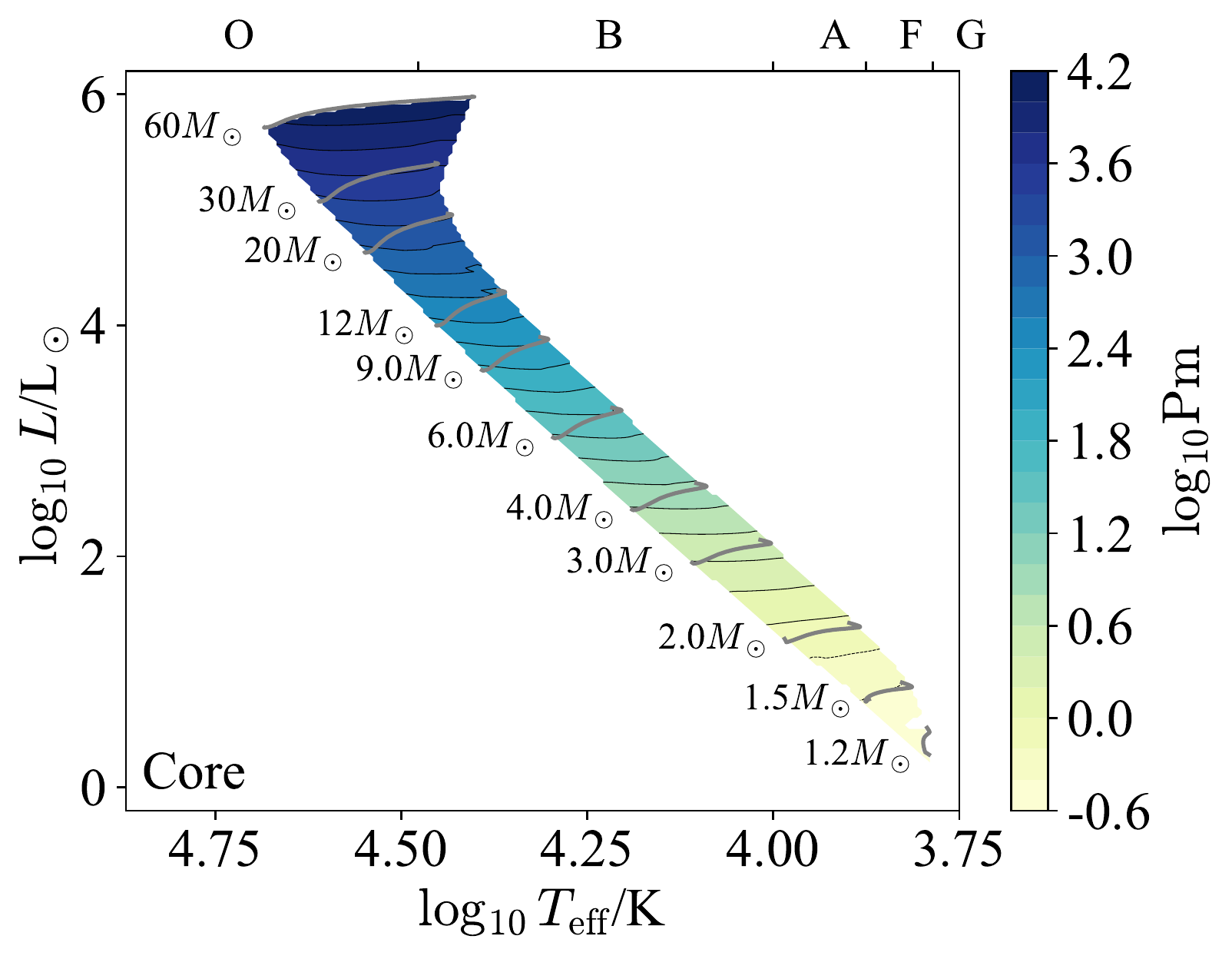}
\end{minipage}
\hfill

\caption{The Prandtl number $\mathrm{Pr}$ (left) and magnetic Prandtl number $\mathrm{Pm}$ (right) are shown in terms of $\log T_{\rm eff}$/spectral type and $\log L$ for stellar models with Core CZs and Milky Way metallicity $Z=0.014$. Note that both $\mathrm{Pr}$ and $\mathrm{Pm}$ are input parameters, and so do not depend on a specific theory of convection. Regions with $\mathrm{Ra} < \mathrm{Ra}_{\rm crit}$ are stable to convection and shaded in grey.}
\label{fig:Core_diffusivities}
\end{figure*}

The Prandtl number $\mathrm{Pr}$ (Figure~\ref{fig:Core_diffusivities}, left) measures the relative importance of thermal diffusion and viscosity, and the magnetic Prandtl number $\mathrm{Pm}$ (Figure~\ref{fig:Core_diffusivities}, right) measures the same for magnetic diffusion and viscosity.
The Prandtl number is always small in these models, so the thermal diffusion length-scale is much larger than the viscous scale.
By contrast, the magnetic Prandtl number varies from order-unity at low masses to large ($10^4$) at high masses.

The fact that $\mathrm{Pm}$ is large at high masses is notable because the quasistatic approximation for magnetohydrodynamics has frequently been used to study magnetoconvection in minimal 3D MHD simulations of planetary and stellar interiors~\citep[e.g.][]{yan_calkins_maffei_julien_tobias_marti_2019} and assumes that $\mathrm{Rm} = \mathrm{Pm} \mathrm{Re} \rightarrow 0$; in doing so, this approximation assumes a global background magnetic field is dominant and neglects the nonlinear portion of the Lorentz force. This approximation breaks down in convection zones with $\mathrm{Pm} > 1$ and future numerical experiments should seek to understand how magnetoconvection operates in this regime.

\begin{figure*}
\centering
\begin{minipage}{0.48\textwidth}
\includegraphics[width=\textwidth]{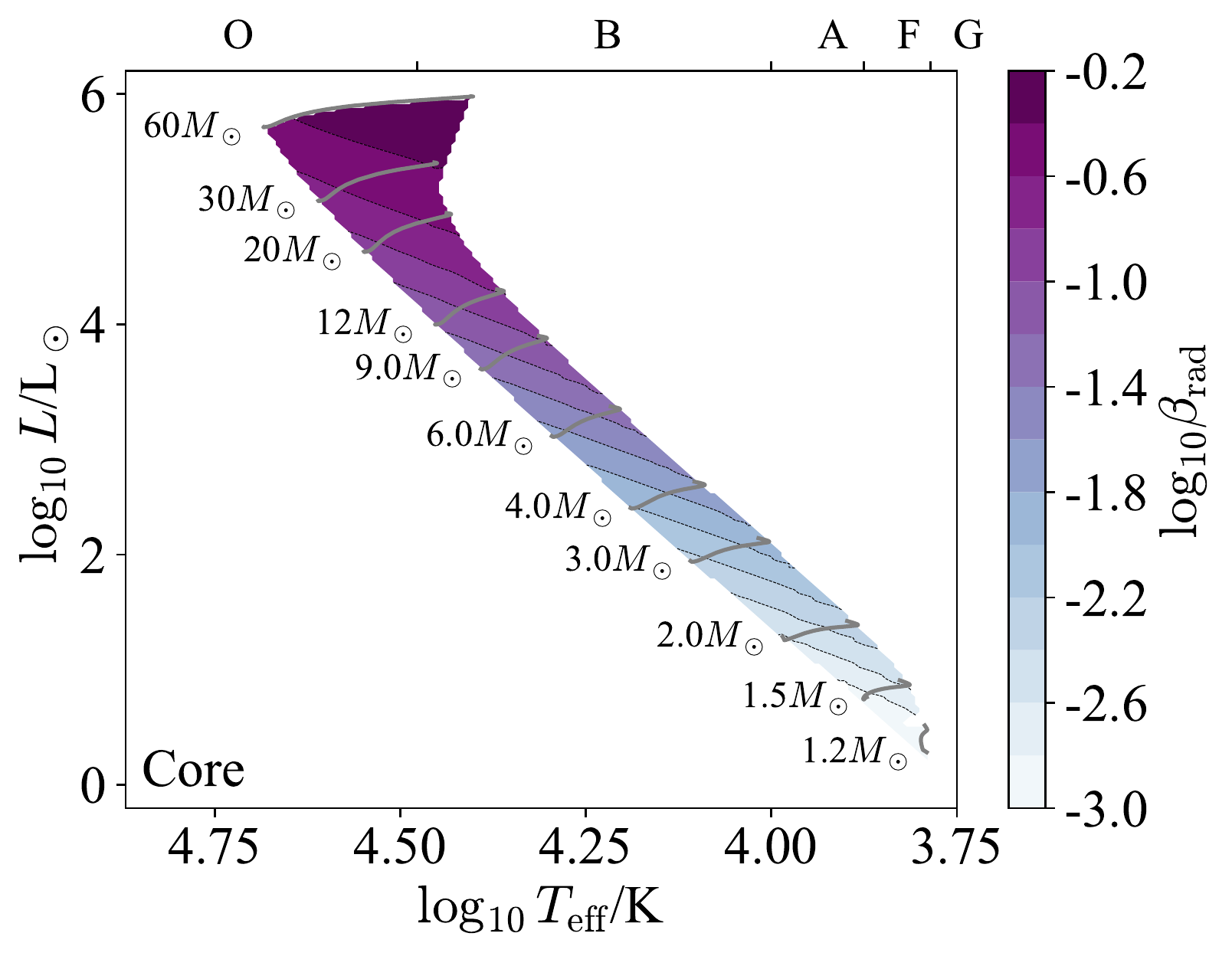}
\end{minipage}
\hfill

\caption{The radiation pressure ratio $\beta_{\rm rad}$ is shown in terms of $\log T_{\rm eff}$/spectral type and $\log L$ for stellar models with Core CZs and Milky Way metallicity $Z=0.014$. Note that this ratio is an input parameter, and does not depend on a specific theory of convection. Regions with $\mathrm{Ra} < \mathrm{Ra}_{\rm crit}$ are stable to convection and shaded in grey.}
\label{fig:Core_beta}
\end{figure*}

The radiation pressure ratio $\beta_{\rm rad}$ (Figure~\ref{fig:Core_beta}) measures the importance of radiation in setting the thermodynamic properties of the fluid.
This is small at low masses ($M_\star \la 9 M_\odot$) but reaches a 30-100\% correction at high masses ($M_\star > 20 M_\odot$) and so radiation pressure is very important to capture at the high mass end.

\begin{figure*}
\centering
\begin{minipage}{0.48\textwidth}
\includegraphics[width=\textwidth]{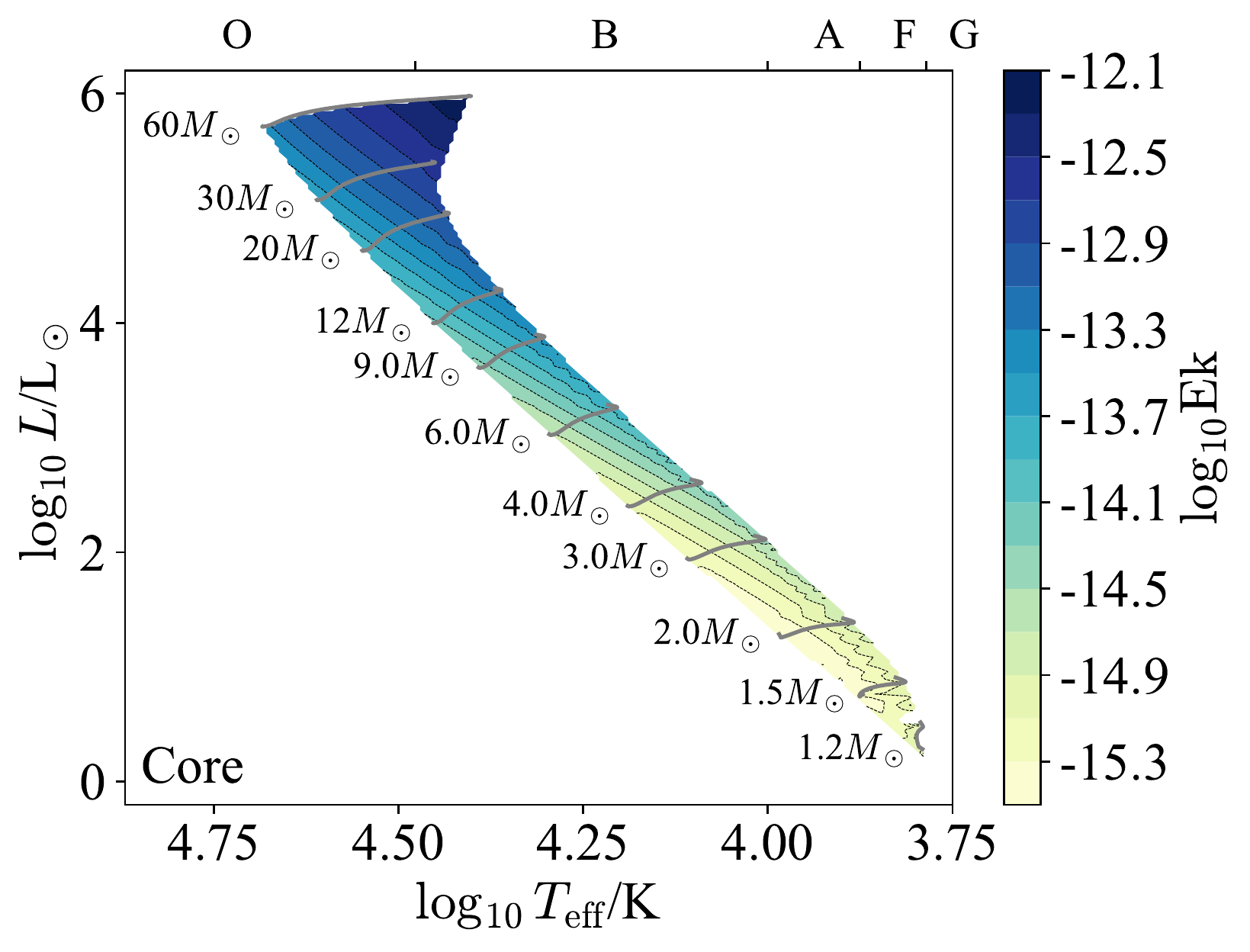}
\end{minipage}

\caption{The Ekman number $\mathrm{Ek}$ is shown in terms of $\log T_{\rm eff}$/spectral type and $\log L$ for stellar models with Core CZs and Milky Way metallicity $Z=0.014$. Note that the Ekman number is an input parameter, and does not depend on a specific theory of convection. Regions with $\mathrm{Ra} < \mathrm{Ra}_{\rm crit}$ are stable to convection and shaded in grey.}
\label{fig:Core_ekman}
\end{figure*}

The Ekman number $\mathrm{Ek}$ (Figure~\ref{fig:Core_ekman}) indicates the relative importance of viscosity and rotation.
This is tiny across the HRD, so we expect rotation to dominate over viscosity, except at very small length-scales.

\begin{figure*}
\centering
\begin{minipage}{0.48\textwidth}
\includegraphics[width=\textwidth]{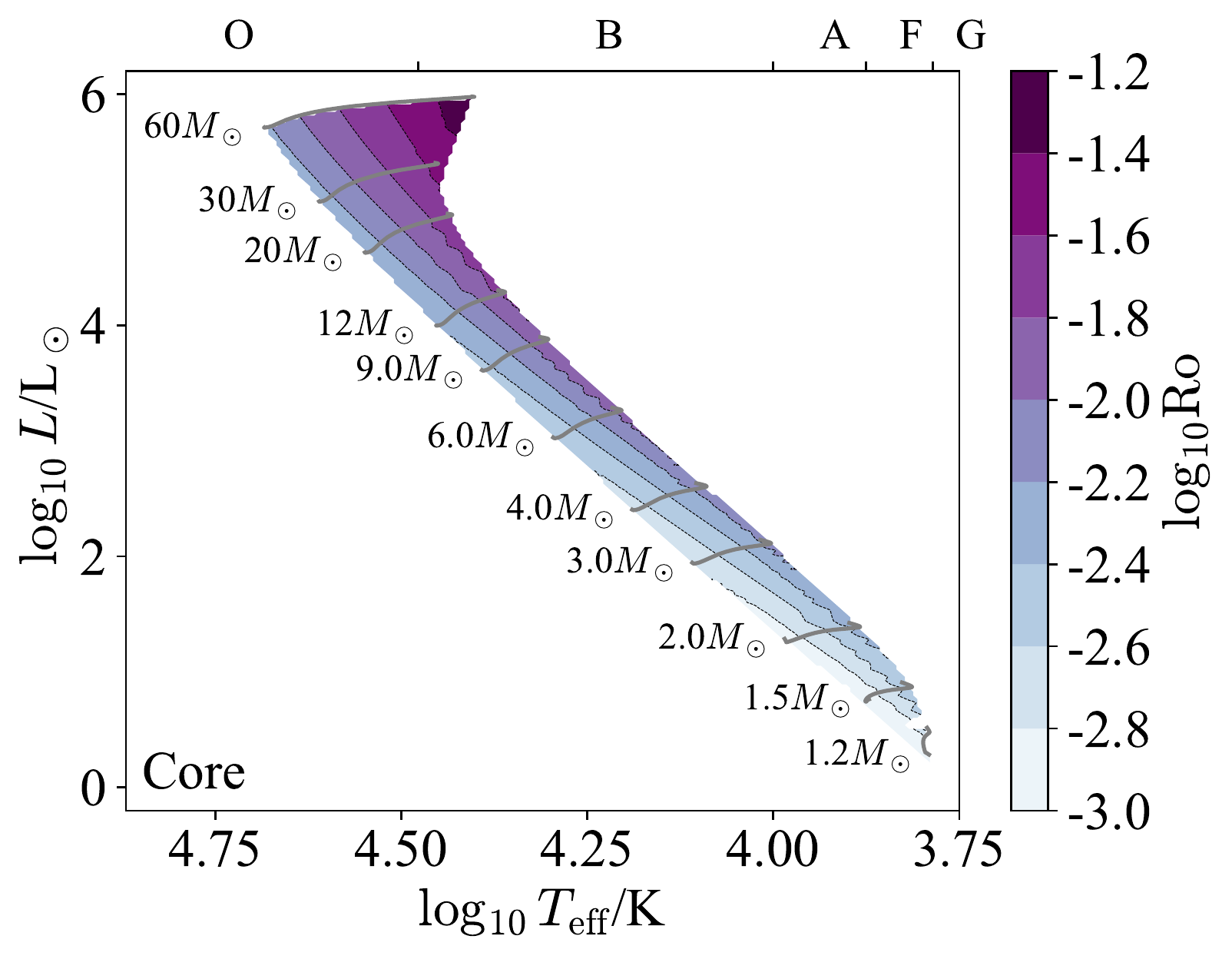}
\end{minipage}
\hfill
\begin{minipage}{0.48\textwidth}
\includegraphics[width=\textwidth]{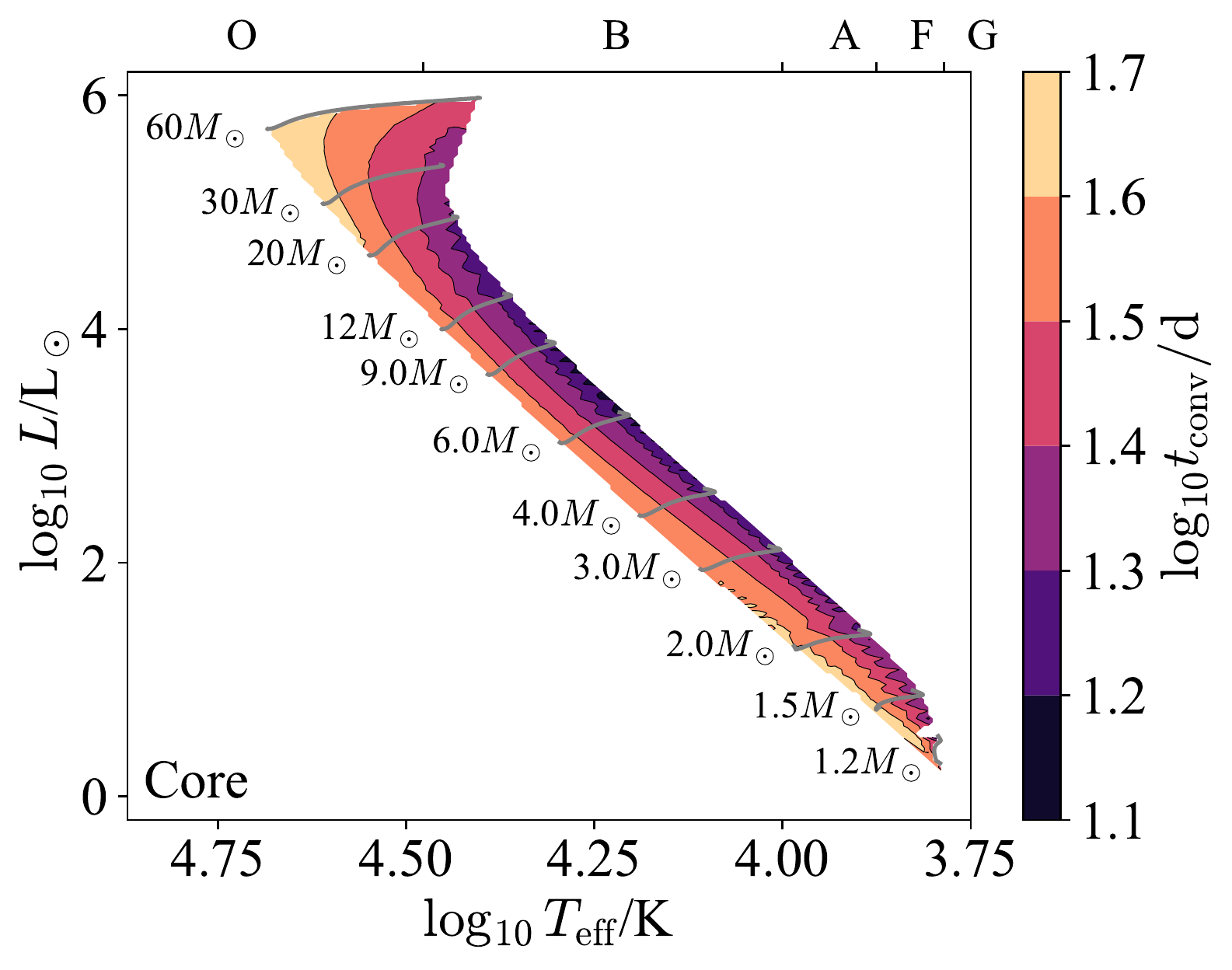}
\end{minipage}
\hfill

\caption{The Rossby number $\mathrm{Ro}$ (left) and turnover time $t_{\rm conv}$ (right) are shown in terms of $\log T_{\rm eff}$/spectral type and $\log L$ for stellar models with Core CZs and Milky Way metallicity $Z=0.014$. Note that both $\mathrm{Ro}$ and $t_{\rm conv}$ are outputs of a theory of convection and so are model-dependent. Regions with $\mathrm{Ra} < \mathrm{Ra}_{\rm crit}$ are stable to convection and shaded in grey.}
\label{fig:Core_rotation}
\end{figure*}

The Rossby number $\mathrm{Ro}$ (Figure~\ref{fig:Core_rotation}, left) measures the relative importance of rotation and inertia.
This is small ($10^{-2}-10^{-3}$), meaning that the Core CZ is rotationally constrained for typical rotation rates~\citep{2013A&A...557L..10N}.

We have assumed a fiducial rotation law to calculate $\mathrm{Ro}$.
Stars exhibit a variety of different rotation rates, so we also show the convective turnover time $t_{\rm conv}$ (Figure~\ref{fig:Core_rotation}, right) which may be used to estimate the Rossby number for different rotation periods.

\begin{figure*}
\centering
\begin{minipage}{0.48\textwidth}
\includegraphics[width=\textwidth]{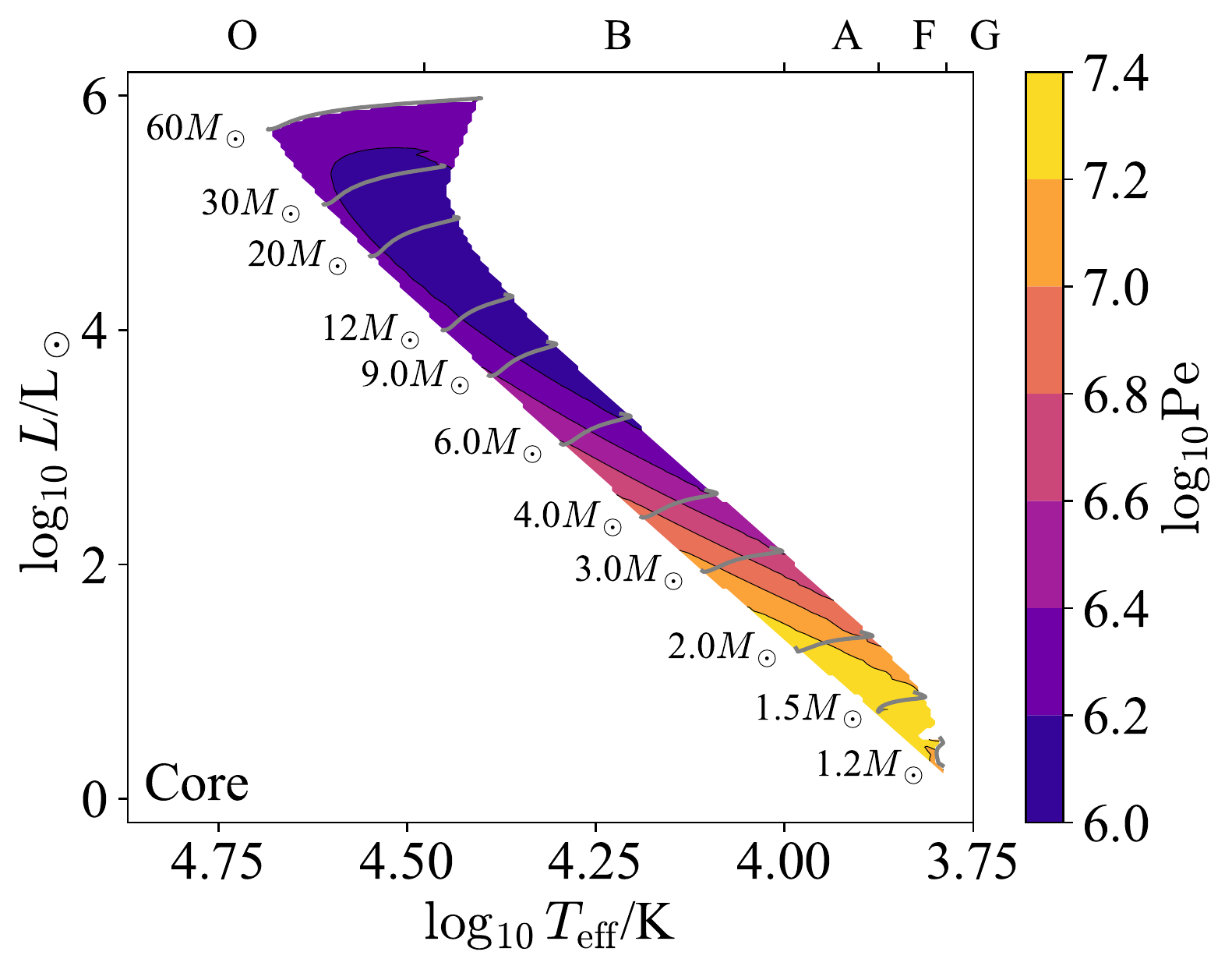}
\end{minipage}
\hfill
\begin{minipage}{0.48\textwidth}
\includegraphics[width=\textwidth]{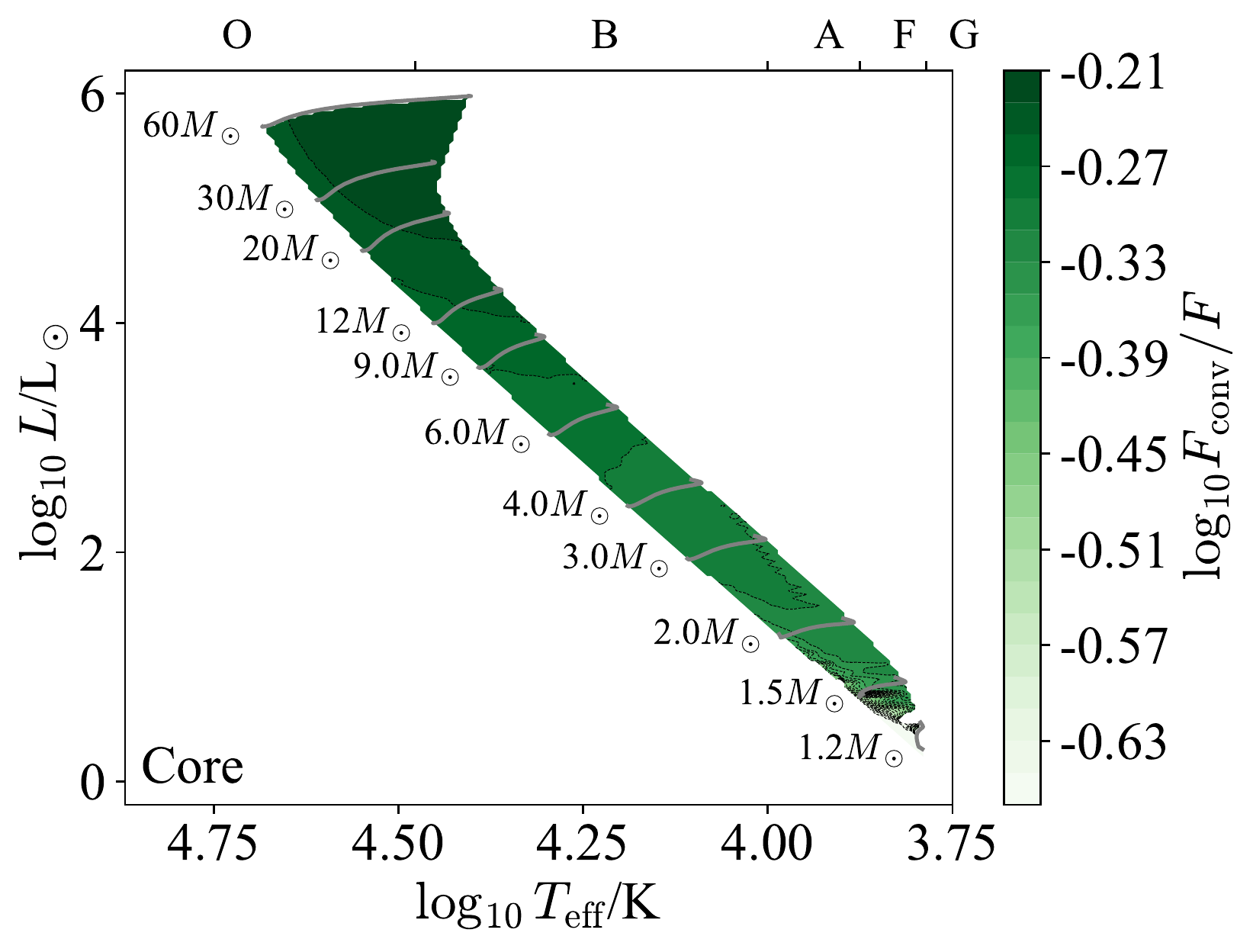}
\end{minipage}

\caption{The P{\'e}clet number $\mathrm{Pe}$ (left) and $F_{\rm conv}/F$ (right) are shown in terms of $\log T_{\rm eff}$/spectral type and $\log L$ for stellar models with Core CZs and Milky Way metallicity $Z=0.014$. Note that both $\mathrm{Pe}$ and $F_{\rm conv}/F$ are outputs of a theory of convection and so are model-dependent. Regions with $\mathrm{Ra} < \mathrm{Ra}_{\rm crit}$ are stable to convection and shaded in grey.}
\label{fig:Core_efficiency}
\end{figure*}

The P{\'e}clet number $\mathrm{Pe}$ (Figure~\ref{fig:Core_efficiency}, left) measures the relative importance of advection and diffusion in transporting heat, and the flux ratio $F_{\rm conv}/F$ (Figure~\ref{fig:Core_efficiency}, right) reports the fraction of the energy flux which is advected.
The P{\'e}clet number is always large ($10^6-10^7$) and the flux ratio varies from $\sim 0.2$ at the low mass end ($M \la 2 M_\odot$) to $\sim 0.6$ for $M \ga 3 M_\odot$.
In general then core convection is very efficient, and carries a substantial fraction of the flux.

\begin{figure*}
\centering
\begin{minipage}{0.48\textwidth}
\includegraphics[width=\textwidth]{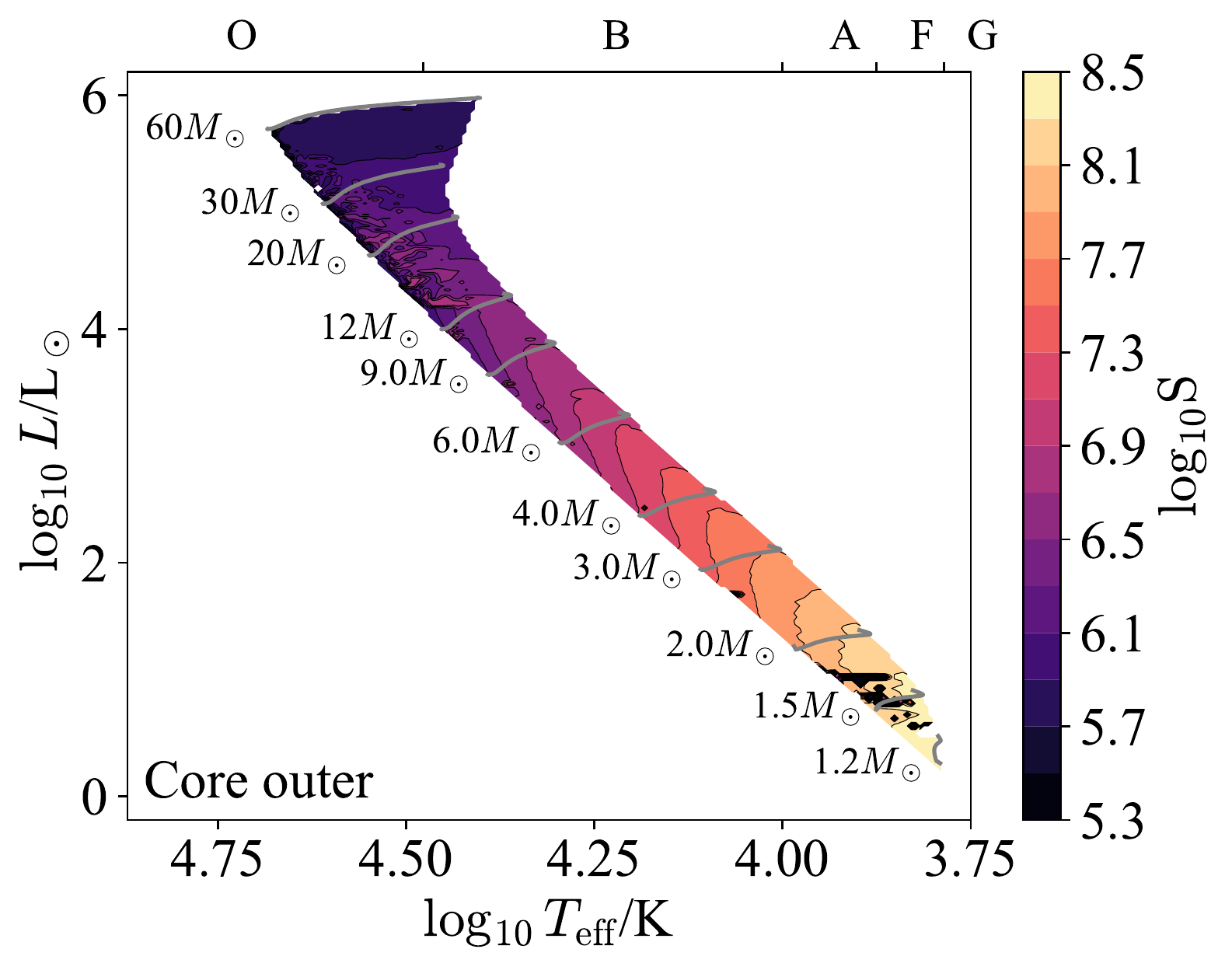}
\end{minipage}

\caption{The stiffness of the outer convective boundary is shown in terms of $\log T_{\rm eff}$ and $\log L$ for stellar models with Core CZs and Milky Way metallicity $Z=0.014$. Note that the stiffness is an output of a theory of convection and so is model-dependent. Regions with $\mathrm{Ra} < \mathrm{Ra}_{\rm crit}$ are stable to convection and shaded in grey.}
\label{fig:Core_stiff}
\end{figure*}

Finally, Figure~\ref{fig:Core_stiff} shows the stiffness of the outer boundary of the core CZ.
Over the whole mass range and main-sequence this boundary is very stiff ($S \sim 10^{5-9}$), so we do not expect much mechanical overshooting, though there could still well be convective penetration~\citep{2021arXiv211011356A}.

\bibliography{refs}
\bibliographystyle{aasjournal}

\end{document}